UNIVERSITAT POLITÈCNICA DE CATALUNYA

DOCTORAL THESIS

---

# Polarization and Index Modulations: a Theoretical and Practical Perspective

---

by
POL HENAREJOS

*Director:*
Prof. Ana Isabel PÉREZ-NEIRA

*A thesis submitted in fulfilment of the requirements for the degree of*

DOCTOR OF PHILOSOPHY

*in the*
Departament de la Teoria del Senyal i Comunicacions

*in collaboration with*
Centre Tecnològic de Telecomunicacions de Catalunya

Barcelona, June 1, 2017

*"My greatest concern was what to call it. I thought of calling it information, but the word was overly used, so I decided to call it uncertainty. When I discussed it with John von Neumann, he had a better idea. Von Neumann told me, ≪You should call it entropy, for two reasons. In the first place your uncertainty function has been used in statistical mechanics under that name, so it already has a name. In the second place, and more important, nobody knows what entropy really is, so in a debate you will always have the advantage.≫"*

Claude Elwood Shannon (1916—2001)

Universitat Politècnica de Catalunya

# *Abstract*

Departament de la Teoria del Senyal i Comunicacions
Centre Tecnològic de Telecomunicacions de Catalunya

Doctor of Philosophy

**Polarization and Index Modulations: a Theoretical and Practical Perspective**

by Pol HENAREJOS

Radiocommunication systems have evolved significantly in recent years in order to meet present and future demands. Historically, time, frequency and more recently, spatial dimensions have been used to improve capacity and robustness. Paradoxically, radiocommunications that leverage the polarization dimension have not evolved at the same pace. In particular, these communications are widely used by satellites, where several streams are multiplexed in each orthogonal polarization.

Current communication trends advocate for simplifying and unifying different frameworks in order to increase flexibility and address future needs. Due to this, systems that do not require channel information are progressively gaining traction, as they help to improve the overall quality of the network instead of that of specific users only.

The search for new paradigms aimed at improving the quality of wireless communications is unstoppable. In order to increase the capacity of current communications systems, new horizons and physical dimensions must be explored.

This dissertation aims at challenging this perspective and promoting the use of polarization in new radiocommunication systems. Consequently, the goal of this thesis is twofold: first, we aim at increasing the current capacity of point-to-point and point-to-multipoint links. Secondly, we introduce new mechanisms to increase the robustness of communications in particularly hostile environments. In this context, this thesis advocates for the use of polarization as a dimension to be exploited in radiocommunications.

In addition to the use of polarization, index modulations help increase transmission rates whilst improving robustness against errors and imperfections with a low computational complexity. Thus, the study of polarization in these systems is essential. This dissertation explores primordial aspects in this area, such as channel capacity, transmitter and receiver design and performance benchmarking with current systems. Finally, we identify and discuss various characteristic aspects of polarization.

In this thesis, the reader will navigate the mathematical foundations of the proposed concepts as well as their implementation in real-life scenarios. After all, engineering excels at the intersection of the underlying physical principles with their real-life implementation.

Universitat Politècnica de Catalunya

# *Resum*

Departament de la Teoria del Senyal i Comunicacions
Centre Tecnològic de Telecomunicacions de Catalunya

Doctor of Philosophy

**Polarization and Index Modulations: a Theoretical and Practical Perspective**

by Pol HENAREJOS

Els sistemes de radiocomunicacions han evolucionat significativament els últims anys per cobrir les exigents demandes del present i el futur. Històricament s'han utilitzat les dimensions temporal i freqüencial, i més recentment la espacial, per dotar les radiocomunicacions de més capacitat i robustesa. Paradoxalment, les radiocomunicacions que utilizen la dimensió de polarització no han evolucionat al mateix ritme. Concretament, s'utilitzen significativament en radiocomunicacions per satèl·lit, multiplexant canals en cada polarització ortogonal.

Addicionalment, la tendència actual és la de simplificar i unificar diferents marcs de treball per flexibilitzar i millorar les adaptacions a les futures necessitats. Per aquesta raó, els sistemes robustos que no requereixen informació de canal estan guanyant importància, ja que ajuden a millorar la qualitat global de la xarxa i no només per a determinats usuaris.

La cerca de nous paradigmes que milloren la qualitat de les comunicacions inalàmbriques és constant i imparable. Per aquesta raó es fa necessari buscar nous horitzons i noves dimensions físiques que permetin incrementar la capacitat actual de les comunicacions.

La present dissertació vol canviar aquesta perspectiva i potenciar l'ús de la polarització en els nous sistemes de radiocomunicacions. L'objectiu d'aquesta tesi doctoral és doble: primer, incrementar la capacitat actual dels enllaços punt a punt i punt a multipunt i, segon, dotar-los de millor robustesa davant d'entorns hostils. Per això, aquesta tesi presenta la polarització com a una dimensió a explotar en les radiocomunicacions.

Paral·lelament, les modulacions per índex incrementen la velocitat de transmissió, alhora que proporcionen una major robustesa davant errors i imperfeccions, tot mantenint una baixa complexitat computacional. Així doncs, l'estudi de la polarització en aquests sistemes es fa necessari i primordial. El present manuscrit estudia aspectes com la capacitat de canal en aquests àmbits, el seu disseny tant del transmissor com del receptor, així com comparacions amb sistemes existents. Finalment, també s'analitzen diferents aspectes característics i únics de la polarització i es proposa una nova modulació en 3 dimensions, únicament possible utilitzant la polarització.

En aquesta dissertació el lector podrà trobar els fonaments matemàtics que sustenten el que es presenta així com implementacions realistes de les diferents propostes en entorns reals. L'èxit de l'enginyeria es produeix quan s'aconsegueix amalgamar els fonaments teòrics amb la implementabilitat en entorns reals.

Universitat Politècnica de Catalunya

# *Resumen*

Departament de la Teoria del Senyal i Comunicacions
Centre Tecnològic de Telecomunicacions de Catalunya

Doctor of Philosophy

**Polarization and Index Modulations: a Theoretical and Practical Perspective**

by Pol HENAREJOS

Los sistemas de radiocomunicaciones han evolucionado significativamente en los últimos años para cubrir las exigentes demandas del presente y el futuro. Históricamente se han utilizado las dimensiones temporal y frecuencial, y más recientemente la espacial, para dotar las radiocomunicaciones de más capacidad y robustez. Paradójicamente, las radiocomunicaciones que utilizan la dimensión de la polarización no han evolucionado al mismo ritmo. Concretamente, se utilizan significativamente en enlaces con satélites, multiplexando canales en cada polarización ortogonal.

Adicionalmente, la tendencia actual es la de simplificar y unificar diferentes marcos de trabajo para flexibilizar y mejorar las adaptaciones a las futuras necesidades. Por esta razón, los sistemas robustos que no requieren información de canal están ganando importancia, ya que ayudan a mejorar la calidad global de la red y no sólo para determinados usuarios.

La búsqueda de nuevos paradigmas que mejoran la calidad de las comunicaciones inalámbricas es constante e imparable. Por este motivo se hace necesario buscar nuevos horizontes y nuevas dimensiones físicas que permitan incrementar la capacidad actual de las comunicaciones.

La presente disertación quiere cambiar esta perspectiva y potenciar el uso de la polarización en los nuevos sistemas de radiocomunicaciones. El objetivo de esta tesis doctoral es doble: primero, incrementar la velocidad de transmisión actual de los enlaces punto a punto y punto a multipunto y, segundo, dotarlos de una mayor robustez en entornos hostiles. Por ello, esta tesis presenta la polarización como una dimensión a explotar en las radiocomunicaciones.

Paralelamente, las modulaciones por índice incrementan las velocidades de transmisión, a la par que proporcionan una mayor robustez ante errores e imperfecciones, logrando mantener la complejidad computacional baja en todo momento. Así pues, el estudio de la polarización en estos sistemas se hace necesario y primordial. El presente manuscrito estudia aspectos como la capacidad de canal en estos ámbitos, su diseño tanto del transmisor como del receptor, así como comparaciones con sistemas existentes. Finalmente, también se analizan diferentes aspectos característicos y únicos de la polarización y se propone una nueva modulación en 3 dimensiones, únicamente posible utilizando la polarización.

En esta disertación el lector podrá encontrar los fundamentos matemáticos que sustentan lo que se presenta, así como implementaciones realistas de las distintas propuestas en entornos reales. El éxito de la ingeniería se produce cuando se consigue amalgamar los fundamentos teóricos con la implementabilidad en entornos reales.

# *Preface*

This thesis originates from the curiosity driving an engineer to discover new ways to improve current communication systems in his spare time. Juggling work, family and personal passions is not always an easy task. Contrary to the common belief, the key ingredient of a doctoral thesis is not technical knowledge nor academic skills, but a high degree of perseverance and patience. Conducting science is not always a grateful task. Experiments sometimes do not turn out as expected and this, in turn, results on contradictions that can easily infuse a sense of defeat. In very few occasions, unique discoveries are made - revolutions that truly have an impact on the world as we know it. However, the vast majority of experiment outcomes often fail to provide an answer to the very questions they seek to address.

True science is based precisely on this premise: to depart from personal beliefs in order to find objective, evidence-based truths. One of the situations that made a particular impact in my professional career was when a student knocked on the door of a retired professor and showed him the results of an experiment based on a theory in which he had worked all his life. The experiment actually proved the existence of his theory. What I found particularly impressive was not the magnitude of the discovery, but the fact that he was able to detach himself from his feelings and follow the path of science, although that meant departing from his personal beliefs.

With modesty and humility, this thesis aims at providing a small contribution to the field of communications by polarization. I still remember when I first heard the word polarization. I was twelve when my mother bought a parabolic antenna and a receiver as a gift to watch foreign TV channels. What stroke me the most was that two different channels could use the same frequency simultaneously just by changing their polarization. Who would have thought that 20 years later I would be writing my doctoral thesis on communication systems built upon the very same concept of polarization.

This thesis only bears my name but it would not have been possible without the contribution of many other people who have contributed with their point of view. From great discussions, technical arguments, ideas and informal talks this dissertation has finally emerged. Therefore, this thesis also serves as testimony of gratitude to the countless contributions from many different people.

The first acknowledgement is for my family. My wife, Sara, has been the main breadwinner during all these years. Thanks to her love, patience and understanding this thesis has finally been able to see the light. And to my children, Lluc and Mateu, for being the best thing that has ever happened to me.

The second thanks go to my parents, Àngels and Àngel. Educating a child comes with a great responsibility and this is not always easy. Thanks to them I am who I am. From them I have learned to take responsibility for my commitments and to have true determination and grit. And, especially, my mother, because her unconditional love has taught me to love everything I do.

The third thanks goes to my thesis director, Ana. We met 10 years ago while I was still finishing my university studies. She saw something in me and from that moment we began to work together. Thanks to her I discovered the world of academic research and that the best way to achieve my objectives is to never stop persevering.

I would also like to extend a huge thanks to my great friend, Jaume. Not only have we spent countless hours discussing about fascinating topics, but we have also helped each other in many occasions. Also, I would like to express my special gratitude for helping me carry out the arduous task of reviewing this thesis.

Finally, I would like to thank all my colleagues at the Centre Tecnològic de Telecomunicacions de Catalunya for being not only remarkable human beings, but also extremely competent professionals. You make working here a fantastic experience. In addition, I would also like to thank its entire management board for allowing me to pursue this thesis and help me to combine my professional obligations with this divertimento.

And lastly, I also want to thank you, reader, for showing interest and appreciation in the work that I hereby present. I hope you find it useful and helpful.

# Contents

# List of Figures

# List of Tables

# List of Abbreviations

| | |
|---|---|
| **AMC** | Adaptive Modulation and Coding |
| **AWGN** | Additive White Gaussian Noise |
| **BER** | Bit Error Rate |
| **BGAN** | Broadband Global Area Network |
| **bpcu** | bits per channel use |
| **BPSK** | Binary Phase-Shift Keying |
| **CASTLE** | Cloud Architecture for Standards development |
| **CDF** | Cumulative Density Function |
| **CSI** | Channel State Information |
| **CSIT** | Channel State Information at Transmitter |
| **DVB** | Digital Video Broadcasting |
| **ESM** | Effective S(I)NR Mapping |
| **ETSI** | European Telecommunications Standard Institute |
| **ETU** | Extended Typical Urban |
| **FEC** | Forward Error Correction |
| **FMod** | Frequency index Modulation |
| **FSK** | Frequency Shift Keying |
| **GIM** | Generalized Index Modulations |
| **GSMod** | Generalized Spatial Modulation |
| **HARQ** | Hybrid Automated Repeat Request |
| **IM** | Index Modulations |
| **LHCP** | Left Hand Circular Polarization |
| **LLR** | Log-Likelihood Ratio |
| **LoS** | Line of Sight |
| **LTE** | Long Term Evolution |
| **MI** | Mutual Information |
| **MIESM** | Mutual Information Effective S(I)NR Mapping |
| **MIMO** | **M**ultiple-**I**nput **M**ultiple-**O**utput |
| **MISO** | **M**ultiple-**I**nput **S**ingle-**O**utput |
| **ML** | **M**aximum **L**ikelihood |
| **MMSE** | **M**inimum **M**ean **S**quare **E**rror |
| **MODCOD** | **Mod**ulation and **Cod**e |
| **OPTBC** | **O**rthogonal **P**olarization-**T**ime **B**lock **C**odes |
| **OSTBC** | **O**rthogonal **S**pace-**T**ime **B**lock **C**odes |
| **PDF** | **P**robability **D**ensity **F**unction |
| **PER** | **P**acket **E**rror **R**ate |
| **PLA** | **P**hysical **L**ayer **A**bstraction |
| **PMod** | **P**olarized **Mod**ulation |
| **PolSK** | **Pol**arization **S**hift **K**eying |
| **QAM** | **Q**uadrature **A**mplitude **M**odulation |
| **QoS** | **Q**uality **o**f **S**ervice |
| **QPSK** | **Q**uadrature **P**hase-**S**hift **K**eying |
| **RB** | **R**esource **B**lock |
| **RF** | **R**adio**f**requency |
| **RHCP** | **R**ight **H**and **C**ircular **P**olarization |
| **RBIR** | **R**eceived **B**it Mutual **I**nformation **R**ate |
| **RTT** | **R**ound **T**rip **T**ime |
| **RV** | **R**andom **V**ariable |
| **SK** | **S**hift **K**eying |
| **SINR** | **S**ignal to **I**nterference and **N**oise **R**atio |
| **SNR** | **S**ignal to **N**oise **R**atio |
| **SE** | **S**pectral **E**fficiency |
| **SER** | **S**ymbol **E**rror **R**ate |
| **SSK** | **S**patial **S**hift **K**eying |
| **SIMO** | **S**ingle-**I**nput **M**ultiple-**O**utput |
| **SISO** | **S**ingle-**I**nput **S**ingle-**O**utput |
| **SMod** | **S**patial **Mod**ulation |
| **UE** | **U**ser **E**quipment |
| **V-BLAST** | **V**ertical **B**ell **LA**boratories Layer **S**pace-**T**ime |
| **XPD** | Cross-**P**olarization **D**iscrimination |
| **ZF** | **Z**ero **F**orcer |

# Physical Constants

| | |
|---|---|
| Speed of Light | $c = 2.997\,924\,58 \times 10^{8}\ \mathrm{m\,s^{-1}}$ |
| Charge of an electron | $q = -1.602\,176\,62 \times 10^{-19}\ \mathrm{C}$ |
| Vacuum permittivity | $\varepsilon_0 = 8.854\,187\,817 \times 10^{-12}\ \mathrm{F\,m^{-1}}$ |
| Mass of electron | $m_e = 9.109\,383\,561\,1 \times 10^{-31}\ \mathrm{kg}$ |

# Notation

| | |
|---|---|
| $a$ | scalar |
| $\mathbf{a}$ | column vector |
| $\mathbf{A}$ | matrix |
| $\mathbb{R}$, $\mathbb{C}$ | set of real and complex numbers, respectively |
| $\mathbb{N}_0$ | set of natural numbers, including the zero number |
| $\mathbb{R}^r$, $\mathbb{C}^r$ | set of $r$-dimensional vectors with real and complex entries, respectively |
| $\mathbb{R}^{r\times t}$, $\mathbb{C}^{r\times t}$ | set of $r \times t$ matrices with real and complex entries, respectively |
| $\|a\|$ | magnitude of $a$ |
| $\|\mathbf{a}\|$ | $\ell_2$ norm of $\mathbf{a}$, $\|\mathbf{a}\|= \sqrt{\mathbf{a}^H\mathbf{a}}$ |
| $a_{ij}$ | entry of $\mathbf{A}$ at $i$th row and $j$th column |
| $\mathrm{tr}\,(\mathbf{A})$ | trace of $\mathbf{A}$, $\mathrm{tr}(\mathbf{A}) = \sum_i a_{ii}$ |
| $\det(\mathbf{A})$ | determinant of $\mathbf{A}$ |
| $\mathbf{a}^H\mathbf{b}$ | inner product |
| $\mathbf{a}\mathbf{b}^H$ | outer product |
| $\mathbb{E}_X\{x\}$ | expectation of $x$ over the $X$ RV, $\mathbb{E}_X\{x\} = \int_{-\infty}^{\infty} f_X(x)x\,\mathrm{d}x$ |
| $\mathbf{A}^H$ | conjugate and transpose of $\mathbf{A}$ |
| $\mathbf{A}^T$ | transpose of $\mathbf{A}$ |
| $\sqrt{\mathbf{A}}$ | square root of matrix such that $\sqrt{\mathbf{A}}\sqrt{\mathbf{A}} = \mathbf{A}$ |
| $\mathcal{N}(\boldsymbol{\mu}, \mathbf{\Sigma})$ | multivariate Gaussian vector distribution, with mean $\boldsymbol{\mu}$ and covariance |
| $\mathcal{CN}(\boldsymbol{\mu}, \mathbf{\Sigma})$ | multivariate complex circularly symmetric Gaussian vector distribution, with mean $\boldsymbol{\mu}$ and covariance $\mathbf{\Sigma}$ |
| $\hat{a}$ | estimation of $a$ |
| $\arg\max_a f(a)$ | value of $a$ that maximizes $f(a)$ |
| $\arg\min_a f(a)$ | value of $a$ that minimizes $f(a)$ |
| $\max_a f(a)$ | maximum value of $f(a)$ |
| $\min_a f(a)$ | minimum value of $f(a)$ |
| $\Re(a),\ \Im(a)$ | real and imaginary parts of $a$, respectively |
| $\log(a)$ | natural logarithm |
| $\log_b(a)$ | base-$b$ logarithm |
| $\nabla$ | vector differential operator, $\nabla = \begin{pmatrix} \frac{\partial}{\partial x_1} & \cdots & \frac{\partial}{\partial x_N} \end{pmatrix}^T$ |
| $\nabla f$ | gradient of $f$, $\nabla f = \begin{pmatrix} \frac{\partial f}{\partial x_1} & \cdots & \frac{\partial f}{\partial x_N} \end{pmatrix}^T$ |
| $\nabla \cdot \mathbf{f}$ | divergence of $\mathbf{f}$, $\nabla \cdot \mathbf{f} = \sum_n \frac{\partial f_n}{\partial x_n}$ |
| $\nabla \times \mathbf{f}$ | curl of $\mathbf{f}$ |
| $\nabla^2 f$ | laplacian of $f$, $\nabla^2 f = \sum_n \frac{\partial^2 f}{\partial x_n^2}$ |
| $Q(a)$ | Q-function at $a$, $Q(a) = \frac{1}{\sqrt{2\pi}} \int_a^\infty e^{-\frac{x^2}{2}}\, \mathrm{d}x$ |

# List of Symbols

| | | |
|---|---|---|
| $j$ | imaginary unit, $\sqrt{-1}$ | |
| $\mathcal{E}$ | energy | J |
| $f$ | frequency | Hz ($\mathrm{s}^{-1}$) |
| $\gamma$ | S(I)NR | |
| $\omega$ | angular frequency | $\mathrm{rad\,s^{-1}}$ |
| $P$ | power | W ($\mathrm{J\,s^{-1}}$) |
| $\mathbf{x} \in \mathbb{C}^t$ | transmitted vector | |
| $\mathbf{H} \in \mathbb{C}^{r \times t}$ | channel matrix | |
| $\mathbf{y} \in \mathbb{C}^r$ | received vector | |
| $\mathbf{w} \in \mathbb{C}^r$ | noise vector | |
| $C$ | Capacity | |
| $l$ | hopping index | |
| $r$ | MIMO output dimension | |
| $s$ | transmitted symbol | |
| $t$ | MIMO input dimension | |
| $\sigma^2$ | variance | |
| $\mathbf{I}_n$ | identity $n \times n$ matrix | |
| $\lambda$ | wavelength | $\mathrm{m}^{-1}$ |

*To my wife and children.*
*You are my light.*

# Chapter 1

# Introduction

Science is not only a disciple of reason but, also, one of romance and passion.

S. Hawking

## 1.1 Presentation, Motivation and Objectives

Polarization was discovered many centuries ago, even before the discovery of Maxwell's equations. Paradoxically, the polarization dimension is not a relevant actor in communication systems. Although it is proven that it provides high diversity gains and flexibility, it is still restricted to minor scenarios.

This dissertation aims at spreading the use of the polarization dimension by analyzing and studying polarization fundamentals. It has been applied to communication systems as well as benchmarked against existing schemes. In order to provide a substantiated answer to the technical questions laid out in this dissertation, the following objectives have been defined:

- To understand how polarization can be used in future communication systems.
- To study use cases where the usage of polarization offers advantages compared to existing schemes.

- To analyze from both a theoretical and practical viewpoint the gains of Index Modulation (IM) in the polarization dimension.
- To introduce new insights that can only be realized with the polarization dimension.
- To raise awareness of the polarization limitations as well as to identify scenarios that are not recommended for polarization dimension.

## 1.2 Structure of the Dissertation

In this dissertation we explore different aspects of polarization as a new dimension in communication systems, specially focused on IM. First, we introduce the fundamentals and analyze the complexity of polarization from a theoretical perspective. Secondly, we extend the analysis of polarization applied to IM in terms of mutual information. Once the theoretical fundamentals and analysis have been introduced, we proceed to describe the properties, algorithms of detection and implementation of IM in the polarization domain. Then, we move beyond the classic IM approach in order to introduce the concept of 3D Polarized Modulation, where the constellation is designed in a 3D (rather than a 2D) space. Afterwards, we continue with the introduction of adaptive polarization modulation schemes. Finally, we conclude with a discussion about open issues and future research lines, followed by final thoughts and conclusions.

**Chapter 2**

This chapter describes the fundamentals of polarization and its common effects, as well as current communication systems without Channel State Information (CSI) where polarization is employed. In this chapter, IM, which motivates the present dissertation, is also introduced. For a detailed description of the Faraday

Rotation effect, the reader is referred to Appendix C.

#### Chapter 3

The third chapter describes an analysis on the mutual information of IM. The chapter departs from the mathematical formulation of mutual information and presents a closed-form expression of the solution, with approximations of 2nd and 4th orders. In addition to a static channel the study is also particularized for Rayleigh, Rice and Nakagami-$m$ statistical channel distributions. In the chapter, the proposed IM scheme is benchmarked against other current approaches such as OSTBC or V-BLAST in terms of mutual information. Finally, the chapter also studies different applications of IM in three different domains (frequency, spatial and polarization) and discusses different use cases.

The contribution of this chapter has been submitted to the following journal publication:

- P. Henarejos, A. I. Pérez-Neira, "Capacity Analysis of Index Modulations over Spatial, Polarization and Frequency Dimensions," IEEE Transactions on Communications, under Major Revision.

#### Chapter 4

The fourth chapter describes how the IM concept is applied to polarization dimension. From the communication systems perspective, the transmitter is described in detail and 4 different receivers are proposed, depending on the accuracy and computational complexity.

The presented PMod scheme is implemented with the specifications of the ETSI BGAN standard [ETS]. This chapter is particularly interesting from the real-life communications point of view, since it embeds the PMod in a transmitter chain where channel coding is employed. In this case, different aspects have to be taken into account in order to perform a reliable reception.

In this chapter, the proposed scheme is also implemented in a real-life scenario using an existing multimedia standard (BGAN). Hence, different signal processing blocks such as channel coding, scrambling, puncturing, pilot pattern generation, framing structure are also involved in the communication system, thus providing more accuracy and precision to the final results. Imperfections such as XPD or imperfect channel estimation are also analyzed.

The contributions of this chapter are integrated into the CASTLE Platform®. See Appendix E for more details.

Further contributions of this chapter have also been published in the following journal publication [HP15b]:

- P. Henarejos and A. I. Pérez-Neira, "Dual Polarized Modulation and Reception for Next Generation Mobile Satellite Communications," in IEEE Transactions on Communications, vol. 63, no. 10, pp. 3803-3812, Oct. 2015,

the following conference proceedings [HP15a; Hen+13]:

- P. Henarejos and A. I. Pérez-Neira, "Dual Polarized Modulation and receivers for mobile communications in urban areas," 2015 IEEE 16th International Workshop on Signal Processing Advances in Wireless Communications (SPAWC), Stockholm, 2015, pp. 51-55.

- P. Henarejos, M. Á. Vázquez, G. Cocco, A. I. Pérez-Neira, "Forward Link Interference Mitigation in Mobile Interactive Satellite Systems," in Proceedings of AIAA International Communications Satellite Systems Conference (ICSSC), 14-17 October 2013, Florence (Italy).

and the following patent [HP14]:

- Patent No.: WO2015113603 A1, EP3100371A1

  Name: Method and System for providing diversity in polarization of antennas

Inventors: P. Henarejos, A. I. Pérez-Neira

International Application No.: PCT/EP2014/051801

Submission Date: 30 January 2014

Publication Date: 6 August 2015

**Chapter 5**

The fifth chapter describes a new method for PMod. Classic modulation constellations are designed in two dimensions, where each dimension corresponds to the in-phase and quadrature baseband complex model. The chapter introduces a new constellation mapping which takes place in a three-dimensional sphere. Exploiting the invariant property of differential phase we are able to apply the concept of PMod in 3D.

The chapter introduces various theoretical concepts and mathematical expressions that allow the implementation of the proposed modulation scheme. In addition, it also discusses some benchmarking results compared with other existing classical implementations.

The main contributions from this chapter have been submitted to the following journal publication:

- P. Henarejos, A. I. Pérez-Neira, "3D Polarized Modulation," submitted to IEEE Transactions on Communications.

**Chapter 6**

In the previous chapters we have introduced different polarization communication schemes. Each scheme has been studied in detail by benchmarking its performance against others mechanisms. In the sixth chapter, we analyze the proposed polarization communication schemes in the context of adaptive transmissions. Depending on the scenario, it might be more beneficial to adapt not only between modulation and coding schemes but also between polarization schemes. Therefore, we analyze the impact of adapting three degrees of freedom instead of the classical approach of Adaptive Modulation and Coding (AMC) Schemes. The results show a substantial gain compared to current deployments.

This chapter is multidisciplinary in nature, as it introduces many different concepts: first, we present the Physical Layer Abstraction (PLA) framework applied to the polarization dimension on top of the BGAN standard. In addition, we describe a methodology for generating a reliable channel modeling that accepts multiple polarizations, time-correlated and with the presence of interfering patterns from other sources. Finally, we also consider realistic impairments such as delayed feedback and beamforming power patterns.

The main contributions from this chapter have been published in the following conference proceeding [Hen+16]:

- P. Henarejos, A. I. Pérez-Neira, N. Mazzali and C. Mosquera, "Advanced signal processing techniques for fixed and mobile satellite communications," 2016 8th Advanced Satellite Multimedia Systems Conference and the 14th Signal Processing for Space Communications Workshop (ASMS/SPSC), Palma de Mallorca, 2016, pp. 1-8.

- A. Tato, P. Henarejos, C. Mosquera, A. I. Pérez-Neira, "Link Adaptation Algorithms for Dual Polarization Mobile Satellite Systems", 9th EAI International Conference on Wireless and Satellite Systems (WiSATs), Oxford, 2017.

and the following project deliverable Technical Note [Hen16]:

- P. Henarejos, "Advanced signal processing techniques for fixed and mobile satellite communications," Technical Note 1 Work Item 2 Call 1 of Order 1. Satellite Network of Experts IV (SatNEx 4), founded by European Space Agency (ESA).

**Chapter 7**

In the final chapter we discuss some open issues and lay out some lines for future research. In particular, we initiate the discussion of 3D PMod with amplitude modulation and modulations using Orbital Angular Moments, jointly with polarization. Finally, the chapter is finalized with this dissertation's conclusions.

# Chapter 2

# State of the Art and Fundamentals

> We live in a society exquisitely dependent on science and technology, in which hardly anyone knows anything about science and technology.
>
> C. Sagan

In the first part of this chapter we introduce common communication systems used in this dissertation as benchmarks and we derive the need of polarization from the Channel Capacity point of view. In the second part of this chapter, we introduce all the mathematical and physic principles of electromagnetic polarization, as well as the impairments that break the diagonality of polarization.

## 2.1 Communication Systems without CSI

Channel capacity is a fundamental metric in every communication system since it is an upper bound of which is the maximum transmission rate that the system can achieve with a reliable reception. Beyond this bound, the reception cannot succeed. Hence, before discussing different communication systems, we describe the fundamentals of Capacity of MIMO Communication Systems [Gol+03]. Recent developments in novel transmission schemes present revolutionary mechanisms aimed at increasing the channel capacity. In particular, the use of MIMO architectures enables a large number of new challenging schemes, introduced as a promising way to notably increase the spectral efficiency (SE) [Sch+08b; Sch+08a; Ara+11].

MIMO Signal Processing is a mathematical model where vectors and matrices represent different components of different dimensions involved in a communication system. Particularly, MIMO Signal Processing takes an important relevance when multiple inputs and/or outputs are modeled in a single system. For instance, MIMO Signal Processing is widely used to model spatial systems, where each input corresponds to a radiating antenna and each output to a receiving antenna. However, the spatial dimension is not the only one that can be modeled by MIMO. Time-delayed systems that convolve channel impulse responses, Fast Fourier Transforms, frequency-time or space-time can also be modeled with MIMO Signal Processing. Therefore, polarization can be modeled using MIMO Signal Processing, where each component corresponds to each polarization. Generally speaking, the system model with $t$ inputs and $r$ outputs can be described as

$$\mathbf{y} = \mathbf{H}\mathbf{x} + \mathbf{w}, \tag{2.1}$$

where $\mathbf{y} \in \mathbb{C}^r$ is the received vector and contains the output components, $\mathbf{H} \in \mathbb{C}^{r\times t}$ is the transition matrix, $\mathbf{x} \in \mathbb{C}^t$ is the transmitted vector and contains the input components, $\mathbf{w} \in \mathbb{C}^r$ is the noise vector. In this model, $\mathbf{H}$ is the transition matrix from inputs to outputs and is often referred to as *channel matrix*. Each entry of this matrix describes the statistics of the environment that affect the transmitted signal. The channel matrix characterizes the scenario and the physical dimension used in the communication system. In other words, depending on the dimension used, the statistics of each entry may be radically distinct. Finally, the noise vector describes the statistics of perturbations that are added to the output.

First, we discuss what each MIMO component describes physically in these dimensions:

- Spatial Dimension: each input corresponds to the voltage of each transmitting antenna using the baseband model. Each output is the voltage measured by each receiving antenna. Hence, inputs are not orthogonal and have a physical measurement in a precise spatial position.
- Polarization Dimension: each input and output corresponds to each component of the voltage in the polarization ellipse at transmission and reception, respectively. Hence, the components are orthogonal by definition, since they correspond to the orthogonal basis chosen by convenience and the number of inputs and outputs is equal to 2, i.e., $t = r = 2$.

The capacity of a communication system is defined as the maximum transmission rate with an error probability arbitrary small and is described by

$$C = \max_{f(x)} I(X, Y) \tag{2.2}$$

$$= \max_{\mathbf{Q}:\mathrm{tr}(\mathbf{Q})=P} \log_2 \left|\mathbf{I}_r + \mathbf{Q}\mathbf{H}^H\mathbf{R}_w^{-1}\mathbf{H}\right|$$

$$= \max_{\mathbf{Q}:\mathrm{tr}(\mathbf{Q})=P} \log_2 \left|\mathbf{I}_r + \mathbf{Q}\mathbf{R}_H\right|, \tag{2.3}$$

where $\mathbf{x} \sim \mathcal{CN}(\mathbf{0}, \mathbf{Q})$, $\mathbf{w} \sim \mathcal{CN}(\mathbf{0}, \mathbf{R}_w)$ and $\mathbf{R}_H = \mathbf{H}^H\mathbf{R}_w^{-1}\mathbf{H}$. Hence, the capacity expression is achieved when the product $\mathbf{Q}\mathbf{R}_H$ is diagonal. In the spatial dimension the $\mathbf{R}_H$ is not diagonal, since each wave is radiated omnidirectionally and interferes with the other. In order to achieve it, transmitter and receiver perform a pre-processing (precoder) and post-processing (postcoder) operations to diagonalize the channel. The optimum processing is to implement the Singular Value Decomposition (SVD). This decomposition is characterized as

$$\mathbf{H} = \mathbf{U}\mathbf{\Sigma}\mathbf{V}^H \tag{2.4}$$

where $\mathbf{U} \in \mathbb{C}^{r\times r}$, $\mathbf{U}^H\mathbf{U} = \mathbf{I}_r$, $\mathbf{\Sigma} \in \mathbb{C}^{r\times t}$ is diagonal, and $\mathbf{V} \in \mathbb{C}^{t\times t}$, $\mathbf{V}^H\mathbf{V} = \mathbf{I}_t$. Transmitter and receiver perform the precoding and postcoding stages as follows:

$$\begin{aligned} \mathbf{r} &= \mathbf{A}\mathbf{y} \\ &= \mathbf{A}\mathbf{H}\mathbf{x} + \mathbf{A}\mathbf{w} \\ &= \mathbf{A}\mathbf{H}\mathbf{B}\mathbf{s} + \mathbf{A}\mathbf{w} \end{aligned} \tag{2.5}$$

where $\mathbf{A} \in \mathbb{C}^{r\times r}$ is the postcoder, $\mathbf{B} \in \mathbb{C}^{t\times t}$ is the precoder, and $\mathrm{s} \in \mathbb{C}^{t}$ is the vector of transmitted symbols. Transmitter and receiver can diagonalize the channel by using $\mathbf{A} = \mathbf{U}^H$ and $\mathbf{B} = \mathbf{V}\sqrt{\mathbf{P}}$, where $\mathbf{P}$ is a diagonal matrix to adjust the power constraint to each eigenvalue using the waterfilling algorithm [Gol+03]. Thus, (2.5) is reduced to

$$\begin{aligned} \mathbf{r} &= \mathbf{U}^H\mathbf{y} \\ &= \mathbf{U}^H\mathbf{H}\mathbf{V}\sqrt{\mathbf{P}}\mathbf{s} + \mathbf{U}^H\mathbf{w} \\ &= \mathbf{\Sigma}\mathbf{s} + \mathbf{U}^H\mathbf{w}. \end{aligned} \tag{2.6}$$

The covariance matrix of transmitter is $\mathbf{Q} = \mathbf{V}\mathbf{P}\mathbf{V}^H$ and therefore, expression (2.3) is reduced to

$$\begin{aligned} C &= \max_{\mathbf{Q}:\mathrm{tr}(\mathbf{Q})=P} \log_2 \left|\mathbf{I}_r + \mathbf{P}\mathbf{\Sigma}^H\mathbf{U}^H\mathbf{R}_w^{-1}\mathbf{U}\mathbf{\Sigma}\right| \\ &\overset{(1)}{=} \max_{\mathbf{Q}:\mathrm{tr}(\mathbf{Q})=P} \log_2 \left|\mathbf{I}_r + \frac{1}{N_0}\mathbf{P}\mathbf{\Sigma}^H\mathbf{\Sigma}\right|, \end{aligned} \tag{2.7}$$

where the last step (1) is possible when AWGN is considered. As we stated previously, this is the maximum possible capacity since $\mathbf{Q}\mathbf{R}_H$ is diagonal.

However, the transmitter needs a-priori knowledge of the channel matrix $\mathbf{H}$ in order to perform the operation, which is not always feasible, specially when $\mathbf{H}$ contains a strong fast fading component. This presents a major drawback in terms of capacity in the spatial dimension since it is difficult to achieve it. By the time the transmitter receives the feedback parameters, the channel has varied and the CSI becomes outdated.

Under these circumstances where the transmitter cannot exploit the knowledge of the channel, the best that it can be done is to employ diversity. *Diversity* concept implies that the same information is conveyed through different ways to exploit the characteristics of the channel. For instance, when we refer to *time diversity*, it implies that the information is transmitted in different time slots. Assuming the channel varies between time slots, the same information experiences different channel conditions at different instants. As a result, the robustness of the communication system is increased due to two main reasons: 1) there are multiple copies of the same information at the receiver (redundancy), and 2) the information is affected by different channel realizations, thus maximizing the probability to experience a favourable channel realization. This is analogous when we refer to frequency, code, spatial or polarization dimension.

The key remark is that diversity works properly when the paths[1] are uncorrelated. Hence, given two zero-mean paths, $h(i)$ and $h(j)$, we define the correlation as $\mathbb{E}\{h(i)h(j)^*\}$, for $i \neq j$[2]. If the correlation is zero, the paths are uncorrelated. Otherwise, diversity is reduced as the correlation increases.[3] Depending on the dimension, the diversity is achieved as follows:

- Frequency dimension: uncorrelated paths are achieved when

[1]We are referring to *path* with a general perspective. A *path* is an abstraction at information level, where the transmitter and receiver are linked by this *path*.

[2]Note that $\mathbb{E}\{h(i)h(i)^*\} = E_{h(i)} > 0$ is the energy of $h(i)$.

[3]In practice, we assume that both are uncorrelated if the correlation is less than $0.4$.

two frequencies experience uncorrelated channel realizations. This is achieved when the receiver is surrounded by objects that introduce a reflection on the incident electromagnetic wave with a delayed version of the transmitted information. Hence, the terminal receives the same information delayed in time, with different amplitudes and phases at different frequencies. The resulting wave is a superposition of different replicas, whose spectrum is not flat and becomes frequency-selective. The bandwidth that is considered flat is the *coherence bandwidth* and is inversely proportional to the maximum delay spread of the channel. Hence, we can assume that the frequencies separated by the coherence bandwidth or more are uncorrelated.

Frequency diversity can be also achieved by using orthogonal frequencies. Note that orthogonality implies uncorrelation and this implies that the transmitter and receiver may use several frequency subcarriers to carry the same information. In this case, the terminal receives the same information through different frequencies.

- Time dimension: having uncorrelated channel realizations implies that the channel has varied sufficiently in such a way that are completely different. Channel realizations in the time domain are different if the object is moving across a scenario where multipath is present. Hence, the channel varies inversely proportional to the maximum Doppler frequency. In other words, the higher the receiver speed, the faster the channel varies.

- Spatial dimension: having more than one antenna creates a new scenario. The transmitter can transmit the same information in all antennas and each stream experiences different channel conditions in certain circumstances. As in the previous dimensions, two spatial paths are uncorrelated if the antennas' separation is larger than $\lambda/2$.

- Code dimension: in this case, codes are designed artificially in such a way that they are always orthogonal. This approach is particularly interesting in spreading scenarios in order to increase the security of the communication. The spectrum is spread and masked below the noise threshold.

- Polarization dimension: this case is particularly interesting because both polarizations are orthogonal and thus, both experience independent spatial paths. In contrast to spatial dimension where all waves are superposed and are self-interfered, in the polarization dimension two waves do not interfere with each other. This case is equivalent to the spatial dimension where the channel is diagonalized and we demonstrated previously that a communication system achieves its full capacity when the channel is diagonalized. However, polarization dimension can provide only two uncorrelated paths, since all polarizations are described as a linear combination of two orthogonal polarizations. It is worth to mention that this diversity is degraded if the cross-polarization is increased or, in other words, when the diagonality of the channel is broken. If it is not zero, there is cross-talk between both polarizations and, thus, the diversity is degraded.

If the full diversity is achieved, the capacity expression of (2.3) is reduced to

$$C = \log_2 \left( 1 + \frac{\|\mathbf{H}\|_F^2 P}{N_0} \right) \tag{2.8}$$

and it is achieved when orthogonal block codes are used, which are described in the next section.

### 2.1.1 Orthogonal Block Codes

Orthogonal Block Codes (OBC) are a special case of Linear Block Codes. OBC present the advantage that allow a coherent detection,

decoupling all streams without cross-interference. Orthogonal Space-Time Block Codes (OSTBC) [PPL04; JSB06] were proposed by Alamouti in 1998 [Ala98]. Originally, this scheme was proposed using space and time dimensions, but it was extended to many other dimensions rapidly. Thus, there are schemes with space-frequency dimensions, such as in LTE [211], or with polarization-time [HP15a; HP15b; HP14]. Depending on the channel constraints, some dimensions provide more benefits in front the others. For instance, in satellite links, it is shown that using polarization provides higher gain than spatial dimension [Ara+10; Ara+11].

In summary, OBCs convey information in one dimension and shift this information in the other dimension. Hence, having a set of $N$ symbols, a transmission scheme can be described by

$$\mathbf{Y} = \mathbf{H}\mathbf{X} + \mathbf{W}, \tag{2.9}$$

where $\mathbf{Y} \in \mathbb{C}^{r\times M}$ is the received matrix, $\mathbf{H} \in \mathbb{C}^{r\times t}$ is the channel matrix, which lies in the first dimension, $\mathbf{X} \in \mathbb{C}^{t\times M}$ is the transmitted matrix, which bridges the information between the second dimension to the first, and $\mathbf{W} \in \mathbb{C}^{r\times M}$ is the additive noise. In this case, $M$ refers to the cardinality of the second dimension, and $t$, $r$ refer to the cardinality of the first dimension, at transmission and reception, respectively. $M$ has to be the same number at transmission and reception, but $t$ and $r$ can be different. Hence, we have

- OSTBC: $t$ is the number of transmitting antennas, $r$ is the number of receiving antennas, and $M$ is the number of time accesses.

- OSFBC: $t$ and $r$ are the number of antennas at transmission and reception, respectively, and $M$ is the number of frequencies.

- OPTBC: $t$ and $r$ are the number of polarizations at transmission and reception, respectively, and $M$ is the number of time accesses. It is usual that $t = r$.

A Linear Block Code, with a transmission rate of $R = N/M$, is considered orthogonal when

$$\mathbf{X}\mathbf{X}^H = \mathbf{I}\sum_{n=1}^{N} |s_n|^2 , \tag{2.10}$$

where $s_n$ is the complex symbol mapped from the constellation. The solution of these codes holds only if $\mathbf{X}$ can be decoupled in a set of $N$ amicable orthogonal designs (AOD).

In general, there are many designs for different $t$. However, there is a unique special case where the full rate is achieved, i.e., $R = 1$. This case holds when $t = 2$. For higher values, solutions provide lower rates. In summary, we can conclude that OBC provide full diversity but do not increase the rate[4]. OSTBC exploits the full channel diversity at the expense of sacrificing capacity.

### Other Block Codes

Although we focus on specific block codes such as OSTBC, there are many other block codes:

- Golden Codes: proposed by [BRV05], these codes aim at achieving full diversity but with higher rate compared to OBC, at expenses of breaking the orthogonality and the increase of the computational complexity at receiver. See B.3 for the design.

- Silver Codes: these codes were proposed by [TH01] and outperform Golden Codes [MOJ10] in optical channels. They have the advantage of a reduced computational complexity at receiver compared with Golden Codes. See B.3 for the design.

- Perfect Codes: these codes are used to detect and correct errors during the transmission and have the particularity that

[4]Note that the rate can be increased by means of channel coding techniques or higher constellation orders, although they require higher SNR

every pair of codewords in the alphabet has a minimum distance of $d = 2e + 1$ at least. Considering an encoding system with $k$ input and $n$ output symbols, there are finite classes of perfect codes:

- Codes with no redundancy: $k = n$.
- Repetition block codes with odd block length: $n = 2m + 1$, $k = 2m$, $d = m$.
- Hamming codes [Mac03]: binary $n = 2^m - 1$, $n - k = m$ and non-binary $n = (q^m - 1)/(q - 1)$, $n - k = m$, $q > 2$.
- Binary Golay code [Gol49]: $q = 2$, $n = 23$, $k = 12$, $d = 3$.
- Ternary Golay code: $q = 3$, $n = 11$, $k = 6$, $d = 2$.

### 2.1.2 Data Multiplexing

Data Multiplexing (DM) is a special case of Linear Block Codes. In this scheme, the block code is not orthogonal and thus, cross-interference is produced. These schemes are intended to multiplex different data streams to increase the transmission rate, at expenses of increasing the error metrics due to the cross-interference. Schemes such as Vertical Bell Laboratories Layered Space-Time (V-BLAST) [Wol+98; Gol+99] or Diagonal BLAST (D-BLAST) were developed in 1998 to transmit different data streams using different number of antennas. Although they are not new, V-BLAST is still used in most scenarios where the transmitter does not have CSI.

Although V-BLAST offers higher capacity, it requires high power budget. Amongst the different approaches that do not use CSI, V-BLAST scheme and successive improvements present a simple way to increase the achievable rate with a relative increase of the processing complexity [Wol+98; Gol+99; She+03; XK03] in the absence of CSI. However, V-BLAST introduces interference between the streams since all signals are transmitted through all antennas without any interference pre-cancellation. In consequence, the signals must be transmitted with higher amplitude to obtain the same error rates compared with the single stream case [Hen+13].

The reception of such schemes is based on the principle of SiC: the strongest stream is decoded first, subtracted from the received signal, and the next stream is decoded recursively. This is due to the fact that these schemes introduce cross-interference, which is necessary to mitigate in reception.

### 2.1.3 Index Modulations

OBC are designed to exploit the diversity of the channel by transmitting different replicas of the message by different paths. This produces an enhancement of the SNR but constraints the transmission rate. For instance, in the case of OSTBC, only a single stream can be transmitted. On the contrary, DM multiplexes several streams of data, increasing the transmission rate. However, the SNR is penalized severely and the probability error is often higher. Index Modulations (IM), term employed by [Bas15; Bas16], aim at filling this gap.

Traditionally, the information is conveyed by mapping bits onto the electric field. By estimating the electric field at reception it is possible to reconstruct the original message and extract the information bits. But this method is not the only one that can be leveraged. IM places the information not only in the radiated waveform but also in the hops of the channel.

IM combines two sources for conveying information:

1. By placing the information in the radiated waveform (first part of message).

2. By placing the information in the hops of different components of dimension (second part message). For instance, in the case of spatial dimension, depending on the second part of the message, one antenna is selected at transmission to radiate the waveform that corresponds to the first part of the message. Hence, the receiver can recover the second message by estimating which antenna radiated the first message. In other words, $t$ channel vectors are considered as a $t$-sized constellation diagram and used to convey information.

Fig. 2.1 illustrates the concept of IM. The message is splitted into two sources. The first is modulated as usual, using the in-phase / quadrature baseband model. The second denotes the activation of different radiating elements. Note that $L$ dimension may not be equal to $t$. For instance, in Chapter 5, we impose $L > t = 2$.

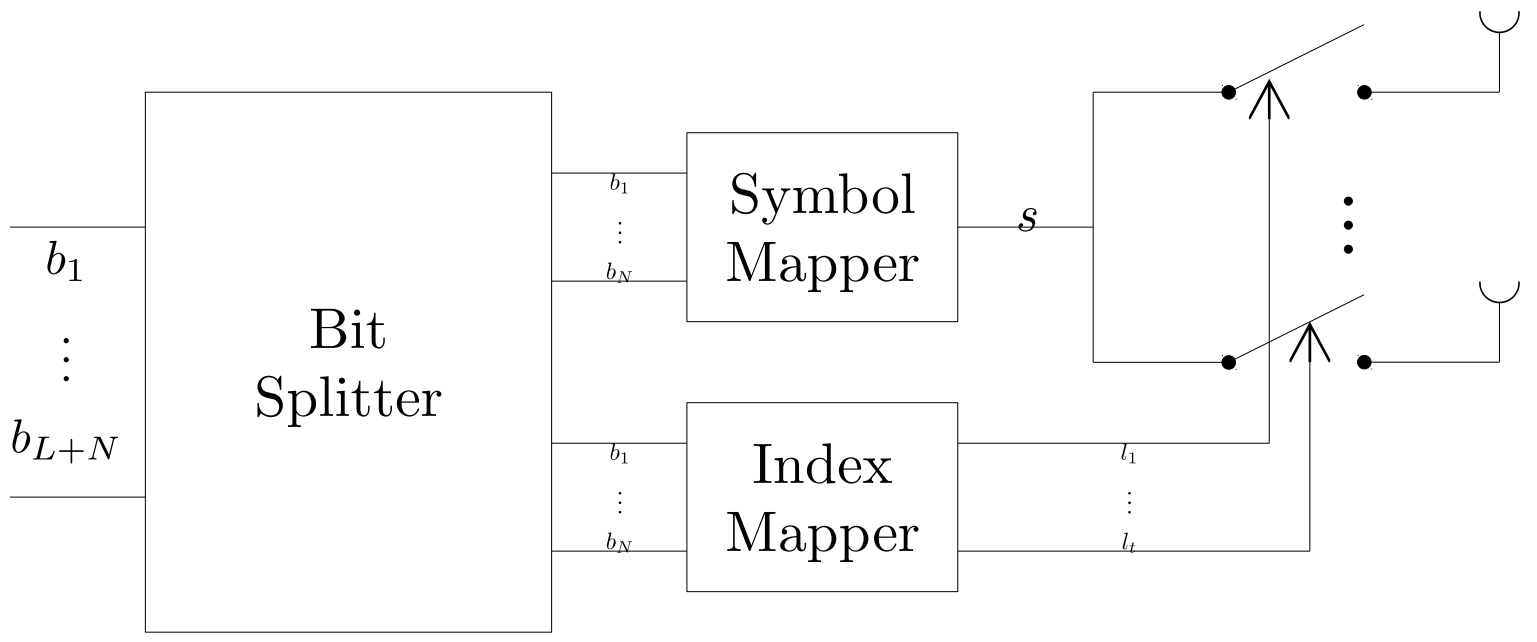

FIGURE 2.1: Index Modulation Diagram.

In the case of frequency, spatial or polarization dimension, the selection is performed on different carriers, antennas or polarizations, respectively.

The vector $\mathbf{l} \in \{0, 1\}^t$ denotes the index vector, whose entries enable or disable the selection of that channel. In IM, the vector $\mathbf{l}$ contains one 1 and the remaining are 0, meaning that only a single element is activated. Generalized Index Modulations (GIM) do not have this constraint and allow to activate multiple radiating elements simultaneously. The number of permutations is defined by $\binom{t}{k}$, where $k$ is the number of activated elements simultaneously. Hybrid schemes such as Quadrature Spatial Modulation (QSM) combine the activation of two antennas simultaneously, such as in GIM, each radiating the real and imaginary parts, respectively, [MIA15; MY17].

In particular, the polarization dimension does not allow the use of GIM, since only two polarizations are available. Activating two polarizations transforms the system into the data multiplexing scheme. In the next chapter, the capacity of IM is studied in detail.

**Example 1.** *We consider a system with* $t = 4$ *antennas and a* 16*-QAM constellation. The achievable rate is* $\log_2(t) + \log_2(N) = \log_2(4) + \log_2(16) = 6$ *bpcu. Considering the following input sequence* [101101]*, the first four bits are mapped in the QAM constellation. Thus,* $s = -3 + 3j$*. The last two bits are mapped to the antenna index, i.e.,* $l = 1$ *and the second antenna (starts counting from* 0*) is selected to radiate the symbol* $s$*.*

Note that IM is an extension of Shift Keying (SK) scheme. With SK, information is located only in the shifts between different elements of a particular dimension. For instance, Spatial Shift Keying (SSK) conveys the information only by selecting the transmitter antenna [Jeg+09]. Additionally, in multi-user scenarios, SSK and reconfigurable antennas can be used to select which user is intended, pointing a beam to it [Bou+15; Bou+16].

### Applications to Spatial Dimension

In opposition to V-BLAST and OSTBC, Spatial Modulation (SMod) [KPB03; Mes+08; JGS08; Yan12] appeared recently to increase the SE [Mes+08; CY01; Mes+06]. SMod strikes a balance between V-BLAST and OSTBC by increasing the channel capacity that OSTBC offers with less power requirements and is a trade-off between V-BLAST and OSTBC. In addition, it also provides more flexibility than OSTBC in the achievable rate that is obtained by increasing the number of transmitting antennas.

As it has been already introduced in Fig. 2.1, if there are multiple antennas in the transmitter, IM can select one of them depending on the sequence of bits to be transmitted. This specific implementation of IM is called SMod.

The receiver can extract the information from the radiated symbol by detecting which channel is being used. Nevertheless, this approach is very sensitive to channel variations and requires an accurate channel estimation as well as spatially uncorrelated channels [DH12; DHG11; MGH07]. The receiver can enhance the detection if the channel's correlation is low and exploits the fact the signal can be received from individual paths. Thus, although the terminal is moving, it may be able to detect from which antenna the information is being conveyed.

It is important to remark that SSK can be interpreted as a particular case of SMod that uses the same radiation pattern for all shifts. SMod uses the same SSK principle, but the radiation pattern is modulated according to additional information.

Although SMod and SSK only transmit through a single channel for a time instant, this can be generalized to an arbitrary number of streams, the so-called Generalized SMod (GSMod). Hence, by activating different channels simultaneously, different streams can be multiplexed and the data rate is also increased. This is well explained in [IKL16]. Additionally, there are studies about the achievable rate for GSMod [DC13] and for GIM [DEC16]. These generalized modulations are attractive from the computational complexity point of view and achievable rate. However, the capacity analysis is still open.

SMod is currently studied in detail to be proposed to standardization bodies. In [Di +14] and [Yan+15], implementation and challenge aspects for standardization are studied as well as comparisons with other spatial techniques deployed in standards such as LTE or Wi-Fi. In this sense, patents such as [Moh+15] and [MKA17] describe the transmitter and receiver of spatial modulation being used in the next Wi-Fi generation, defined in the 802.11ax specifications.

Finally, it is worth to mention that in satellite scenarios, due to the Line of Sight (LoS), the spatial components become correlated at the receiver side although the transmitting antennas may be separated at half wavelength. Hence, in the absence of scatterers, the receiver only discovers a single transmission path and the sensitivity of the terminal is not enough to distinguish the different spatial signatures and detect the antenna indices. Because of this, SMod does not seem suitable as it does not provide sufficient diversity in satellite scenarios.

### Applications to Frequency Dimension

In the frequency domain, analogous consideration can be pursued [EN02; Bas16]. In this case, the transmitter alternates different carriers, where an information symbol is radiated in the selected carrier frequency. Whereas Frequency Shift Keying (FSK) only conveys information in the selected frequency hop, Frequency Index Modulation (FMod) conveys information in the frequency hop as well as in the radiated symbol.

Note that FMod is different from Frequency Hopping, where the hop sequence is determined by the spreading code that should identify the user in a multi-user scheme. We note that FMod cannot be used as multi-user scheme, unless the FMod scheme is applied to only a group of frequencies and each user is identified by a different group.

In [DEC16], a design of a system where spatial and frequency domains coexist jointly is introduced and the achievable rate as the number of maximum bits that this technique can transmit, regardless the channel capacity, is presented.

### Applications to Polarization Dimension

Exploiting the spatial or frequency domains is not the only possible dimension. Polarization domain can also be used. In the case of polarization, Polarized Modulation (PMod) shifts between different orthogonal polarizations following a sequence of bits in order to radiate a different symbol in each shift [HP15b; HP15a].

In this case, Polarization Shift Keying (PolSK) is a particular case of PMod that uses the same radiation pattern for all shifts [BP92].

In contrast to spatial or frequency dimensions, the polarization satellite channel provides more diversity and may be used for these kind of schemes [HF06]. Although dual polarized antennas were used for broadcasting, where subscribers only tuned a single polarization, recent studies unveil that a dual-polarized MIMO channel is richer in terms of diversity [Lio+10]. Additionally, the use of dual polarized antennas is increasingly motivated by the new possibilities arising, together with the newest standards including dual polarized MIMO, such as Digital Video Broadcasting-Next Generation broadcasting system to Hand-held (DVB-NGH [DVBa]). Finally, research projects such as [Hen+13; Gal+14] reported that the throughput can be increased as in a conventional MIMO system if more antennas, and the consequent radio frequency (RF) chains [Vas00; Ara+10; Zor+08], are added in order to multiplex polarizations. The price to pay is that the complexity of the satellite payload increases since interference among polarizations appears. For instance, extending the V-BLAST strategy to dual polarized schemes requires higher transmit power to maintain the same Quality of Service (QoS) in point-to-point clients [Hen+13] compared with OPTBC or PMod.

## 2.2 Polarization as a Natural Diagonalizer

In the previous section we emphasized that polarization presents unique benefits, such as capacity increase. In this section we discuss the benefits of using the polarization dimension in front of spatial dimension in terms of MIMO Capacity.

The polarization domain has the inherent property that the polarization components are orthogonal by definition. Hence, the matrix $\mathbf{H}$ is usually diagonal. In this case, the transmitter can achieve capacity without additional pre-processing operations. However, there are some cases where this cannot always be fulfilled:

- If the transmitter and receiver are not perfectly aligned. This implies that the transmitter and receiver have to share the same orthogonal basis, i.e., $\mathbf{x}$ and $\mathbf{y}$ are the same in both sides. If this does not hold, each component of electric field is projected onto the basis of the receiver and, thus, the channel matrix is not diagonal.
- If the channel is aggressive and rotates the polarization. The result is the same as in the previous point: transmitter and receiver are not perfectly aligned.
- Some surfaces change the phase of some of the components. In this case, there is a second ray reflected with the same polarization with the inverted phase. Hence, the receiver experiments a rotation of the transmitted waveform polarization.

In contrast to the spatial dimension, where a pre-processing operation needs to be performed, the polarization dimension performs this task by definition. Hence, we can describe the polarization dimension *as a natural diagonalizer*.

The fact that the polarization is characterized by an orthogonal basis is the main advantage but, simultaneously, the major limitation. As explained in next section, the solution to the wave equations of the Maxwell's equations is in the form of planar waves. Hence, this basis is always limited by two components. There are some works that define basis with an arbitrary number of components, such as [Kob+12; Yof+13; Web+13]. However, only two of them are orthogonal and the rest are lineal combinations. This implies that the channel always has a rank of 2, regardless of the number of polarizations used. This is the major limitation factor since the capacity expression is constrained to $\min(t, r)$, which is always 2 in the polarization dimension.

As we explained above, the optimal transmission scheme if the channel is diagonal is to use V-BLAST scheme with the power loading using the waterfilling algorithm [Gol+03]. However, the intrinsic diagonal aspect of polarization is not always fulfilled. Depending on the surrounding environment each polarization component may be rotated randomly. Additionally, impairments such as cross-talk may degrade the diagonal shape of the matrix. Thus, at the receiver side the channel matrix may not be diagonal.

In these cases, the V-BLAST is not the optimal since the channel is not diagonal and cannot be diagonalized. The present dissertation demonstrates that, in these cases, Index Modulations with polarization present important gains compared with other schemes. In the next section we introduce the fundamentals of electromagnetic polarization, as well as the physical impairments that may break the diagonality of polarization.

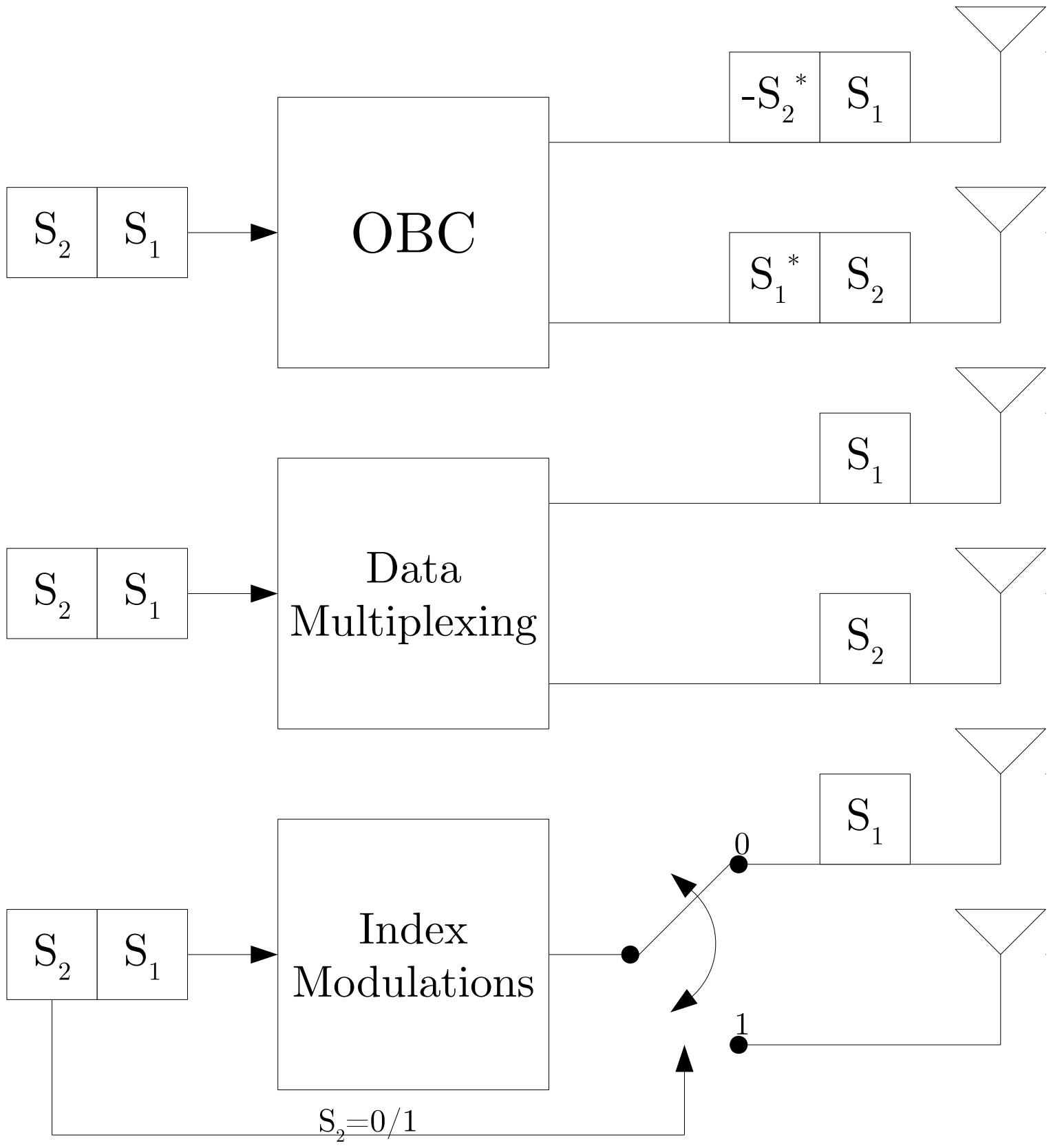

FIGURE 2.2: Index Modulation Diagram.

Finally, to conclude this section we summarize all the described communication schemes in Fig. 2.2.

## 2.3 Fundamentals of Electromagnetic Polarization

In 1672, Christian Huygens was the first scientist to suggest that the light was not a scalar but rather a vectorial magnitude. This behaviour was denoted as *polarization*. This theory, however, could not be validated until James Clerk Maxwell published the four equations that govern all electromagnetic forces.

Maxwell's Equations define the electromagnetic force by means of four well-known equations in the vacuum and homogeneous free space (no charges, $\rho = 0$, and no currents, $\mathbf{j} = \mathbf{0}$). These equations are summarized as

$$\nabla \cdot \mathbf{E} = 0 \tag{2.11}$$

$$\nabla \cdot \mathbf{B} = 0 \tag{2.12}$$

$$\nabla \times \mathbf{E} + \frac{\partial \mathbf{B}}{\partial t} = 0 \tag{2.13}$$

$$\nabla \times \mathbf{B} - \frac{1}{c^2}\frac{\partial \mathbf{E}}{\partial t} = 0. \tag{2.14}$$

The wave equations are obtained by taking the curl of (2.13) and (2.14) as follows

$$\begin{aligned} \frac{1}{c^2}\frac{\partial^2 \mathbf{E}}{\partial t^2} - \nabla^2 \mathbf{E} = 0 \\ \frac{1}{c^2}\frac{\partial^2 \mathbf{B}}{\partial t^2} - \nabla^2 \mathbf{B} = 0. \end{aligned} \tag{2.15}$$

The solution of the previous wave equations is a sinusoidal plane wave in the form

$$\begin{aligned} \mathbf{E}(\mathbf{r}, t) &= \Re\left\{\mathbf{E}_0 e^{j\left(\omega t - \mathbf{k}^T \mathbf{r}\right)}\right\} \\ \mathbf{B}(\mathbf{r}, t) &= \Re\left\{\mathbf{B}_0 e^{j\left(\omega t - \mathbf{k}^T \mathbf{r}\right)}\right\}, \end{aligned} \tag{2.16}$$

where $\mathbf{r}$ is the position vector in the spatial reference system, $\omega = 2\pi f$ is the angular frequency, $\mathbf{k} = k\mathbf{n}$ is the wave vector with the relationship $k = \omega/c = 2\pi/\lambda$, $\mathbf{n}$ is the normal of the propagation plane, $\lambda = c/f$ is the wavelength and $\mathbf{E}_0$ and $\mathbf{B}_0$ are constant electric and magnetic fields in time and space, respectively. Placing the three-dimensional Cartesian plane $(\mathbf{x}, \mathbf{y}, \mathbf{z})$ oriented in such a way that $\mathbf{n} = \mathbf{z}$, then

$$\begin{aligned} \mathbf{E}(\mathbf{r}, t) &= \Re\left\{\mathbf{E}_0 e^{j(\omega t - kz)}\right\} \\ \mathbf{B}(\mathbf{r}, t) &= \Re\left\{\mathbf{B}_0 e^{j(\omega t - kz)}\right\}. \end{aligned} \tag{2.17}$$

The static electric field $\mathbf{E}_0$ can be parametrized in the x-y plane as

$$\mathbf{E}_0 = \begin{pmatrix} E_x \\ E_y \end{pmatrix} = \begin{pmatrix} E_{0x} e^{j\varphi_x} \\ E_{0y} e^{j\varphi_y} \end{pmatrix}. \tag{2.18}$$

Hence,

$$\mathbf{E}(z, t) = \begin{pmatrix} E_{0x} \\ E_{0y} \end{pmatrix} \Re\begin{Bmatrix} e^{j(\omega t - kz + \varphi_x)} \\ e^{j(\omega t - kz + \varphi_y)} \end{Bmatrix} = \begin{pmatrix} E_{0x} \cos(\omega t - kz + \varphi_x) \\ E_{0y} \cos(\omega_t - kz + \varphi_y) \end{pmatrix}. \tag{2.19}$$

Note that $E_{0x}$ and $E_{0y}$ can be parametrized using polar coordinates and, thus, the expression (2.19) becomes

$$\begin{aligned} \mathbf{E}(z, t) &= \begin{pmatrix} E_0 \cos\alpha \\ E_0 \sin\alpha \end{pmatrix} \Re\begin{Bmatrix} e^{j(\omega t - kz + \varphi_x)} \\ e^{j(\omega t - kz + \varphi_y)} \end{Bmatrix} = \begin{pmatrix} E_0 \cos\alpha \cos(\omega t - kz + \varphi_x) \\ E_0 \sin\alpha \cos(\omega t - kz + \varphi_y) \end{pmatrix} \\ &= E_0 \cos(\omega t - kz) \begin{pmatrix} \cos\alpha \cos\varphi_x \\ \sin\alpha \cos\varphi_y \end{pmatrix}. \end{aligned} \tag{2.20}$$

Equation (2.20) is particularly interesting since it allows to express the electric field as a single amplitude $E_0$ with a tilt of $\alpha$.

The vector $\mathbf{E}_0$ is the Jones Vector and represents the amplitude and phase of the electric field in the $\mathbf{x}$ and $\mathbf{y}$ basis. The Jones vector is frequently normalized to the unitary norm, i.e., $E_{0x}^2 + E_{0y}^2 = 1$.

By applying trigonometric identities and after some mathematical manipulations, the following equality can be written

$$\frac{E_x^2}{E_{0x}^2} + \frac{E_y^2}{E_{0y}^2} - 2\frac{E_x}{E_{0x}}\frac{E_y}{E_{0y}}\cos\delta = \sin^2\delta, \tag{2.21}$$

where $\delta = \varphi_y - \varphi_x$. Equation (2.21) is the general parametric form of an ellipse and demonstrates that, at any time, the polarization describes an ellipse, with an orientation equal to $\delta$. The ellipse is also characterized by the parameter $\chi$. This parameter is defined as $\tan\chi = b/a$, where $a$ and $b$ are the major and minor semi-axis of the ellipse, and satisfy $a^2 + b^2 = E_{0x}^2 + E_{0y}^2$.

### 2.3.1 Stokes Vector

In 1852, Sir George Gabriel Stokes discovered that the polarization state of any electromagnetic wave can be characterized by four parameters, now known as the Stokes parameters. These parameters can be represented with the Stokes Vector and are defined as

$$\begin{aligned} S_0 &= |E_x|^2 + |E_y|^2 = E_{0x}^2 + E_{0y}^2 \\ S_1 &= |E_x|^2 - |E_y|^2 = E_{0x}^2 - E_{0y}^2 \\ S_2 &= E_x E_y^* + E_x^* E_y = 2E_{0x}E_{0y}\cos\delta \\ S_3 &= j\left(E_x E_y^* - E_x^* E_y\right) = 2E_{0x}E_{0y}\sin\delta. \end{aligned} \tag{2.22}$$

Hence, the Stokes vector, denoted as $\mathbf{S}$, is represented as

$$\mathbf{S} = \begin{pmatrix} S_0 \\ S_1 \\ S_2 \\ S_3 \end{pmatrix}. \tag{2.23}$$

Stokes parameters are real quantities and not only characterize the state of the polarization but also indicate the degree of polarization with the following inequality

$$S_0^2 \geq S_1^2 + S_2^2 + S_3^2, \tag{2.24}$$

which is fulfilled with equality when the electromagnetic wave is completely polarized. These four parameters describe the following states of the polarization:

- $S_0$ describes the total intensity of the wave.
- $S_1$ describes the amount of vertical / horizontal polarization.
- $S_2$ describes the rotation of the wavefront with respect to $\pi/4$.
- $S_3$ describes the direction of rotation.

The polarization degree defines the quantity of polarization of the electromagnetic wave and is defined as

$$P = \frac{\sqrt{S_1^2 + S_2^2 + S_3^2}}{S_0},\ 0 \leq P \leq 1, \tag{2.25}$$

where $P = 0$ means that the wave is totally unpolarized and $P = 1$ is fully polarized.

Stokes parameters can also be represented using the parameters of the polarization ellipse ($\delta$ and $\chi$):

$$\begin{aligned} S_0 &= I \\ S_1 &= IP \cos 2\delta \cos 2\chi \\ S_2 &= IP \sin 2\delta \cos 2\chi \\ S_3 &= IP \sin 2\chi. \end{aligned} \tag{2.26}$$

$S_1$, $S_2$ and $S_3$ span a three-dimensional polarization space and therefore all polarization states can be expressed as a linear combination of these parameters. Hereinafter, we assume that the electromagnetic wave is always fully polarized. Hence, the equality in (2.24), these parameters describe a sphere of radius $S_0$ centred at the origin. This sphere was proposed by Henri Poincaré in 1892 and now it is called Poincaré Sphere.

Note that all polarized electromagnetic waves, even if they are only partially polarized, can be represented by a superposition of unpolarized and fully polarized waves. Thus,

$$\mathbf{S} = \begin{pmatrix} S_0 \\ S_1 \\ S_2 \\ S_3 \end{pmatrix} = (1-P)\begin{pmatrix} S_0 \\ 0 \\ 0 \\ 0 \end{pmatrix} + P\begin{pmatrix} S_0 \\ S_1 \\ S_2 \\ S_3 \end{pmatrix}, \ 0 \leq P \leq 1. \tag{2.27}$$

### 2.3.2 Reference System

Often we distinguish several types of linear polarization: vertical, horizontal or slant. However, it is not clear when a given polarization is vertical or horizontal. Originally, the terms horizontal and vertical polarizations were defined according to the Earth surface. Decades later, with the commercial deployment of satellite communications, the reference system was defined in a more general context to include terrestrial and Earth-space communications.

From [ITU86], the polarization reference plane is defined as the plane composed by the centre of the Earth, the transmitter and receiver. Hence, the component of the electric field normal to this plane is the horizontally polarized component; and the component of the electric field parallel to this plane is the vertically polarized component.

### 2.3.3 Common Polarizations

In general, linear polarizations are obtained if $\varphi_x = \varphi_y \pm k\pi,\ k \in \mathbb{N}_0$. If $E_{0x} = E_{0y} = E$, the polarization becomes $45°$ slant if $k$ is even, or $-45°$ slant if $k$ is odd. Vertical or horizontal polarizations are achieved when $E_{0x} = 0$ or $E_{0y} = 0$. In this case, the electromagnetic wave oscillates only in a single plane. Finally, circular polarizations are obtained when $E_{0x} = E_{0y} = E$ if $\varphi_x = \varphi_y \pm k\frac{\pi}{2},\ k \in \mathbb{N}_0$ (plus sign produces RHCP and minus sign produces LHCP). Otherwise, the wave is elliptically polarized. It is straightforward to show that the circular polarization can be decoupled into two linear polarizations, with a phase difference of $\pi/2$. Figs. 2.3, 2.4, 2.5, 2.6 and 2.7 illustrate the electric field in a region of the space for a particular time instant. Note that the electric field is oscillating along the time at the $\omega$ frequency.

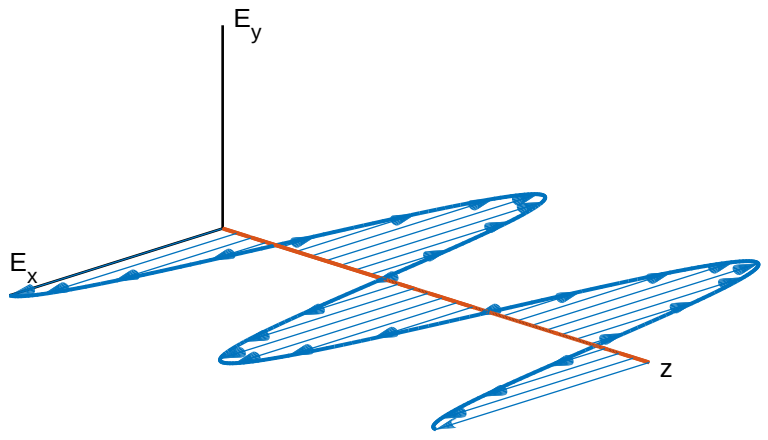

(A) Horizontal Polarization

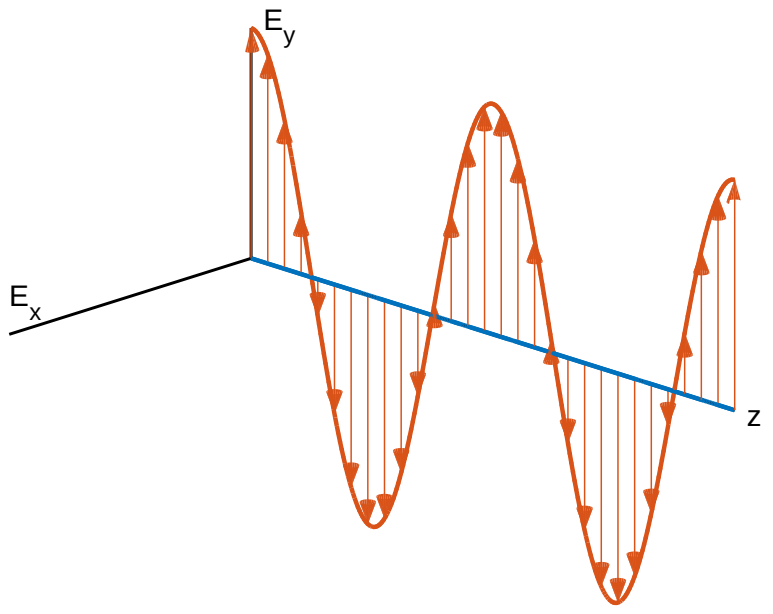

(B) Vertical Polarization

FIGURE 2.3: Electric field of horizontal and vertical polarizations for a particular time instant.

The Poincaré sphere is a representation of each possible polarization state. Each point on the sphere describes a type of polarization. The four points at the equator, $(1, 0, 0)$, $(0, 1, 0)$, $(-1, 0, 0)$ and $(0, -1, 0)$ correspond to the horizontal, $45°$ slant, vertical and $-45°$ slant polarizations, respectively. Top point $(0, 0, 1)$ and bottom point $(0, 0, -1)$ on the sphere correspond to Left Hand Circular Polarization (LHCP) and Right Hand Circular Polarization (RHCP). Tables 2.1, 2.2, 2.3, 2.4, 2.5 and 2.6 depict the Jones and Stokes vectors, the polarization ellipse and the point on the Poincaré sphere of most common used polarization states.

TABLE 2.1: Horizontal Polarization

<table>
<tr><td>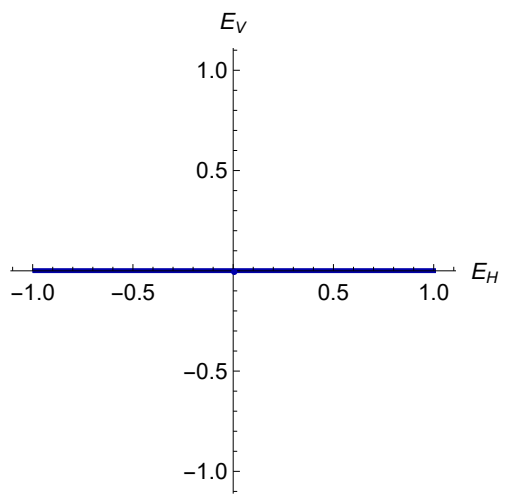

Polarization Ellipse</td><td>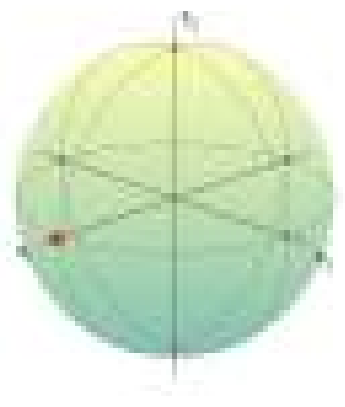
Poincaré Sphere</td></tr>
<tr><td>$\begin{pmatrix}1\\0\end{pmatrix}$<br>Jones vector</td><td>$\begin{pmatrix}1\\1\\0\\0\end{pmatrix}$<br>Stokes vector</td></tr>
</table>

TABLE 2.2: Vertical Polarization

<table>
<tr><td>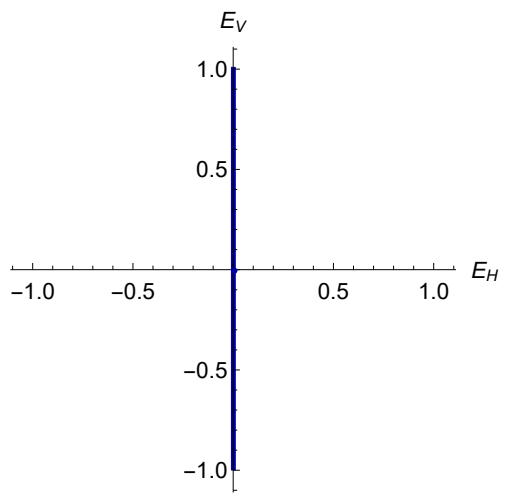

Polarization Ellipse</td><td>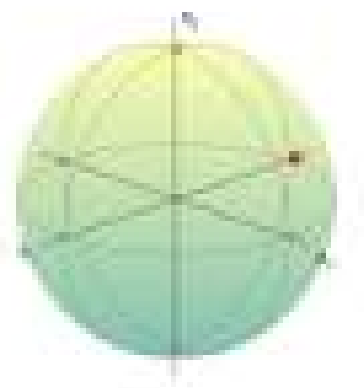
Poincaré Sphere</td></tr>
<tr><td>$\begin{pmatrix}0\\1\end{pmatrix}$<br>Jones vector</td><td>$\begin{pmatrix}1\\-1\\0\\0\end{pmatrix}$<br>Stokes vector</td></tr>
</table>

TABLE 2.3: $45°$ Slant Polarization

| | |
|---|---|
| 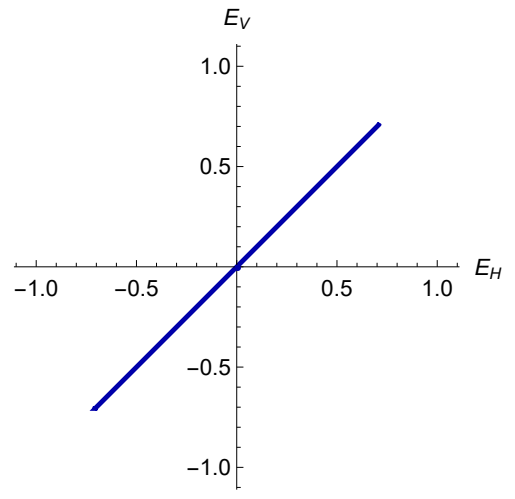<br>Polarization Ellipse | 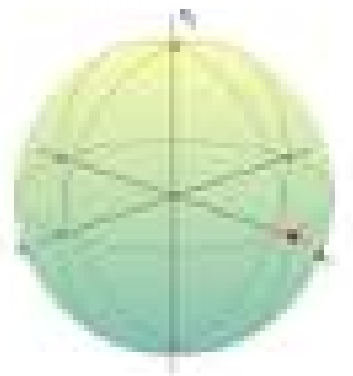<br>Poincaré Sphere |
| $\frac{1}{\sqrt{2}}\begin{pmatrix}1\\1\end{pmatrix}$<br>Jones vector | $\begin{pmatrix}1\\0\\1\\0\end{pmatrix}$<br>Stokes vector |

TABLE 2.4: $-45°$ Slant Polarization

| | |
|---|---|
| 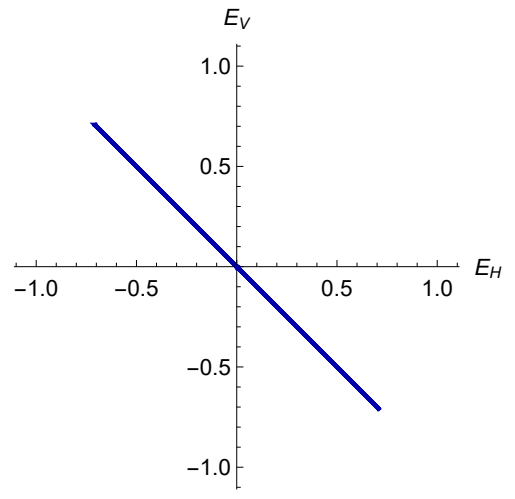<br>Polarization Ellipse | 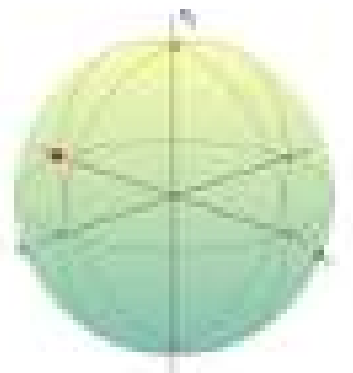<br>Poincaré Sphere |
| $\frac{1}{\sqrt{2}}\begin{pmatrix}1\\-1\end{pmatrix}$<br>Jones vector | $\begin{pmatrix}1\\0\\-1\\0\end{pmatrix}$<br>Stokes vector |

TABLE 2.5: Righ-Hand Circular Polarization

<table>
<tr><td>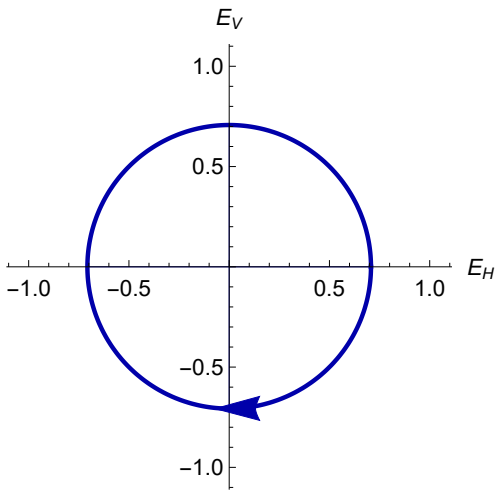

Polarization Ellipse</td><td>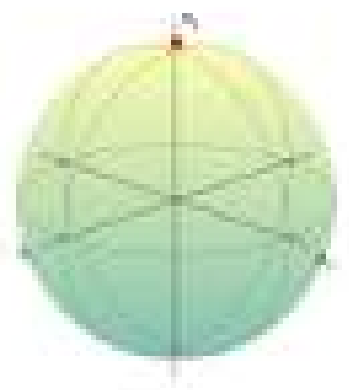
Poincaré Sphere</td></tr>
<tr><td>$\frac{1}{\sqrt{2}}\begin{pmatrix}1\\-j\end{pmatrix}$<br>Jones vector</td><td>$\begin{pmatrix}1\\0\\0\\1\end{pmatrix}$<br>Stokes vector</td></tr>
</table>

TABLE 2.6: Left-Hand Circular Polarization

<table>
<tr><td>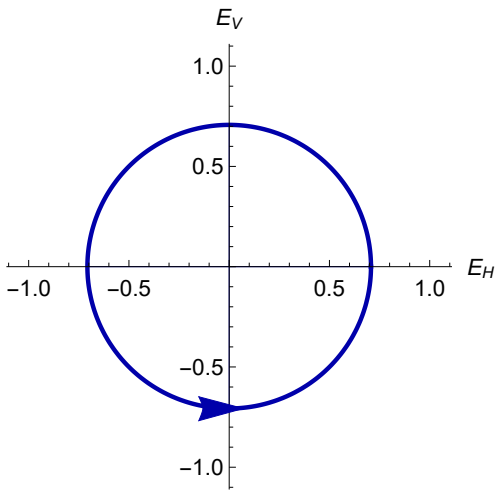

Polarization Ellipse</td><td>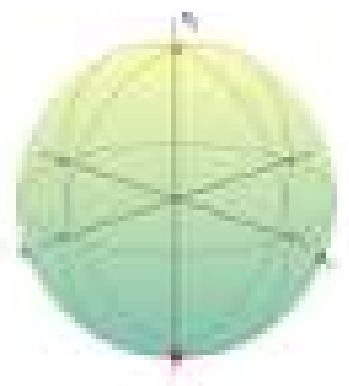
Poincaré Sphere</td></tr>
<tr><td>$\frac{1}{\sqrt{2}}\begin{pmatrix}1\\j\end{pmatrix}$<br>Jones vector</td><td>$\begin{pmatrix}1\\0\\0\\-1\end{pmatrix}$<br>Stokes vector</td></tr>
</table>

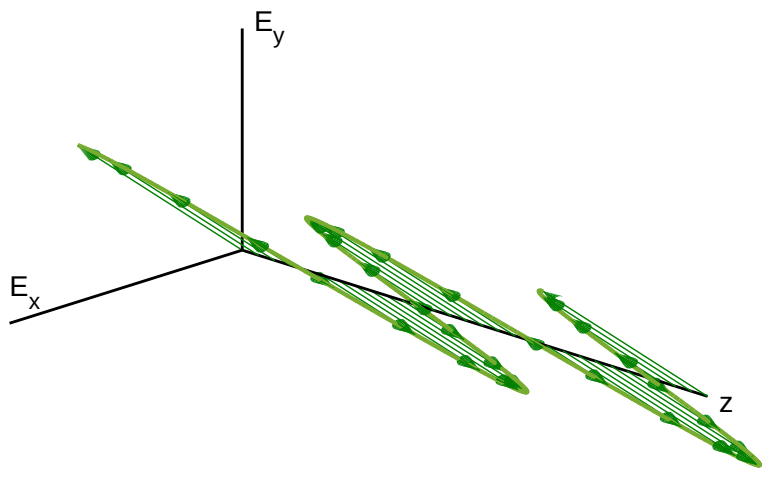

(A) 45° Slant Polarization

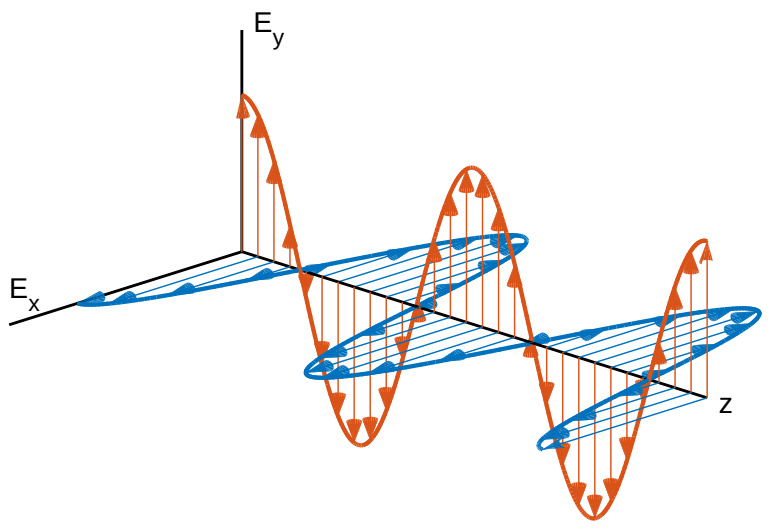

(B) 45° Slant Polarization Components

FIGURE 2.4: Electric field of 45° slant polarization for a particular time instant.

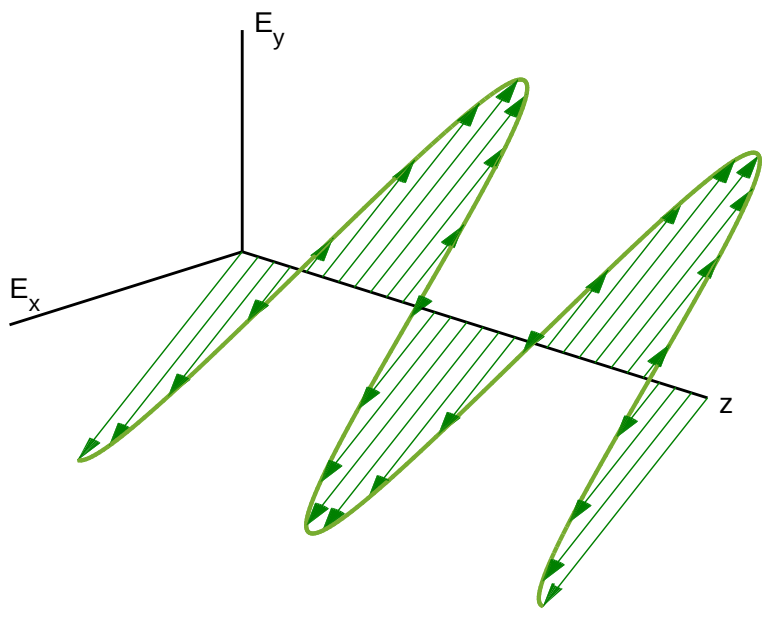

(A) −45° Slant Polarization

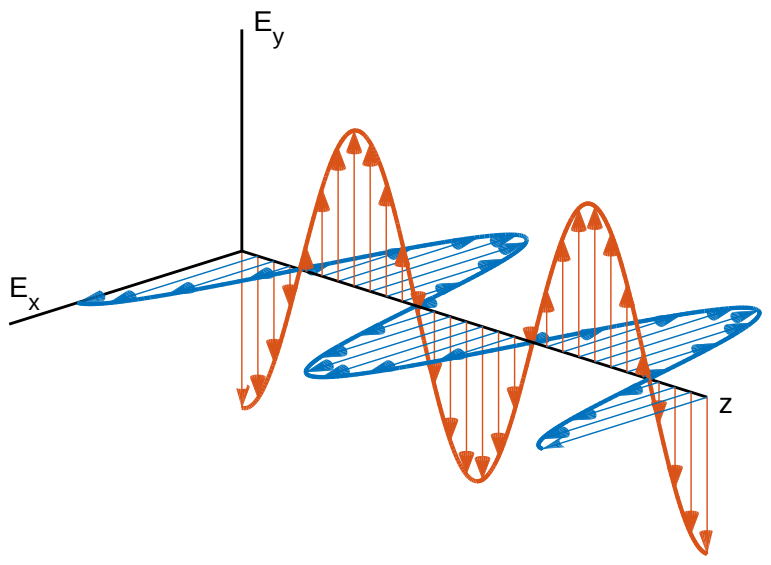

(B) −45° Slant Polarization Components

FIGURE 2.5: Electric field of −45° slant polarization for a particular time instant.

### 2.3.4 Reflection on Surfaces: Specular Component

Polarized electromagnetic waves may change the polarization state when are reflected on a surface. Depending on the characteristics of the surface, the reflection may modify the polarization ellipse or even reverse the direction of rotation.

We define the *specular component* as the electromagnetic wave reflected from a surface such that the incident angle is equal to the angle of reflection. The amplitude of the reflected wave is equal to the amplitude of the incident wave multiplied by the modulus of the reflection coefficient, $|R|$, which is defined as

$$R = \frac{\sin\phi - \sqrt{C}}{\sin\phi + \sqrt{C}}, \tag{2.28}$$

where

$$C = \begin{cases} \eta - \cos^2\phi & \text{for horizontal polarization} \\ \left(\eta - \cos^2\phi\right)/\eta^2 & \text{for vertical polarization} \end{cases} \tag{2.29}$$

and $\eta = \varepsilon(f) - j60\lambda\sigma(f)$, $\varepsilon(f)$ is the relative permittivity of the surface at frequency $f$, $\sigma(f)$ is the conductivity of the surface at frequency $f$ and $\lambda$ is the wavelength of incident wave defined as $\lambda = c/f$.

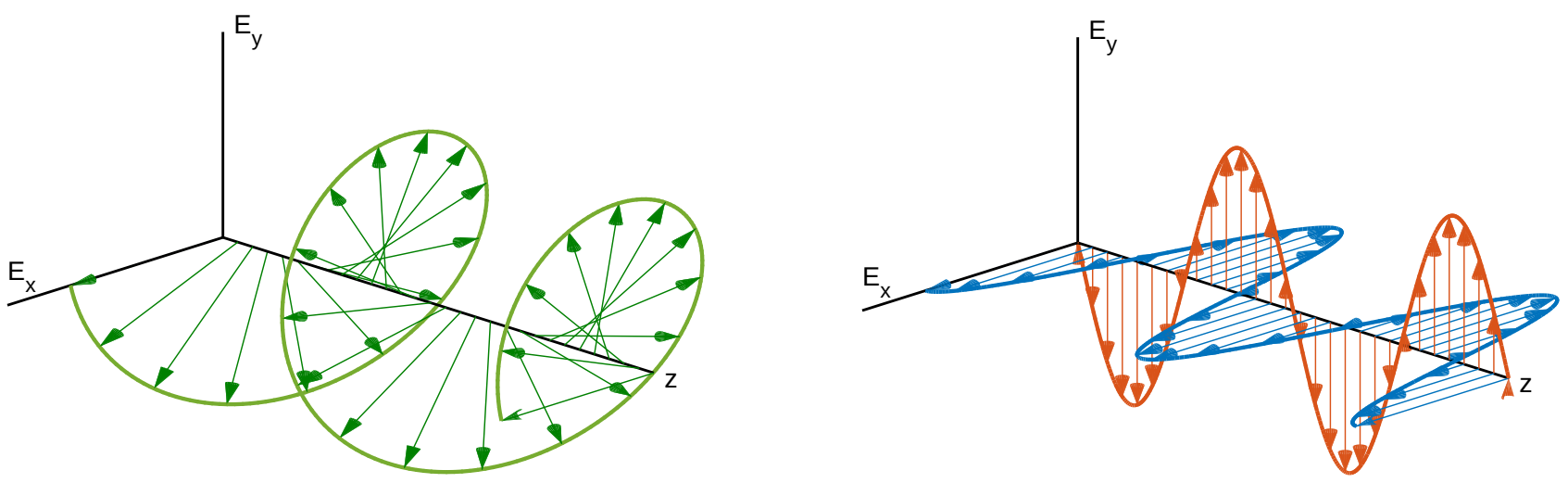

(A) RHCP

(B) RHCP Components

FIGURE 2.6: Electric field of RHCP for a particular time instant.

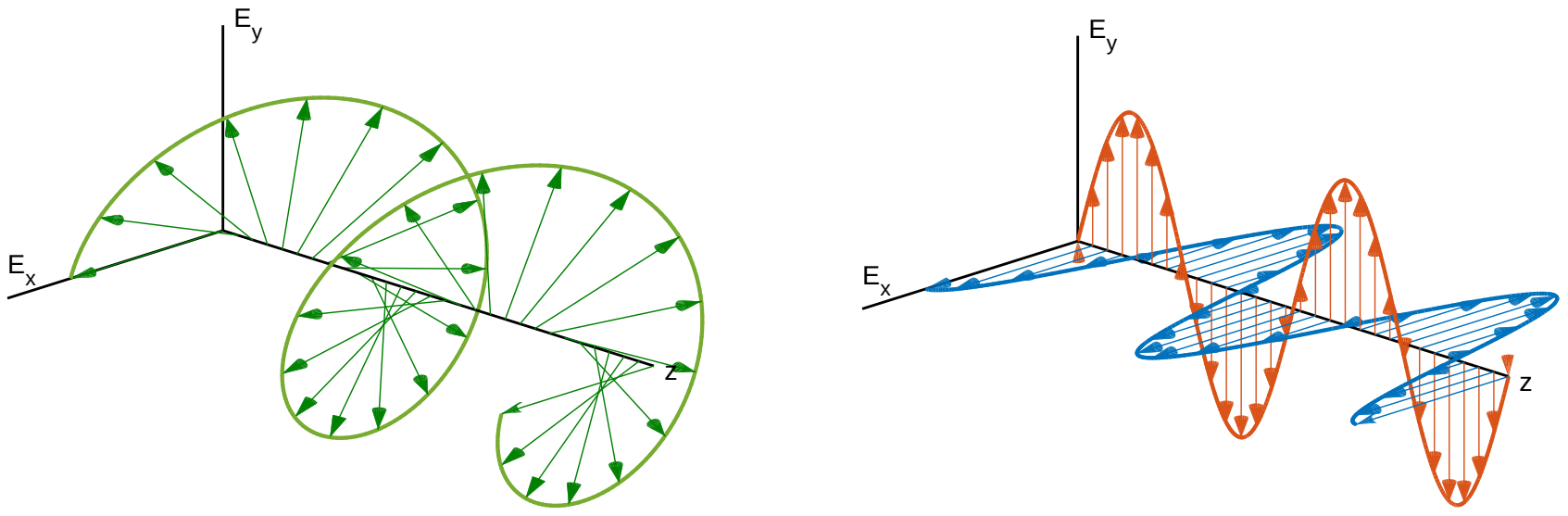

(A) LHCP

(B) LHCP Components

FIGURE 2.7: Electric field of LHCP for a particular time instant.

It can be shown that there is an angle such that the reflection coefficient for vertical polarization is minimum. This angle is called the Brewster angle and satisfies the following expression

$$\sin\phi_B = \frac{1}{\sqrt{|\eta|}}. \tag{2.30}$$

When the incident angle is greater than Brewster angle, the direction of rotation of the reflected wave is reversed.

Finally, the minimum incident angle is defined as

$$\phi_{\min} = \sqrt[3]{\frac{21}{10f}}. \tag{2.31}$$

Below this angle, diffraction phenomenon is more relevant and the transmission cannot succeed.

### 2.3.5 Faraday Effect

The Faraday effect is a phenomenon where the polarization rotates in the presence of an external magnetic field. A linearly polarized wave can be decoupled in two circular polarized waves, with opposite direction of rotation. See Annex C for a complete demonstration.

When an electromagnetic wave enters into a plasma of free moving electrons, such as the Ionosphere, and with a presence of an external magnetic field, such the Earth's magnetic field, the electric field of the wave induces a force in each electron. This force is described by the Lorentz's force and is proportional to the flux of the external magnetic field and the motion of the electron. Since the electric field of each polarization component is circular, the movement of electrons is also circular. According to Faraday's law, a circular electric movement induces a magnetic field across the motion. The induced magnetic field from one circular polarization is added to the external magnetic field; the induced magnetic field from the other circular polarization is added destructively to the external magnetic field. The different interaction depending on the direction of rotation causes that one circular component is delayed with respect to the other. Hence, when the incident wave leaves the medium, one circular polarization component is delayed with respect to the other and, therefore, the total linear polarization is rotated. This phenomena is also known as *birefringence*.

The Faraday rotation is proportional to the square of the wavelength. Hence, for the microwave wavelengths used by satellites, this rotation is negligible. However, for low frequencies such as S or L bands, this effect can have a significant impact. For this reason, the polarization used in these bands is circular instead of linear, as it is more robust to rotation effects. Since the circular polarization is composed by two linear components, these two components are also rotated in the same manner and the resulting polarization is still circular.

## 2.4 Polarization in Index Modulations Implementation

In this section we describe how to implement dual polarization communication schemes. In general, when we employ MIMO Signal Processing $t$ and $r$ RF chains are needed at transmitter and receiver. But one of the major advantages of Index Modulations is the fact that only one RF chain is needed. The selection of the polarization can be achieved by selecting the correspondent antenna.

Obviously, if only one RF chain is available at transmission, we need to use as many polarized antennas as polarizations we consider. But, due to the properties of polarization, the transmitter can implement two RF chains to use all polarizations, since all can be decomposed into two orthogonal polarizations.

However, at receiver side, since the selected polarization is not known a priori, it is necessary to receive both polarizations at same time to decide which polarization is being used. In this case, the receiver has to implement two RF chains.

To radiate and receive different polarizations, we can employ antennas with specific polarizations or dual polarization antennas.

### 2.4.1 Specific Polarizations

These antennas radiate with specific polarization. Using these antennas implies that the transmitter and receiver have to implement all antennas corresponding to the considered polarizations. Depending on the frequency, a guide-wave or dish may be necessary behind the radiating element.

Linear polarizations are obtained by using dipoles or monopoles. In particular, vertical and horizontal polarizations are obtained by facing the poles vertically or horizontally, respectively. Analogously, LHCP and RHCP are achieved by using specific antennas. Fig. 2.8 illustrates two antennas that resonate with circular polarizations.

Clover leaf antennas are commonly used for First Person View communications with drones. For instance, [Tao] provides clover leaf antennas with a peak gain of $3$ dBiC at $5.8$ GHz band. Electrical specifications and radiation patterns can be found in its website.

On the contrary, helical antennas are intended for professional users. For instance, [Wir] provides helical antennas with gains of $14$ dBi in the frequency range $460-900$ MHz and they have higher directivity compared with clover leaf antennas.

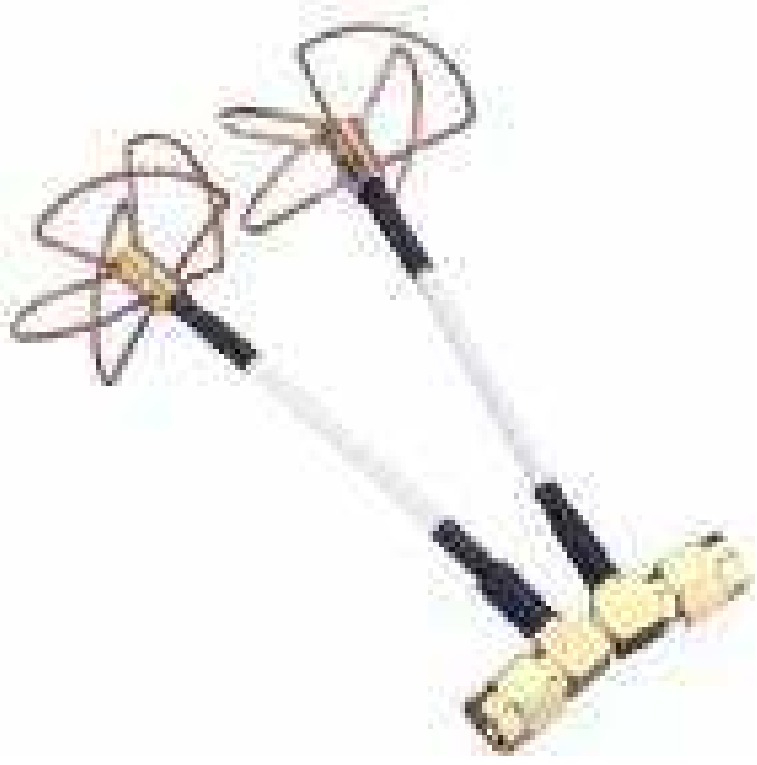

(A) Clover Leaf Antennas

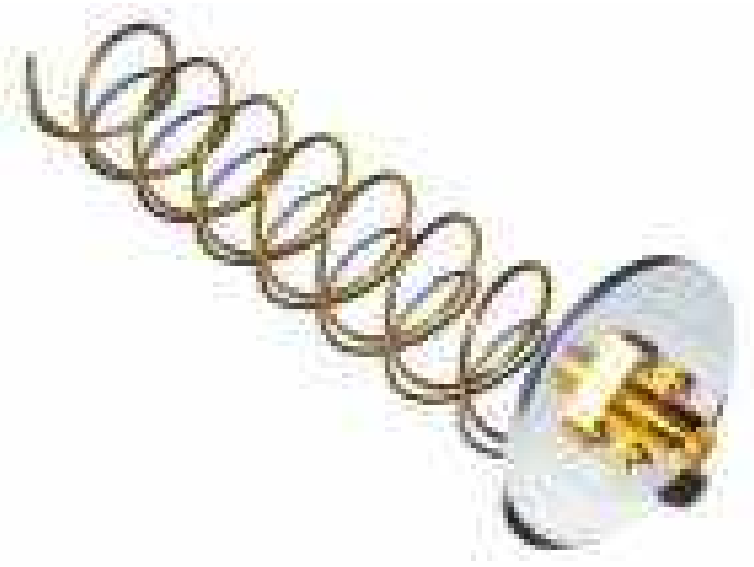

(B) Helical Antenna

FIGURE 2.8: Circular polarization antennas.

### 2.4.2 Dual Polarized Antennas

Dual polarized antennas contain two orthogonal radiating elements. With these antennas, all types of polarizations can be processed, since all polarizations can be decoupled into two orthogonal polarizations. These antennas have two ports (one per each polarization) and require two RF chains to feed both polarizations. At the same time, both polarizations can be received simultaneously. Fig 2.9 illustrates two types of polarized antennas. In particular, Fig. 2.9a is a Low Noise Block-Feedhorn disassembled where both orthogonal monopoles are appreciable. These LNBFs are used jointly with dishes to transmit or receive satellite signals. In particular, [Tec] provides specifications of dual polarized LNBFs with gains of 65 dBi at C band.

Fig. 2.9b is a log-periodic antenna with crossed orthogonal dipoles. These antennas are used for [Lin] in the VHF/UHF band and provide gains of 7 dBi.

In contrast to Fig. 2.9 where horizontal and vertical polarization are received simultaneously, dual polarized antennas with circular polarizations are also employed. This is is achieved by using the Septum polarizer [AS83; LK10; Fra11; Lim+14]. The

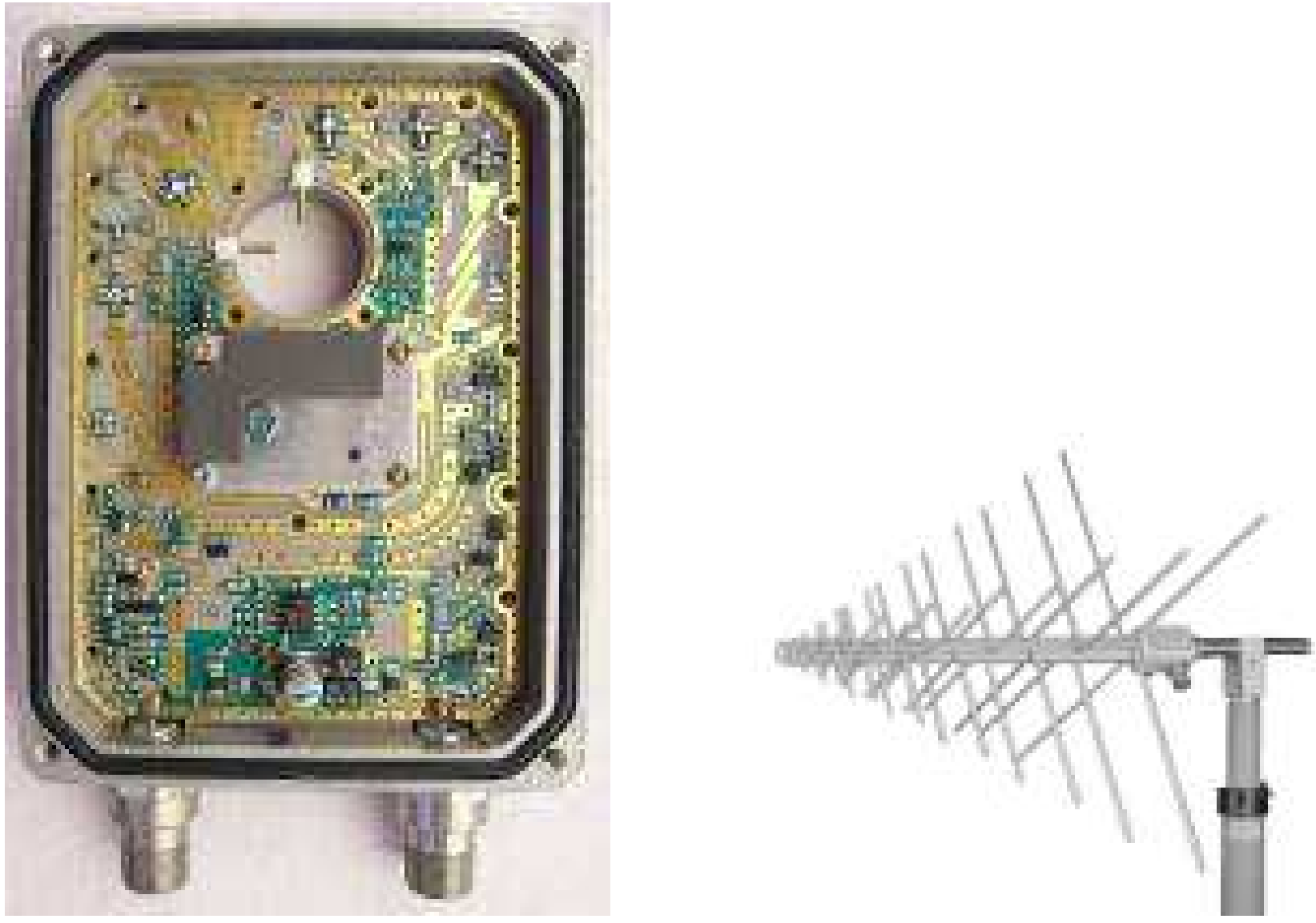

(A) LNBF with two orthogonal monopoles.

(B) Dual Polarized Log Period Antenna

FIGURE 2.9: Dual polarized antennas.

Septum polarizer is placed inside a wave-guide and both RHCP and LHCP can be received in each side of polarizer. Thus, this antenna have two ports, one per each circular polarization. Fig. 2.10 illustrates the septum polarizer.

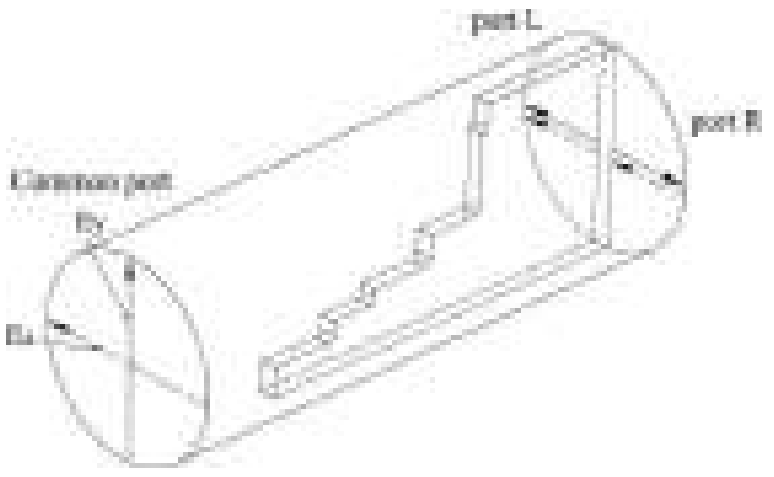

(A) Septum polarizer described in [CCT14]

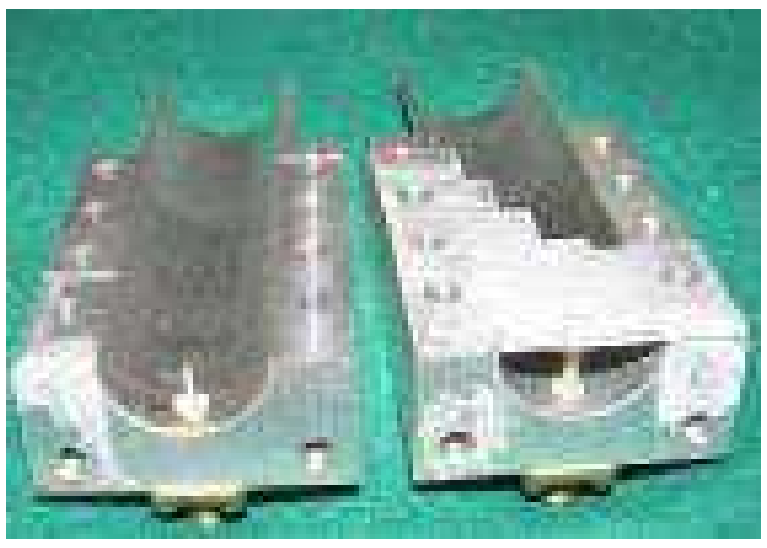

(B) Real horn with Septum polarizer disassembled.

FIGURE 2.10: Septum polarizer.

# Chapter 3

# Capacity Analysis of Index Modulations

> We have no right to assume that any physical laws exist, or if they have existed up to now, that they will continue to exist in a similar manner in the future.
>
> M. Planck

This chapter studies de capacity of IM in order to better understand them. The study consists of a providing a novel and accurate closed-form expression for the capacity. Index Modulations are usually presented as a technique to convey information using two sources: the radiation of an electromagnetic wave and hopping between different radio channels. The capacity analysis of IM can be performed regardless the physical domain of the applied scheme, e.g., frequency, polarization, etc.

Providing closed-form analytical expressions of fundamental metrics of IM (such as capacity) is a challenging task. The first work on channel capacity for the previous schemes is introduced in [Son+04]. Later works such as [YJ08a; YJ08b] introduce the capacity expression in its integral-based expression and formulate the instantaneous capacity for a single receiver antenna. However, manipulating the integral-based expression is rather complicated. In [RHH14], the authors extend the previous work to an arbitrary dimension at the receiver. In this case, although the capacity is also expressed in its integral-based expression, upper and lower bounds based on the Jensen inequality are introduced. Finally, in [ZYH15] another integral-based expression of the capacity is provided.

In this chapter we focus on the capacity of IM. Authors of [BRH16] and [NC16] have also analyzed the mutual information of SMod, however those expressions are in the integral-based form. In order to address this issue, [FEH16] presents a first approximation of the integral-based expression using the Meijer G function. However, this approximation is only valid for MISO systems.

This chapter provides a closed-form analytic expression for an accurate approximation of IM capacity. In addition, since we do not constrain the channel to a particular distribution, we obtain the expression for a generalized fading channel. Later on, we compute the ergodic capacity based on different channel distributions. Finally, to illustrate the applicability of the proposed formulas, we perform the capacity analysis of IM applied to three physical domains: spatial, polarization and frequency. Although the analyzed expression is the same for all results, its consequences are different depending on the physical dimension. Thus, the usefulness of the obtained expression is provided.

## 3.1 System Model and Capacity

Given a discrete time instant, the IM over an arbitrary MIMO channel realization, with $t$ inputs and $r$ outputs, is defined as

$$\mathbf{y} = \sqrt{\gamma}\mathbf{H}\mathbf{x} + \mathbf{w}, \tag{3.1}$$

where $\mathbf{y} \in \mathbb{C}^r$ is the received vector, $\gamma$ is the average SNR, $\mathbf{x} = \mathbf{l}s$, $\mathbf{l}$ is the all-zero vector except at position $l$ that is 1, $\mathbf{H} = [\mathbf{h}_1 \, \ldots \, \mathbf{h}_t] \in \mathbb{C}^{r \times t}$ is the channel matrix, $l \in [1, t]$ is the hopping index, $s \in \mathbb{C}$ is the complex symbol from the constellation $\mathcal{S}$. The AWGN noise is modeled as vector $\mathbf{w} \in \mathbb{C}^r \sim \mathcal{CN}(\mathbf{0}, \mathbf{I}_2)$. In other words, $\mathbf{x}$ has only one component different from zero (component $l$) and its value is $s$; thus, the ed symbol hops among the different channels. Note that MIMO channel can be obtained by simultaneous channels in any of the dimensions.

In this section we do not analyze the statistics of $\mathbf{H}$ yet, as we are only interested in the instantaneous capacity given a channel realization. $\mathbf{H}$ models effects and specific impairments of the employed domain (spatial, polarization, frequency, etc.). For instance, in the case of SMod, $\mathbf{H}$ models the channel and the antennas imperfections; in the case of PMod, $\mathbf{H}$ includes all imperfections of polarization dipoles. Note that the aspects of using reconfigurable antennas can also be reflected in $\mathbf{H}$. In following sections we will study the capacity for the different channel statistics.

Since the transmitted vector is determined by $(l, s)$, it is possible to rewrite (3.1) as

$$\mathbf{y} = \sqrt{\gamma}\mathbf{h}_l s + \mathbf{w}. \tag{3.2}$$

Thus, the symbol $s$ as well as the hopping index $l$ transmit information. The capacity can be expressed as

$$C = \max_{f_S(s), p_L(l)} I(\mathbf{y}; s, l) \text{ [bpcu]} \tag{3.3}$$

where $f_S(s)$ and $p_L(l)$[1] are the probability density functions (pdf) of the random variables (RV) of the complex symbol $s$ and hopping index $l$, respectively, and $I(X, Y)$ is the mutual information (MI) between RV $X$ and $Y$. Applying the chain rule [CT12], the

[1] Note that $p_L(l)$ is restricted to a discrete probability function since the channel hops are a discrete RV.

MI can be decomposed as

$$I(\mathbf{y}; s, l) = I(\mathbf{y}; s|l) + I(\mathbf{y}; l) \triangleq I_1 + I_2, \tag{3.4}$$

where $I(\mathbf{y}; s|l)$ is the MI between the received vector $\mathbf{y}$ and the transmitted symbol $s$ conditioned to the hopping index $l$, and $I(\mathbf{y}; l)$ is the MI between the received vector $\mathbf{y}$ and the hopping index $l$.

As described in [CT12], $I_1$ is maximized when $s$ presents zero mean complex Gaussian distribution. Thus, for a fixed $l$ index, we obtain

$$I_1 = I(\mathbf{y}; s|l) = \frac{1}{t} \sum_{l=1}^{t} \log_2 \left(\sigma_l^2\right), \tag{3.5}$$

where $\sigma_l^2 = 1 + \gamma \|\mathbf{h}_l\|^2$. Note that $\sigma_l^2$ depends on which channel is selected by the $l$ index.

Using the sufficient statistics transformation $Y = \frac{\mathbf{h}_l^H}{\|\mathbf{h}\|} \boldsymbol{Y}$, the second term of (3.4) can therefore be expressed as

$$\begin{aligned} I(\mathbf{y}; l) &\equiv I(y; l) \\ &= -H(Y|L) + H(Y), \end{aligned} \tag{3.6}$$

where $y = \sqrt{\gamma} \|\mathbf{h}_l\| s + w$, $w \sim \mathcal{CN}(0, 1)$. Therefore $Y$ is a RV such that, for a given $l$, $Y \sim \mathcal{CN}\left(0, \sigma_l^2\right)$.

Hence,

$$\begin{aligned} I_2 &= I(\mathbf{y}; l) \\ &= -\sum_{l=1}^{t} H(Y|L = l)\, p_L(l) - \int_{\mathcal{Y}} f_Y(y) \log_2 \left(f_Y(y)\right) \mathrm{d}y \qquad (3.7) \\ &= -\frac{1}{t} \sum_{l=1}^{t} \log_2 \left(\pi e \sigma_l^2\right) \\ &\quad - \frac{1}{t} \sum_{l=1}^{t} \int_{\mathcal{Y}} f_{Y|L}(y|l) \log_2 \left( \frac{1}{t} \sum_{l'=1}^{t} f_{Y|L'}\left(y|l'\right) \right) \mathrm{d}y, \qquad (3.8) \end{aligned}$$

where $\mathcal{Y}$ is the domain of $y$.

In (3.8), the conditioned pdf $f_{Y|L}(y|l)$ is described by the pdf of the zero mean valued complex Gaussian distribution, which is expressed as

$$f_{Y|L}(y|l) = \frac{1}{\pi\sigma_l^2} e^{-\frac{|y|^2}{\sigma_l^2}}. \tag{3.9}$$

The integral-based term (3.8) can be expressed as

$$\int_{\mathcal{Y}} f_{Y|L}(y|l) \log_2\left(\frac{1}{t}\sum_{l'=1}^{t} f_{Y|L'}(y|l')\right) \mathrm{d}y \tag{3.10}$$

$$= -\log_2(t) + \mathbb{E}_{Y|L}\left\{\log_2\left(\sum_{l'=1}^{t}\frac{1}{\pi\sigma_{l'}^2} e^{-\frac{|y|^2}{\sigma_{l'}^2}}\right)\right\}. \tag{3.11}$$

In order to evaluate (3.11), we decompose the expectation function into its multivariate Taylor series expansion near the mean [FTW92]. Note that whereas [IKL16] expands the logarithm function, we are decomposing the expectation function. Given a sufficiently differentiable function $g$, the Taylor series for the $g(\mathbf{x})$ function with $\mathbf{x} = \begin{pmatrix} x_1 & \dots & x_N \end{pmatrix}^T$ in the proximity of $\mathbf{a} = \begin{pmatrix} a_1 & \dots & a_N \end{pmatrix}$ is described by the multi-index notation [Sai91] as[2]

$$T(g, \mathbf{x}, \mathbf{a}) = \sum_{n=0}^{\infty}\sum_{|\alpha|=n}\frac{1}{\alpha!}\partial^{\alpha} g(\mathbf{a})(\mathbf{x}-\mathbf{a})^{\alpha}, \tag{3.12}$$

where $|\alpha| = \alpha_1 + \ldots + \alpha_N$, $\alpha! = \alpha_1! \ldots \alpha_N!$, $\mathbf{x}^{\alpha} = x_1^{\alpha_1} \ldots x_N^{\alpha_N}$ and $\partial^{\alpha} g = \partial^{\alpha_1} \ldots \partial^{\alpha_N} = \frac{\partial^{|\alpha|} g}{\partial x_1^{\alpha_1} \ldots \partial x_N^{\alpha_N}}$.

Thus, given a RV $\boldsymbol{X}$ with finite moments and such that all of its components are uncorrelated, i.e. $\mathbb{E}\{X_i X_j\} = \delta_{ij}$, the expectation of $g(\boldsymbol{X})$ can be expressed as

$$\mathbb{E}_{\boldsymbol{X}}\{T(g, \boldsymbol{X}, \mathbf{a})\} = \sum_{n=0}^{\infty}\frac{1}{n!}\sum_{m=1}^{N}\frac{\partial^n g}{\partial x_m^n}(\mathbf{a})\,\mathbb{E}\{(X_m - a_m)^n\}. \tag{3.13}$$

[2] $\sum_{|\alpha|=n}$. This corresponds to the sum of all possible combinations such that $|\alpha| = n$. For example, for $N = 3$, $\sum_{|\alpha|=2}\mathbf{x}^{\alpha} = x_1x_2 + x_1x_3 + x_2x_3 + x_1^2 + x_2^2 + x_3^2$.

By considering the Taylor series expansion near the expected value of $\boldsymbol{X}$, $\mathbf{a} = \mu_{\boldsymbol{X}}$, then (3.13) becomes

$$\mathbb{E}_{\boldsymbol{X}}\left\{T\left(g, \boldsymbol{X}, \mu_{\boldsymbol{X}}\right)\right\} = \sum_{n=0}^{\infty} \frac{1}{n!} \sum_{m=1}^{N} \frac{\partial^n g}{\partial x_m^n}\left(\mu_{\boldsymbol{X}}\right) \vartheta_{X_m}^n, \tag{3.14}$$

where $\vartheta_{X_m}^n$ is the centred $n$th moment of $X_m$.

The $n$th moment can be computed by deriving $n$ times the Moment-generating function $M_{\boldsymbol{X}}\left(t\right)$ [GS12] and equating it to zero, which for the multivariate normal case is defined as

$$M_{X_i}(t) = e^{\mu_{X_i} t + \frac{1}{2}\sigma_{X_i}^2 t^2}, \tag{3.15}$$

where $\mu_{X_i}$ and $\sigma_{X_i}^2$ are the mean and variance of $X_i$, respectively.

Since the received signal is a complex normal RV, which is a uncorrelated bivariate normal RV, such that $y = (\Re(y), \Im(y)) = (y_1, y_2)$, $N = 2$, the mean and variance of real and imaginary parts are defined by

$$\mu_{y_i} = \mathbb{E}\left\{y_i\right\} = 0,\ i = 1, 2 \tag{3.16}$$

$$\sigma_{y_i}^2 = \mathbb{E}\left\{\left(y_i - \mathbb{E}\left\{y_i\right\}\right)^2\right\} = \frac{\sigma_l^2}{2},\ i = 1, 2, \tag{3.17}$$

where the variance of the real and imaginary parts is the half of the variance of the received signal constraint to $l$, $\sigma_l^2$. Therefore, the $n$th moment of $y$ is expressed as

$$\vartheta_{y_i}^n = \begin{cases} (n-1)!!\,\frac{\sigma_l^n}{2^{\frac{n}{2}}} & \text{if } n \text{ is even} \\ 0 & \text{if } n \text{ is odd} \end{cases} \tag{3.18}$$

where $n!! = n(n-2)(n-4)...1$. Assuming that $g(y)$ is symmetric in its derivatives, $\frac{\partial^n g}{\partial y_1^n}(a) = \frac{\partial^n g}{\partial y_2^n}(a)$, then (3.14) can be reduced to

$$\mathbb{E}_{Y}\left\{T\left(g,y,0\right)\right\} \tag{3.19}$$

$$= g(0) + \sum_{n=1}^{\infty} \frac{\sigma_l^{2n}}{2^{2n-1} n!} \frac{\partial^{2n} g}{\partial y_1^{2n}}(0). \tag{3.20}$$

Assuming that

$$g\left(y\right) = \log_2\left(\sum_{l'=1}^{t} \frac{1}{\pi\sigma_{l'}^2} e^{-\frac{y_1^2+y_2^2}{\sigma_{l'}^2}}\right), \tag{3.21}$$

the first term $g(0)$ is expressed as

$$g(0) = \log_2\left(\sum_{l=1}^{t} \frac{1}{\pi\sigma_l^2}\right). \tag{3.22}$$

Therefore (3.11) is described as

$$\int_{\mathcal{Y}} f_{Y|L}\left(y|l\right) \log_2\left(\frac{1}{t}\sum_{l'=1}^{t} f_{Y|L}\left(y|l'\right)\right) \mathrm{d}y$$

$$= -\log_2\left(t\right) + \log_2\left(\sum_{l'=1}^{t} \frac{1}{\pi\sigma_{l'}^2}\right) + \sum_{n=1}^{\infty} \frac{\sigma_l^{2n}}{2^{2n-1} n!} \frac{\partial^{2n} g}{\partial y_1^{2n}}(0). \tag{3.23}$$

Combining (3.23), (3.8) can be expressed as

$$I_2 = I(\mathbf{y}; l) = -\frac{1}{t}\sum_{l=1}^{t} \log_2\left(\pi e \sigma_l^2\right) + \log_2\left(t\right) - \log_2\left(\sum_{l=1}^{t} \frac{1}{\pi\sigma_l^2}\right)$$

$$-\frac{1}{t}\sum_{n=1}^{\infty} \frac{1}{2^{2n-1} n!} \frac{\partial^{2n} g}{\partial y_1^{2n}}(0) \sum_{l=1}^{t} \sigma_l^{2n}. \tag{3.24}$$

Finally, combining (3.5) and (3.24), we can describe the capacity of IM as

$$C = \log_2\left(e^{-1} H\left(\boldsymbol{\sigma}^2\right)\right) - \sum_{n=1}^{\infty} \frac{A\left(\boldsymbol{\sigma}^{2n}\right)}{2^{2n-1} n!} \frac{\partial^{2n} g}{\partial y_1^{2n}}(0), \tag{3.25}$$

where $\boldsymbol{\sigma}^n = \begin{pmatrix} \sigma_1^n & \ldots & \sigma_t^n \end{pmatrix}^T$, and $A(\cdot)$ and $H(\cdot)$ are the arithmetic and harmonic mean operators, respectively.

In order to get some insight into (3.25), we consider its second and fourth order approximations. Taking the second order, $n = 1$, the second derivative of $g(y)$ at zero is expressed as

$$\frac{\partial^2 g}{\partial y_1^2}(0) = -\frac{2}{\log(2)} \frac{H\left(\boldsymbol{\sigma}^2\right)}{H\left(\boldsymbol{\sigma}^4\right)}. \tag{3.26}$$

Hence, we can obtain a 2nd order approximation of (3.25) by

$$C \simeq \log_2\left(H\left(\boldsymbol{\sigma}^2\right)\right) - \frac{1}{\log(2)}\left(1 - A\left(\boldsymbol{\sigma}^2\right) \frac{H\left(\boldsymbol{\sigma}^2\right)}{H\left(\boldsymbol{\sigma}^4\right)}\right). \tag{3.27}$$

Similarly, by taking the fourth order, $n = 2$, the fourth derivative of $g(y)$ at zero can be expressed as

$$\frac{\partial^4 g}{\partial y_1^4}(0) = \frac{12}{\log(2)}\left(\frac{H\left(\boldsymbol{\sigma}^2\right)}{H\left(\boldsymbol{\sigma}^6\right)} - \frac{H^2\left(\boldsymbol{\sigma}^2\right)}{H^2\left(\boldsymbol{\sigma}^4\right)}\right) \tag{3.28}$$

and therefore a 4th order approximation of (3.25) can be obtained by

$$\begin{aligned} C \simeq \log_2\left(H\left(\boldsymbol{\sigma}^2\right)\right) - \frac{1}{\log(2)}\Bigg(1 - A\left(\boldsymbol{\sigma}^2\right) \frac{H\left(\boldsymbol{\sigma}^2\right)}{H\left(\boldsymbol{\sigma}^4\right)} \\ + \frac{3}{4} A\left(\boldsymbol{\sigma}^4\right)\left(\frac{H\left(\boldsymbol{\sigma}^2\right)}{H\left(\boldsymbol{\sigma}^6\right)} - \frac{H^2\left(\boldsymbol{\sigma}^2\right)}{H^2\left(\boldsymbol{\sigma}^4\right)}\right)\Bigg). \end{aligned} \tag{3.29}$$

### 3.1.1 Capacity analysis in high SNR

Under the assumption of high SNR regime, i.e., $\gamma \to \infty$, we can use (3.25) at the limit and write

$$\lim_{\gamma \to \infty} C_{\text{IM}} = \lim_{\gamma \to \infty} \log_2(\gamma) . \tag{3.30}$$

We compare it with the asymptotic capacity of MIMO scheme without CSI. This is expressed as

$$C_{\text{MIMO}} = \log_2 \left| \mathbf{I} + \frac{\gamma}{t} \mathbf{H}^H \mathbf{H} \right| = \sum_{n=1}^{L} \log_2 \left( 1 + \frac{\gamma}{t} \lambda_n \right) , \tag{3.31}$$

where $\lambda_n$ is the $n$th non-zero eigenvalue of $\mathbf{H}^H\mathbf{H}$ and $L$ is the rank of $\mathbf{H}$. At high SNR, we can denote

$$\lim_{\gamma \to \infty} C_{\text{MIMO}} = \lim_{\gamma \to \infty} L \log_2(\gamma) . \tag{3.32}$$

Comparing (3.30) and (3.32) we can conclude that, effectively, any MIMO technique where the transmitter uses all channel modes (e.g. V-BLAST), increases the capacity by a factor of $L$. Instead, by using IM, the use of channel matrix is constrained to a single column and, therefore, the capacity formula is not multiplied by $L$, i.e. (3.30).

## 3.2 Remainder Analysis

In this section we analyze the expectation of the remainder of the 2nd order approximation (3.27). From (3.14), the Taylor series expansion with the remainder term can be written as follows

$$\mathbb{E}_{\boldsymbol{X}} \left\{ T_k \left( g, \boldsymbol{X}, \mu_{\boldsymbol{X}} \right) \right\} = \sum_{n=0}^{k} \frac{1}{n!} \sum_{m=1}^{N} \frac{\partial^n g}{\partial x_m^n} \left( \mu_{\boldsymbol{X}} \right) \vartheta_{X_m}^n + R_k \left( g, \boldsymbol{X}, \xi \right) , \tag{3.33}$$

where $k$ is the order of the Taylor series expansion and for some $\xi$ in the segment $[0, X]$. Since the third moment is zero, we analyze the remainder of order $k = 3$. $R_3\left(g, \boldsymbol{X}, \xi\right)$ is found by trunking the sum of (3.20) at $4$th order and evaluated at some point $\xi \in [0, y]$. Thus,

$$\mathbb{E}_Y\left\{R_3(C)\right\} = \frac{A\left(\boldsymbol{\sigma}^4\right)}{32}\left(\frac{\partial^4 g}{\partial y_1^4}(\xi) + \frac{\partial^4 g}{\partial y_2^4}(\xi)\right). \tag{3.34}$$

Expression (3.34) depends on $\xi$, which also depends on $y$. Since $y$ depends on the SNR, we perform the analysis for very low and very high SNR. First, we state the following theorem: First, we state the following theorem:

**Theorem 1.** *The expectation of the remainder $\mathbb{E}_Y\left\{R_3(C)\right\}$ is $0$ when all $\sigma_l^2$ tend to the same value $S$, regardless the value $S$. Hence,*

$$\lim_{\boldsymbol{\sigma}^2 \to S\mathbf{1}} \mathbb{E}_Y\left\{R_3(C)\right\} = 0. \tag{3.35}$$

See Annex G for the proof.

Based on the Theorem (1) we can formulate the following corollary:

**Corollary 1.** *The expectation of the remainder tends to $0$ for low SNR, i.e., when $\gamma \to 0$.*

$$\lim_{\gamma \to 0} \mathbb{E}_Y\left\{R_3(C)\right\} = 0. \tag{3.36}$$

*Proof.* If $\gamma \to 0$, then $\boldsymbol{\sigma}^2 \to \mathbf{1}$, which is the condition of Theorem 1 for $S = 1$ and therefore this concludes the proof. □

For high values of SNR, i.e., $\gamma \to \infty$, we introduce the following theorem:

**Theorem 2.** *The expectation of the remainder $\mathbb{E}_Y\left\{R_3(C)\right\}$ is $o(\gamma)$ when $\gamma \to \infty$*

$$\lim_{\gamma \to \infty} \mathbb{E}_Y\left\{R_3(C)\right\} = o(\gamma). \tag{3.37}$$

See Annex G for the proof.

Finally, we can state that:

- For low SNR, the expected error tends to zero.
- For high SNR, the expected error tends to a constant that does not depend on the SNR, but on the channel realization instead.

In the next section, the 2nd order approximation is used to obtain detailed expressions for the ergodic capacity.

## 3.3 Ergodic Capacity

In the previous sections, we studied the capacity for an arbitrary realization of the channel matrix. In this section we analyze the ergodic capacity for different channel statistics. The ergodic capacity is defined as

$$\bar{C} = \mathbb{E}_{\mathbf{H}}\{C\} \tag{3.38}$$

where $\mathbb{E}_{\mathbf{H}}$ is the expectation over all channel realizations. For the sake of clarity, hereinafter we omit the sub-index in the expectation operator, referring to the expectation over the channel statistics. For fast fading channels or when interleaving is carried out, ergodic capacity is a useful bound.

Although the equation (3.38) does not provide a closed-form expression, we exploit the property of harmonic mean of two RV, $H\left(\boldsymbol{\sigma}^2\right) = \frac{2\sigma_1^2\sigma_2^2}{\sigma_1^2+\sigma_2^2}$. This case corresponds to the IM applied to polarization dimension, i.e., PMod [HP15b], and the SMod with $2$ transmitting antennas. It is important to remark that there is no restriction on $r$.

Expanding (3.27) and after simplifying, we obtain

$$\begin{aligned} C = {} & 1 + \log_2\left(\sigma_1^2\right) + \log_2\left(\sigma_2^2\right) \\ & - \log_2\left(\sigma_1^2 + \sigma_2^2\right) - \frac{1}{\log(2)}\left(1 - \frac{1}{2}\left(\frac{\sigma_1^2}{\sigma_2^2} + \frac{\sigma_2^2}{\sigma_1^2}\right)\right). \end{aligned} \tag{3.39}$$

Assuming that all $\sigma_l^2$ follow the same distribution with the same parameters, we apply the statistical average of (3.38) to obtain

$$\bar{C} = 1 + 2\mathbb{E}\left\{\log_2\left(\sigma_1^2\right)\right\} \\ - \mathbb{E}\left\{\log_2\left(\sigma_1^2 + \sigma_2^2\right)\right\} - \frac{1}{\log(2)}\left(1 - \mathbb{E}\left\{\frac{\sigma_1^2}{\sigma_2^2}\right\}\right). \tag{3.40}$$

After mathematical manipulations, Table 3.1 summarizes the ergodic capacities of Nakagami-$m$, Rice and Rayleigh channel distributions, where

$$\begin{aligned} \beta &= \frac{m}{\gamma\Omega} \\ \Upsilon(r,\beta) &= \sum_{j=1}^{r-1}\sum_{k=1}^{j}\left(\frac{(k-1)\,!}{j!}(-\beta)^{j-k}\right) - \mathrm{E}_s(r,\beta) \\ \mathrm{E}_s(r,\beta) &= e^{\beta}\mathrm{E}_i\left(-\beta\right)\sum_{j=0}^{r-1}\frac{(-\beta)^j}{j!} \\ \mathrm{E}_i(-x) &= -\Gamma(0,x) \end{aligned} \tag{3.41}$$

and $\mathrm{E}_i(-x)$ is the Exponential Integral function for negative argument and $\Gamma(s,x) = \int_x^\infty t^{s-1}e^{-t}\,\mathrm{d}t$ is the Incomplete Upper Gamma Function. Note that the Rayleigh ergodic capacity can be obtained by particularizing the Nakagami-$m$ distribution with $m = 1$ and $\Omega = 2\varrho^2$, where $\varrho$ is the standard deviation of the Rayleigh RV. The derivations can be found in the Appendix D. It is important to remark that channel phase distribution does not affect capacity analysis, since only $\sigma_l^2$ are used in the expressions.

## 3.4 Results

In this section we present some results stemming from the work described in the previous sections. First, we compare the proposed approximations (3.27) and (3.29). In order to compute the system capacity we first generate $N$ channel realizations. Then, we compute the instantaneous capacity for each channel realization following different approaches. Finally, we average the obtained results amongst all realizations. For very large $N$, this procedure is equivalent to calculating the ergodic capacity. Unless explicitly stated otherwise, both the transmitter and the receiver have two inputs and two outputs, $t = r = 2$.

TABLE 3.1: Ergodic capacity of the Nakagami-$m$, Rice and Rayleigh channels

| Channel distribution | Ergodic capacity |
|---|---|
| **Nakagami-$m$** | $\bar{C} = \frac{1}{\log(2)} \left(2\Upsilon\left(mr, \beta\right) - \Upsilon\left(2mr, 2\beta\right) + \left(1 + mr\beta^{-1}\right)\beta^{mr}e^{\beta}\Gamma\left(1 - mr, \beta\right) - 1\right)$ |
| **Rice** | $\bar{C} = \frac{1}{\log(2)} \sum_{k=0}^{\infty} \frac{e^{-\frac{\lambda}{2}}\left(\frac{\lambda}{2}\right)^k}{k!} \left(2\Upsilon\left(r+k, \beta\right) - \Upsilon\left(2r+k, 2\beta\right) + \left(1 + \frac{r+k}{\beta}\right)\beta^{r+k}e^{\beta}\Gamma\left(1 - r - k, \beta\right) - 1\right)$ |
| **Rayleigh** | $\bar{C} = \frac{1}{\log(2)} \left(2\Upsilon\left(r, \beta\right) - \Upsilon\left(2r, 2\beta\right) + \left(1 + r\beta^{-1}\right)\beta^{r}e^{\beta}\Gamma\left(1 - r, \beta\right) - 1\right)$ |

The following subsections describe two studies. In the first one, the proposed approximations are validated and compared with the integral-based expressions introduced in [YJ08b] and [RHH14]. In the second, we employ the proposed approximations to compare and analyze different applications of IM. Concretely, we counterpose IM applied to the frequency, spatial and polarization domains.

### 3.4.1 Analytical Results

This section studies analytical results using an arbitrary channel matrix. Fig. 3.1 depicts the computed ergodic capacity of the proposed approximations (3.27) and (3.29) (2nd and 4th orders) and its integral-based expression, described in [YJ08b]. The results are obtained by generating matrices following the Rayleigh distribution and averaging the result. This picture shows how the approximation of the integral-based expression evolves. This is described by Taylor's Theorem [GS12] and is validated in the picture. As the order is increased, the error between the approximation and the integral-based expression decreases notably.

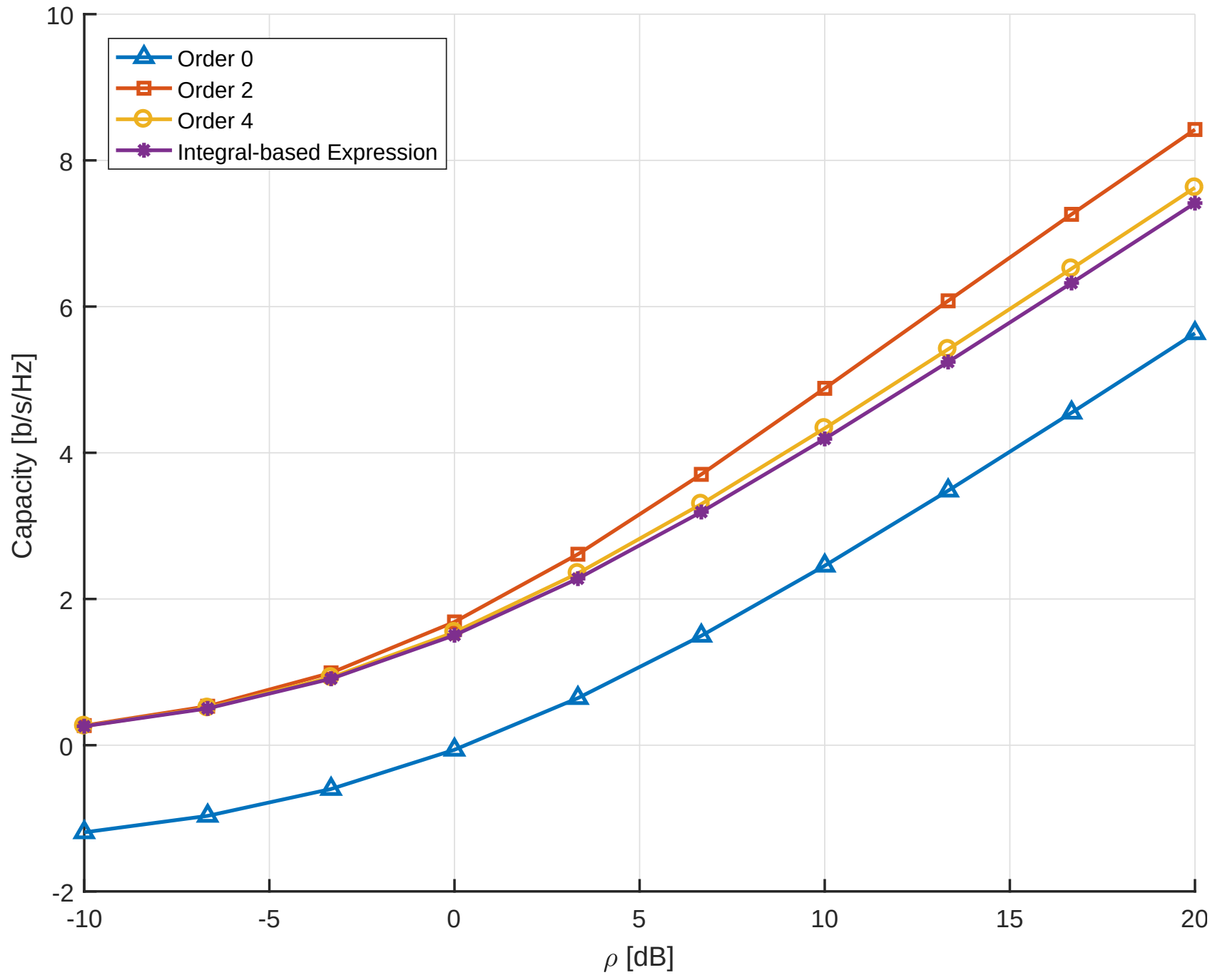

FIGURE 3.1: Average instantaneous capacity approximations for different orders compared to the integral-based expression for $N_t = 2$ and $N_r = 2$.

Fig. 3.2 illustrates the normalised error of each approximations relative to the integral-based expression. Note that this is defined as a function of the approximation's order as

$$E(o) = \frac{\left|\sum_n C_{\text{Order=o}}(n) - C_{\text{Integral}}(n)\right|^2}{\left|\sum_n C_{\text{Integral}}(n)\right|^2}. \tag{3.42}$$

With this figure, first we show that the normalised error tends to zero as we increase the order. Second, the figure validates theorems (1) and (2), which state that the error tends to zero for low SNR and to a constant for high SNR, respectively.

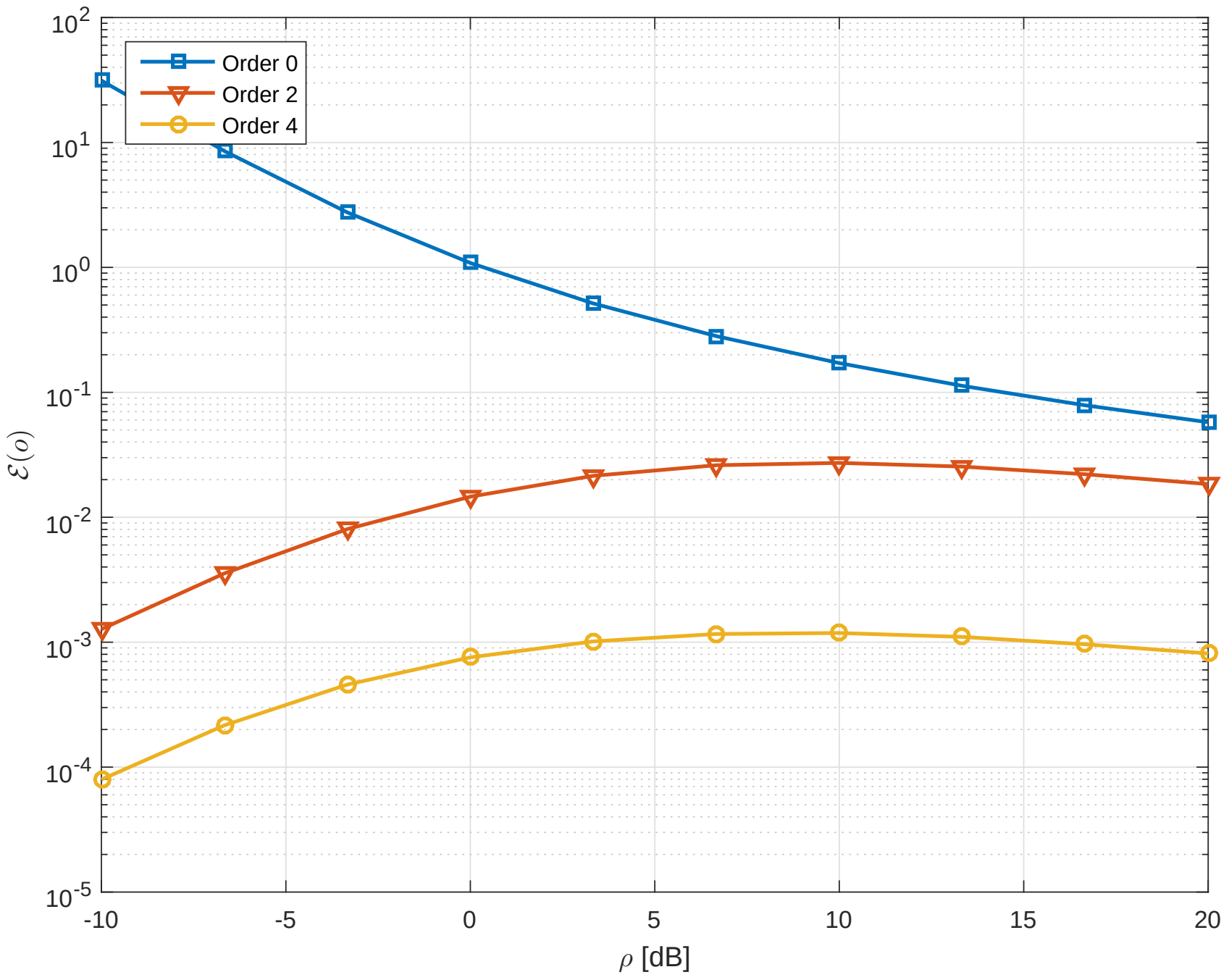

FIGURE 3.2: Normalized error of the different approximation orders for $N_t = 2$ and $N_r = 2$.

Fig. 3.3 shows the second and fourth order approximations and the upper and lower bounds described in [RHH14]. With this figure, we demonstrate that the proposed approximations are placed between both bounds. The MIMO capacity in absence of CSI is also depicted, which is described in (3.31). This picture reflects the effect of the rank of the channel, $L$. For instance, whereas the slope of IM is $2/5$, the slope of MIMO is $4/5$, increased by $L = 2$ with respect to IM.

Finally, a last analysis is performed in terms of computational complexity. This analysis is particularly interesting due to the a-priori higher precision of the integral-based expression. However, its computational complexity is much higher compared with the proposed approximations. Table 3.2 describes the average computational complexity in terms time consumption and its relative increment with respect to the Order $0$ approximation. As shown in the table, the proposed approximations can reduce time consumption by more than $4$ times with respect to the integral-based expression. Therefore, the proposed approximations represent a fair trade-off between precision and computational complexity.

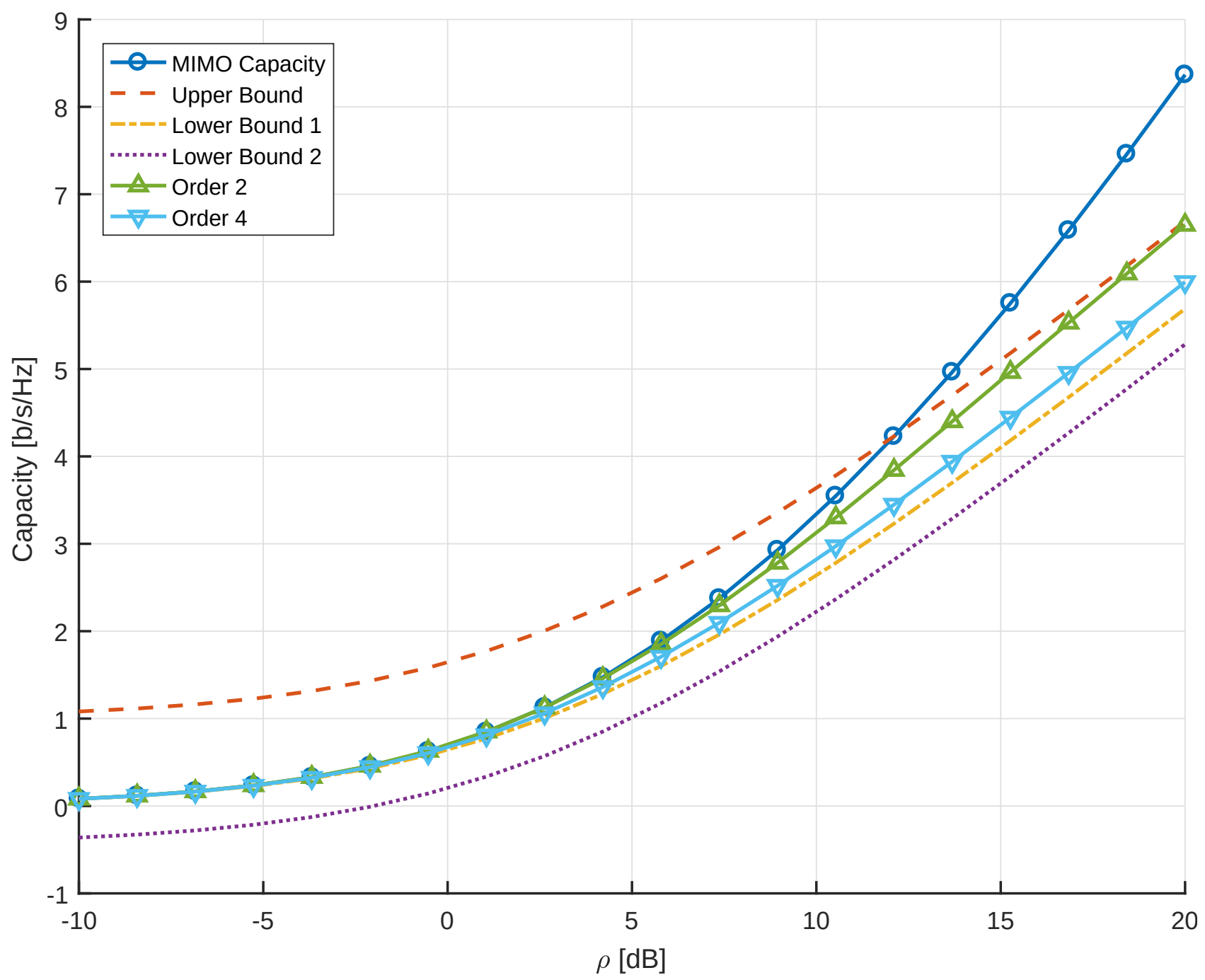

FIGURE 3.3: Approximations of the capacity for different orders compared to the upper and lower bounds described in [RHH14].

### 3.4.2 Applications of Index Modulations

In this section we discuss the applicability of IM to different domains: spatial, polarization and frequency. We note that, in all three cases, the capacity in bits per channel use is the one shown (i.e., spectral efficiency). Therefore, the impact of more bandwidth in FMod does not come up in the study that follows.

TABLE 3.2: Computational Complexity

| | **Average Time Consumption** [$\mu s$] | **Relative increment** [%] | **Precision (MSE)** [$10^{-3}$] |
|---|---|---|---|
| **Order** 0 | 27.758 | – | 26 |
| **Order** 2 | 31.466 | 13.36 | 5.5 |
| **Order** 4 | 38.802 | 26.43 | 0.3 |
| **Integral-based expression** | 151.115 | 404.62 | – |

### Spatial Modulation

SMod consists in applying IM to the spatial domain. Using several antennas at transmission, the transmitter can modulate additional information deciding which antenna uses for transmission. Assuming that the channels are uncorrelated, the receiver can obtain the additional information by detecting which antenna is being used at transmission. This scheme is specially interesting where transmitters are equipped with many antennas, such as Long Term Evolution (LTE) or Wi-Fi (IEEE 802.11n and future releases).

To evaluate the capacity of SMod under realistic scenarios we employ the channel model described by 3GPP [101; 104]. The channel profile corresponds to the one described by the Extended Typical Urban model (ETU), with independent realizations. This implies that consecutive channel realizations are not correlated and, thus, do not depend on the Doppler frequency shift. Spatial channels are uncorrelated if the separation between antennas is greater than $\lambda/2$, which is desirable when SMod is used. However, due to imperfections of the transmitter and receiver, antennas can be correlated in different levels. We use 2 grades defined in the specifications: no correlation and high correlation. Antenna correlation matrices are defined by Table 3.3 and Table 3.4.

Fig. 3.4a and 3.4b depict the capacity of SMod under ETU

TABLE 3.3: Antenna Correlation Matrices

| One antenna | Two antennas | Four antennas |
|---|---|---|
| $R_{TX} = 1$ | $R_{TX} = \begin{pmatrix} 1 & \alpha \\ \alpha^* & 1 \end{pmatrix}$ | $R_{TX} = \begin{pmatrix} 1 & \alpha^{1/9} & \alpha^{4/9} & \alpha \\ \alpha^{*1/9} & 1 & \alpha^{1/9} & \alpha^{4/9} \\ \alpha^{*4/9} & \alpha^{*1/9} & 1 & \alpha^{1/9} \\ \alpha & \alpha^{*4/9} & \alpha^{*1/9} & 1 \end{pmatrix}$ |
| $R_{RX} = 1$ | $R_{RX} = \begin{pmatrix} 1 & \beta \\ \beta^* & 1 \end{pmatrix}$ | $R_{RX} = \begin{pmatrix} 1 & \beta^{1/9} & \beta^{4/9} & \beta \\ \beta^{*1/9} & 1 & \beta^{1/9} & \beta^{4/9} \\ \beta^{*4/9} & \beta^{*1/9} & 1 & \beta^{1/9} \\ \beta & \beta^{*4/9} & \beta^{*1/9} & 1 \end{pmatrix}$ |

TABLE 3.4: Antenna Correlation Parameters

| | $\alpha$ | $\beta$ |
|---|---|---|
| No correlation | 0 | 0 |
| Medium correlation | 0.3 | 0.9 |
| High correlation | 0.9 | 0.9 |

channel conditions for no correlation and high correlation of antennas at transmission and reception, and for different number of antennas. From these pictures, the following observations arise:

1. Increasing the number of antennas at transmission increases the capacity in SMod. For instance, the highest capacity is achieved for the $4 \times 4$ mode.

2. As expected, the capacity of SMod decreases when antenna correlation is introduced. SMod exploit spatial diversity inherently by hopping between spatial channels. If antennas are correlated, spatial channels are also correlated, diversity is not fully exploited and the capacity decreases.

3. The presence of antenna correlation may underperform other modes. For instance $2 \times 4$ underperforms $2 \times 2$ in the presence of high correlation.

4. As expected, $1 \times 2$ obtains the lower capacity.

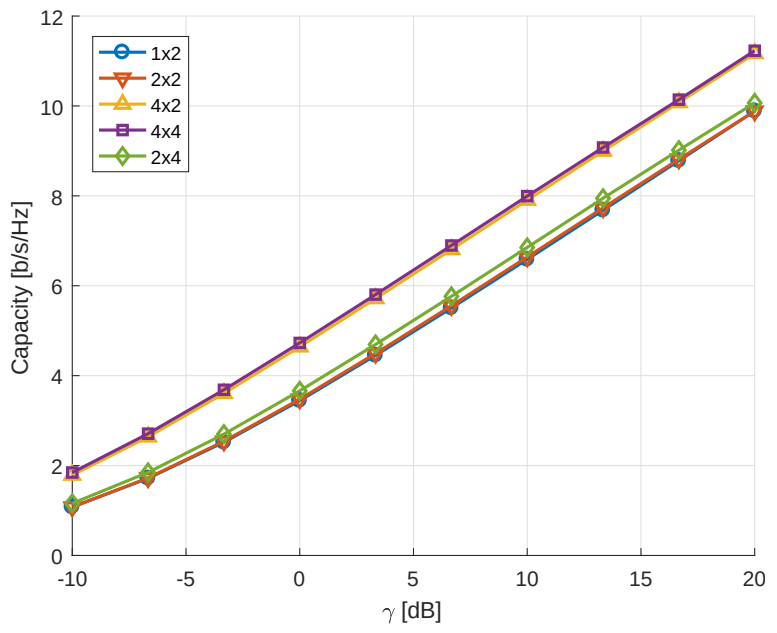

(A) Low Spatial Correlation

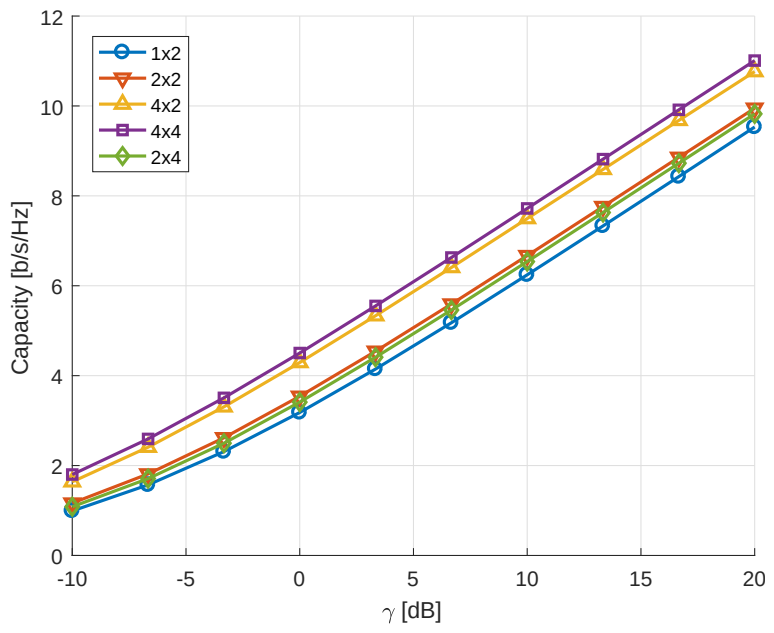

(B) High Spatial Correlation

FIGURE 3.4: Capacity evaluation for different antennas at transmission and reception applying IM to the spatial domain of LTE ETU channel for low and high antenna correlations.

### Polarized Modulation

In contrast to the previous section, polarization domain is not widely used in mobile radio communications. Mobile terminals are handed in different ways with different physical orientations, without respecting the polarization direction. Nevertheless, it is still possible to employ the polarization domain with fixed terminals, such as those generally used in satellite services. Moreover, in satellite communications it is not possible to exploit the spatial diversity due to the correlation between spatial paths. Hence, in these scenarios the polarization dimension takes an important relevance and becomes more challenging.

We use the channel model proposed in [SGL06], which describes different scenarios for land mobile satellite communications. It incorporates parameters such as correlation between rays, direct, specular, and diffuse rays; as well as cross-talk between inputs and other features. By tuning these parameters, different scenarios such as urban, suburban or maritime environments can be modeled. Fig. 3.5 depicts the IM capacity in different scenarios using the polarization domain. With this picture, we aim to employ the 2nd order approximation to compare different satellite channels in terms of capacity. Thanks to it, we are able to classify which environmental conditions are more suitable for IM.

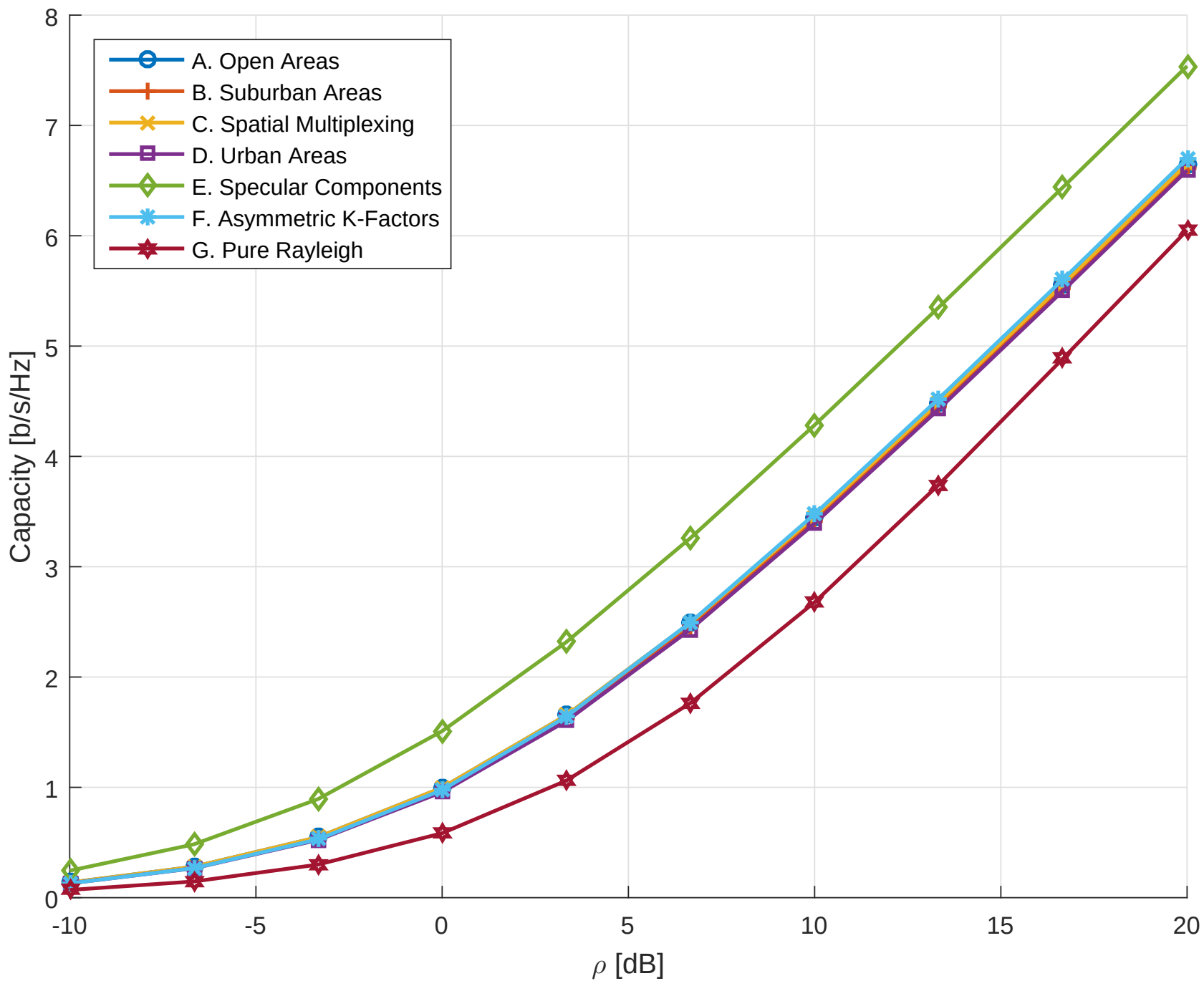

FIGURE 3.5: Capacity evaluation for different scenarios applying IM to the polarization domain of Land Mobile Satellite channel.

Specifically, we consider PMod activating only vertical or horizontal polarization in each hop. We use typical parameters such as a sampling frequency of $F_s = 33.6$ kHz, carrier centred at L-band and a mobility of 5 m/s. Whilst scenarios such as open areas, suburban areas, spatial multiplexing, urban areas, and Rice channels with asymmetric $K$-factors attain the same capacity, scenarios with specular components increase the capacity by an additional 1 b/s/Hz with respect to the others. These scenarios achieve a better performance, as specular components can be added to the direct ray. Note that, in general, specular components are present in scenarios where there is a strong reflection, such as, for instance, the maritime scenario, due to the strong reflection of the sea.

#### Frequency Index Modulation

In the frequency dimension, the index modulation is achieved by hopping between available subcarriers. On one hand, flat fading channels imply that all subcarriers are affected by the same channel magnitude and phase and therefore the receiver has to estimate which subcarrier is used by the transmitter. This approach requires high frequency isolation and power budget. These channels are typical in scenarios where there is a strong LoS component. Note that FMod complements FSK, where the information is placed only in the shifts, but differs from Frequency Hopping (FH). In the latter case, no information is placed in the hops and its objective is to exploit frequency diversity and increase security at physical level.

On the other hand, frequency selective channels generate rich frequency diversity since the subcarriers are affected by different channel magnitudes and phases. In this case, the frequency isolation is not critical as previously since the receiver can exploit the CSI to estimate the used subcarrier more accurately. These channels are typical in scenarios with multipath.

In order to exploit frequency selectivity property, frequency hops cannot be adjacent. Intuitively we could think that the more separated subcarriers are, the better capacity the system will achieve. But this is not true. Frequency selective channels present frequency fading randomly at different subcarriers. Fig. 3.6 depicts an example of snapshot of ETU channel. It can be appreciated that choosing too separated subcarriers may not be the best strategy.

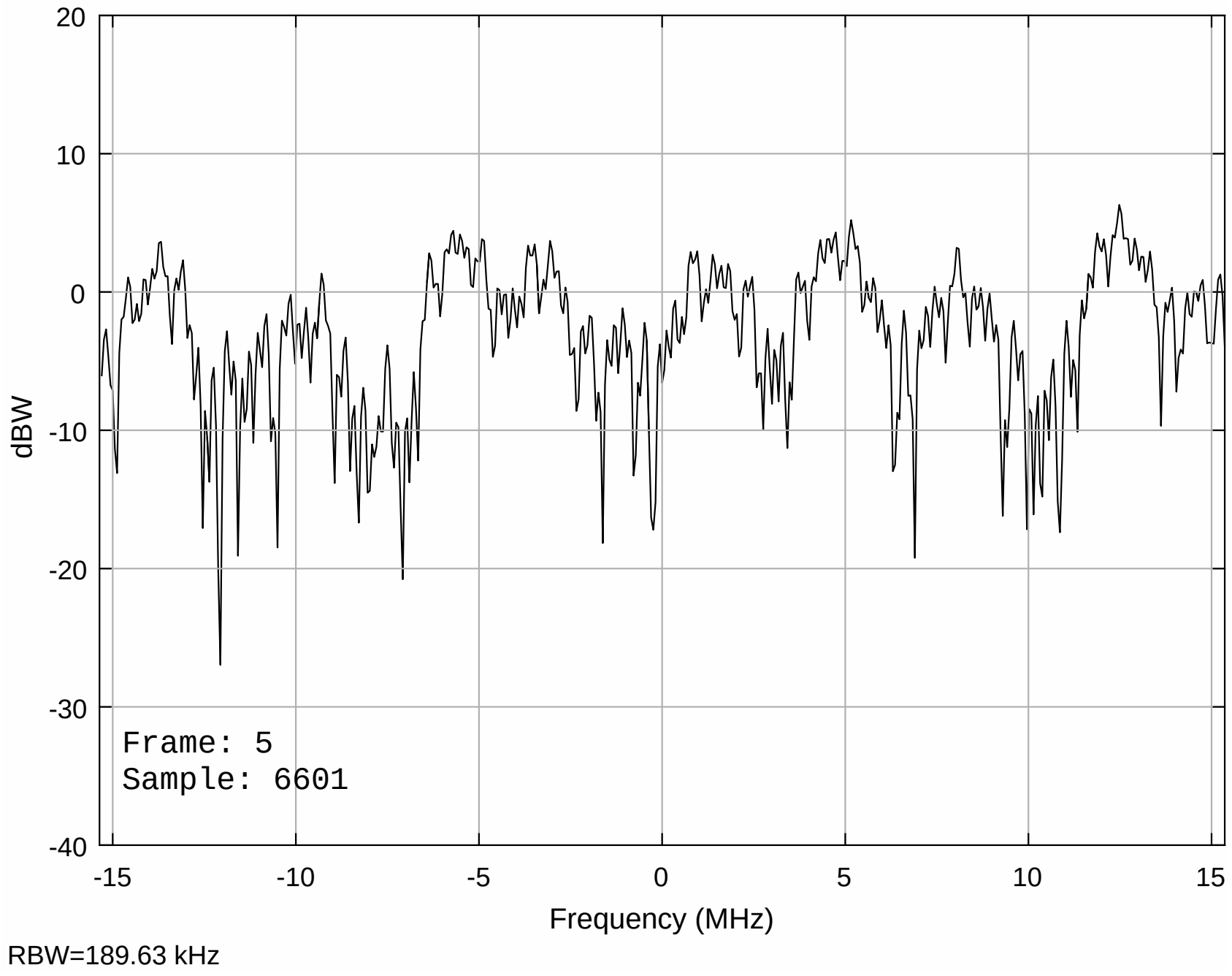

FIGURE 3.6: Snapshot of the spectrum of the ETU channel model.

Fig. 3.7 depicts the capacity of FMod for different separations, in Resource Blocks units (1 RB = 180 kHz). Clearly, 1 RB of separation achieves the lowest capacity. However, the maximum separation (99 RB) does not achieve the highest capacity. In this case, the maximum capacity is achieved when the separation is 40 RB.

An additional important aspect is that, in contrast to SMod or PMod, the performance in the capacity is the same for low SNR regime, regardless the separation of frequency subcarriers. This means that in low SNR regime, the separation of subcarriers is not relevant and does not affect the performance. Also, FMod occupies more bandwidth when number of hops is increased.

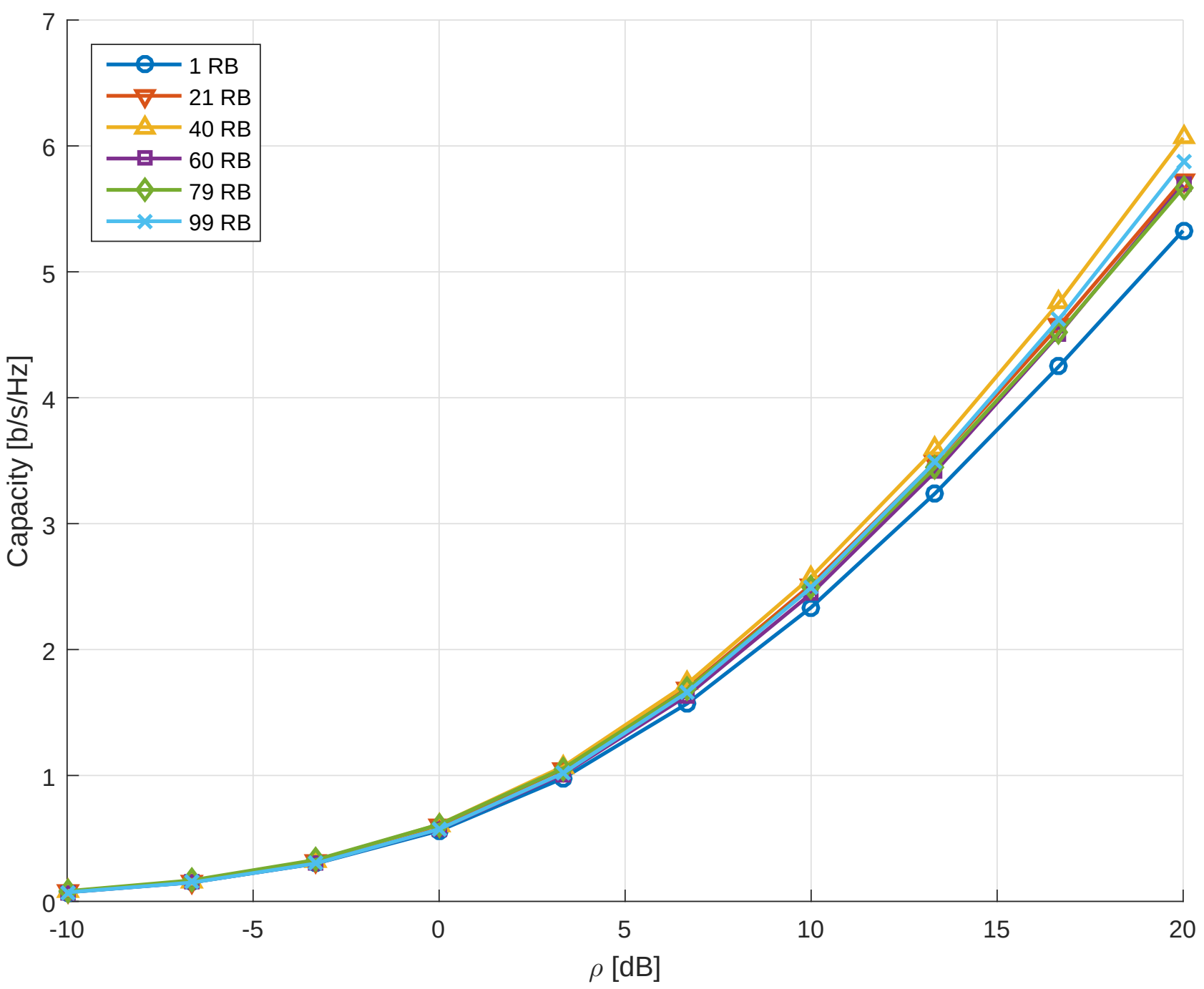

FIGURE 3.7: Capacity evaluation for several frequency separations applying IM to the frequency domain of LTE ETU channel.

## 3.5 Conclusions

In this chapter we present a closed-form expression of the IM capacity, (3.25), as well as two closed-forms of its 2nd and 4th order approximations, which are (3.27) and (3.29), respectively. These expressions are valid for different channel distributions, and provide an approximation to the integral-based expression. We analytically demonstrate that the expectation of the error of the 2nd and 4th order approximations tends to zero for low SNR and is $o(\mathrm{SNR})$. This fact is illustrated with several simulations. We also compute the ergodic capacity for Rayleigh, Rice, and Nakagami-$m$ channels based on its 2nd order approximation, summarized in Table 3.1. These expressions allow to find the ergodic channel capacity without computing the instantaneous capacity over many channel realizations. Finally, we apply the capacity analysis of IM to three physical properties: spatial, polarization and frequency. With SMod, the number of antennas at transmitter and receiver increases the capacity, as well as the correlation between antennas; with PMod, the maximum capacity is achieved when the channel contains specular components; with FMod, the separation between subcarriers affects directly the capacity of the system only in medium and high SNR regimes.

# Chapter 4

# Polarized Modulation

One never notices what has
been done; one can only see
what remains to be done.

M. Curie

In the previous chapter, we studied the Capacity metric of Index Modulations, where a single hop is used in each channel access. The primary motivation of this chapter is to focus Index Modulations to dual polarized communications and contextualize this in the challenging scenario of mobile satellite channels. Hereinafter, this scheme is referred to as Polarized Modulation (PMod). Indeed, it is not polarization multiplexing since only one polarization is activated at a time and, therefore, precludes the presence of interference. Although this work has been conceived as an attempt to apply a simple diversity technique to the satellite scenario, the chapter could also be seen as the extension to satellite communications of the PMod idea that has been reported previously in optical communications [KA09]. However, to the best of our knowledge, there is no literature describing the PMod demodulation scheme (detection and decoding) for optical communications in detail, being polarization multiplexing far more common.

The proposed scheme provides the following contributions:

- The proposed PMod technique exploits the polarization diversity in satellite scenarios, where the spatial diversity is highly penalized.
- PMod does not require CSI and increases the throughput maintaining the robustness based on the polarization diversity.
- Satellite systems that operate with dual polarized antennas are able to employ PMod with the minimum requirement of using a dual polarized feeder.
- The success of this scheme lies not only on the simplicity of the transmission technique, but also on the receiver design, which is also one main contribution of the present work, together with the performance evaluation. Note that the information is conveyed not only in the transmitted bit stream, but also in the polarization.
- Finally, as we demonstrate in a maritime mobile satellite L-band scenario, the result is an increase of the overall performance in terms of throughput whilst guaranteeing a minimum QoS and requiring a minimum increase in power usage. The best performance is obtained for low order modulations, where the proposed method achieves a gain of 2 when it is compared with a basic system without PMod.

## 4.1 System Model

Let us consider a MIMO system where transmitter and receiver are equipped with a single antenna with dual polarization, and a Rice frequency flat channel. Each symbol contains $b + 1$ bits of information, where $b$ bits are mapped within the constellation $\mathcal{S}$ and the remaining bit is used for polarization selection, among the two possible orthogonal polarizations. This remaining bit is denoted as $l \in \{0, 1\}$ and the modulated bits as $s \in \mathbb{C}$. It must be noted that the information is conveyed through the symbol $s$ as well as bit $l$. For the sake of clarity, we identify the polarizations as polarizations 0 and 1. The channels across the polarizations 0 and 1 are denoted $h_{00} \in \mathbb{C}$ and $h_{11} \in \mathbb{C}$, respectively, and their respective cross-channels as $h_{10} \in \mathbb{C}$ and $h_{01} \in \mathbb{C}$. Note that polarizations 0 and 1 are orthogonal and can be either vertical/horizontal, RHCP/LHCP, slant $\pm 45°$, etc. All channel coefficients $h_{ij}$ follow a Rice statistical distribution with ($K$,$\sigma_h$)[1] parameters. The received signals for polarizations 0 and 1 are denoted as $y_0 \in \mathbb{C}$ and $y_1 \in \mathbb{C}$, respectively, and $\mathbf{w} \in \mathbb{C}^2$ follows Additive White Gaussian Noise (AWGN), $\mathbf{w} \sim \mathcal{CN}(\mathbf{0}, \mathbf{I}_2)$.

Depending on the value of the bit $l$, $s$ symbol is conveyed using one polarization or the other. Hence, we can formulate the system model as:

$$\begin{aligned} \mathbf{y} &= \sqrt{\gamma}\mathbf{h}_l s + \mathbf{w} \\ &= \sqrt{\gamma}\mathbf{H}\mathbf{l} s + \mathbf{w}, \end{aligned} \tag{4.1}$$

where $\gamma$ is the signal to noise ratio (SNR), $\mathbf{h}_l = \begin{pmatrix} h_{0l} & h_{1l} \end{pmatrix}^T$ is the channel corresponding to the $l$th polarization, $\mathbf{H} = \begin{pmatrix} \mathbf{h}_0 & \mathbf{h}_1 \end{pmatrix}^T$ is the MIMO channel and $\mathbf{l} = \begin{pmatrix} 1-l & l \end{pmatrix}^T$.

Since this scheme adds an additional bit to the transmission by keeping the same power budget, the achievable gain of PMod with respect to the conventional Single-input Multiple-output (SIMO) case is

$$G = \frac{b+1}{b} = 1 + \frac{1}{b}. \tag{4.2}$$

For higher order modulations, (4.2) is asymptotically 1 and thus the proposed PMod scheme increases the gain for low order modulations. For instance, the gain is 2 for Binary Phase-Shift Keying (BPSK) modulation or 1.5 for Quadrature Phase-Shift Keying (QPSK) modulation. Since low order modulations are used in low SNR regime, it is clear that PMod increases significantly the throughput gain $G$ in low SNR systems. This is exactly the scenario for mobile satellite communications where shadowing, fading and power limitations cause low SNR.

[1] Note that Rayleigh distribution can be obtained by imposing $K = 1$.

## 4.2 Demodulation Schemes

In this section we study different classes of receivers. First, we analyze the optimal receiver, where a joint decoding is performed. Later, due to the high computational complexity, we study different suboptimal approaches, with different types of optimizations.

### 4.2.1 Optimal Receiver

Assuming AWG noise and equiprobable symbols, the optimal receiver is the Maximum Likelihood (ML) receiver, whose mathematical expression is characterized by

$$\hat{\mathbf{x}} = \arg\min_{\mathbf{x}\in\mathcal{X}} \|\mathbf{y} - \sqrt{\gamma}\mathbf{H}\mathbf{x}\| \,, \tag{4.3}$$

where $\mathbf{x} = \mathbf{l}s$ and $\mathcal{X}$ is the joint constellation from the Cartesian product of variables $l$ and $s$. Note that there is no restriction on the characteristics of the channel matrix. Thus, we make no assumptions on the statistical independence of $\mathbf{H}$.

One of the particularities of PMod (and Index Modulations) is the property that the index $l$ is decoupled from the modulated symbol $s$. Hence, (4.3) can be simplified as

$$\left(\hat{l}, \hat{s}\right) = \arg\min_{l,s} \|\mathbf{y} - \sqrt{\gamma}\mathbf{h}_l s\| \tag{4.4}$$

This expression performs a joint estimation of the index $l$ and the transmitted symbol $s$. The computational complexity of this receiver is $o(S^2)$, where $S$ is the number of symbols in the constellation $\mathcal{S}$.

### 4.2.2 Suboptimal Receivers

Due to the lack of implementability of the optimal receiver, we explore different suboptimal receivers, with much less computational complexity. As we mentioned, $l$ and $s$ are conveyed independently.

The implementation of the receiver derives into several approaches depending on the scenario constraints. Since PMod transmits a single stream, we aim to extract this stream to be processed into a SIMO decoder. This scheme offers two main advantages:

- Reduces the complexity drastically since the signal processing is one dimensional.
- Can be combined with existing SIMO decoders, maintaining the compatibility with the current standards.

The reception scheme is illustrated in Fig. 4.1.

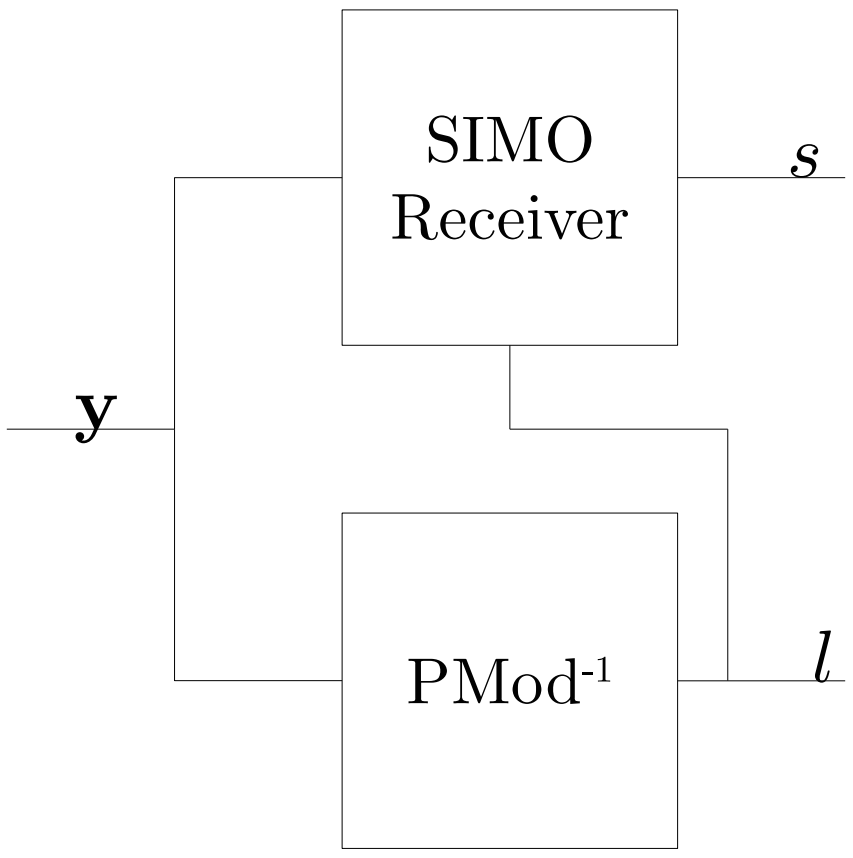

FIGURE 4.1: Reception scheme. $\text{PMod}^{-1}$ applies one of the following demodulation schemes to estimate the bit $l$.

SIMO receivers are widely described in the literature. Although this is beyond the scope of this dissertation, we describe the most used ones:

- Selection Diversity (SD): only a single antenna is selected at reception, whose gain is maximum. This is the simplest

receiver and has the lowest computational complexity. Additionally, it is the less costly receiver, since it only contains one single RF chain.

- Equal Gain Combining (EGC): the receiver compensates only the phase of the channel and preserves the magnitude. Assuming a-priori knowledge of the channel impulse response, the equalized signal $r$ is described by

$$r = y_0 e^{-j\theta_{0l}} + y_1 e^{-j\theta_{1l}} = \sqrt{\gamma}\left(|h_{0l}|+|h_{1l}|\right) s + \tilde{w}, \tag{4.5}$$

where $\tilde{w} = w_0 e^{-j\theta_{0l}} + w_1 e^{-j\theta_{1l}}$. This scheme exploits the gains of the channel to enhance the signal.

- Maximum Ratio Combining (MRC): this receiver weights the signal in each component by the channel component. Hence,

$$r = \frac{\mathbf{h}_l^H \mathbf{y}}{\|\mathbf{h}_l\|^2} = \sqrt{\gamma} s + \tilde{w}, \tag{4.6}$$

where $\tilde{w} = \frac{\mathbf{h}_l^H \mathbf{w}}{\|\mathbf{h}_l\|^2}$

Note that, in all schemes, the SIMO receiver need a-priori knowledge of the $l$ index before the detection of $s$.

### Linear Receiver

The first linear receiver under consideration is obtained by finding the x that minimizes the ML equation, i.e.,

$$\begin{aligned} &\frac{\mathrm{d}f}{\mathrm{d}\mathbf{x}^H} = 0 \\ &f(\mathbf{x}) = \|\mathbf{y} - \sqrt{\gamma}\mathbf{H}\mathbf{x}\|^2. \end{aligned} \tag{4.7}$$

The solution to this problem is

$$\begin{aligned} \hat{\mathbf{x}} &= \left(\sqrt{\gamma}\mathbf{H}^H\mathbf{H}\right)^{-1}\mathbf{H}^H\mathbf{y} \\ &= \mathbf{A}_{\text{ZF}}\mathbf{y} \end{aligned} \tag{4.8}$$

and is known as Zero Forcing (ZF).

We recall that $\hat{\mathbf{x}} = \hat{\mathbf{l}}\hat{s}$. Hence, in the absence of noise, $\hat{\mathbf{x}} = \begin{pmatrix}\hat{s} & 0\end{pmatrix}$ if $l = 0$, or $\hat{\mathbf{x}} = \begin{pmatrix}0 & \hat{s}\end{pmatrix}$ if $l = 1$. Therefore, to decide on $l$ we propose a power detector, described as

$$\hat{l} = \arg\max_{l} |\hat{x}_l|^2 , \tag{4.9}$$

where $x_l$ is the $l$th component of $\mathbf{x}$.

However, this receiver presents a major disadvantage. If $\mathbf{H}^H\mathbf{H}$ is badly conditioned, it may produce an excessive noise enhancement, making the demodulation impossible.

Another well known receiver is the Minimum Mean Square Error (MMSE). The MMSE filter is a post-processing filter $\mathbf{A}_{\text{MMSE}}$ that minimizes the expectation of the square error. If the error is defined as $\mathbf{e} = \hat{\mathbf{x}} - \mathbf{x}$, where $\hat{\mathbf{x}} = \mathbf{A}_{\text{MMSE}}\mathbf{y}$, then

$$\begin{aligned} \frac{\mathrm{d}f}{\mathrm{d}\mathbf{A}^H} &= 0 \\ f(\mathbf{x}) &= \mathbb{E}\left\{\|\mathbf{A}_{\text{MMSE}}\mathbf{y} - \mathbf{x}\|^2\right\} , \end{aligned} \tag{4.10}$$

with the solution

$$\mathbf{A}_{\text{MMSE}} = \mathbf{H}^H \left(\sqrt{\gamma}\mathbf{H}\mathbf{H}^H + \mathbf{I}_2\right)^{-1} \tag{4.11}$$

As with ZF, we can estimate the index $l$ by using (4.9) with $\hat{\mathbf{x}}$.

As stated in the previous section, applying the $\mathbf{A}$ filter may introduce important distortions. Since the solution have to lie in the subset $\mathcal{X}$, the solution of this approach is suboptimal.

### Likelihood Ratio with Hard Decision

In the presence of coded information, as it can be seen in [FJM12], soft decoding outperforms the previous ML implementation. Usually, to deal with channel impairments, the transmitted bits are channel coded. The channel decoder computes the metrics based on likelihood ratios of the received signal and is able to estimate the uncoded bits.

Following this approach, the $l$ bit is estimated based on the likelihood ratio of the signals received in each polarization. We define the likelihood ratio defined as

$$\Lambda\left(\mathbf{y}\right) = \frac{P_1}{P_0} = \frac{P\left(l=1|\mathbf{y}\right)}{P\left(l=0|\mathbf{y}\right)} = \frac{\sum\limits_{\tilde{s}\in\mathcal{S}} \exp\left(-\|\mathbf{y}-\sqrt{\gamma}\mathbf{h}_1\tilde{s}\|^2\right)}{\sum\limits_{\tilde{s}\in\mathcal{S}} \exp\left(-\|\mathbf{y}-\sqrt{\gamma}\mathbf{h}_0\tilde{s}\|^2\right)}. \tag{4.12}$$

Hence, $l=1$ if $\Lambda\left(\mathbf{y}\right) > 1$, and $l=0$ if $\Lambda\left(\mathbf{y}\right) < 1$. This decision rule depends only on the sign of $\log\left(\Lambda\left(\mathbf{y}\right)\right)$. Thus,

$$\hat{l} = \frac{1+\mathrm{sign}\left(\log\left(\Lambda\left(\mathbf{y}\right)\right)\right)}{2}. \tag{4.13}$$

However, this scheme introduces a non-linearity with the $\mathrm{sign}(\cdot)$ function.

### Likelihood Ratio with Soft Decision

The three receivers described above perform hard decision for the estimation of bit $l$. However, they can introduce errors if the system conveys coded information as it was mentioned. The soft version of bit $l$ corresponds to the log-likelihood, exactly as the bits $b$. That is

$$\hat{l} = \log\left(\Lambda\left(\mathbf{y}\right)\right). \tag{4.14}$$

Later on, the bit $l$ is soft and can be passed to the soft decoder. The bit $l$ contains the information about the probability of which is the most probable polarization is used for the transmission of the symbol $s$. Since it takes values in $\mathbb{R}$, it is not possible to use the reception scheme described in fig. 4.1. In order to address this, we propose a cascade demodulation scheme. In the first stage we equalize the channel matrix $\mathbf{H}$. Note that, in contrast to the previous receivers, a-priori knowledge of the $l$ index is not needed, since we are equalizing the full channel matrix. This equalizer can be based on ZF or MMSE, for instance. After this stage, the equalized signal $\tilde{\mathbf{y}}$ contains the symbol and a modified version of the noise, i.e.,

$$\begin{aligned}\tilde{\mathbf{y}} &= \mathbf{A}\mathbf{y} \\ &= \mathbf{l}s + \mathbf{A}\mathbf{w},\end{aligned} \tag{4.15}$$

where $\mathbf{A}$ is the equalizer filter. Fig. 4.2 illustrates this concept.

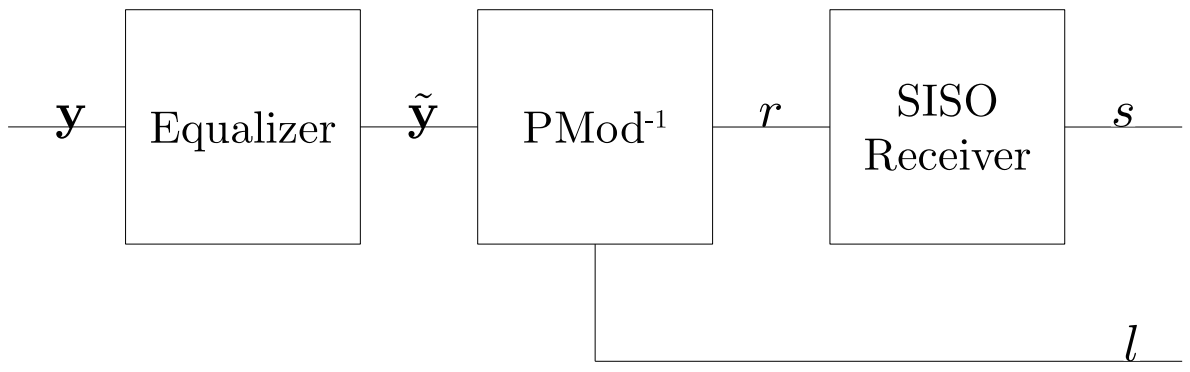

FIGURE 4.2: Reception scheme. $\text{PMod}^{-1}$ applies the approach described in this section.

Although the channel contribution is cleared from the signal at reception, both polarizations contain noise but only one contains the symbol $s$. Instead of choosing one polarization over the other, we weight each polarization component by its probability $P_0$ or $P_1$. Hence, the average received signal is described as

$$r = P_0\tilde{y}_0 + P_1\tilde{y}_1. \tag{4.16}$$

Using the likelihood ratio $\Lambda(\mathbf{y})$ computed as in (4.12), and using

$$P_1 = P(l = 1|\mathbf{y}) = 1 - P(l = 0|\mathbf{y}) = 1 - P_0, \tag{4.17}$$

we can rewrite

$$P_1 = P(l = 1|\mathbf{y}) = \frac{\Lambda(\mathbf{y})}{1 + \Lambda(\mathbf{y})}. \tag{4.18}$$

Therefore, the receiver can recover the signal by weighting the received signals from both polarizations by $P_0 = 1 - P_1$ and $P_1$,

respectively. Hence, the averaged received signal takes the following form:

$$r = (1 - P_1)\,\tilde{y}_0 + P_1 \tilde{y}_1 = \frac{1}{1 + \Lambda(\mathbf{y})} \left(\tilde{y}_0 + \tilde{y}_1 \Lambda(\mathbf{y})\right). \tag{4.19}$$

Finally, the combined signal $r$ is passed to the SISO decoder in order to obtain the symbol $s$. It must be noted that the proposed cascade scheme maintains soft decoding in all signals, which is necessary when channel encoding is employed. Note that we use the likelihood ratio $\Lambda(\mathbf{y})$ and not $\Lambda(\tilde{\mathbf{y}})$ since we avoid channel equalization inaccuracies.

## 4.3 Numerical Results for Uncoded BER

In this section we analyze the results of the proposed schemes. In order to compare them, we deploy a system conveying QPSK symbols in addition to the switching bit $l$. For this purpose, we only examine the uncoded bit error rate (BER). The channel model used corresponds to the Rice maritime mobile channel model described in the experiment V in [SGL06] with a correlation factor of $\rho_{ij}$. All parameters are summarized in Table 4.1.

In all results the following labels are used:

1. *Reference* denotes the reference scenario, i.e. the scenario where single polarization is used.
2. *VBLAST* is the polarization multiplexing V-BLAST coding scheme.
3. *PMod ZF* is the first approach described in Section 4.2.2.
4. *PMod ML* is the second approach described Section 4.2.1.
5. *PMod HD* is the third approach described in Section 4.2.2.
6. *PMod SD* is the fourth approach described in Section 4.2.2.

TABLE 4.1: Scenario Main Parameters

| | |
|---|---|
| **Profile** | Maritime |
| **Channel model** | Rice flat fading |
| **Rice K factor** | 10 |
| **Doppler shift** | 2 Hz |
| **Doppler spectrum** | Jakes |
| **Stream correlation** | $\rho_{ij} = 0.5$ |
| **Path distance** | 35786 km |
| **Path loss** | 187.05 dB |
| **Bandwidth** | 200 kHz |
| **Terminal G/T** | $-12.5$ dB/K |
| **Carrier band** | L (1.59 GHz) |
| **Code rate** | 0.625 |
| **Bitrate** | 40 kbps |

7. *OSTBC* corresponds to the Orthogonal Space-Time Block Codes applied to polarization instead of spatial diversity [Pér+08].

In Fig. 4.3, we compare the BER of the four PMod schemes. The four curves are labelled in the same order that they have been introduced (from the first to the fourth approach).

As stated in Section 4.2.1, the ML solution provides the lowest error rate, immediately followed by the fourth solution. As expected, in the absence of channel coding, the ML receiver becomes the optimal solution. Although we observe that ML uses a reduced search space of order $o\left(S^2\right)$, the computational complexity is sensibly higher with respect to the other solutions.

Next to the curve of ML is the pure soft scheme (i.e., the fourth mechanism). By examining the magnified area we observe that the gap between the ML solution and the pure soft is tight. Hence, we conclude that the fourth demodulation scheme stays very close to the optimal solution.

Finally, the third approach (which does not use the conditional mean of the signal) performs very similarly to PMod SD. Instead,

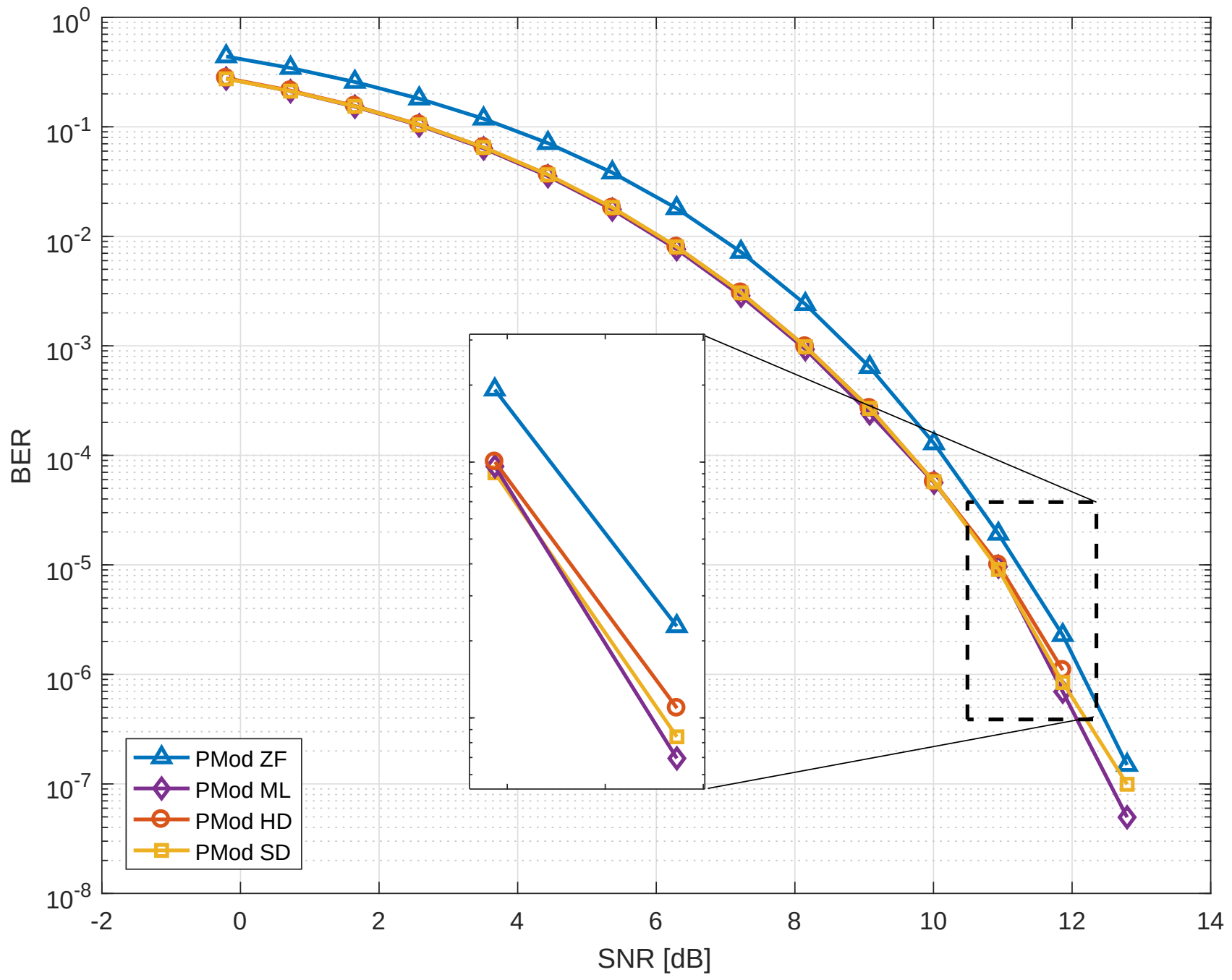

FIGURE 4.3: Comparison of the uncoded BER of the four proposed PMod techniques conveying a QPSK constellation.

the first approach PMod ZF obtains the highest BER.

In the following sections we select the fourth approach (PMod SD) as a benchmark baseline with other schemes different to PMod due to:

- A near-optimal ML solution performance, with a small gap of $0.05$ dB of SNR for a fixed BER of $10^{-6}$.
- A lower computational complexity than ML.

Fig. 4.4 compares the PMod SD solution with the conventional OSTBC, V-BLAST and reference scenario using a QPSK constellation for all schemes. Note that, even though we use the same constellation for all schemes, the total SE is different for each scheme. Therefore, although different schemes with different SE are being compared, the most remarkable observation is that PMod is bounded by OSTBC (lower SE) and V-BLAST (higher SE) and, therefore, PMod achieves a trade-off between OSTBC and V-BLAST in terms of BER and SE.

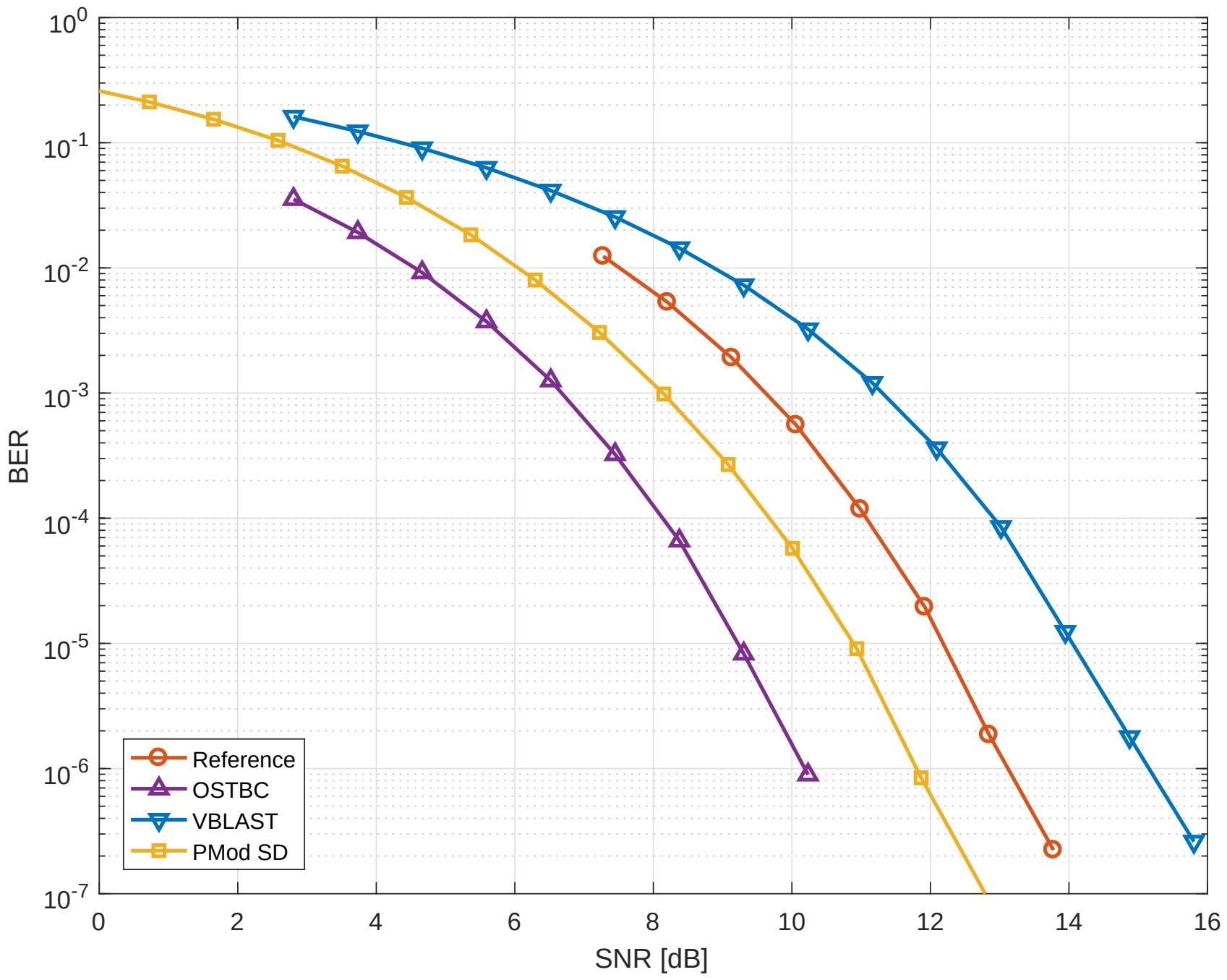

FIGURE 4.4: Comparison of the uncoded BER of the PMod SD with other existing techniques conveying a QPSK constellation.

As expected, OSTBC obtains lowest BER, followed by PMod and V-BLAST. However, OSTBC does not allow to increase the granularity of the adaptive bitrate. In other words, there is no choice to transmit 3 bpcu. The next step is to transmit a 16QAM with OSTBC, which is 4 bpcu use. Newest standards, such as DVB-S2X [DVBb], aim to include new modulation schemes to refine the rate adaptation curve.

### 4.3.1 Equal SE Analysis

In contrast to the previous section, where benchmarking is performed whilst keeping the same constellation, in this section we analyze the performance of PMod compared with the other schemes but constrained to the same SE. In order to do this, we use the following transmission schemes:

- PMod SD with BPSK constellation.
- V-BLAST with BPSK constellation.
- OSTBC with QPSK constellation.
- Reference with QPSK constellation.

In all these schemes, $2$ bpcu are conveyed. Fig. 4.5 describes the curves of the different throughputs, which is defined as

$$T = 64000(1 - \text{SER}), \tag{4.20}$$

where $64000$ is the rate reference of Symbol Error Rate (SER), defined in [ETS].

By observing the figure, it is clear that all curves tend to the same throughput for high SNR.

Fig. 4.6 depicts the BER for the different techniques. In this case, *OSTBC* obtains the lowest BER, followed by *PMod SD*, *Reference* and *VBLAST*, respectively. As expected, OSTBC exploits the full diversity of the channel and is closely followed by PMod. However, one of the main advantages of PMod compared with OSTBC is the ability to increase the granularity of the throughput adaptation. Whilst OSTBC increases the throughput by powers of two, PMod can increase the throughput by small fractions, as seen in (4.2).

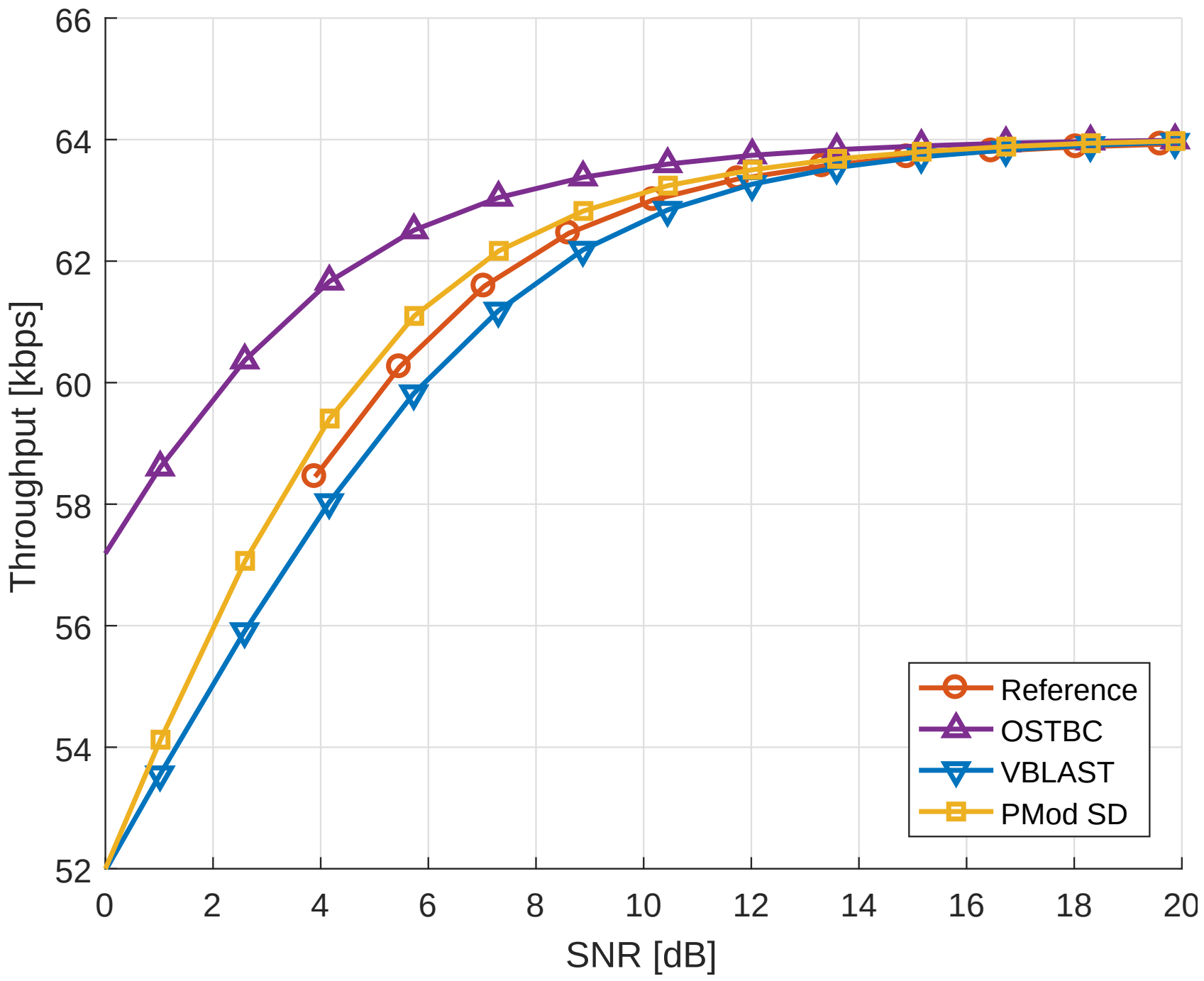

FIGURE 4.5: Comparison of the throughput of the PMod SD with other existing techniques constraint to the same SE.

## 4.4 Results in a Realistic System Context

In this section we describe the implementation of the PMod solution in the context of the Broadband Global Area Network (BGAN) standard. Amongst other procedures, this part of the standard defines the scrambling, turbo coding and mapping stages. In order to offer flexibility in terms of data rate, several bearers and subbearers are detailed. Put briefly, they are different profiles with many combinations of coding rate and constellations. Focusing in the downlink part, the symbol rate is 33.6 ksps and the frame length is 80 ms, where the blocks of coded symbols are not interleaved. In order to simplify the model, QPSK bearers will be used in all simulations. In detail, we employ the bearer F80T1Q4B-L4. Note that BGAN standard contains the specifications of the feedback channel, where the SINR is conveyed to transmitter. This value is used for the adaptive modulation and coding scheme mechanism, despite of PMod does not use its value.

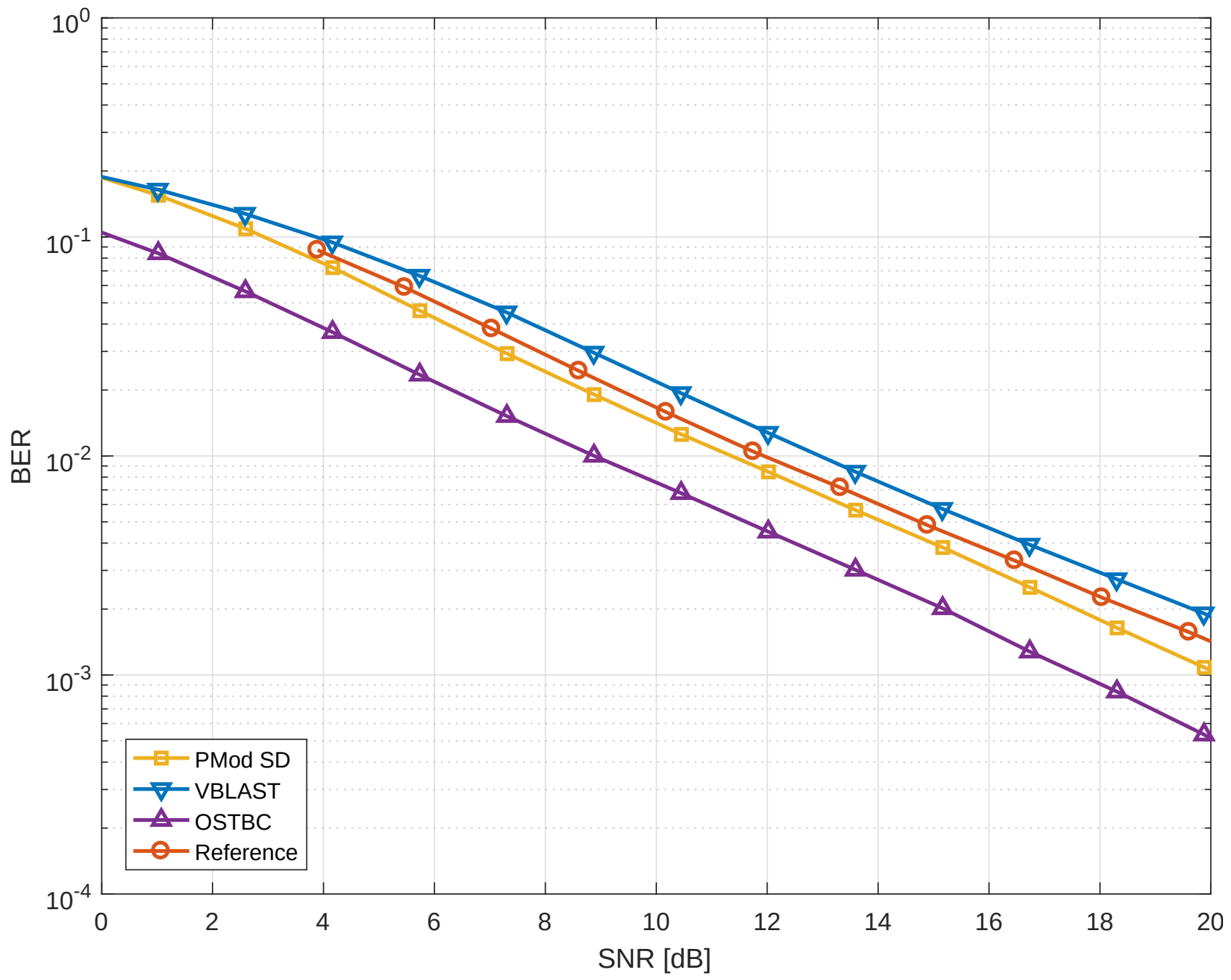

FIGURE 4.6: Comparison of the Uncoded BER of the PMod SD with other existing techniques constraint to the same SE.

### 4.4.1 Next Generation Satellite Communications Simulation Framework

We simulate a L-band geostationary satellite with 7 beams (one desired beam and six interfering beams) and dual polarization. Since these beams are not perfectly orthogonal, we consider six adjacent beams at the same frequency as interferences, as well as the cross polarization couplings. All these values are summarized in Table 4.2 and are obtained via realistic multibeam antenna pattern in the context of the Next Generation Waveforms for Improved Spectral Efficiency (NGW) ESA ARTES 1 project, whose results are summarized in [Hen+13]. In more detail, Fig. 4.7 illustrates the beam pattern, where the working beam is marked with a red circumference and the interfering beams as yellow circumferences. It must be noted that not all beams induce the same levels of interference. Depending on the position of the satellite and the geometry of the reflectors, the interference power varies across beams. In more detail, Fig. 4.8 and Fig. 4.9 illustrate the co-polar and cross-polar coverage for the forward link with contours at 3 dB (blue lines) and 4.5 dB (red lines). One of key aspect is the asymmetry of the co-polar and cross-polar gains in each beam. From these figures, it emerges that gains are different for each beam spot. Finally, Fig. 4.10 shows the block diagram used for the simulations described hereafter.

TABLE 4.2: Data Coupling Polarization Matrix + Interference Matrices

| | Index | Interference matrix (dB) |
|---|---|---|
| Data | 0 | $\begin{pmatrix} 40.8 & -11.6 \\ -11.6 & 40.8 \end{pmatrix}$ |
| Interference | 1 | $\begin{pmatrix} 3.7 & -12.3 \\ -12.3 & 3.7 \end{pmatrix}$ |
| | 2 | $\begin{pmatrix} 8.7 & -13 \\ -13 & 8.7 \end{pmatrix}$ |
| | 3 | $\begin{pmatrix} 3.6 & -6.7 \\ -6.7 & 3.6 \end{pmatrix}$ |
| | 4 | $\begin{pmatrix} 13.4 & -8.9 \\ -8.9 & 13.4 \end{pmatrix}$ |
| | 5 | $\begin{pmatrix} 8.9 & -4.7 \\ -4.7 & 8.9 \end{pmatrix}$ |
| | 6 | $\begin{pmatrix} 11.6 & -3.7 \\ -3.7 & 11.6 \end{pmatrix}$ |

In Fig. 4.10 the identified blocks are:

- Forward Error Correction (FEC) Encoder: encodes the bit

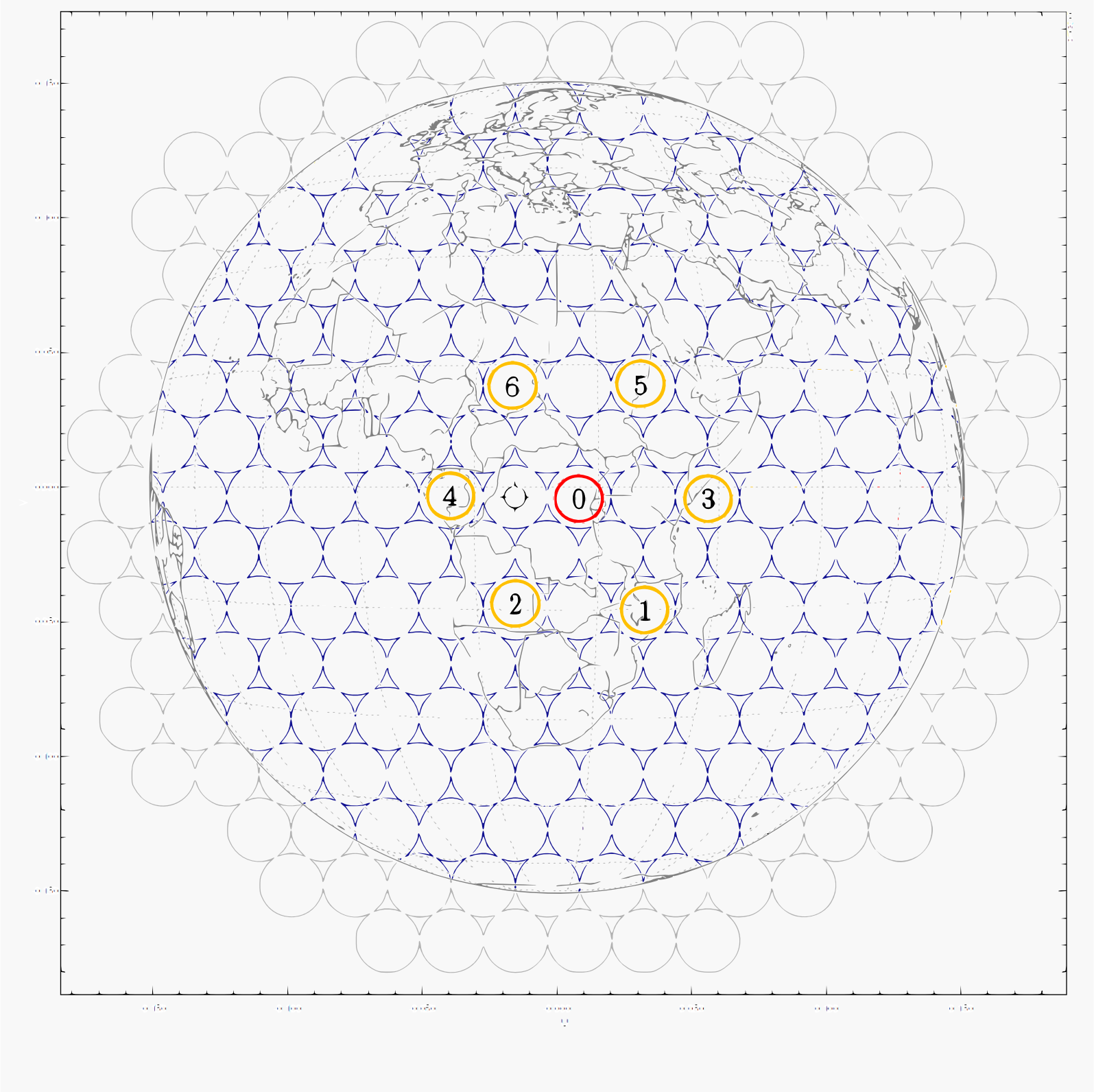

FIGURE 4.7: Considered beam pattern to perform realistic simulations. Working beam is marked with a red circumference and interfering beams as yellow circumferences.

stream using the specifications of [ETS]. Fig. 4.11 describes a detailed view of this block.

- $PMod$: groups the bits in blocks of size $b+1$, maps the bits to symbol $s$ and uses the $l$ bit to select the polarization for each symbol.

- Framing: encapsulates the symbols of each polarization in a frame defined in [ETS]. It inserts pilots for channel estimation, a preamble for synchronization and a header for modulation-code identification.

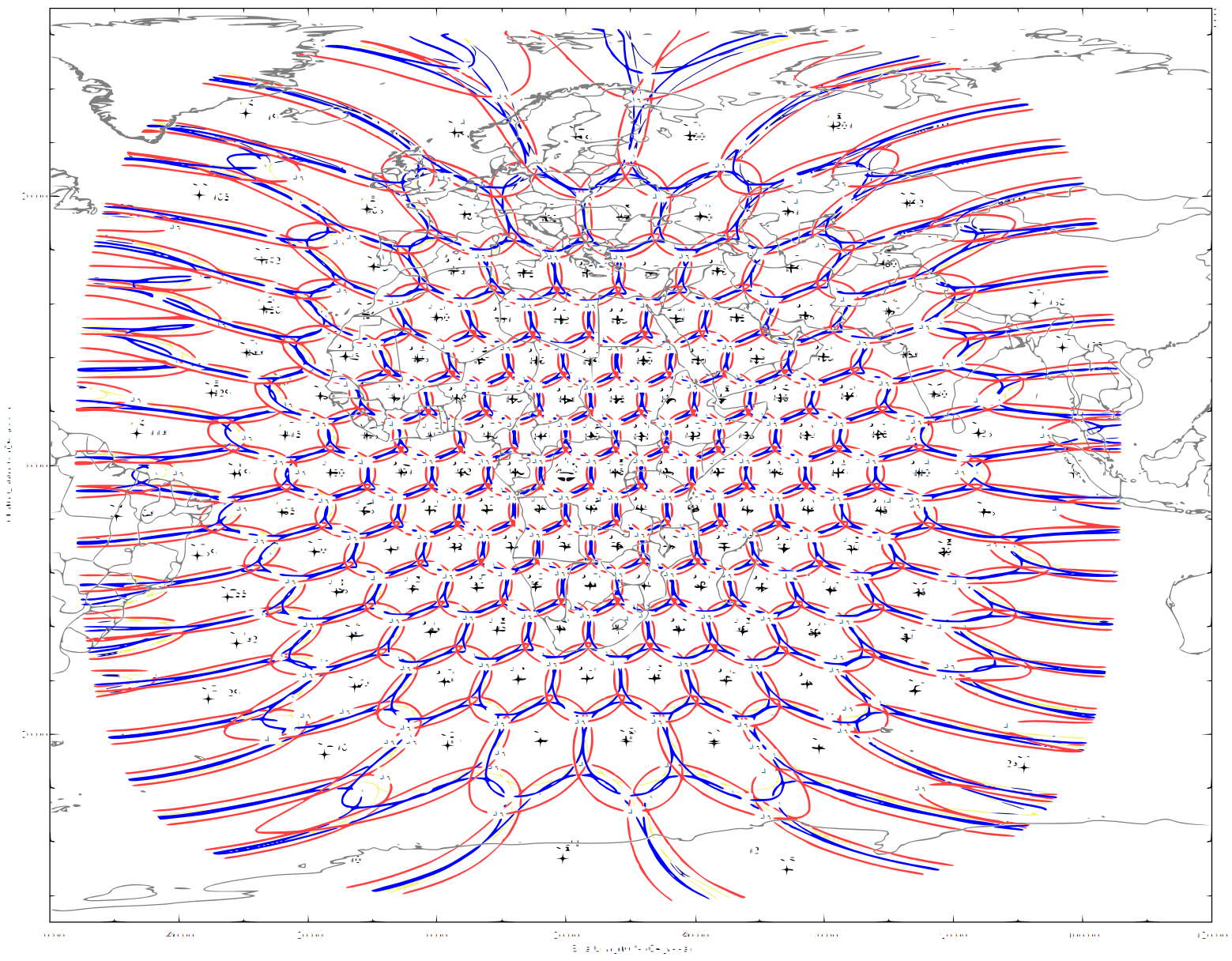

FIGURE 4.8: Co-polar coverage for the forward link with contours at 3 dB (blue lines) and 4.5 dB (red lines).

- Interference matrix $\mathbf{B}_i$: models the cross polarization by a factor defined in Table 4.2. $B_0$ corresponds to the cross-polarized matrix of intended data and $\mathbf{B}_1, \dots, \mathbf{B}_6$ correspond to the cross-polarized matrices of interfering beams.

- $P$: the signal is amplified by a factor of $\sqrt{P}$. It is important to remark that this is possible due to the fact that, for each symbol, only a single polarization is active and thus all power budget $P$ is available, whereas in the case of *VBLAST* and *OSTBC* this factor is divided by 2.

- $L_i$, $i = 0, 1$: equivalent path-loss for each polarization.

- $H_i$, $i = 0, 1$: convolves the signal using the Rice fast fading channel model.

- Noise: adds the AWGN.

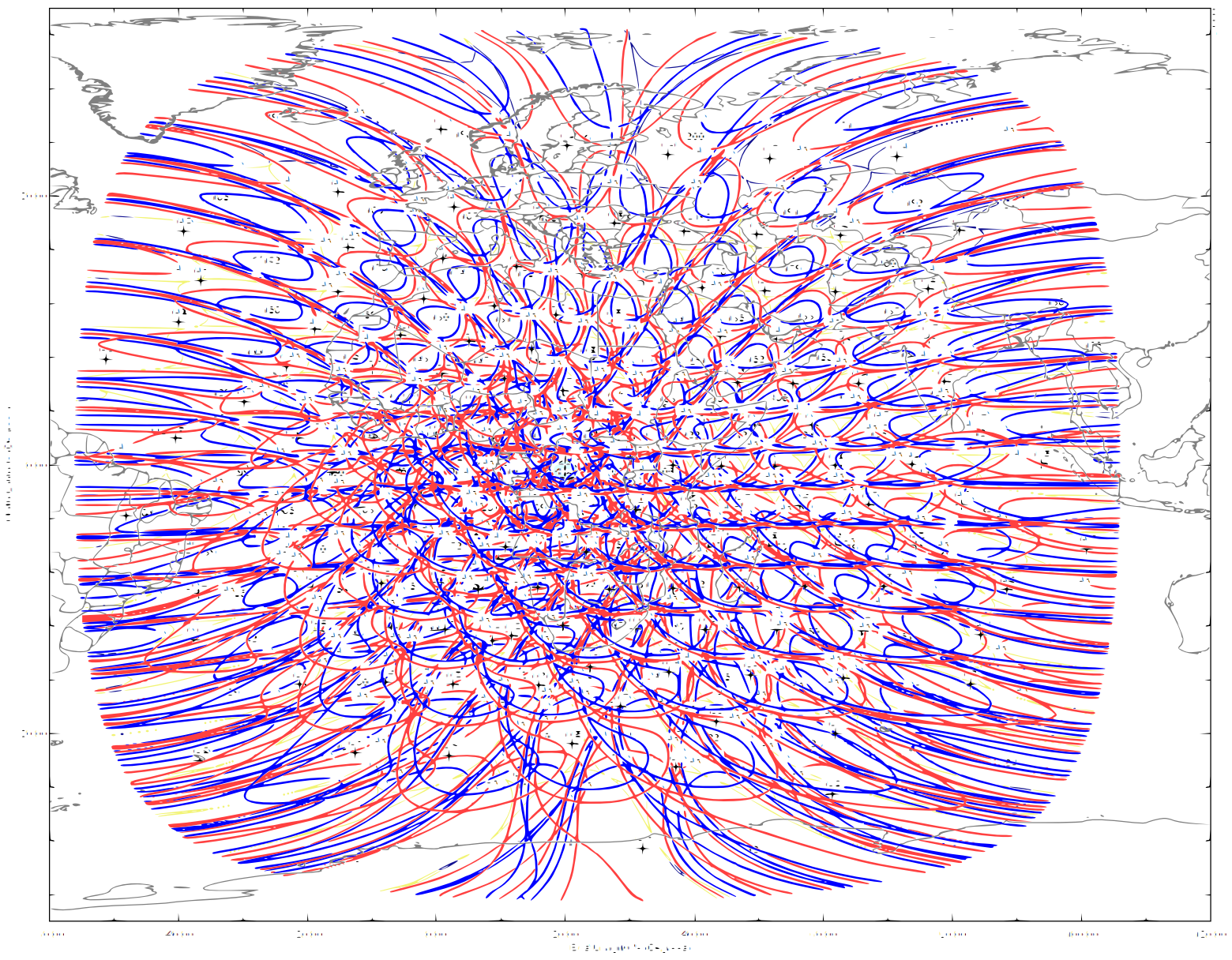

FIGURE 4.9: Cross-polar coverage for the forward link with contours at 3 dB (blue lines) and 4.5 dB (red lines).

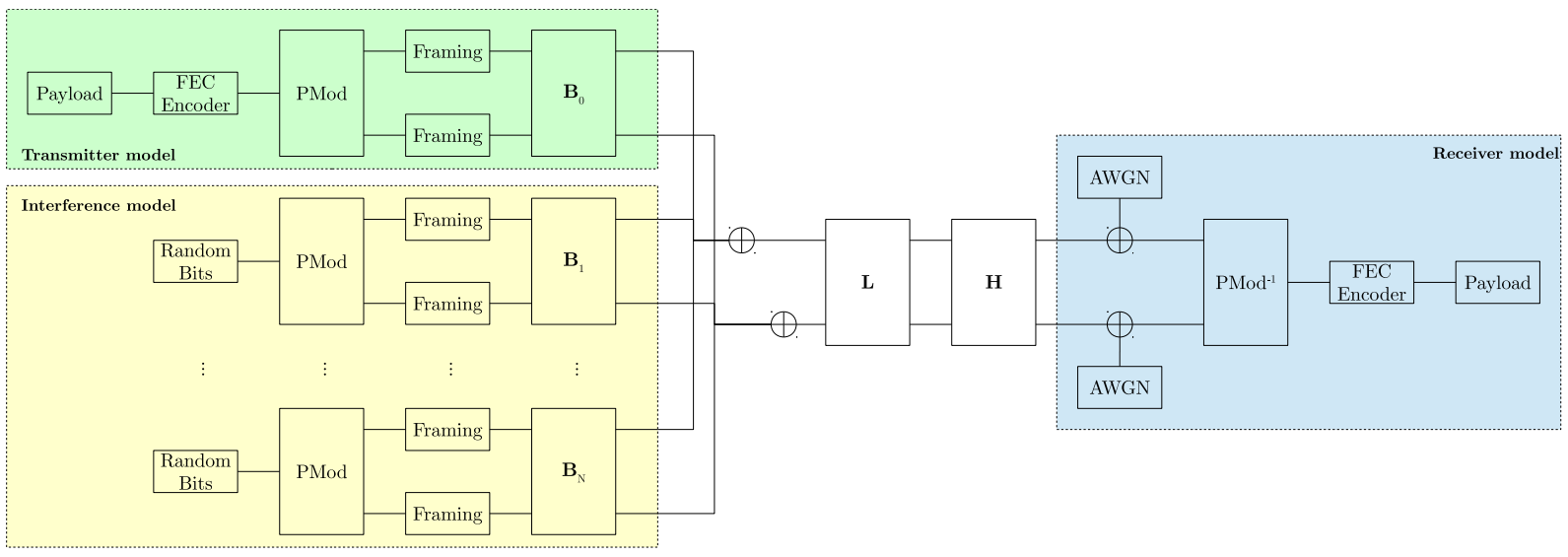

FIGURE 4.10: Block diagram of the simulation framework.

- $PMod^{-1}$: implements one of the schemes.
- FEC Decoder: performs the inverse operation of FEC Encoder. It implements a Turbo Decoder. Fig. 4.12 depicts the detailed view of this block. BCJR implements the algorithm for maximum a posteriori decoding, based on the algorithm discovered by Bahl et al. [Bah+74].

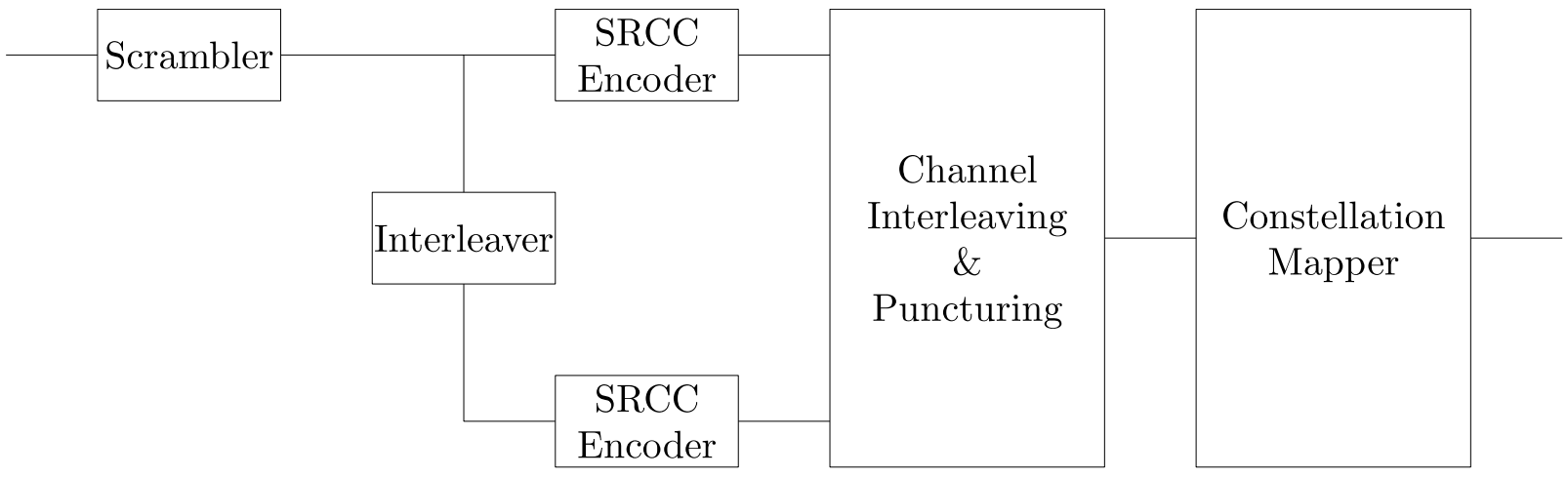

FIGURE 4.11: Detailed view of FEC Encoder.

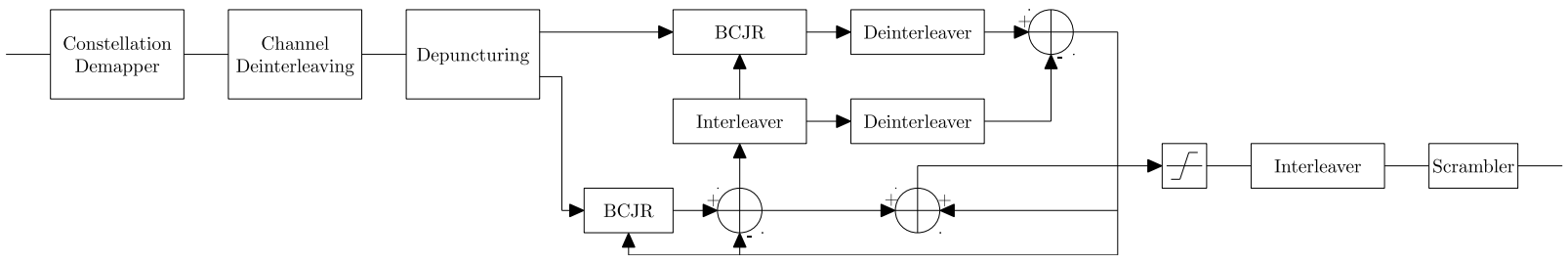

FIGURE 4.12: Detailed view of FEC Decoder.

We consider the Rice maritime mobile channel model described in the experiment V in [SGL06] and the parameters described in Table 4.1.

The aim is to evaluate the basic transmission and reception concepts and schemes. Thus, in this work we assume perfect synchronization at the receiver side, as well as perfect channel estimation. Prior to detecting symbol $s$, one of the four approaches is performed in order to estimate the bit $l$ and to filter the received signal.

We observe that this scenario includes non-Gaussian interference. Thus, since we formulated the PMod solution under this assumption, we need to cope with the interference to minimize it. In order to achieve this, the receiver implements a MMSE linear filter. This configuration mitigates the interferences from the other beams, as well as the other polarization for the detection of symbol $s$.

One important aspect is the Faraday Rotation (FR), which appears in the L-band. This effect is caused by the free electrons in the ionosphere and causes a rotation of the polarization (see Annex C). Since it changes the polarization, FR may be critical in order to estimate the bit $l$. Fortunately, this effect can be reduced using a circular polarization or performing an estimation and assuming that the FR remains invariant during the time slot. An estimation of FR is described in [MN08] and it can be applied using the pilot symbols used by the channel estimation. Nevertheless, for our simulations, we assume that this effect is compensated.

Finally, in the next stage, the demodulated soft bits are conveyed to the turbo decoder and scrambled to obtain the information bits. As we consider interferences in this scenario, we use the signal to interference plus noise ratio (SINR) in the x-axis rather than SNR.

### 4.4.2 Comparing PMod Solutions

We compare the four proposed demodulation schemes. In contrast to Fig. 4.13, although the ML solution is the optimal in absence of channel coding, this is not the case in the presence of coded information. Certainly, the PMod SD scheme produces the lowest BER, followed by PMod HD. Both schemes use soft bits and, thus, their performance is better than the hard solutions (PMod ZF and PMod ML).

In order to benchmark the proposed schemes against the existing ones, we compare the performance in terms of throughput, defined in (4.21). This corresponds to the average rate of successful information delivery and is defined as

$$T = R(1 - \text{PER})G. \tag{4.21}$$

This is equivalent to the bitrate ($R$) of the particular bearer weighted by the probability of no error in the whole block ($1 - \text{PER}$) and the throughput gain ($G$), defined in 4.2. In all simulations, a fixed modulation-code (F80T1Q4B-L4) is simulated with coding rate of $0.625$ ($R = 40$ kbps without frame overhead). PER is obtained by simulations and corresponds to the number of erroneous blocks divided by the total number of blocks. If a single bit in the block is erroneous, the entire block is marked as erroneous block.

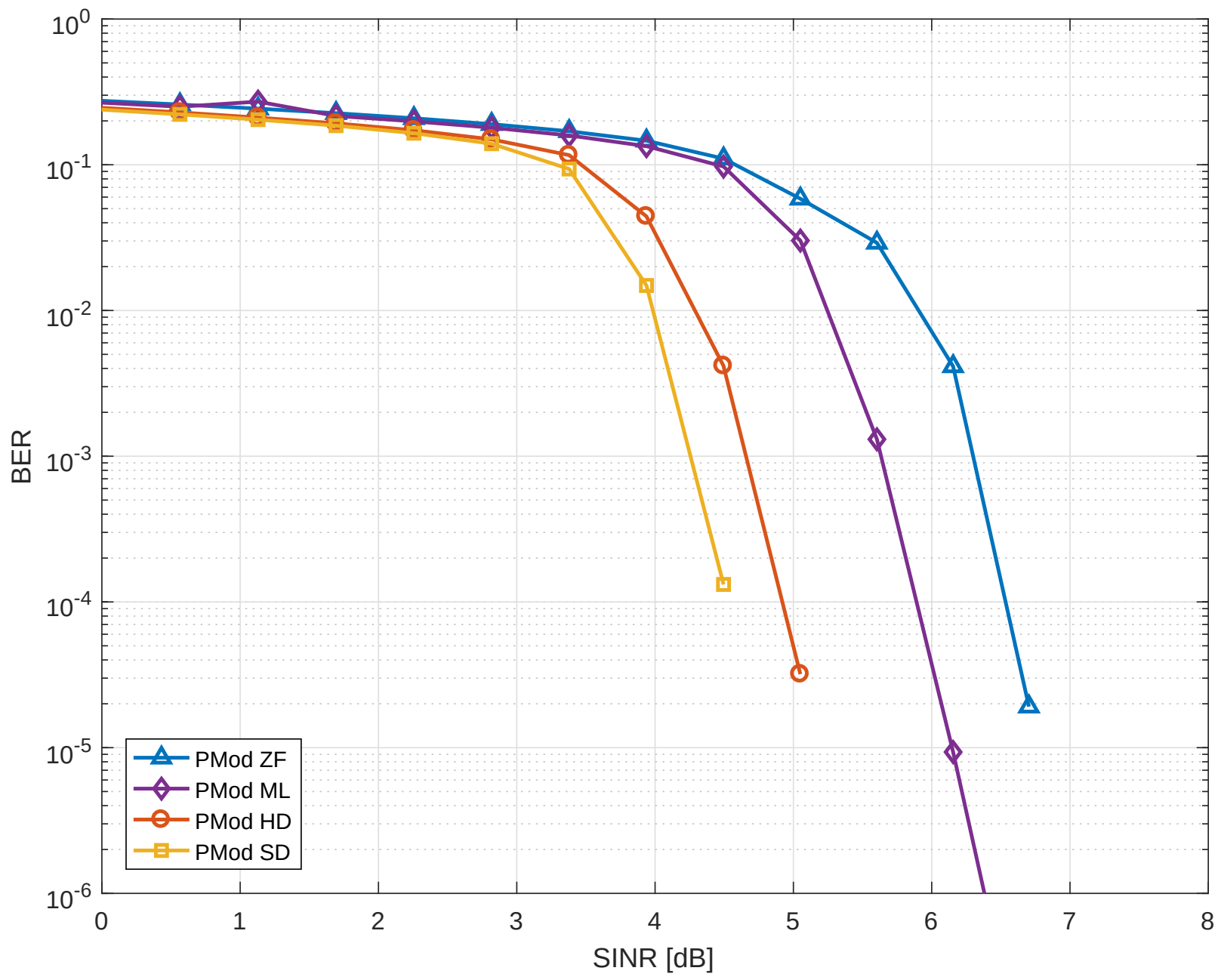

FIGURE 4.13: Comparison of the coded BER of the four proposed PMod techniques conveying a QPSK constellation.

Fig. 4.14 describes the throughput achieved using the four schemes. We observe that the four curves are grouped in the soft and hard receivers. In contrast to the previous section (where the gap between the solutions is tight), in this case the gap increases notably, thus revealing the performance of the PMod SD/HD.

### 4.4.3 Comparing PMod SD with Other Solutions

In this section we compare the performance of the PMod SD with OSTBC and V-BLAST in the same interference scenario. By doing so, we examine different strategies to increase the throughput.

Fig. 4.15 illustrates the coded BER for the different schemes. As with the uncoded BER (see Fig. 4.4), in this case, PMod SD lies between OSTBC and V-BLAST. One important aspect is that improves the BER of the Reference scenario. This is positive since PMod increases the SE but also the error rate.

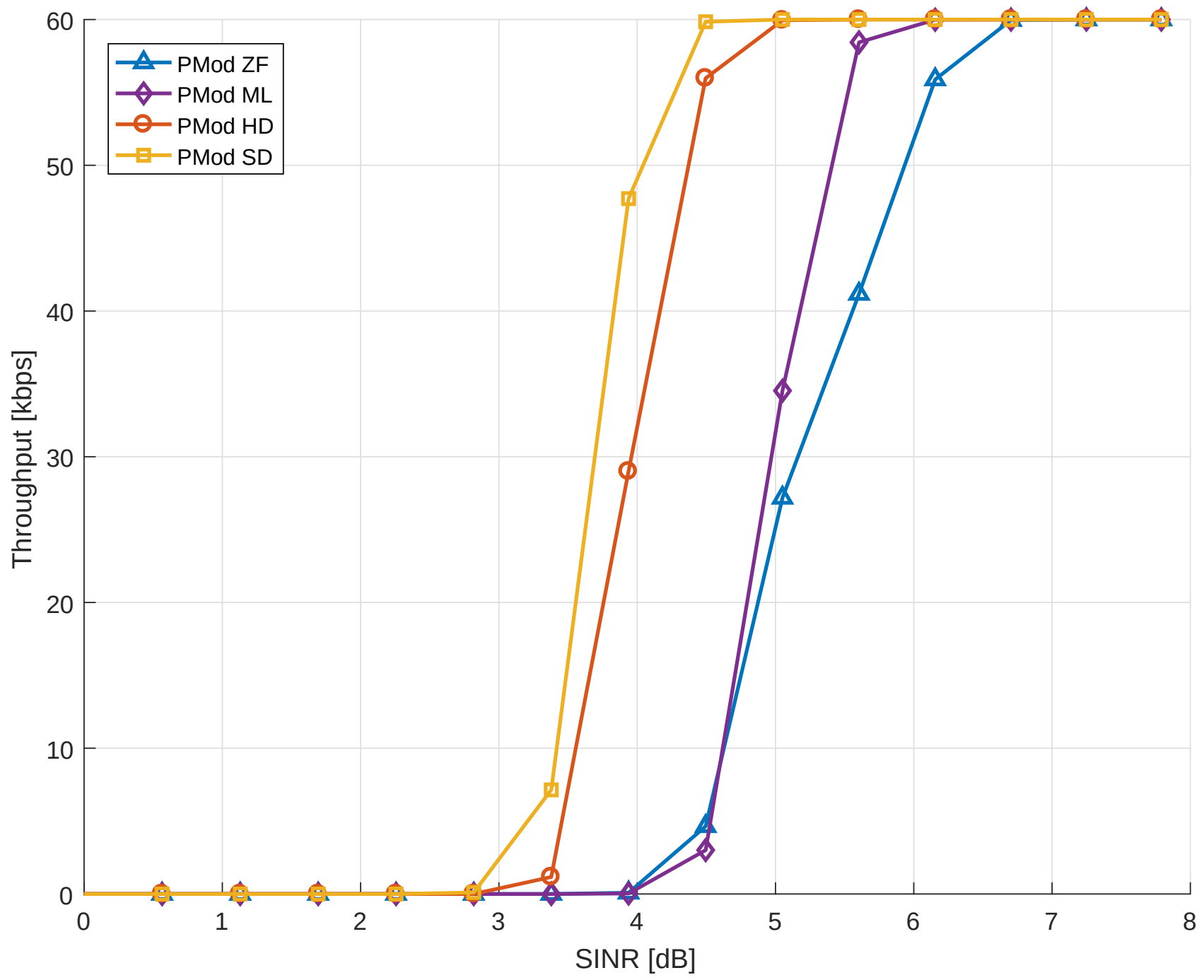

FIGURE 4.14: Comparison of the throughput of the four proposed PMod techniques conveying a QPSK constellation.

Finally, Fig. 4.16 illustrates the throughput achieved by each scheme. The interesting part of this figure is the adaptation of the rate. For very low SNR the most effective scheme is OSTBC. From 3.5 dB, the PMod SD increases the throughput by a factor of 1.5, followed by V-BLAST from 5.5 dB. This motivates the use of PMod in Adaptive Modulation and Coding Schemes (AMC).

### 4.4.4 XPD Analysis

In addition to prior comparisons, we also include a cross-polarization discrimination (XPD) analysis for the PMod. The results are extremely encouraging and reveal that the PMod scheme is robust in front of cross-polarization impairments. The reason for this is twofold:

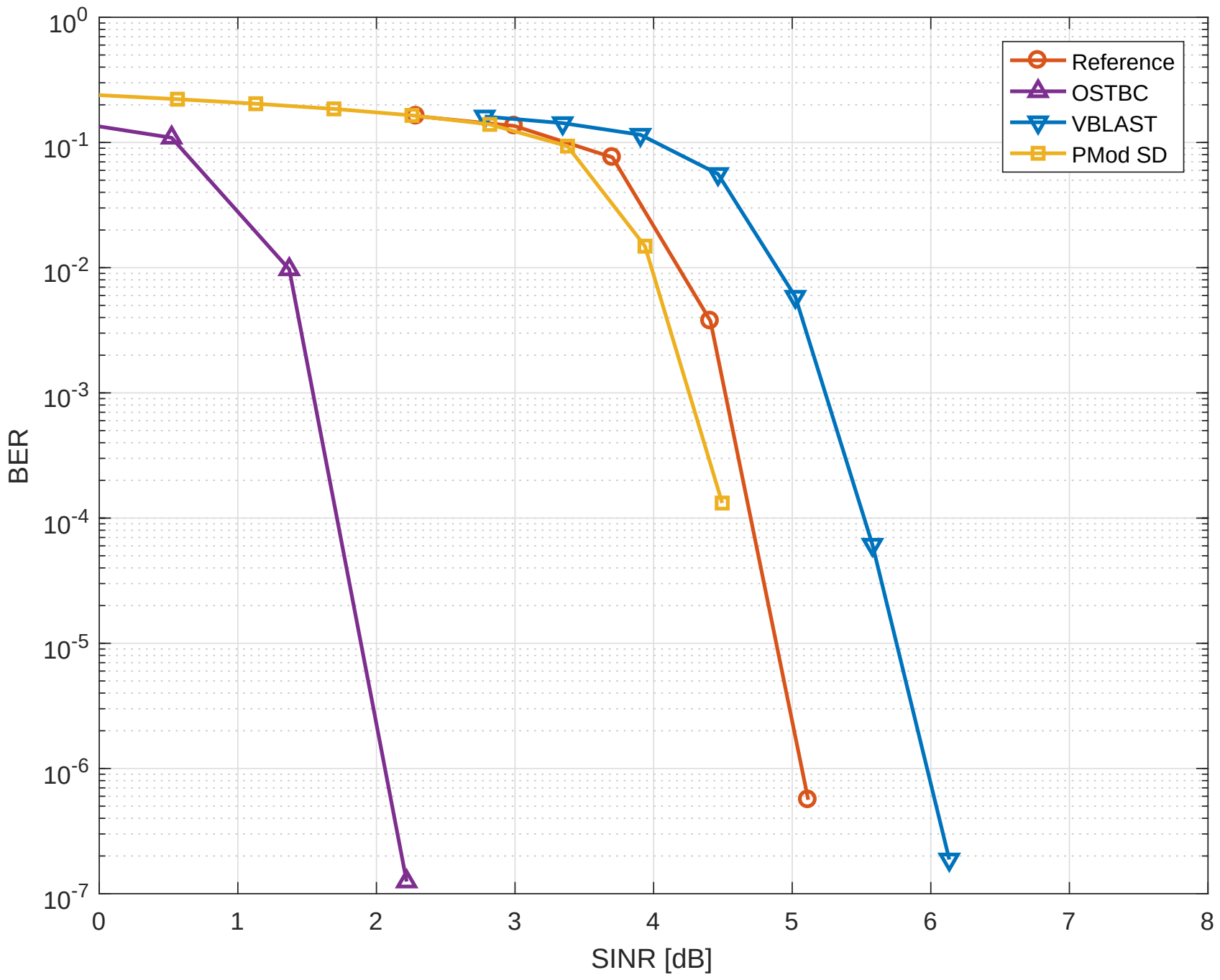

FIGURE 4.15: Comparison of the coded BER of the PMod SD with other existing techniques conveying a QPSK constellation.

- For high XPD values, only one polarization carries the data symbol whereas the other only contains noise. In this case, the system will decode the symbols $s$ and the switching bits $l$ correctly.

- For low XPD values, both polarizations carry the same symbol but only one polarization is decoded. In this case, the probability of error of decoding bit $l$ increases as the XPD decreases but the probability of error of decoding the symbol $s$ remains constant. This is motivated by the fact that, even in the case where the $l$ bit is erroneous and the decoded polarization is the wrong one, it also contains the symbol $s$ and thus, is able to decode the $s$ symbols as if it was decoded from the other polarization.

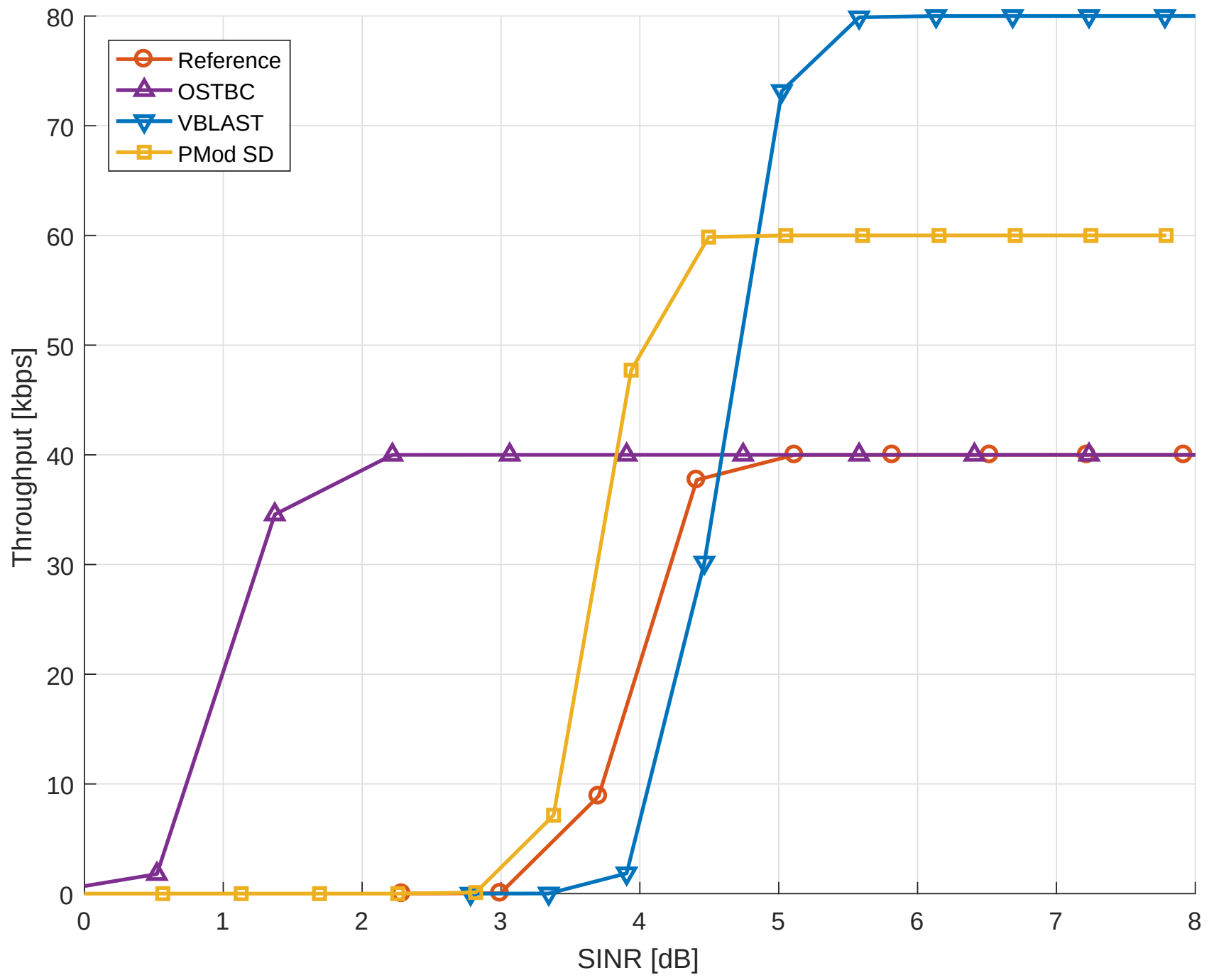

FIGURE 4.16: Comparison of the throughput of the PMod SD with other existing techniques conveying a QPSK constellation.

In order to analyze the XPD of the PMod technique, the XPD is defined as follows:

$$\text{XPD} = 20\log\left(\frac{|y_l|}{|y_{1-l}|}\right), \tag{4.22}$$

where $y_l$ is the signal received at the polarization where the symbol is transmitted and $y_{1-l}$ is the other one.

Fig. 4.17 compares the throughput of the four proposed schemes (*PMod ZF*, *PMod ML*, *PMod HD* and *PMod SD*) for different values to the XPD with the reference (*Reference*). Note that *PMod HD* and *PMod SD* are overlapped in the figure, although the *PMod SD* has slightly higher robustness. Particularly, a fixed SNR of 20 dB was set for this simulation only. In contrast, the value of the remaining parameters was set to that of previous figures. As mentioned before, the PMod technique is robust in front of XPD as it can exploit the fact that the $2/3$ of the bits are transmitted through the both polarizations.

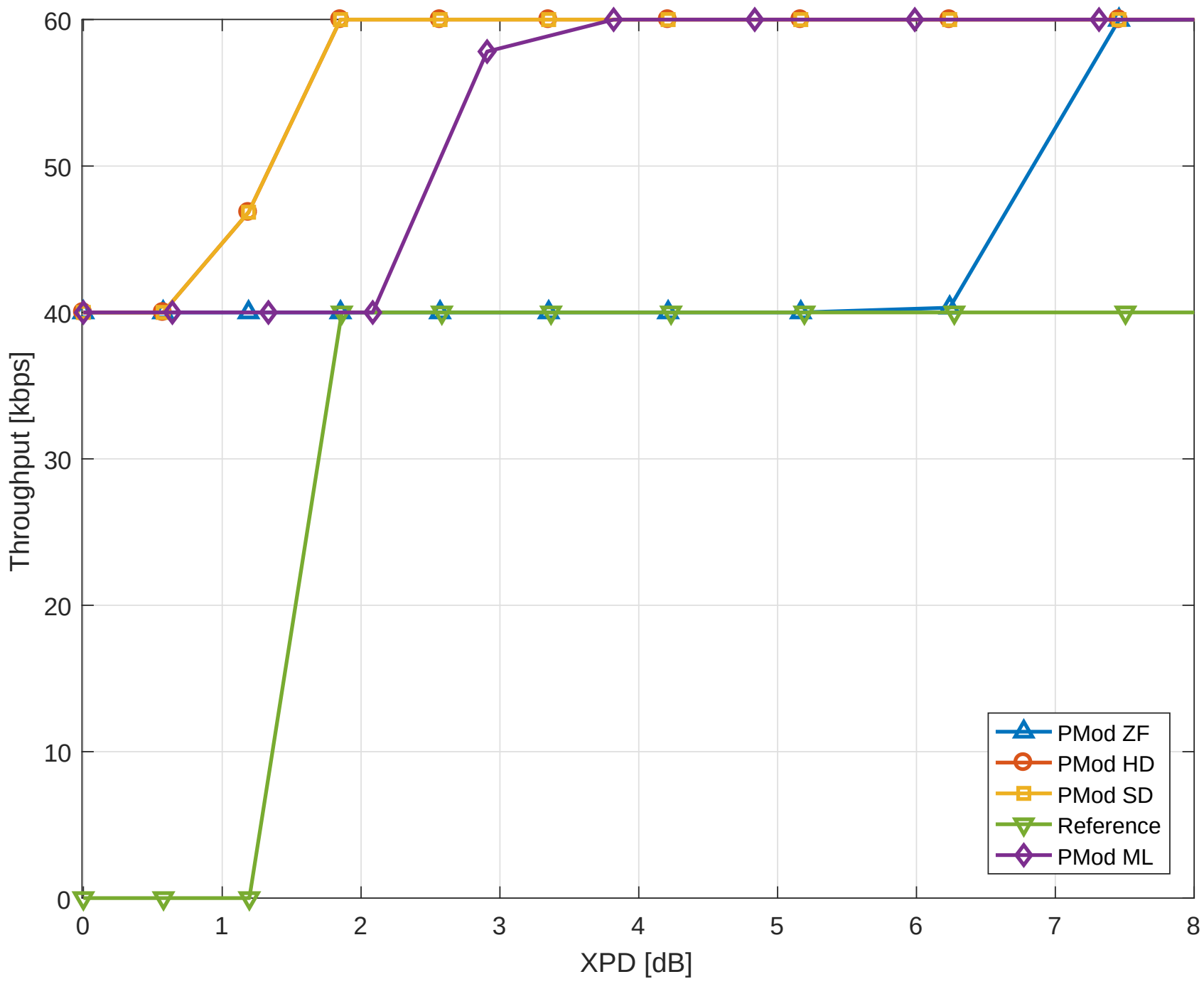

FIGURE 4.17: Comparison of the throughput with respect of XPD of the different techniques conveying a QPSK constellation.

### 4.4.5 Imperfect Channel Estimation

In this section we analyze the impact of an imperfect channel estimation. In order to study this behaviour, we introduced an error $\xi$ into the channel estimation which is normalized by the channel norm. Indeed, the power of the error $\xi$ is defined as follows:

$$|\xi|^2 = \frac{\mathbb{E}\left\{\left|h - \bar{h}\right|^2\right\}}{\mathbb{E}\left\{|h|^2\right\}}, \tag{4.23}$$

where $\mathbb{E}\{x\}$ is the expectation value of the variable $x$ and $\bar{h}$ is the estimated coefficient. The results can be examined in Fig. 4.18.

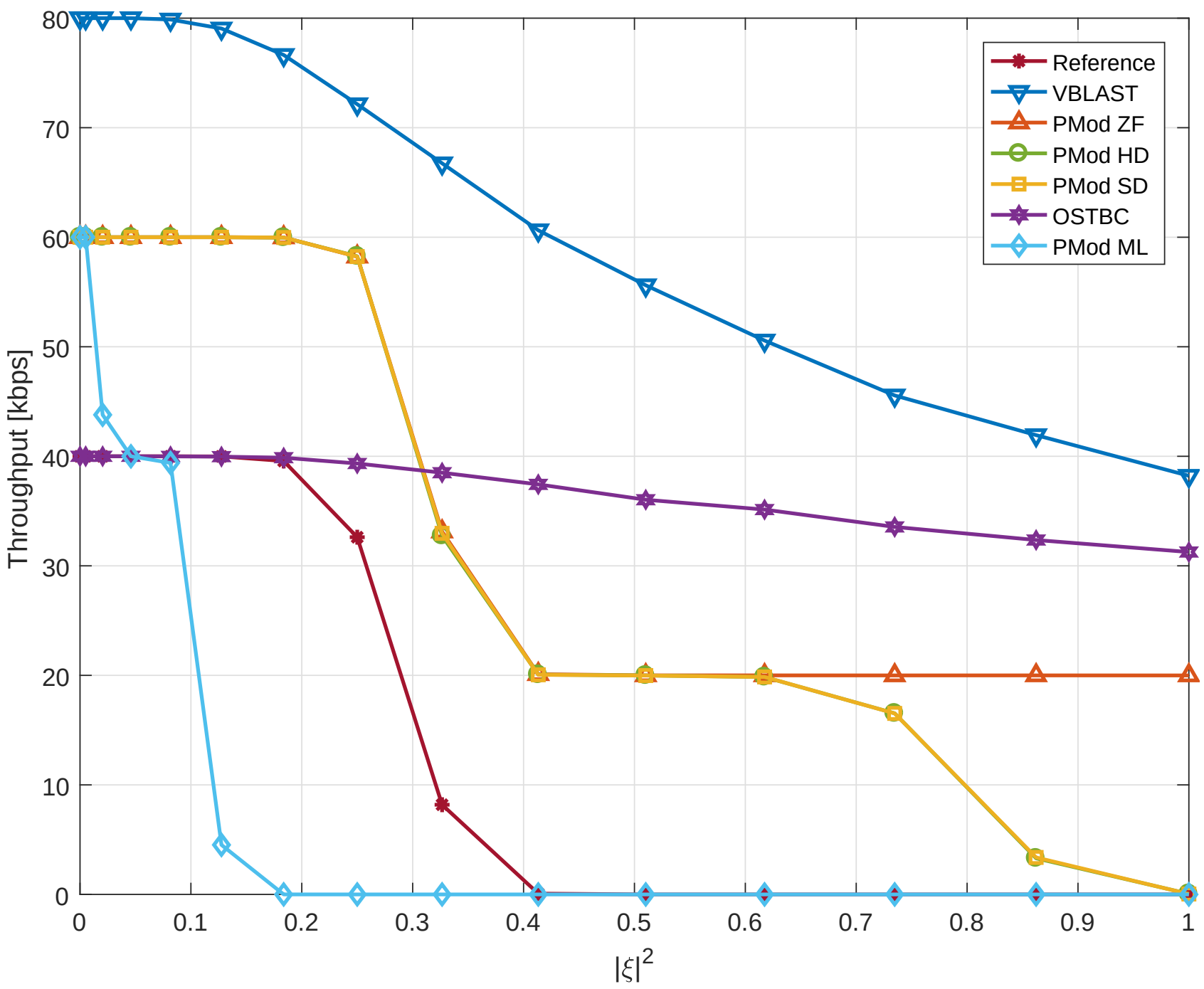

FIGURE 4.18: Impact of the imperfect channel estimation in the different techniques conveying a QPSK constellation.

The PMod scheme becomes more robust than the reference scheme (*Reference*). Particularly, three of the four schemes (*PMod ZF*, *PMod HD* and *PMod SD*) offer the same tolerance. However, the main difference is that *PMod ZF* is able to decode the bit $l$ correctly. This means that, although the scheme may be inaccurate, it is always capable of decoding the bit $l$. This motivates the approach of hierarchical modulations. For example, using the *PMod ZF* we could establish a hierarchical BPSK+QPSK and always succeed on decoding the BPSK scheme at least.

Finally, it is worth mentioning that the PMod technique presents a good trade-off between robust techniques with less througput, such as OSTBC, and more throughput-available techniques (but more power consuming), such as V-BLAST. This is shown in as Fig. 4.18.

## 4.5 Conclusions

This chapter introduces a novel application to mobile satellite communications of the so-called Polarized Modulation. This mechanism is based on dual polarized antennas. The work shows that with dual-polarized modulation throughput can be increased by a factor of $1 + b^{-1}$ in the absence of CSI in low SNR regime and that the transmission results are robust to cross-polarization and imperfect channel estimation. Since performance is highly dependent on the implemented receiver, in this chapter we propose different alternatives that trade-off computational complexity and performance. One of the demodulation schemes is based on probabilities, which involves soft detections and (to the best of the authors' knowledge) it is novel in the context of either spatial or polarized modulation. Finally, the proposed techniques have been thoroughly tested and validated using a maritime mobile satellite scenario and the newest implementation of the novel ETSI's standard TS 102 744 [ETS], known as BGAN. This work validates the PMod scheme and demonstrates the throughput and robustness improvements. Further work comprises the extension of the results and receiver architectures to more than two polarizations, as well as investigating the PMod in aeronautical and urban channels. Since PMod exploits the diversity of the channel, provided that the polarization channel has diversity, the PMod will work as expected. Additionally, although the union bound for Rayleigh channel is provided, an interesting line of work is to study the impact of the averaged probability of error in Rice channels as well as the mutual information and capacity analysis.

# Chapter 5

# 3D Polarized Modulation

> If your experiment needs statistics, you ought to have done a better experiment.
>
> E. Rutherford

In this chapter we introduce a PMod scheme using an arbitrary number of polarizations. We call this scheme 3D Polarized Modulation (3D PMod). By using a 3D sphere as the constellation, we are able to map 3D points with the respective electric field. Hence, compared with the classic 2D I/Q constellation mapping, placing a symbol on a sphere increases the minimum distance between symbols. Therefore, we can reduce the error rate and increase the throughput without requiring additional energy.

## 5.1 Sphere Modulation

In Chapter 2 the Stokes Vector (2.23) and parameters (2.22) are introduced. These parameters are obtained from the electric field measured by each polarization. We use the mapping of Stokes parameters to Poincaré Sphere to produce the constellation in three dimensions. For example, Fig. 5.1 displays the Poincaré Sphere facing $L$ points where the minimum distance is maximized, using the Sloane 3D packs [S+]. It is important to remark that the points proposed in the present manuscript are the corrected version from [S+], produced using exact numbers despite of floating point comma precision.

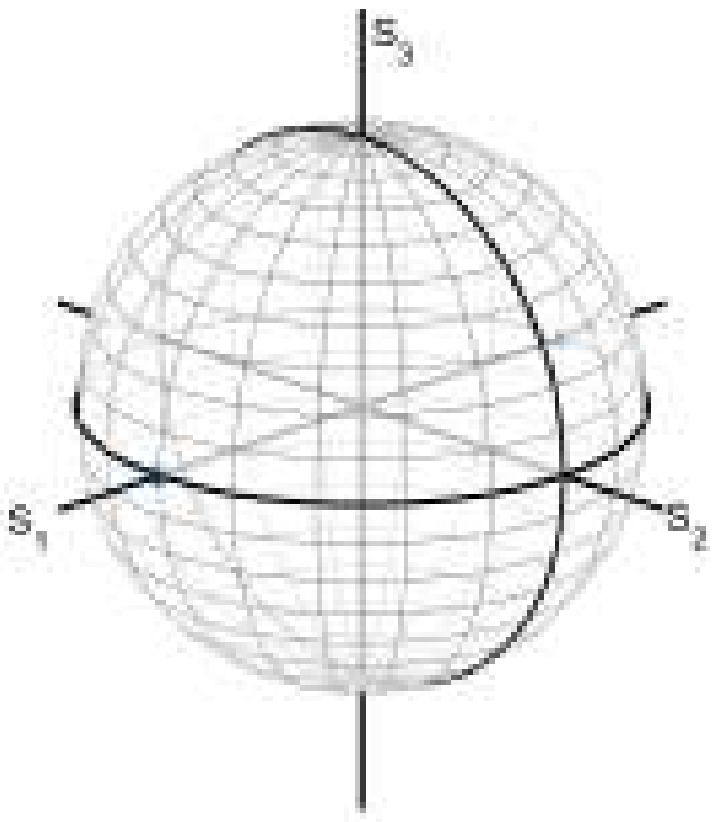

(A) $L = 2$

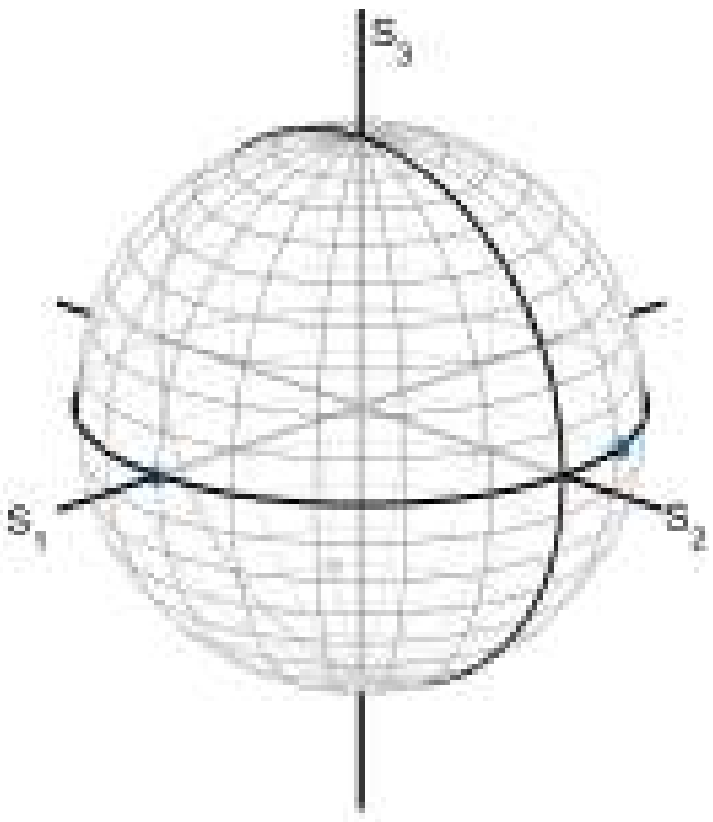

(B) $L = 4$

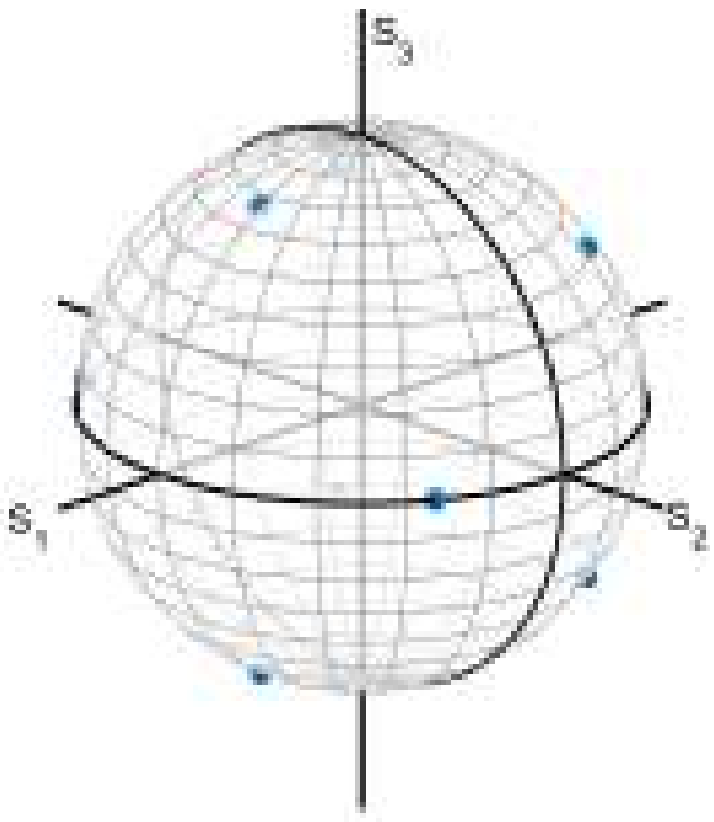

(C) $L = 8$

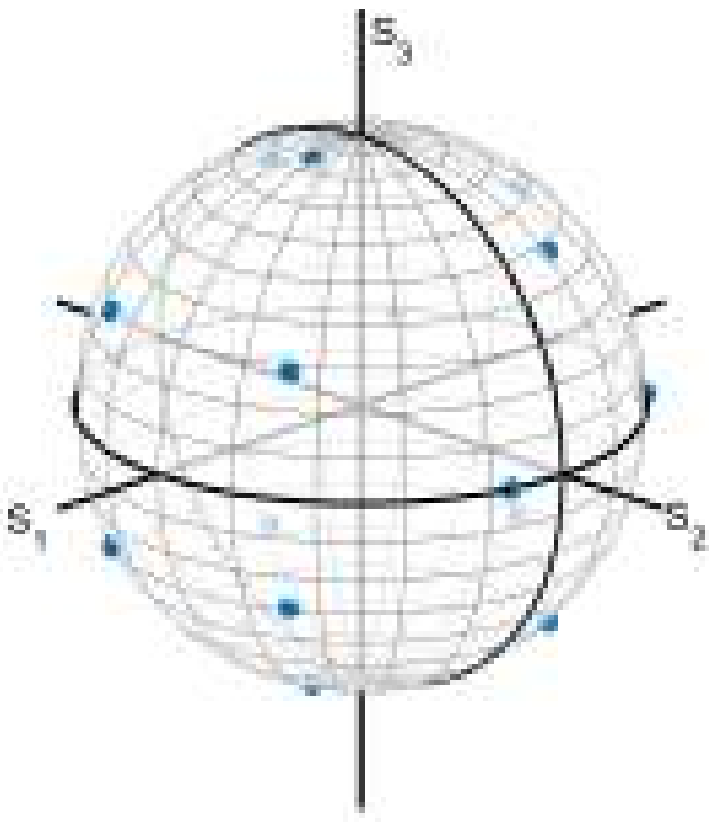

(D) $L = 16$

FIGURE 5.1: Poincaré sphere facing $L$ points in such a way that the minimum distance is maximized.

Expressing the Jones Vector from the Stokes Vector is not straightforward. The Stokes Vector measures the intensities of the polarized wave, whilst the Jones Vector contains information about the complex components, including the magnitude and the phase.

Authors of [Are+16] propose the following transformation

$$\mathbf{E}_0 = \begin{pmatrix} E_x \\ E_y \end{pmatrix} = \begin{pmatrix} \sqrt{\frac{S_0+S_1}{2}} e^{j\xi} \\ \frac{S_2 - jS_3}{\sqrt{2(S_0+S_1)}} e^{j\xi} \end{pmatrix}, \tag{5.1}$$

where $\mathbf{E}_0$ is the Jones Vector. However, (5.1) is not well defined in the entire domain:

$$\lim_{S_1 \to -S_0} E_y = \infty. \tag{5.2}$$

To the best of our knowledge, there is no well-defined, formal proposition of the conversion from the Stokes to the Jones vector. In order to circumvent this problem, [NTK11] proposes to express the Jones vector in the following manner

$$\mathbf{E}_0 = \begin{pmatrix} |E_x| e^{-j\theta} \\ |E_y| e^{j\theta} \end{pmatrix}. \tag{5.3}$$

Using the Stokes Vector we can find the following relationships

$$\begin{aligned} S_0 + S_1 &= 2|E_x|^2 \\ S_0 - S_1 &= 2|E_y|^2 \\ \frac{S_3}{S_2} &= \tan\delta, \ \cos\delta \neq 0. \end{aligned} \tag{5.4}$$

Hence, after some mathematical manipulations, we are able to express the Jones Vector as a function of Stokes parameters by

$$\mathbf{E}_0 = \begin{pmatrix} \sqrt{\frac{S_0+S_1}{2}} e^{-j\theta} \\ \sqrt{\frac{S_0-S_1}{2}} e^{j\theta} \end{pmatrix}, \tag{5.5}$$

where $\theta = \frac{1}{2}\arctan\left(\frac{S_3}{S_2}\right)$. Note that $\arctan\frac{y}{x}$ is the poorer form of the argument since it is not well defined if $x = 0$ and it does not preserve the signs of $x$ and $y$. In order to solve it, we use the

arctan2$(y, x)$ function instead, which is defined as

$$\text{arctan2}(y, x) = \begin{cases} \arctan\left(\frac{y}{x}\right) & \text{if } x > 0 \\ \arctan\left(\frac{y}{x}\right) + \pi & \text{if } x < 0 \text{ and } y \geq 0 \\ \arctan\left(\frac{y}{x}\right) - \pi & \text{if } x < 0 \text{ and } y < 0 \\ \frac{\pi}{2} & \text{if } x = 0 \text{ and } y > 0 \\ -\frac{\pi}{2} & \text{if } x = 0 \text{ and } y < 0 \\ 0 & \text{if } x = 0 \text{ and } y = 0, \end{cases} \tag{5.6}$$

which is well defined in $\mathbb{R}^2$.

The form expressed in (5.5) is further simplified when it is expressed in spherical coordinates, which takes the form

$$\begin{aligned} S_0 &= \mathcal{E} \\ S_1 &= \mathcal{E} \cos \vartheta \\ S_2 &= \mathcal{E} \sin \vartheta \cos \phi \\ S_3 &= \mathcal{E} \sin \vartheta \sin \phi, \end{aligned} \tag{5.7}$$

where $\phi \in [0, 2\pi)$ and $\vartheta \in [0, \pi)$ are the azimuthal and elevation components, respectively.

In the previous representation, by convenience, we place the $S_1$ in the z-axis, and $S_2$ and $S_3$ in the x-axis and y-axis, respectively. Moreover, we consider a sphere with its radius equal to the total energy of the symbol $\mathcal{E}$. Thus, the Jones vector in spherical coordinates is described as

$$\mathbf{E}_0 = \begin{pmatrix} \sqrt{\mathcal{E}} \cos \frac{\vartheta}{2} e^{-j\frac{\phi}{2}} \\ \sqrt{\mathcal{E}} \sin \frac{\vartheta}{2} e^{j\frac{\phi}{2}} \end{pmatrix}. \tag{5.8}$$

Classic digital modulations map $L$ symbols of a finite alphabet to the complex two-dimensional plane I/Q in such a way that the minimum distance is maximized to reduce the bit error rate. Using the Poincaré Sphere, we are able to extend the same concept to the three-dimensional space. However, placing $L$ points on the surface of a sphere is not a straightforward problem. This problem, known as *sphere packing*, is addressed in many works such as [HS95; Slo98; CS13]. Although works such as [S+] published particular solutions for many dimensions and different $L$ values, there is still no closed-form expression for an arbitrary $L$ or dimension. In particular, Sloane provides solutions for $L \in [4 \dots 130]$, i.e., modulation orders from 2 to 7 bits.

Hence, it is possible to transmit information depending on which point on the sphere is used or, in other words, which polarization state is used. This is a more general version of the well known Polarization Shift Keying (PolSK) [BP92], where the information is placed only in the shifts.

## 5.2 Enabling Polarized Modulation

Polarized Modulation [HP15b] combines PolSK with modulated information in the amplitude and phase of the radiated waveform. This concept can also be applied using the described 3D modulation by exploiting the ambiguity of the initial phase. Our proposal focuses on choosing the initial phase by mapping certain number of bits to a PSK constellation. Hence, we are able to use two sources for transmission: the state of polarization and the initial phase.

By packing $L_b$ bits on the sphere surface and $N_b$ bits on the PSK phase we are able to convey $L_b + N_b$ bits in total. Hence, if $2^{L_b} = L$ symbols lay on the sphere and $2^{N_b} = N$ symbols are with the PSK constellation, we are able to describe the transmitted vector as a function of Stokes parameters as follows

$$\begin{aligned}\mathbf{x}[k] &= \mathbf{E}_0[k]e^{j\xi[k]} \\ &= \begin{pmatrix} \sqrt{\frac{S_0[k]+S_1[k]}{2}}e^{-j\theta[k]} \\ \sqrt{\frac{S_0[k]-S_1[k]}{2}}e^{j\theta[k]} \end{pmatrix} e^{j\xi[k]} \\ &\equiv \begin{pmatrix} E_H \\ E_V \end{pmatrix},\end{aligned} \tag{5.9}$$

where $\mathbf{E}_0$ is the Jones Vector of the 3D PolSK contribution, $k$ is the time sample, $\xi[k] = \frac{2\pi}{N}n[k]$ is the modulated initial phase, $n[k]$ is the symbol with the PSK constellation modulated in phase, $S_0[k] = \sqrt{S_1[k]^2 + S_2[k]^2 + S_3[k]^2} = \mathcal{E}$ is the total energy transmitted by the symbol, $\theta[k] = \frac{1}{2}\arctan\left(\frac{S_3[k]}{S_2[k]}\right)$ and $S_1[k]$, $S_2[k]$, $S_3[k]$ are the coordinates of the point on the sphere surface. Note that the Jones vector of 3D PMod is $\mathbf{x}$ and is equal to $\mathbf{E}_0 e^{j\xi}$. The transmitted vector can also be expressed in spherical coordinates as

$$\mathbf{x}[k] = \sqrt{\mathcal{E}}\begin{pmatrix}\cos\frac{\vartheta[k]}{2}e^{-j\frac{\phi[k]}{2}}\\ \sin\frac{\vartheta[k]}{2}e^{j\frac{\phi[k]}{2}}\end{pmatrix}e^{j\xi[k]} \equiv \begin{pmatrix}E_H\\ E_V\end{pmatrix}. \tag{5.10}$$

The expression in spherical coordinates is particularly interesting because it can be seen that $\mathbf{E}_0^H\mathbf{E}_0 = 1,\ \forall(\phi,\vartheta)$.

It is worth mentioning that introducing the $e^{j\xi[k]}$ component does not affect the computation of Stokes parameters, since it is an invariant transformation and, thus, it is independent from the PolSK modulation.

Hence, we can describe the system model as

$$\mathbf{y}[k] = \mathbf{H}[k]\mathbf{x}[k] + \mathbf{w}[k], \tag{5.11}$$

where $\mathbf{y} \in \mathbb{C}^2$ is the received signal, $\mathbf{H} \in \mathbb{C}^{2\times2}$ is the channel matrix and $\mathbf{w} \in \mathbb{C}^2$ is the zero-mean noise vector with a covariance $\mathbf{R_w}$. Fig. 5.2 illustrates the block diagram of 3D Polarized Modulation transmitter, where $E_V$ and $E_H$ denote the vertical and horizontal electric field, respectively, and where $E_H \equiv E_x$ and $E_V \equiv E_y$ for simplicity.

One important advantage of using PSK modulation for the initial phase is that it does not affect the demodulation of Stokes parameters, since they are only affected by the differential phase. Thus, we can decode the grouped bits independently from each other.

Based on that, we introduce two different classes of receivers:

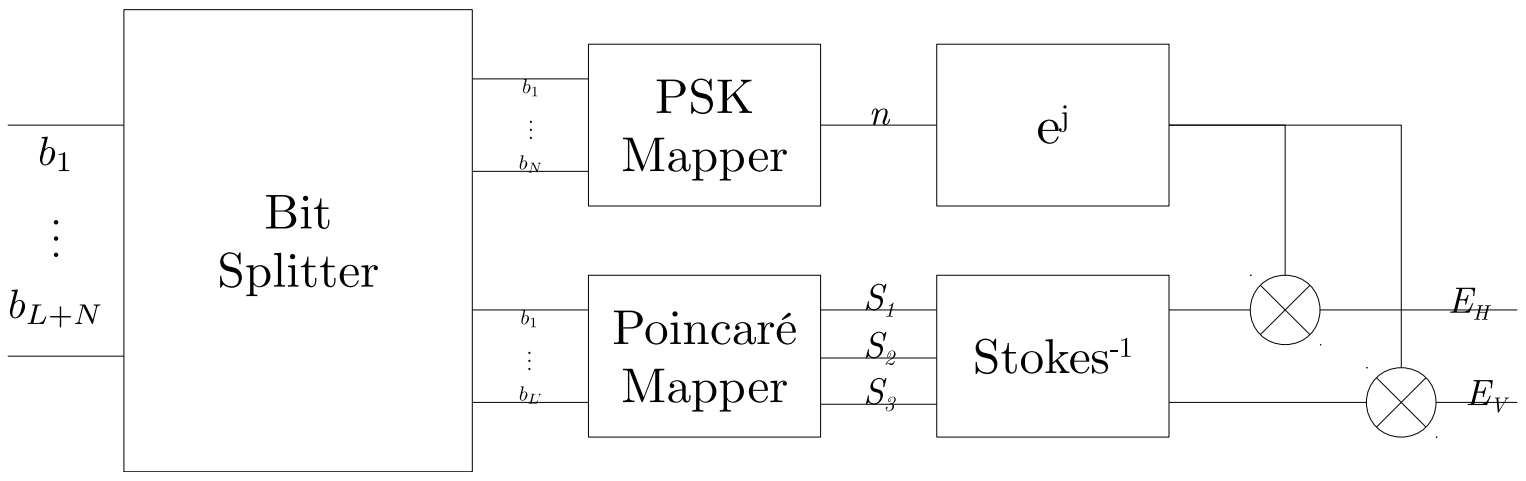

FIGURE 5.2: Block diagram of 3D Polarized Modulation transmitter.

1. Joint receiver: it decodes the symbols from the PSK constellation and the Stokes parameters jointly.

2. Cascade receiver: it is composed by two independent receivers, faced in cascade. First, the Stokes receiver computes the Stokes parameters and second these are used by the PSK receiver.

### 5.2.1 Joint Receiver

This receiver decodes all symbols and bits without decoupling the PSK and 3D PolSK contributions. The optimal receiver implements the Maximum Likelihood (ML) algorithm.

The expression of this receiver is denoted by

$$\begin{aligned}\left(\hat{l}, \hat{n}\right) &= \arg\min_{l,n} \|\mathbf{y}[k] - \mathbf{H}[k]\mathbf{x}[k]\| \\ &= \arg\min_{l,n} \mathbf{y}[k]^H \mathbf{x}_{l,n},\end{aligned} \tag{5.12}$$

where $\|\cdot\|$ is the $\ell^2$-norm, $\mathbf{x}_{l,n}$ is the symbol $\mathbf{x}$ using the $l$th symbol of the Poincaré Sphere and the $n$th symbol of the PSK constellation.

Note that, in this receiver, the search space is $L \times N$ and, hence, its computational complexity is $o(L \times N)$.

### 5.2.2 Cascade Receiver

In order to reduce the complexity next we propose a suboptimal receiver. This receiver decouples the signal into two contributions: the PSK and 3D PolSK. Each contribution is decoded by an independent receiver. The PSK contribution does not affect the 3D PolSK, since the Stokes parameters are obtained using the difference of the phases of each component. Thus, the Stokes parameters can be estimated using the received signal $\mathbf{y}[k]$ straightforwardly. However, the contribution of 3D PolSK affects the PSK contribution. To estimate the phase $\hat{\xi}[k]$ we filter the received signal by a linear filter

$$\hat{\xi}[k] = \arg \hat{r}[k] = \arg \left(\mathbf{a}^{H}[k]\mathbf{y}[k]\right), \tag{5.13}$$

where $\mathbf{a}$ is the linear filter. It can be the Zero Forcer (ZF) filter, which is described by

$$\mathbf{a}_{\text{ZF}} = \frac{\mathbf{H}[k]\hat{\mathbf{E}}_0[k]}{\hat{\mathbf{E}}_0^H[k]\mathbf{H}^H[k]\mathbf{H}[k]\hat{\mathbf{E}}_0[k]}. \tag{5.14}$$

Alternatively, the Minimum Mean Square Error (MMSE) filter is described by

$$\mathbf{a}_{\text{MMSE}} = \left(\mathbf{H}[k]\hat{\mathbf{E}}_0[k]\hat{\mathbf{E}}_0^H[k]\mathbf{H}^H[k] + \mathbf{R}_{\mathbf{w}}\right)^{-1}\mathbf{H}[k]\hat{\mathbf{E}}_0[k]. \tag{5.15}$$

It is important to remark that the PSK estimation depends on the estimation of $\hat{\mathbf{E}}_0$ and thus, the errors of the estimators are propagated.

This receiver has lower computational complexity compared with the Joint Receiver. In particular, the computational complexity is the sum of each sub-receivers, i.e., $o(L) + o(N)$. Fig. 5.3 illustrates the block diagram of this receiver. $\mathbf{a}(\mathbf{H})$ applies the filter operation described by (5.14) or (5.15) and depends on the channel matrix $\mathbf{H}[k]$. In the simulation section this receiver is evaluated and compared with the optimal one; thus, concluding within which SNR range its performance is competitive.

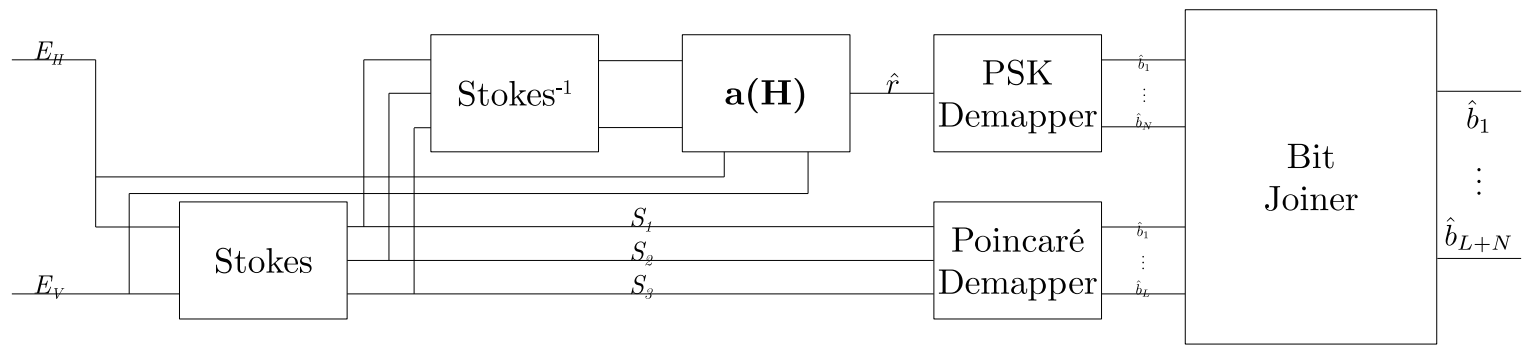

FIGURE 5.3: Block diagram of 3D Polarized Modulation Cascade Receiver.

## 5.3 BER Analysis

We perform the BER analysis of the proposed 3D PMod scheme by means of pairwise error probability (PEP) and union bound, defined in [RH12] as

$$\text{BER} \leq \frac{1}{LN}\frac{1}{L_b N_b}\sum_{l=1}^{L}\sum_{l'=1}^{L}\sum_{n=1}^{N}\sum_{n'=1}^{N}\mathcal{D}\left((l',n')\to(l,n)\right)\text{PEP}\left((l',n')\to(l,n)\right), \tag{5.16}$$

where $\mathcal{D}\left((l',n')\to(l,n)\right)$ is the Hamming distance, i.e., the number of different bits between symbol defined by $(l',n')$ and $(l,n)$. An important remark of [RH12] is the fact that the BER can be decoupled into three contributions. A symbol is decoded erroneously if 1) the polarization is estimated correctly but the initial phase is erroneous, 2) the initial phase estimation is correct but the estimated polarization state is erroneous, 3) neither the polarization state nor the initial phase are estimated correctly. Hence, the BER contributions are described as follows:

- BER obtained by the distance between the symbols belonging to the same PSK constellation, i.e., $\left((l,n')\to(l,n)\right), \forall n' \neq n$.
- BER obtained by the distance between the symbols belonging to the PolSK constellation, i.e., $\left((l',n)\to(l,n)\right), \forall l' \neq l$.

- BER obtained by the distance between the symbols belonging to the PolSK and PSK $((l', n') \rightarrow (l, n))$, $\forall l' \neq l$, $\forall n' \neq n$.

A very interesting observation is that the first BER can be expressed in terms of exact error probability and is widely available in the literature, without incurring into bounding and obtain accurate results. This produces tighter union upper bound. Note that BER depends on the bit mapping. For the sake of homogeneity, we assume Gray bit mapping. Gray mapping is used vastly in the literature and it is proved to be the one that produces the lowest BER. Thus, (5.16) can be grouped as

$$\mathrm{BER} \leq \mathrm{BER}_{\mathrm{Signal}} + \mathrm{BER}_{\mathrm{Index}} + \mathrm{BER}_{\mathrm{Joint}}, \tag{5.17}$$

where

$$\begin{aligned}
\mathrm{BER}_{\mathrm{Signal}} &= \frac{N_b}{L_b + N_b} \mathrm{BER}_{\mathrm{PSK}} \\
\mathrm{BER}_{\mathrm{PolSK}} &= \frac{1}{L} \frac{1}{L_b + N_b} \sum_{l=1}^{L} \sum_{l'=1}^{L} \mathcal{D}\left(l' \rightarrow l\right) Q\left(\sqrt{\frac{d_{l',l}^2}{2N_0}}\right) \\
\mathrm{BER}_{\mathrm{Joint}} &= \frac{1}{LN} \frac{1}{L_b + N_b} \\
&\times \sum_{l=1}^{L} \sum_{n=1}^{N} \sum_{l'=1}^{L} \sum_{n'=1}^{N} \left(\mathcal{D}\left(l' \rightarrow l\right) + \mathcal{D}\left(n' \rightarrow n\right)\right) Q\left(\sqrt{\frac{d_{l',l,n',n}^2}{2N_0}}\right).
\end{aligned} \tag{5.18}$$

The term $\mathrm{BER}_{\mathrm{PSK}}$ can be substituted by the exact BER expression, since it is known in the literature and takes the following expression

$$\mathrm{BER}_{\mathrm{PSK}} = \frac{1}{N_b} \left(1 - \frac{1}{2\pi} \int_{-\frac{\pi}{N}}^{\frac{\pi}{N}} e^{-\gamma \sin^2 \theta} \int_{0}^{\infty} \nu e^{-\frac{(\nu - \sqrt{2\gamma} \cos \theta)^2}{2}} \, \mathrm{d}\nu \, \mathrm{d}\theta \right). \tag{5.19}$$

The previous integral has a closed-form expression only in the case of $N = 2$ and $N = 4$. In these cases, the expression is reduced to

$$\mathrm{BER_{PSK}} = Q\left(\sqrt{2\gamma}\right) \tag{5.20}$$

for $N = 2$ and

$$\mathrm{BER_{PSK}} = Q\left(\sqrt{2\gamma}\right)\left(1 - \frac{Q\left(\sqrt{2\gamma}\right)}{2}\right) \tag{5.21}$$

for $N = 4$, where $\gamma = \mathcal{E}/N_0$.

Before developing the expressions of the distances $d_{l',l}$ and $d_{l',l,n',n}$, for the sake of simplicity, we express the Stokes vector as a function of spherical coordinates $(\phi, \vartheta)$, where $\phi \in [0, 2\pi]$ and $\vartheta \in [0, \pi]$. Hence,

$$\mathbf{S}_l \equiv \mathcal{E}\begin{pmatrix} 1 \\ \cos\vartheta_l \\ \sin\vartheta_l\cos\phi_l \\ \sin\vartheta_l\sin\phi_l \end{pmatrix} \tag{5.22}$$

Using (5.10), the generic distance $d_{l',l,n',n}$ is expressed as the norm of two arbitrary symbols $\|\mathbf{x}_{l',n'} - \mathbf{x}_{l,n}\|$. Thus,

$$\begin{aligned} d^2_{l',l,n',n} = 2\mathcal{E}\Bigg(1 - \Bigg(&\cos\left(\Delta\xi - \frac{\Delta\phi}{2}\right)\cos\frac{\vartheta_{l'}}{2}\cos\frac{\vartheta_l}{2} \\ &+ \cos\left(\Delta\xi + \frac{\Delta\phi}{2}\right)\sin\frac{\vartheta_{l'}}{2}\sin\frac{\vartheta_l}{2}\Bigg)\Bigg) \end{aligned} \tag{5.23}$$

where $\Delta\xi = \xi_{n'} - \xi_n$, $\Delta\phi = \phi_{l'} - \phi_l$, $\xi_n$ is the PSK $n$ symbol, $(\phi_l, \vartheta_l)$ is the 3D PolSK $l$ symbol, composing the $\mathbf{x}_{l,n}$ 3D PMod symbol. This expression is the general version for an arbitrary pair of symbols.

The previous distance is further reduced if both symbols have the same PSK component, i.e., $\Delta\xi = 0$. Then, (5.23) is reduced to

$$d^2_{l',l} = 2\mathcal{E}\left(1 - \cos\left(\frac{\Delta\phi}{2}\right)\cos\left(\frac{\Delta\vartheta}{2}\right)\right), \tag{5.24}$$

where $\Delta\vartheta = \vartheta_{l'} - \vartheta_l$.

Note that if both symbols belong the same PolSK position, i.e., $\Delta\phi = 0$ and $\vartheta_{l'} = \vartheta_l$, the distance expression (5.23) is reduced to

$$d^2_{n',n} = 2\mathcal{E}\left(1 - \cos\Delta\xi\right) = 4\mathcal{E}\sin^2\left(\frac{\pi\Delta n}{N}\right), \tag{5.25}$$

where $\Delta n = n' - n$, which is equivalent to the well known PEP of PSK [Pro].

However, the distance of PolSK cannot be studied analytically, since there is no closed-form expression on the symbols belonging the constellation (they are obtained numerically [S+]). Despite this, we are able to compute this distance numerically for different modulation orders.

The packing problem is not a new problem. Essentially, it aims at finding an answer to the question *How $n$ points should be placed on a sphere surface in such a way that the minimum distance between them is maximized?* This problem is known as *Tammes Problem* [Tam30; Kot91]. Unfortunately, there is no closed solution, although there are known solutions for a small number of points. For instance:

- $L = 1$: the solution is trivial.
- $L = 2$: points at the poles.
- $L = 3$: points at the equator separated $120$ degrees apart.
- $L = 4$: vertices of a regular tetrahedron.
- ...

In particular, for $L = 2, 4, 8, 16$ simple solutions can be found. The z-axis is sliced in few levels with symmetry in the equator and the points are located equispaced in each slice.

The BER analysis can be also performed in terms of minimum distance. In the presence of AWGN, the overall performance of the system is mainly described by the minimum distance. Thus, we compare the minimum distance using different $L \times N$ combinations for several spectral efficiencies. Tables 5.1, 5.2, 5.3, 5.4, 5.5, 5.6 and 5.7 summarize the minimum distance of different schemes for a fixed spectral efficiencies of $2, 3, 4, 5, 6, 7, 8$ bits. The results in boldface denote the mode with the maximum minimum distance. For the sake of clarity, the results are expressed in numeric form instead of using trigonometric functions and fractions.

TABLE 5.1: Minimum distance for spectral efficiency $L_b + N_b = 2$ bits

| $L \times N$ | 3D PMod | Dual QAM | Dual PSK | Single QAM | Single PSK | LAM |
|---|---|---|---|---|---|---|
| $2 \times 2$ | **1.4142** | 1.4142 | 1.4142 | 1.4142 | 1.4142 | 1.4142 |

TABLE 5.2: Minimum distance for spectral efficiency $L_b + N_b = 3$ bits

| $L \times N$ | 3D PMod | Dual QAM | Dual PSK | Single QAM | Single PSK | LAM |
|---|---|---|---|---|---|---|
| $2 \times 4$ | **1.4142** | 1 | 1 | 0.8165 | 0.7654 | 1.4142 |
| $4 \times 2$ | 1 | 1 | 1 | 0.8165 | 0.7654 | 1.4142 |

TABLE 5.3: Minimum distance for spectral efficiency $L_b + N_b = 4$ bits

| $L \times N$ | 3D PMod | Dual QAM | Dual PSK | Single QAM | Single PSK | LAM |
|---|---|---|---|---|---|---|
| $2 \times 8$ | 0.7654 | 0.5774 | 0.5412 | 0.6325 | 0.3902 | 1 |
| $8 \times 2$ | 0.6323 | 0.5774 | 0.5412 | 0.6325 | 0.3902 | 1 |
| $4 \times 4$ | 0.9194 | **1** | **1** | 0.6325 | 0.3902 | 1 |

Figure 5.4 depicts the maximum minimum distance for different spectral efficiencies compared with common schemes, such as Dual QAM, Dual PSK, Single QAM, Single PSK and Lattice Amplitude Modulation (LAM). These schemes are described as follows:

TABLE 5.4: Minimum distance for spectral efficiency $L_b + N_b = 5$ bits

| $L \times N$ | 3D PMod | Dual QAM | Dual PSK | Single QAM | Single PSK | LAM |
|---|---|---|---|---|---|---|
| $2 \times 16$ | 0.3902 | 0.4472 | 0.2759 | 0.4472 | 0.1960 | 0.8165 |
| $16 \times 2$ | 0.5039 | 0.4472 | 0.2759 | 0.4472 | 0.1960 | 0.8165 |
| $4 \times 8$ | **0.7654** | 0.5774 | 0.5412 | 0.4472 | 0.1960 | 0.8165 |
| $8 \times 4$ | 0.6323 | 0.5774 | 0.5412 | 0.4472 | 0.1960 | 0.8165 |

TABLE 5.5: Minimum distance for spectral efficiency $L_b + N_b = 6$ bits

| $L \times N$ | 3D PMod | Dual QAM | Dual PSK | Single QAM | Single PSK | LAM |
|---|---|---|---|---|---|---|
| $2 \times 32$ | 0.1960 | 0.3162 | 0.1386 | 0.3086 | 0.0981 | 0.7559 |
| $32 \times 2$ | 0.3318 | 0.3162 | 0.1386 | 0.3086 | 0.0981 | 0.7559 |
| $4 \times 16$ | 0.3902 | 0.4472 | 0.2759 | 0.3086 | 0.0981 | 0.7559 |
| $16 \times 4$ | 0.5039 | 0.4472 | 0.2759 | 0.3086 | 0.0981 | 0.7559 |
| $8 \times 8$ | **0.6323** | 0.5774 | 0.5412 | 0.3086 | 0.0981 | 0.7559 |

TABLE 5.6: Minimum distance for spectral efficiency $L_b + N_b = 7$ bits

| $L \times N$ | 3D PMod | Dual QAM | Dual PSK | Single QAM | Single PSK | LAM |
|---|---|---|---|---|---|---|
| $2 \times 64$ | 0.0981 | 0.2182 | 0.0694 | 0.2209 | 0.0491 | 0.6324 |
| $64 \times 2$ | 0.2615 | 0.2182 | 0.0694 | 0.2209 | 0.0491 | 0.6324 |
| $4 \times 32$ | 0.1960 | 0.3162 | 0.1386 | 0.2209 | 0.0491 | 0.6324 |
| $32 \times 4$ | 0.3318 | 0.3162 | 0.1386 | 0.2209 | 0.0491 | 0.6324 |
| $8 \times 16$ | 0.3902 | 0.4472 | 0.2759 | 0.2209 | 0.0491 | 0.6324 |
| $16 \times 8$ | **0.4627** | 0.4472 | 0.2759 | 0.2209 | 0.0491 | 0.6324 |

- 3D PMod: the proposed scheme described in (5.9) and (5.10).
- Dual QAM scheme conveys $L$-QAM and $N$-QAM constellations in each horizontal or vertical polarization.

TABLE 5.7: Minimum distance for spectral efficiency $L_b + N_b = 8$ bits

| $L \times N$ | 3D PMod | Dual QAM | Dual PSK | Single QAM | Single PSK | LAM |
|---|---|---|---|---|---|---|
| $2 \times 128$ | 0.0491 | 0.1562 | 0.0347 | 0.1534 | 0.0245 | 0.5443 |
| $128 \times 2$ | 0.1627 | 0.1562 | 0.0347 | 0.1534 | 0.0245 | 0.5443 |
| $4 \times 64$ | 0.0981 | 0.2182 | 0.0694 | 0.1534 | 0.0245 | 0.5443 |
| $64 \times 4$ | 0.2615 | 0.2182 | 0.0694 | 0.1534 | 0.0245 | 0.5443 |
| $8 \times 32$ | 0.1960 | 0.3162 | 0.1386 | 0.1534 | 0.0245 | 0.5443 |
| $32 \times 8$ | 0.3318 | 0.3162 | 0.1386 | 0.1534 | 0.0245 | 0.5443 |
| $16 \times 16$ | 0.3902 | **0.4472** | 0.2759 | 0.1534 | 0.0245 | 0.5443 |

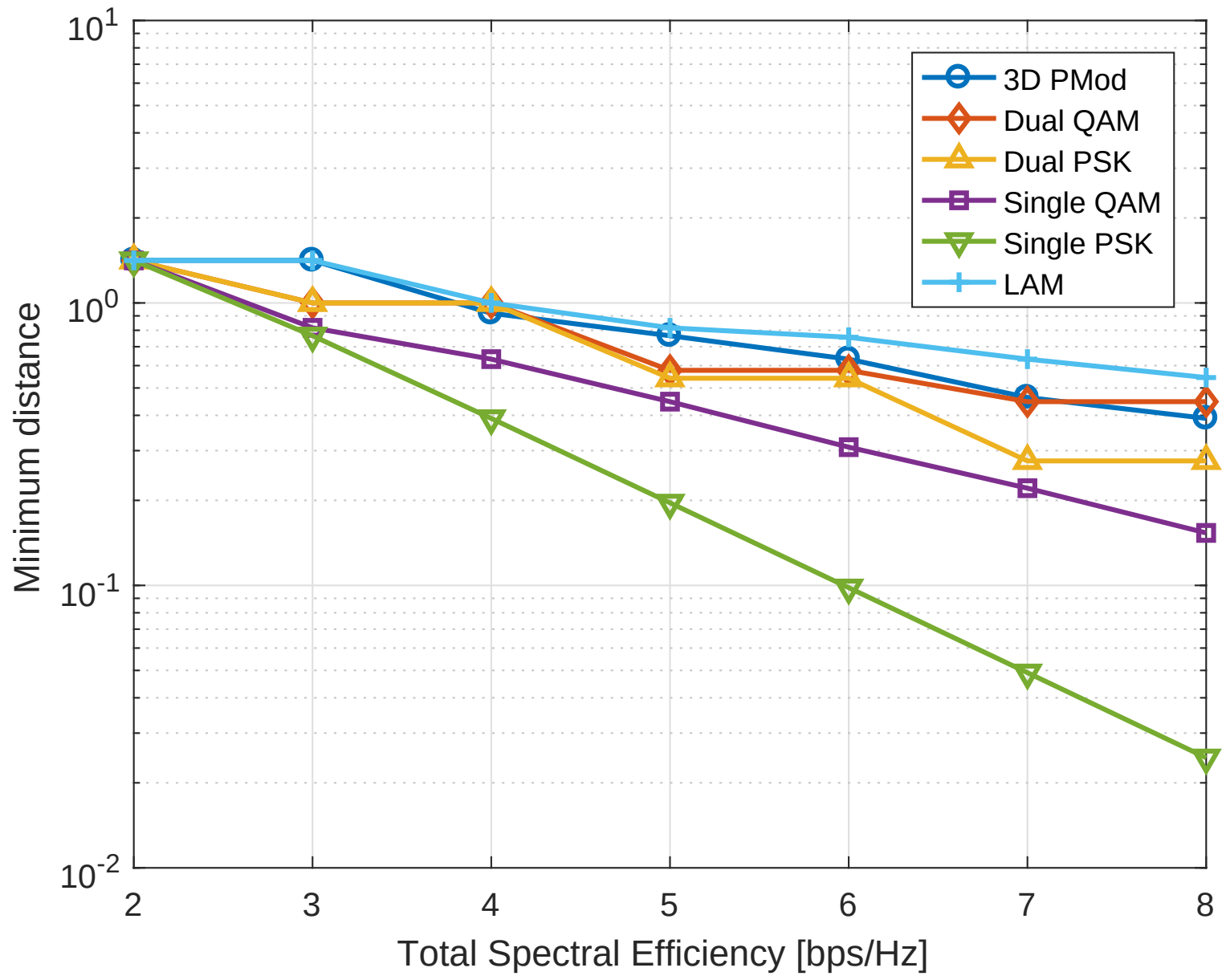

FIGURE 5.4: Maximum minimum distance for different spectral efficiencies.

- Dual PSK scheme conveys $L$-PSK and $N$-PSK constellations in each horizontal or vertical polarization.
- Single QAM conveys a $L \times N$-QAM constellation using a single polarization.

- Single PSK conveys a $L \times N$-PSK constellation using a single polarization.
- LAM conveys a $L \times N$-LAM constellation using both polarizations.

Examining the respective tables, we can conclude the following aspects:

- The minimum distance of $L \times N$ of 3D PMod is determined by the 3D PolSK contribution if $L > N$ or by the PSK contribution if $L < N$. In the case where $L = N$ both contributions are mixed.
- In the cases of Dual QAM and Dual PSK, the minimum distance is determined by the highest modulation order. Thus, it is equivalent to use the minimum distance of the constellations $\max(L, N)$-QAM/PSK, respectively.
- The minimum distance of Single QAM and Single PSK can be computed using the known formulas for a $L \times N$-QAM/PSK constellations [Pro].
- For low modulation orders (below than $8$ bits) 3D PMod achieves the maximum performance, below LAM, except $4 \times 4$.
- For asymmetric $L \times N$ schemes, higher performance if the deviation of $L$ and $N$ is lower. For instance, $4 \times 8$ and $8 \times 4$ have higher minimum distance compared with $2 \times 16$ and $16 \times 2$ for the same spectral efficiency. This is also valid for Dual QAM/PSK. This is because the minimum distance is constraint by the highest order $L$ or $N$.

It is worth mentioning that LAM is designed in such a way that the minimum distance is maximized using both polarizations. The constellation can be envisaged as a 4D constellation, where the points are placed in a hypercube. This constellation is often referred as the optimal, since achieves the highest mutual information [MKM16]. However, LAM constellations present major drawbacks compared to 3D PMod:

- The benefits of LAM are observable for $L \times N > 16$. The benefits of 3D PMod are observable for $L \times N < 256$.
- The bit mapping is not trivial and Gray mapping cannot be always applied. Due to of this, the BER performance is not always the optimal.
- LAM design is based on spherical cuts of a lattice structure. Hence, the PAPR impact is not negligible. 3D PMod has a constant joint envelope, i.e., $\mathbb{E}\{\|\mathbf{x}\|^2\} = \|\mathbf{x}\|^2 = \mathcal{E}$.
- LAM does not allow symbol multiplexing nor codeword, whereas 3D PMod does.
- LAM does not support differentiated modulation order schemes. Both polarizations constitute a single supersymbol.
- LAM receiver is constituted by a ML detector, which is computationally prohibitive.

In [HP15a] and [HP15b] we describe the communication system of 2D PMod. This is a particular case of 3D PMod, where $L = 2$. In detail, we constraint it to H/V or RHCP/LHCP. It is clear that when H/V is used, only one channel is activated, corresponding to the horizontal or vertical polarization. Thus, the BER analysis described in [RH12] can be applied straightforwardly.

In terms of minimum distance, it is determined by the PSK constellation and takes the expression of (5.25). Examining tables 5.1 it is interesting to see that 2D PMod achieves the maximum minimum distance for $2 \times 2$ and $2 \times 4$, i.e., orthogonal polarization + BPSK/QPSK constellations. For higher spectral efficiencies, modes with $L > 2$ achieve higher minimum distance. This is particularly interesting since we demonstrate that 2D PMod obtains an appreciable performance for low modulation order schemes.

## 5.4 Results

In this section we discuss the results obtained when 3D PMod is used. We implement the system described by Fig. 5.2 using different values of $L$ and $N$. We also implemented the Joint Receiver and Cascade Receiver, described in the previous sections. All symbols are encoded using Gray coding and all results are obtained with AWGN channel. The transmission power is normalized to $0$ dBW and the timing and phase synchronization is assumed perfect. The Cascade Receiver uses the MMSE filter expression (5.15). Based on the minimum distance analysis, we evaluate the following $L \times N$ modes to cover spectral efficiencies from $2$ to $7$ bps/Hz: $2 \times 2$, $2 \times 4$, $4 \times 4$, $4 \times 8$, $8 \times 8$, $16 \times 8$. Projections of the real and imaginary parts of these modes in the $H$ and $V$ components are illustrated in figs. 5.5, 5.6, 5.7, 5.8, 5.9, 5.10. The number near each point indicates the number occurrences in the other projection.

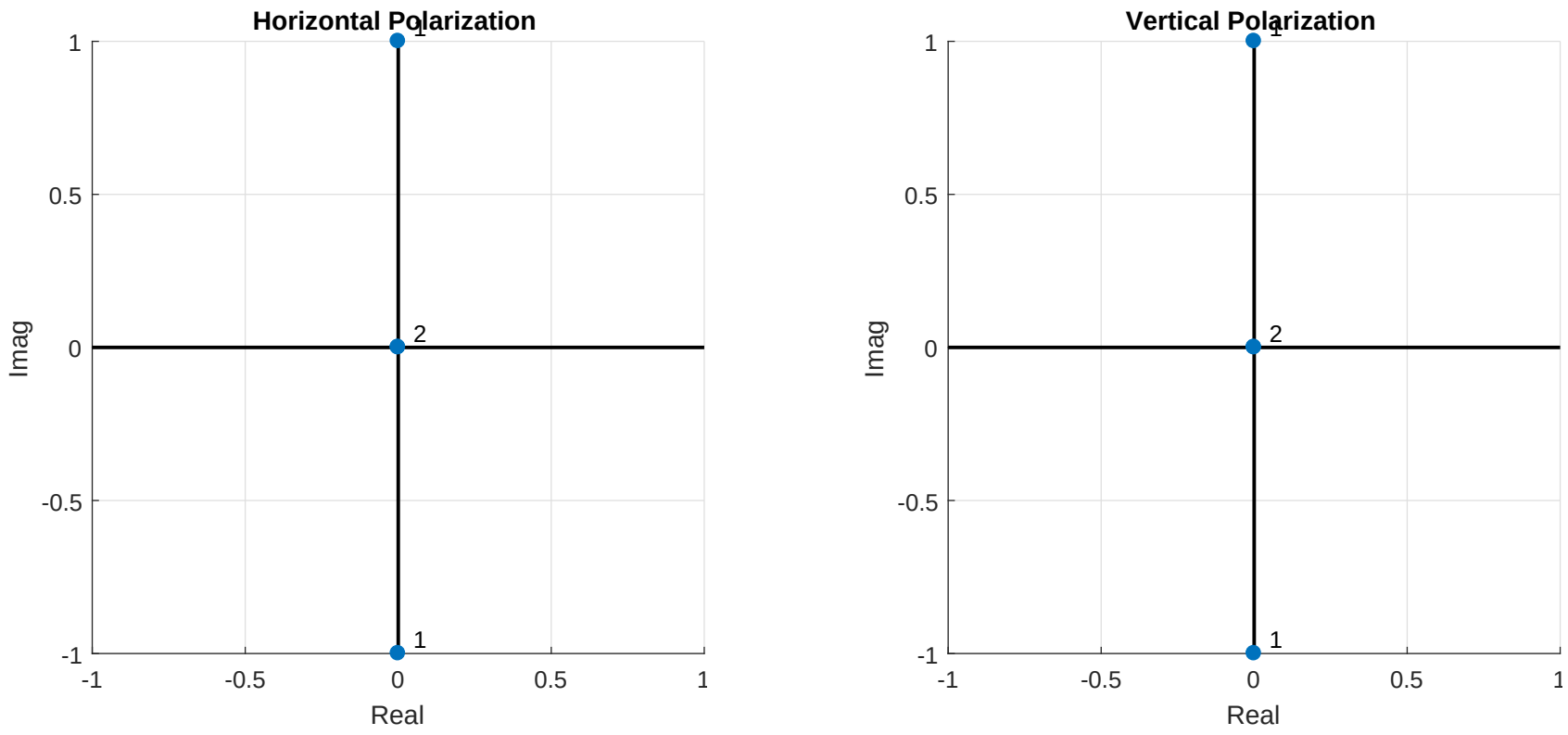

FIGURE 5.5: Projections of real and imaginary parts of $2 \times 2$ (SE 2 bps/Hz) constellation onto horizontal and vertical components.

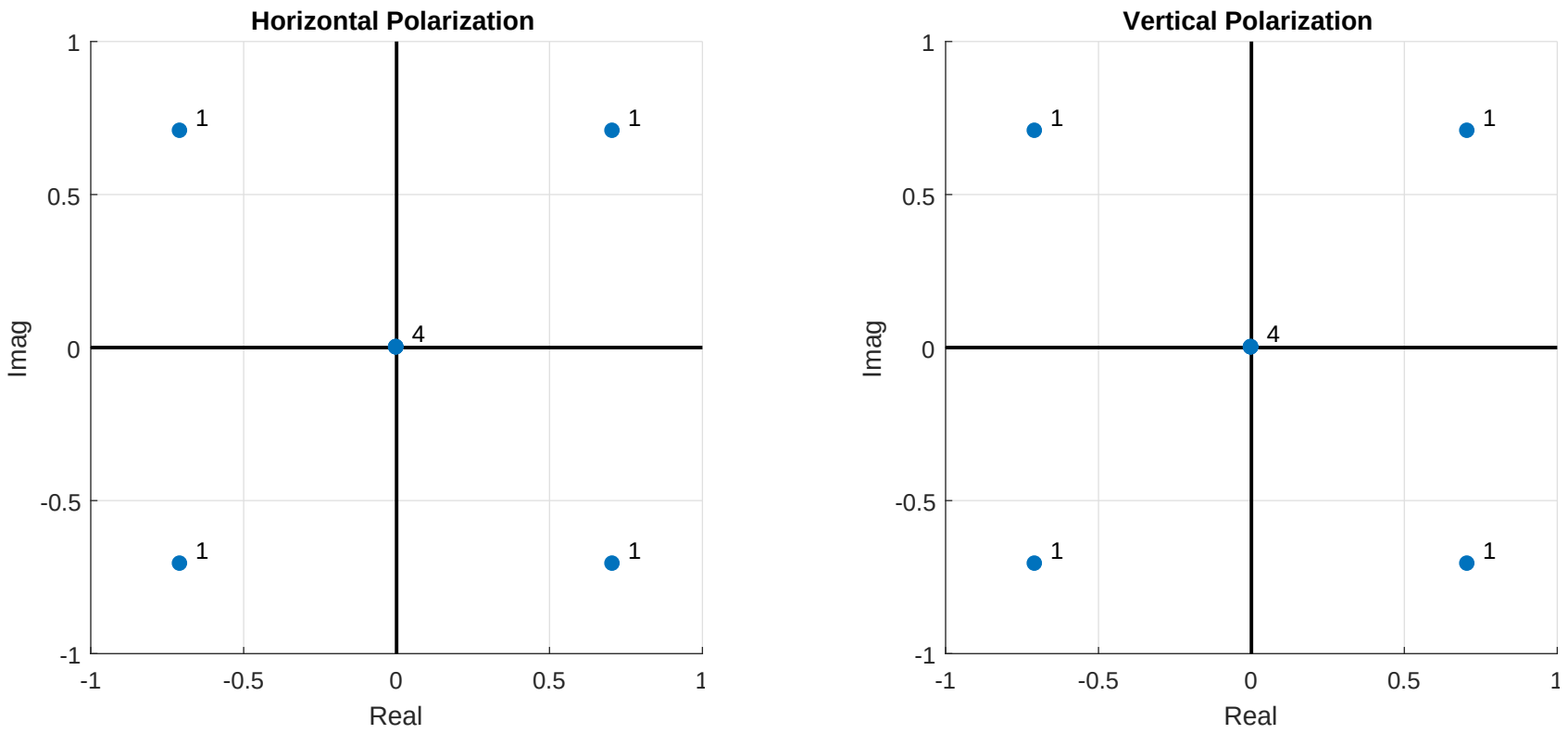

FIGURE 5.6: Projections of real and imaginary parts of $2 \times 4$ (SE 3 bps/Hz) constellation onto horizontal and vertical components.

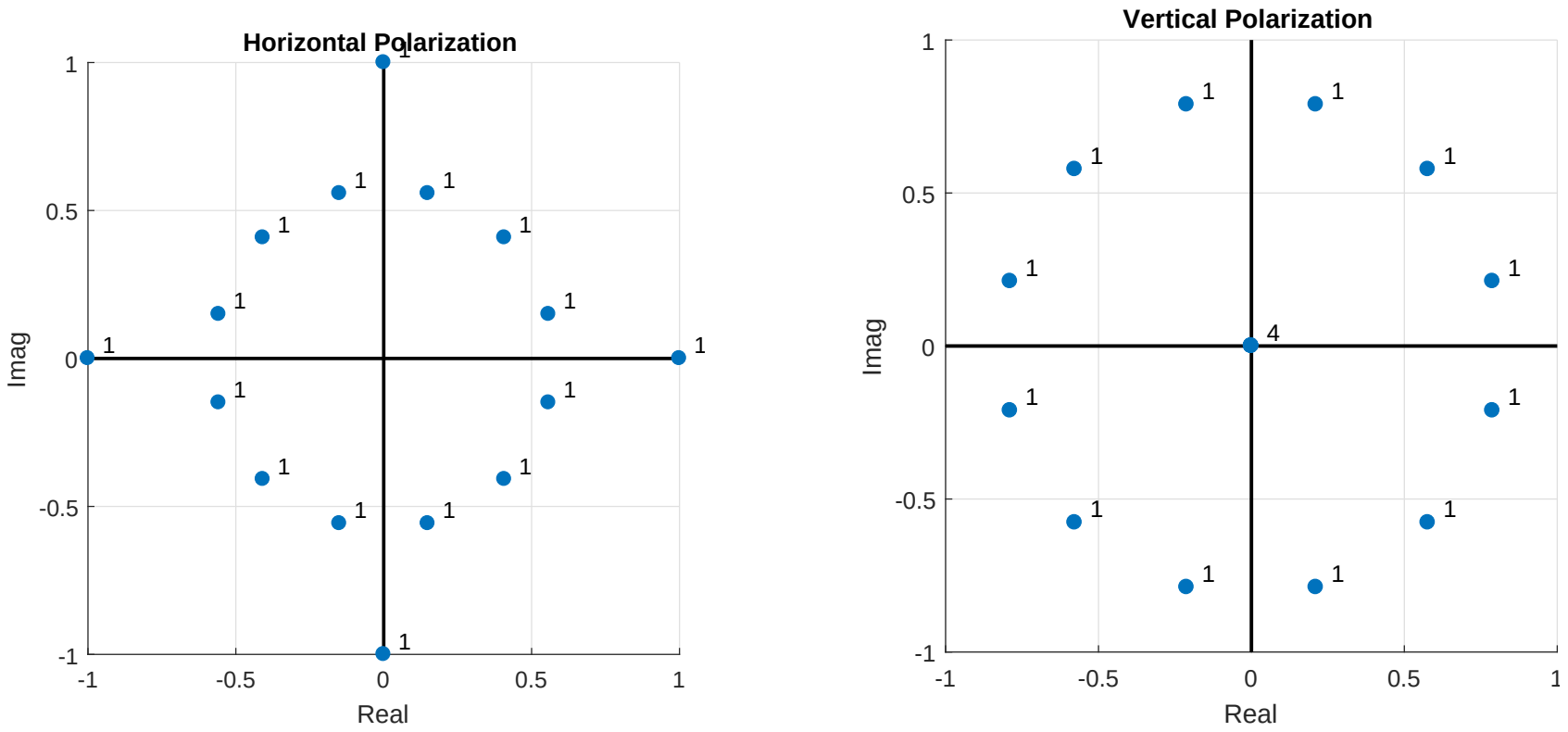

FIGURE 5.7: Projections of real and imaginary parts of $4 \times 4$ (SE 4 bps/Hz) constellation onto horizontal and vertical components.

We first perform an analysis by comparing both receivers, and depict the individual and joint BER and throughput. The throughput is obtained by counting the number of symbols decoded successfully multiplied by the number of bits carried by the symbol. This is equivalent to

$$\text{Throughput} = (L_b + N_b)\,(1 - \text{SER}) \tag{5.26}$$

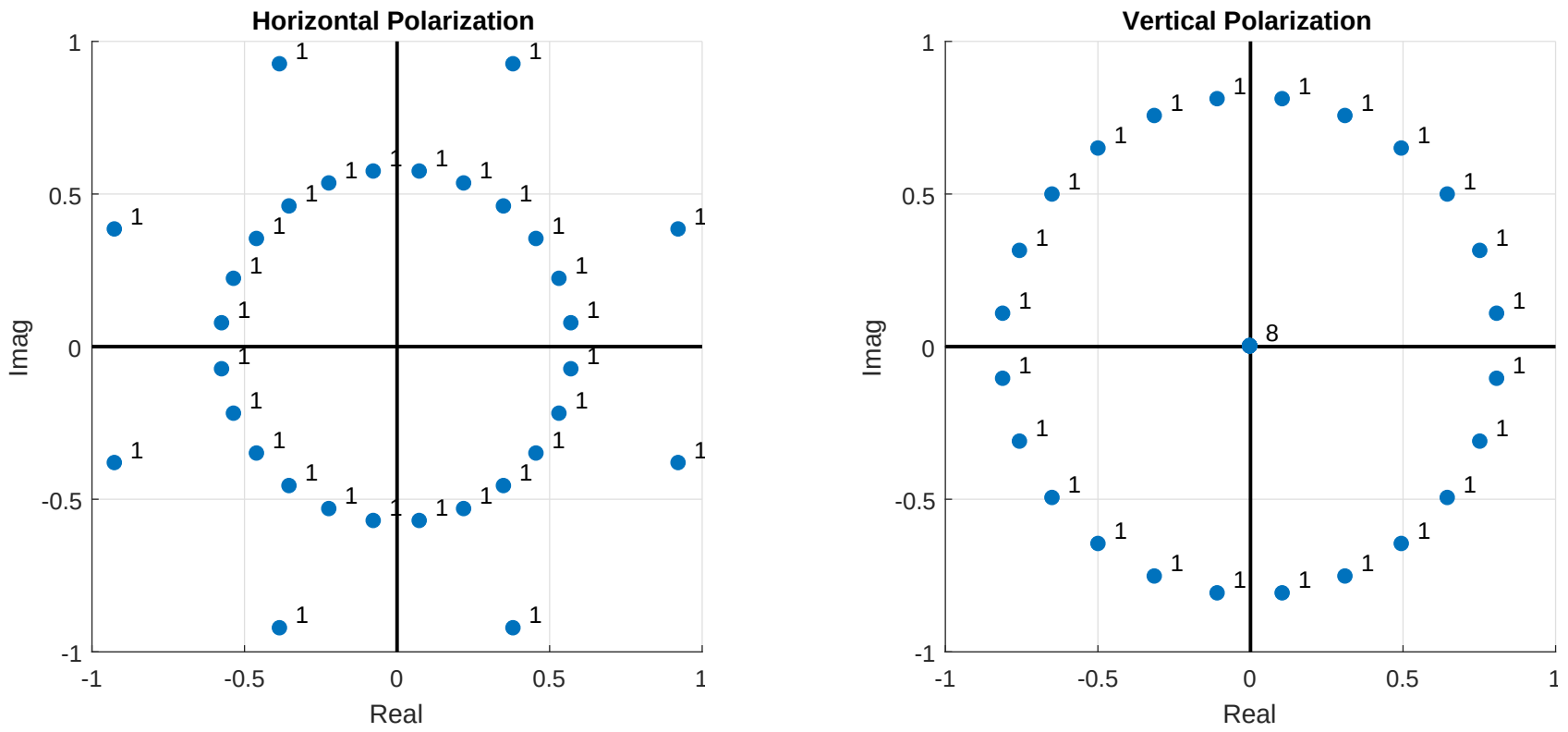

FIGURE 5.8: Projections of real and imaginary parts of $8 \times 4$ (SE 5 bps/Hz) constellation onto horizontal and vertical components.

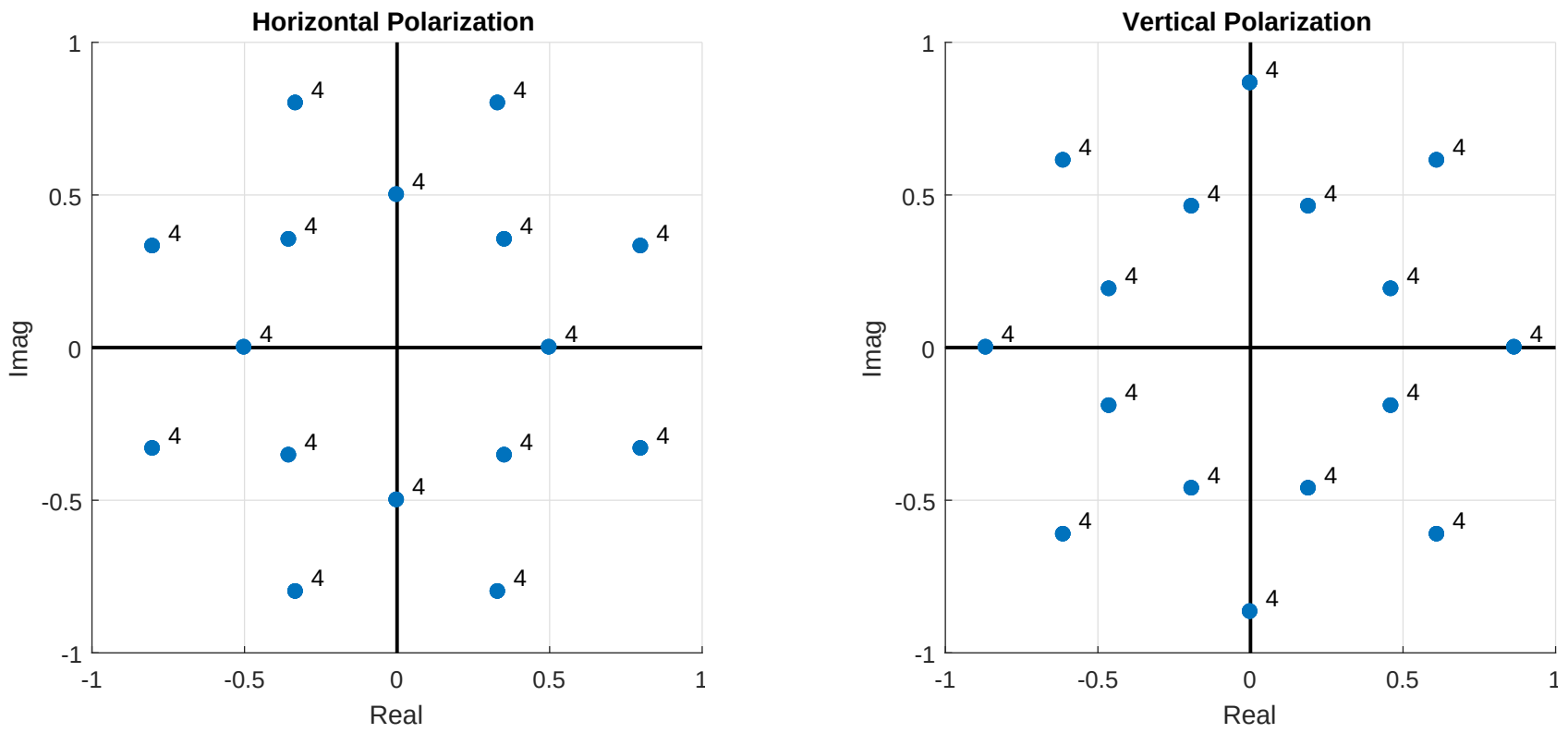

FIGURE 5.9: Projections of real and imaginary parts of $8 \times 8$ (SE 6 bps/Hz) constellation onto horizontal and vertical components.

where SER is the symbol error rate (SER). Note that SER can be computed using the XOR operator as follows

$$\begin{aligned} \text{BER} &= \frac{1}{L_b + N_b} \sum_{n=1}^{L_b+N_b} \left| b_n - \hat{b}_n \right| \\ \text{SER} &= \bigvee_{n=1}^{L_b+N_b} b_n \oplus \hat{b}_n \end{aligned} \tag{5.27}$$

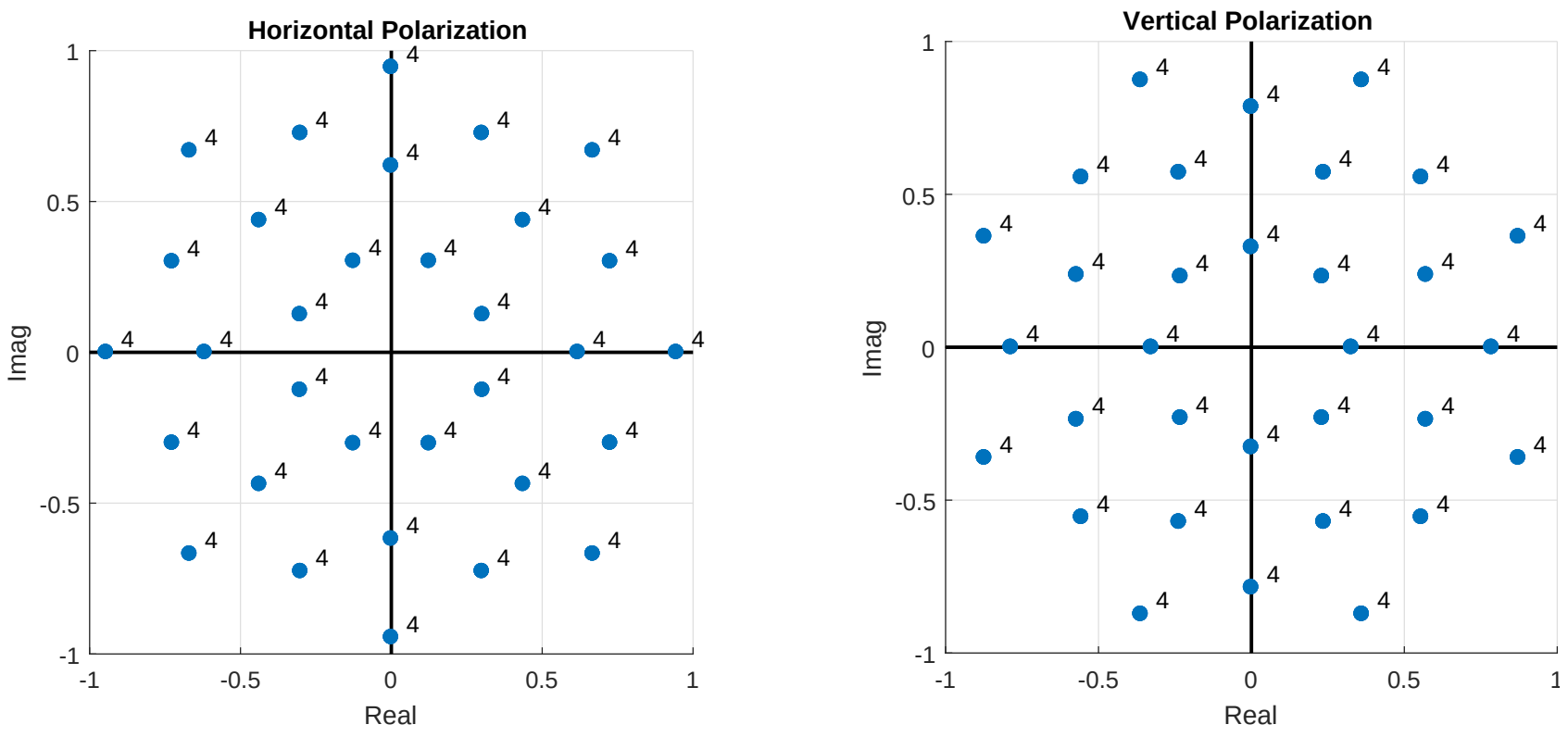

FIGURE 5.10: Projections of real and imaginary parts of $16 \times 4$ (SE 7 bps/Hz) constellation onto horizontal and vertical components.

where $\vee_{n=1}^{N} x_n = x_1 \vee \ldots \vee x_N$ performs the logic OR operation.

### 5.4.1 Comparison of Classes of Receivers

In this section we compare the performance achieved by each class of receivers. Fig. 5.11 illustrates the BER obtained by the different receivers. BERs labelled as *Joint RX*, *PolSK RX* and *PSK RX* are obtained by using the Joint receiver and the Cascade PolSK and PSK sub-receivers. The BER labelled as *PolSK RX + PSK RX* is obtained weighting the BERs of *PolSK RX* and *PSK RX* by the number of bits carried by each one. This figure shows that the Joint receiver outperforms the other schemes. Whilst the Joint receiver obtains lower BER at expenses of a higher computational complexity, the Cascade Receiver reduces drastically the computational complexity of the receiver at the expense of increasing the BER.

Fig. 5.12 compares the throughput obtained by the different classes of receivers for different $L$ and $N$ values. The total throughput of the Cascade Receiver is computed by adding the throughput obtained by each sub-receiver. As with the previous figure, the Joint receiver obtains the maximum throughput within less SNR. However, in low SNR regimes, the Cascade Receiver is able to obtain higher throughput.

### 5.4.2 Comparison of Different Modulation Orders

In this section we compare the different modulation orders in terms of BER and throughput. We use the Joint receiver as an optimal benchmark reference to compute the performance. Fig. 5.13 depicts the BER of 3D PMod using different numbers in the constellation. As expected, as the spectral efficiency increases, the minimum distance decreases, which -in turn- causes the BER to increase.

Fig. 5.14 illustrates the throughput achieved by the same sizes in fig. 5.13. This figure is particularly interesting, as we can appreciate the bitrate adaptation as a function of the SNR. The maximum throughput can be obtained by drawing the envelope of the curves.

### 5.4.3 Comparison with Other Existing Schemes

In this section we compare the performance of the proposed schemes with other existing schemes. In the following figures, all schemes have the same spectral efficiency. We recall that, in the case of asymmetric sizes ($L \neq N$), Dual QAM and Dual PSK convey a $L$-symbol constellation through horizontal polarization and $N$-symbol constellation through vertical polarization. In the case of Single QAM and Single PSK, a $L \times N$-symbol constellation is conveyed through the horizontal polarization. Finally, in the case of LAM, a $L \times N$-symbol constellation is conveyed using both polarizations.

Fig. 5.15 illustrates the BER of 3D PMod compared with the aforementioned schemes for different constellation sizes. As we analyzed previously in terms of minimum distance, 3D PMod outperforms the other conventional schemes except LAM. Compared with LAM, 3D PMod achieves a similar performance, but with a higher degree of flexibility.

Similarly, Fig. 5.16 compares the throughput of 3D PMod compared with other conventional schemes. Following the same criteria as with the previous figure, we observe that 3D PMod always outperforms the other schemes, except LAM, which is near to 3D PMod. Hence, we conclude that 3D PMod is an excellent candidate for medium and high modulation order transmissions.

### Comparison with 2D PMod

In this section we analyze the degradation of 2D PMod in front of 3D PMod. As mentioned above, 2D PMod obtains higher BER for $N > 4$, compared with the optimal case of 3D PMod for the same spectral efficiency. 2D PMod is described in [HP15b] and consists in transmitting a PSK/QAM mapped symbol activating horizontal or vertical polarization depending on the input source. In the previous section, we exposed that, in terms of minimum distance, 2D PMod achieves the higher minimum distance when a BPSK ($N = 2$) or QPSK ($N = 4$) is used as the symbol constellation, compared with other solutions.

Fig. 5.17 depicts the BER of 2D PMod and 3D PMod with optimal mode. As expected, 2D PMod performance is degraded notably when the spectral efficiency is increased. Moreover, since QAM has higher minimum distance than PSK, the performance of 2D-QAM PMod is higher than 2D-PSK PMod, though it is lower when it is compared with 3D PMod. The BER is also compared with Dual-QAM, Dual-PSK and LAM schemes.

Fig. 5.18 performs the same comparison but in terms of throughput. In this case, the throughput is degraded considerably for 2D-PSK PMod, whilst the throughput of 2D-QAM PMod does not decrease as 2D-PSK PMod.

## 5.5 Conclusions

In this chapter we present a new modulation based on the 3D constellation for polarization dimension. This modulation technique maps symbols from a sphere to the respective horizontal and vertical polarizations. This scheme is highly flexible since it allows to place an arbitrary number of symbols on the sphere and presents a low computational complexity. We describe the transmission scheme as well as two classes of receivers, depending on the performance and computational complexity trade-off. We study the analytical BER in terms of minimum distance and Union Bounds. We analyze the performance of the 3D PMod for different constellation sizes in terms of error rate as well in throughput. We compare the proposed classes of receivers and the performance of 3D PMod with other schemes such as Dual Polarization QAM multiplexing, Dual Polarization PSK multiplexing, Single Polarization QAM, Single Polarization PSK and LAM constellations. Finally, we compare the proposed 3D PMod with conventional approaches of 2D PMod. From this, we emphasize that 3D PMod obtains the highest minimum distance for spectral efficiencies below than 8 bps/Hz and in all cases, 3D PMod outperforms the other schemes, except for spectral efficiency of 4 bps/Hz. Hence, we conclude that 3D PMod is an excellent option for medium and high modulation order transmissions.

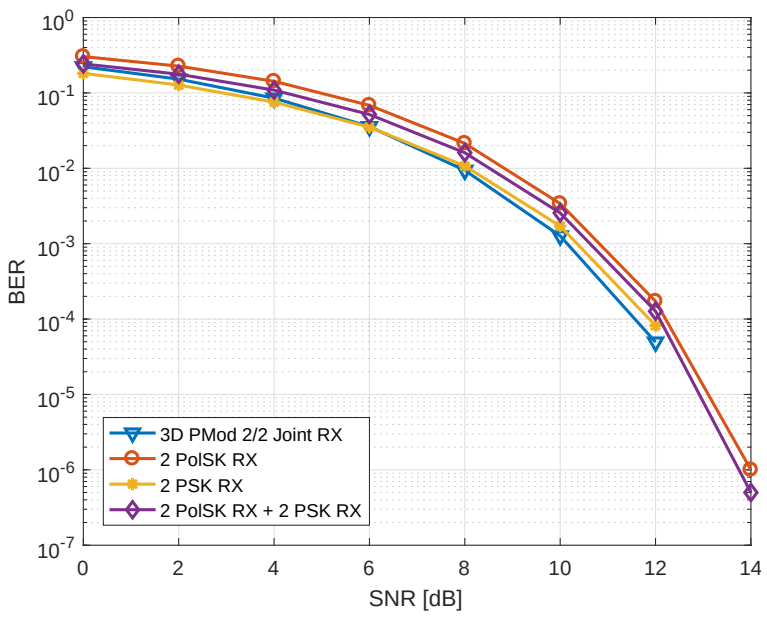

(A) SE 2 bps/Hz $L \times N = 2 \times 2$

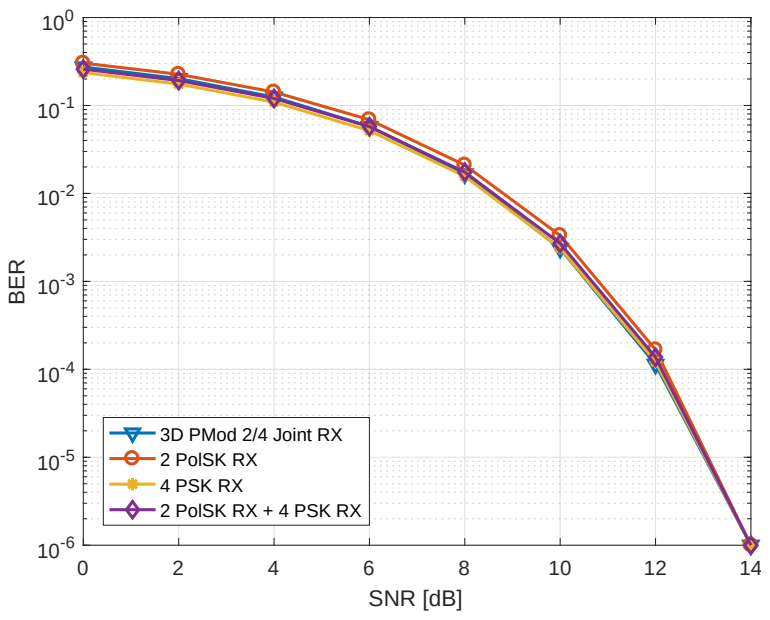

(B) SE 3 bps/Hz $L \times N = 2 \times 4$

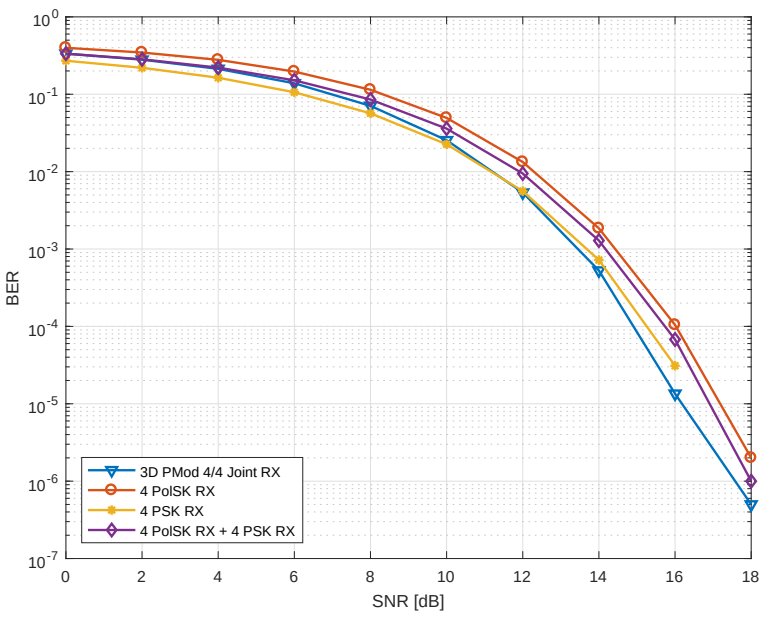

(C) SE 4 bps/Hz $L \times N = 4 \times 4$

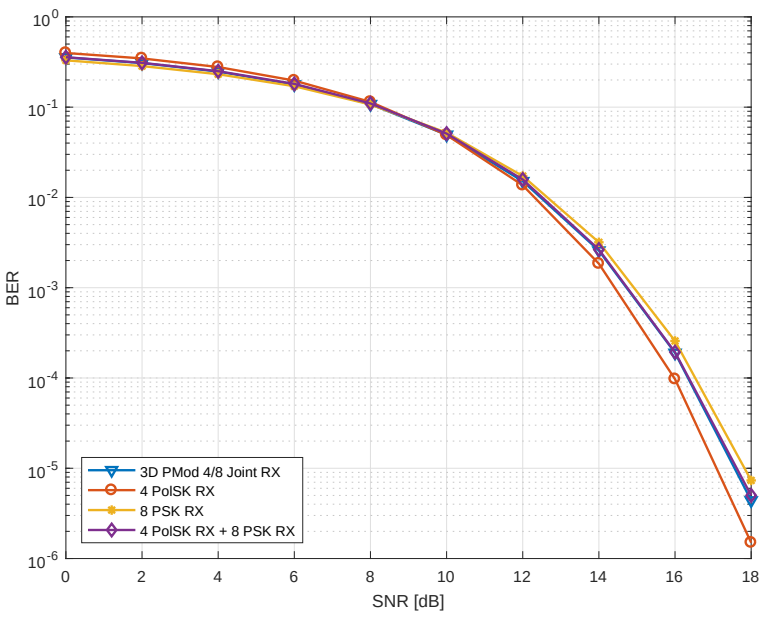

(D) SE 5 bps/Hz $L \times N = 4 \times 8$

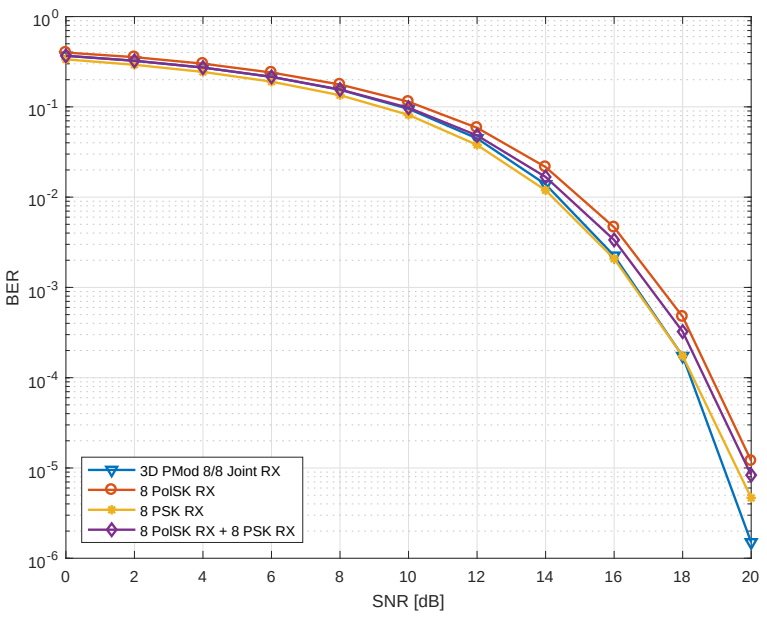

(E) SE 6 bps/Hz $L \times N = 8 \times 8$

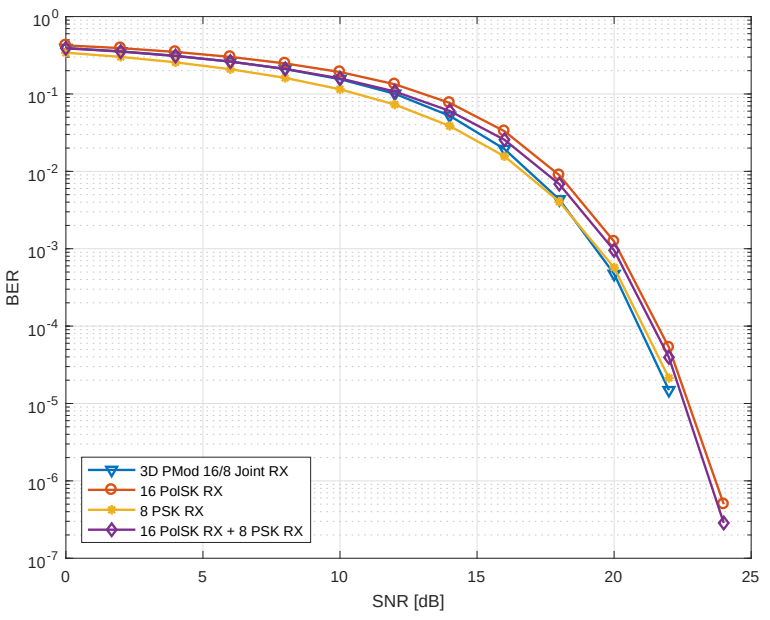

(F) SE 7 bps/Hz $L \times N = 16 \times 8$

FIGURE 5.11: Comparison of the BER of 3D Polarized Modulation for different classes of receivers. The combined BER from the Cascade subreceivers is weighted by the number of bits carried by each modulation.

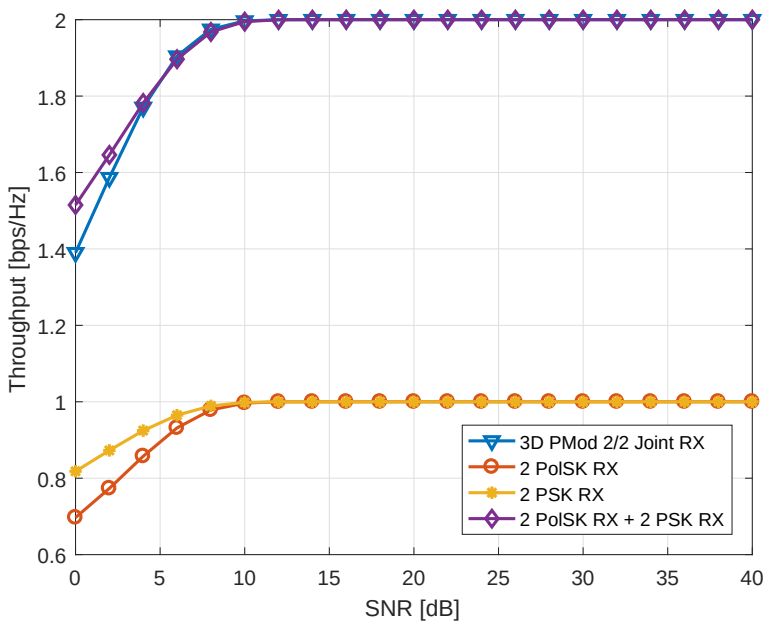

(A) SE 2 bps/Hz $L \times N = 2 \times 2$

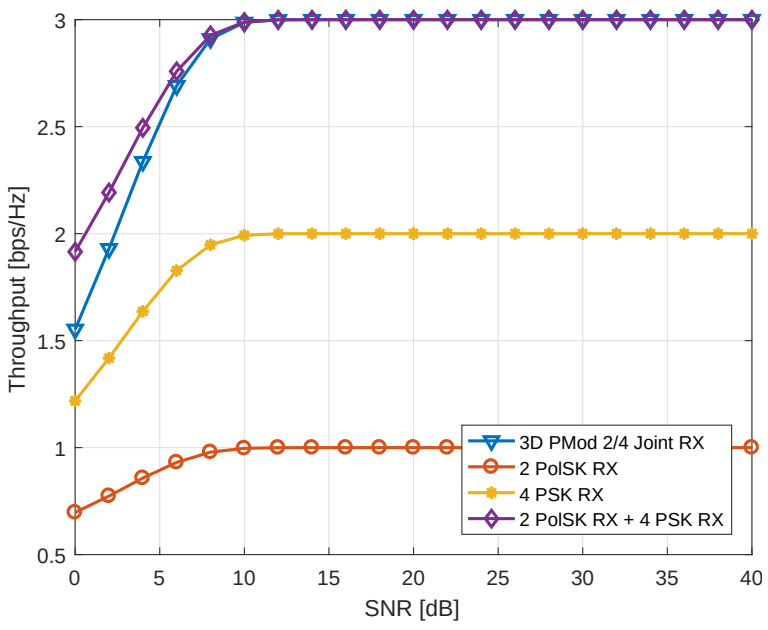

(B) SE 3 bps/Hz $L \times N = 2 \times 4$

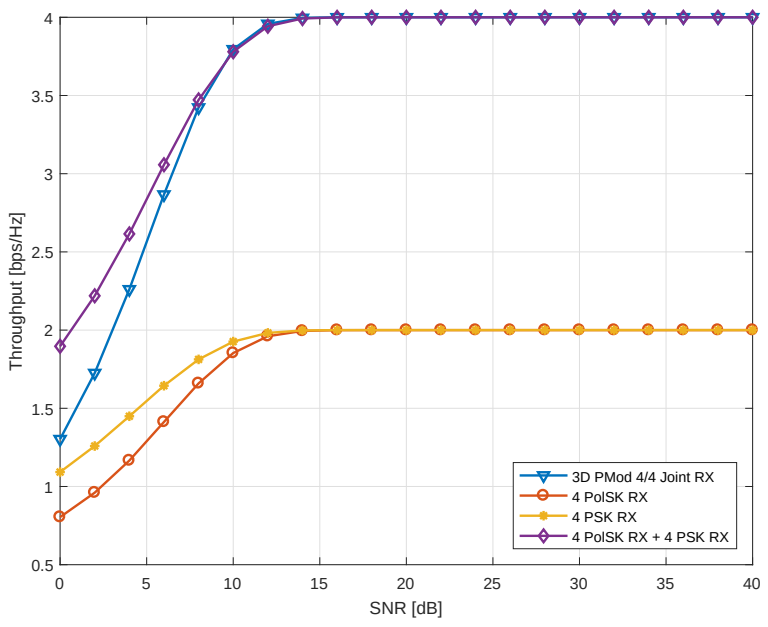

(C) SE 4 bps/Hz $L \times N = 4 \times 4$

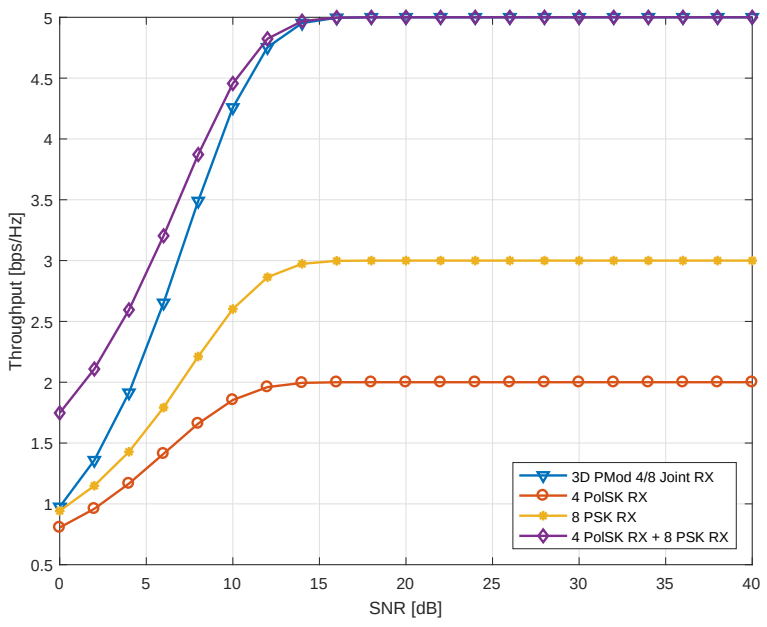

(D) SE 5 bps/Hz $L \times N = 4 \times 8$

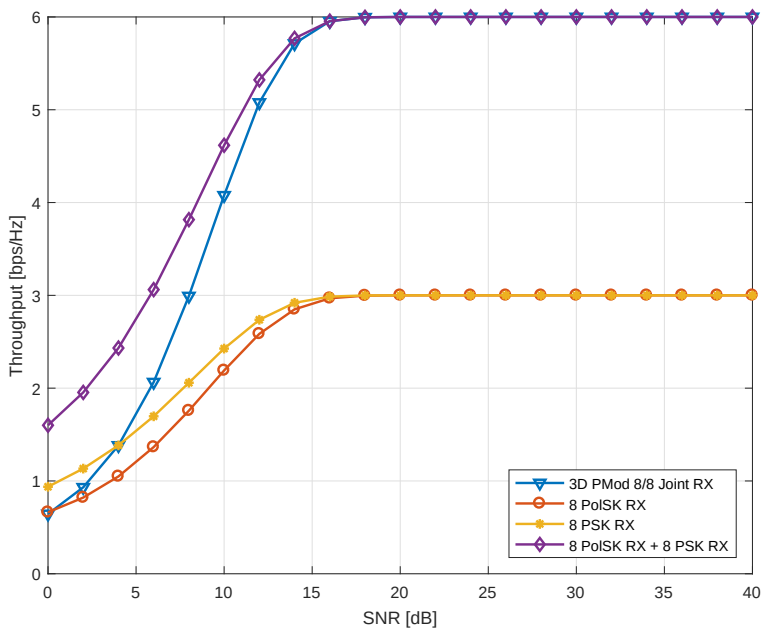

(E) SE 6 bps/Hz $L \times N = 8 \times 8$

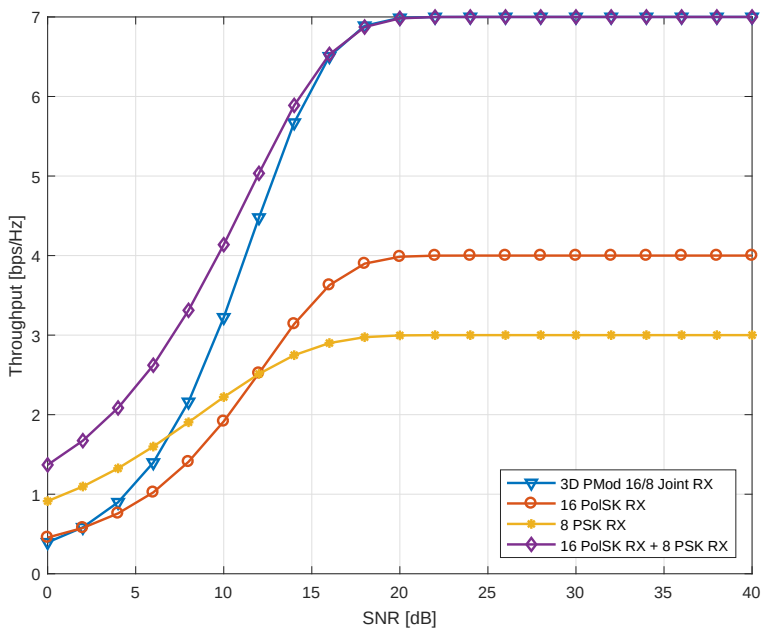

(F) SE 7 bps/Hz $L \times N = 16 \times 8$

FIGURE 5.12: Comparison of the throughput of 3D Polarized Modulation achieved by each class of receiver. The total throughput of the Cascade Receiver is computed by adding the throughput of each sub-receiver.

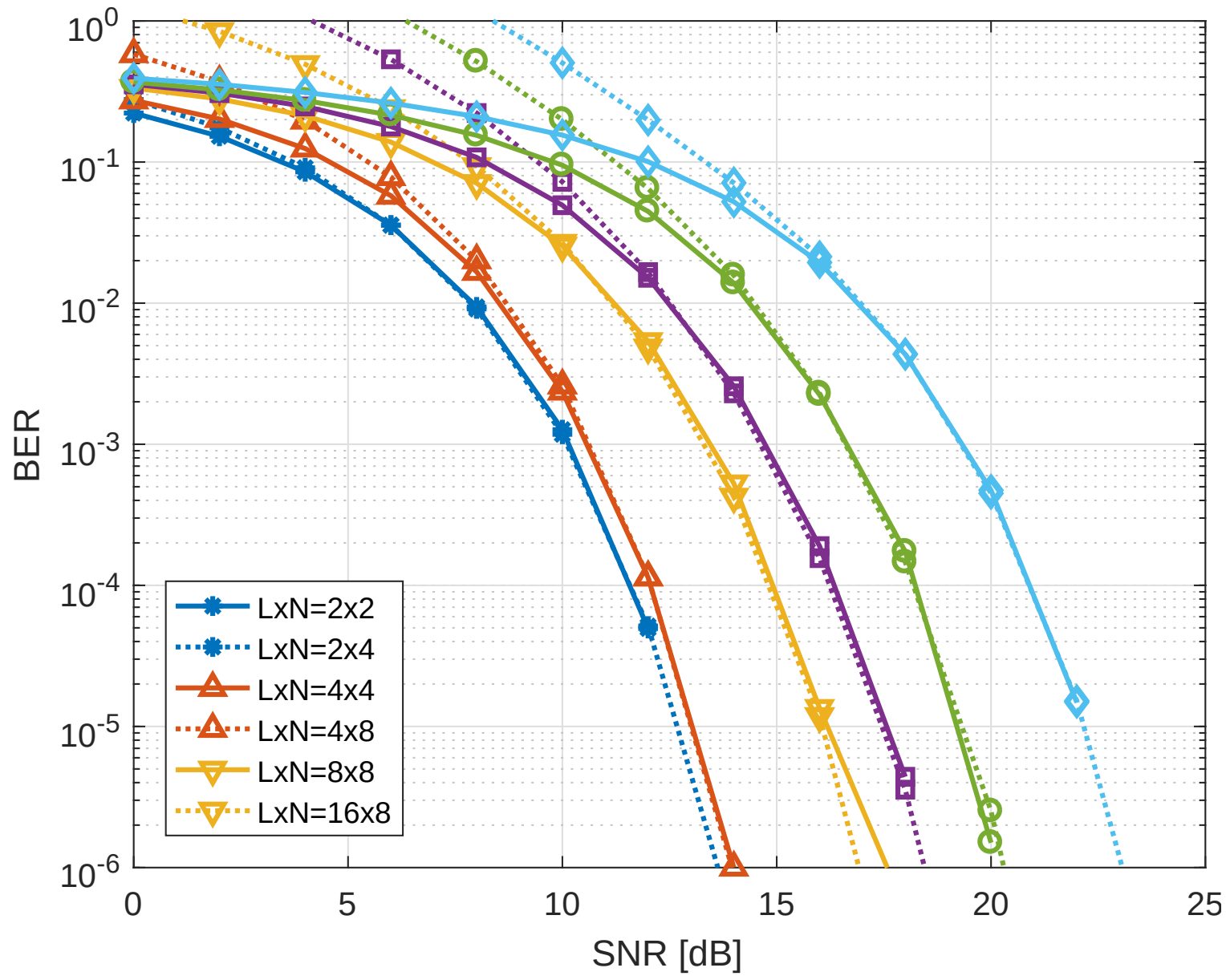

FIGURE 5.13: BER of 3D PMod for the different considered modes. Solid lines are obtained via Monte Carlo simulations. Dashed lines correspond to the Union Bound (5.16).

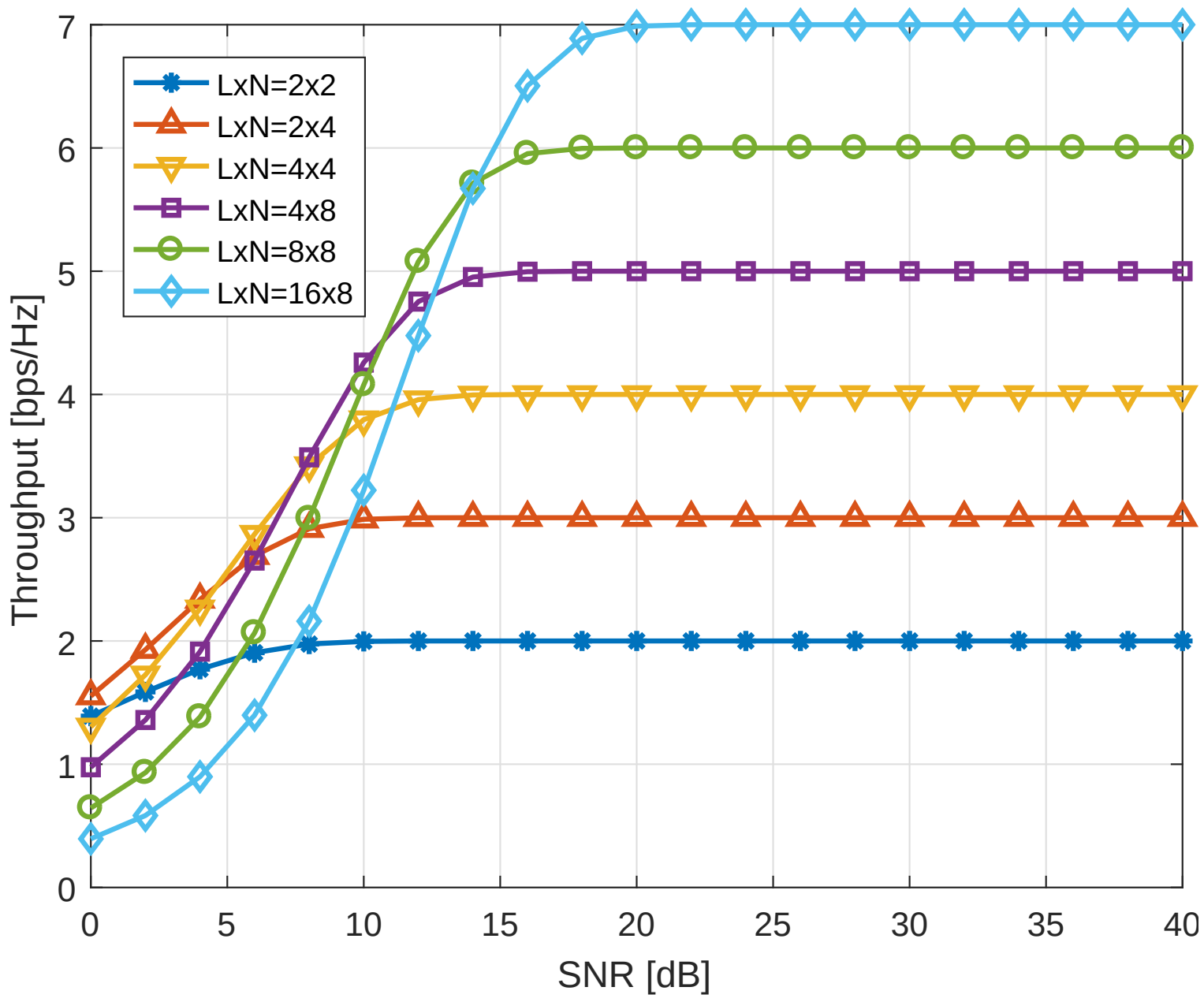

FIGURE 5.14: Throughput of 3D PMod for the different considered modes.

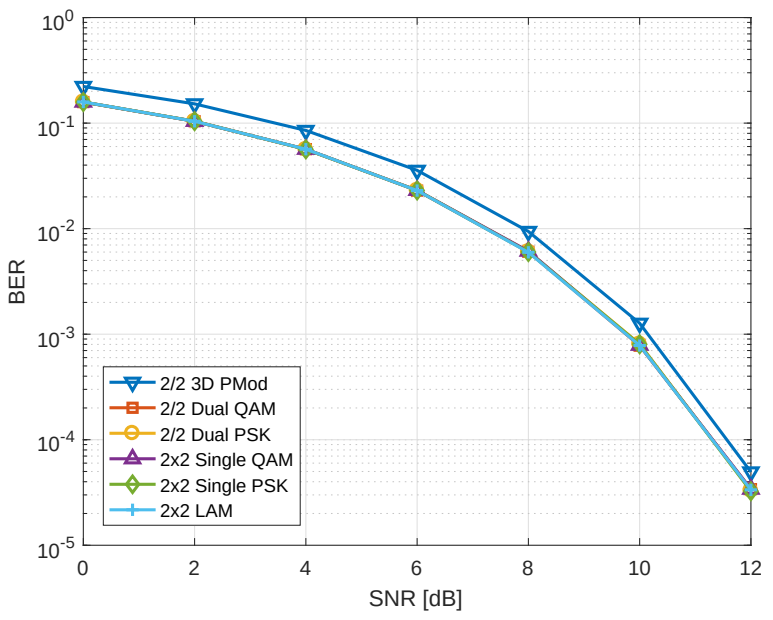

(A) SE 2 bps/Hz $L \times N = 2 \times 2$

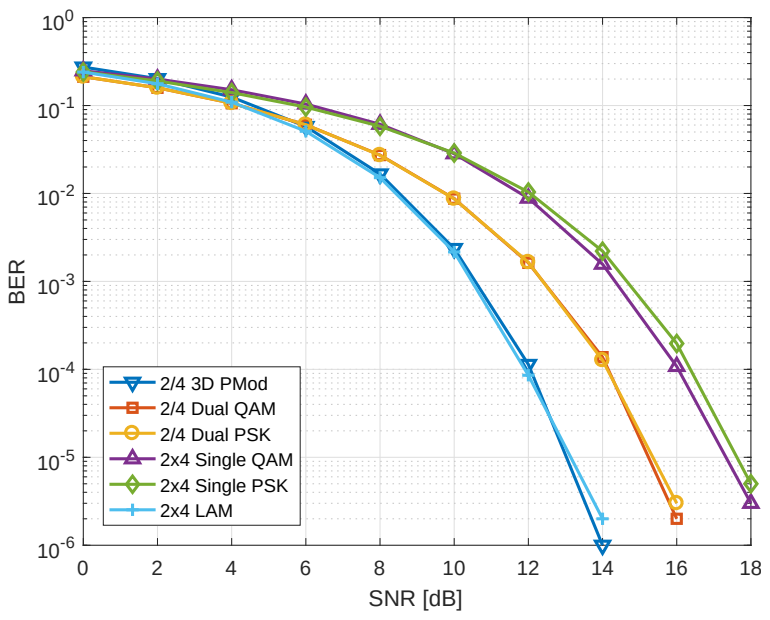

(B) SE 3 bps/Hz $L \times N = 2 \times 4$

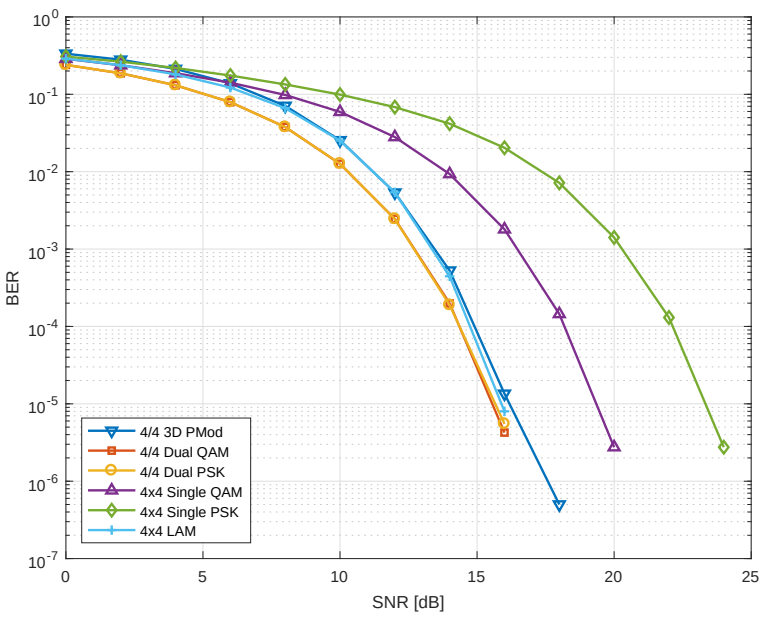

(C) SE 4 bps/Hz $L \times N = 4 \times 4$

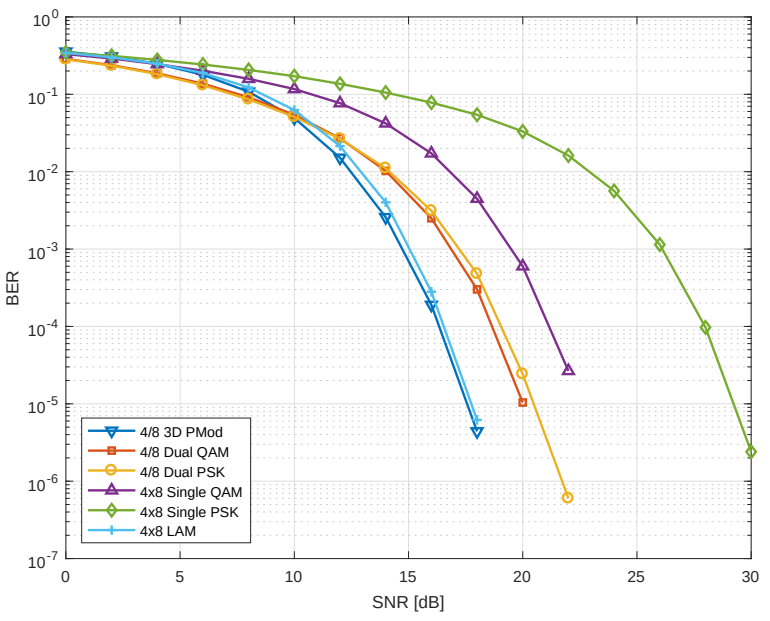

(D) SE 5 bps/Hz $L \times N = 4 \times 8$

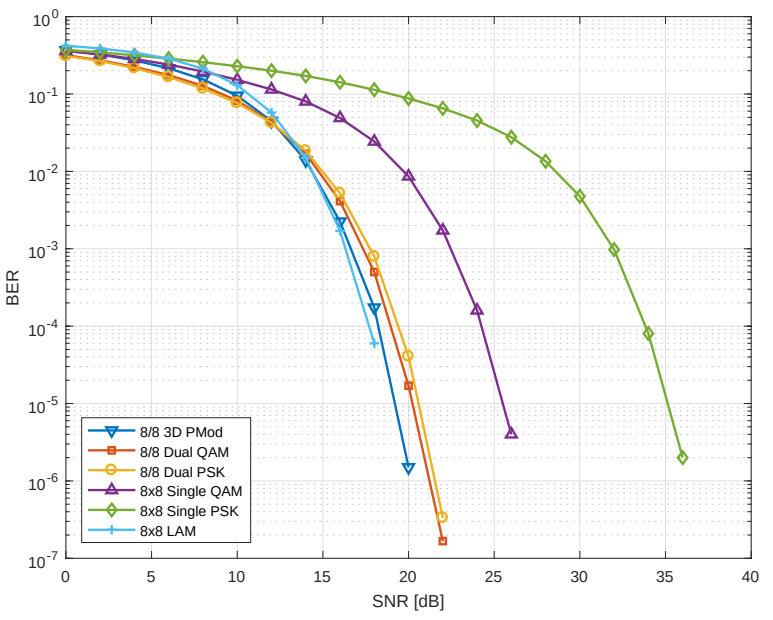

(E) SE 6 bps/Hz $L \times N = 8 \times 8$

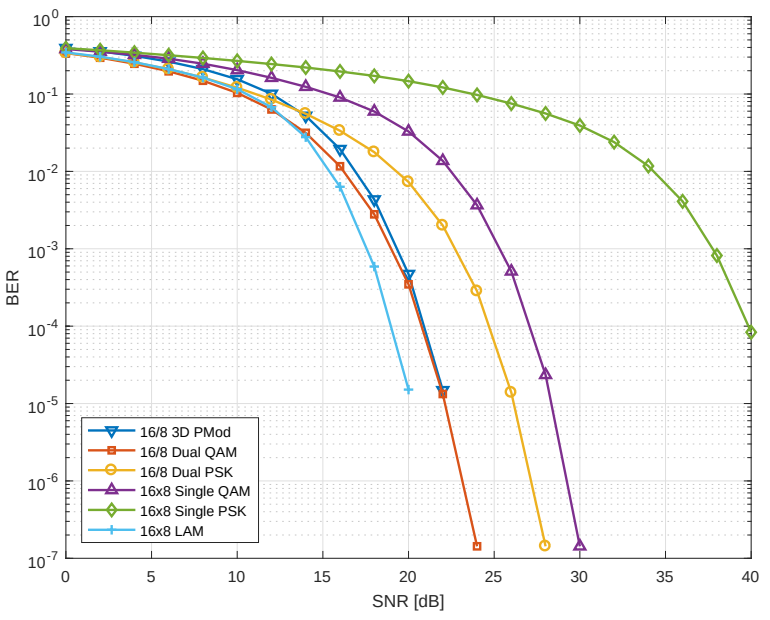

(F) SE 7 bps/Hz $L \times N = 16 \times 8$

FIGURE 5.15: BER of 3D Polarized Modulation compared with other conventional schemes.

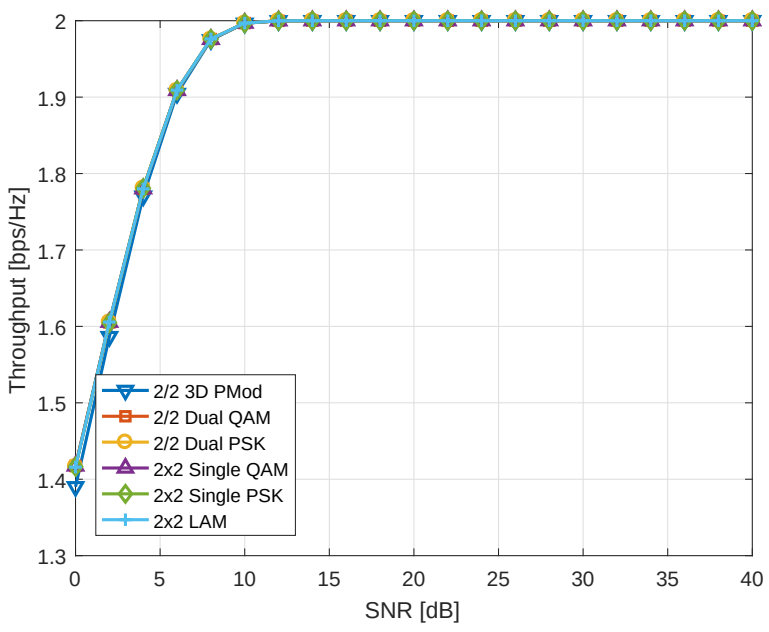

(A) SE 2 bps/Hz $L \times N = 2 \times 2$

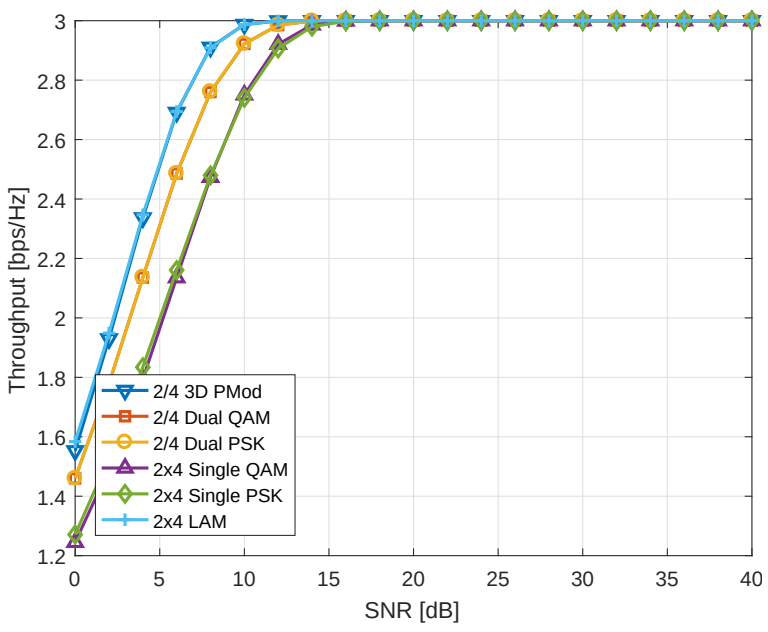

(B) SE 3 bps/Hz $L \times N = 2 \times 4$

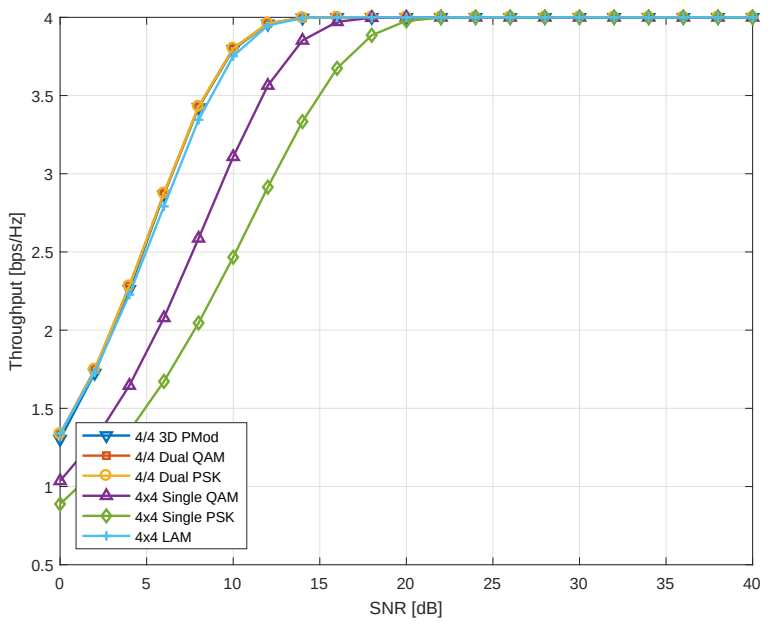

(C) SE 4 bps/Hz $L \times N = 4 \times 4$

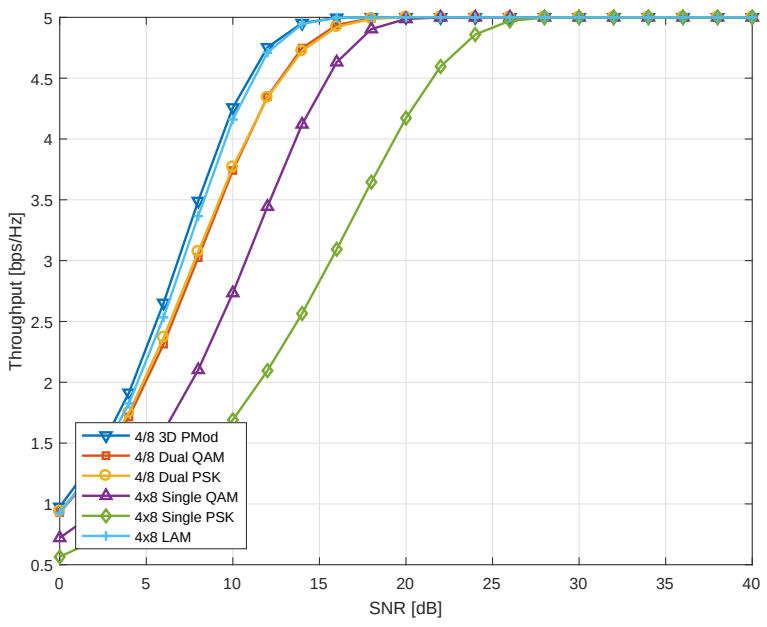

(D) SE 5 bps/Hz $L \times N = 4 \times 8$

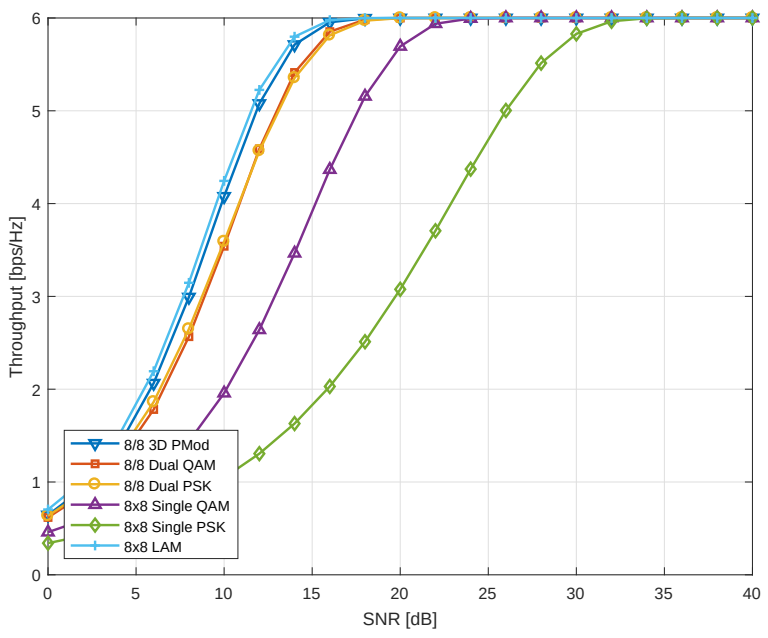

(E) SE 6 bps/Hz $L \times N = 8 \times 8$

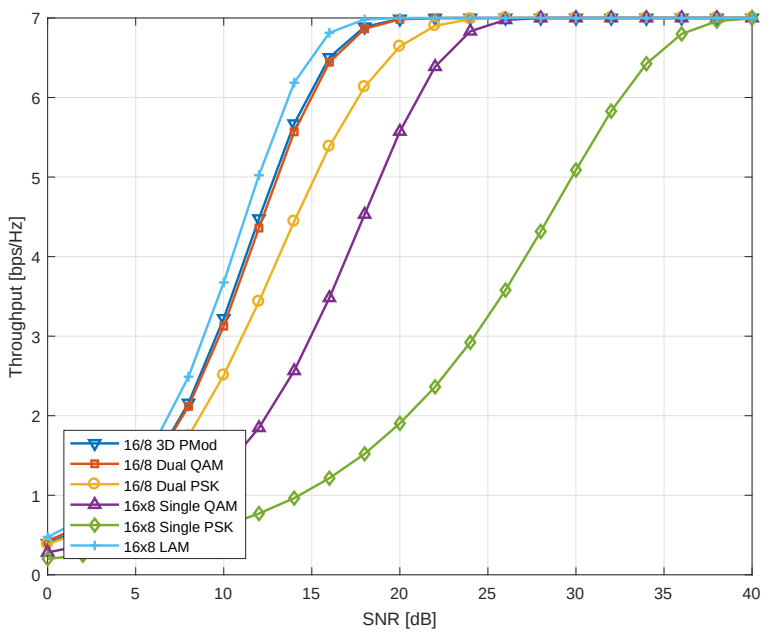

(F) SE 7 bps/Hz $L \times N = 16 \times 8$

FIGURE 5.16: Throughput of 3D Polarized Modulation compared other conventional schemes.

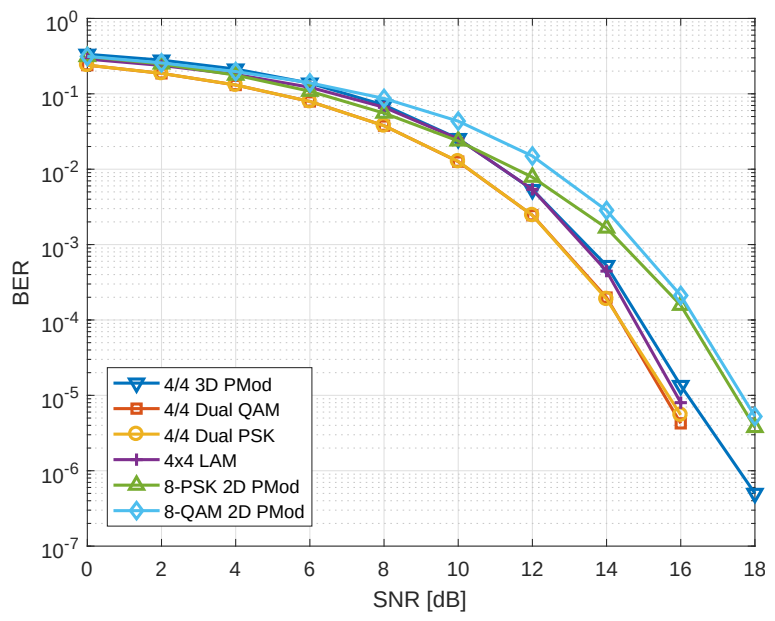

(A) SE 4 bps/Hz

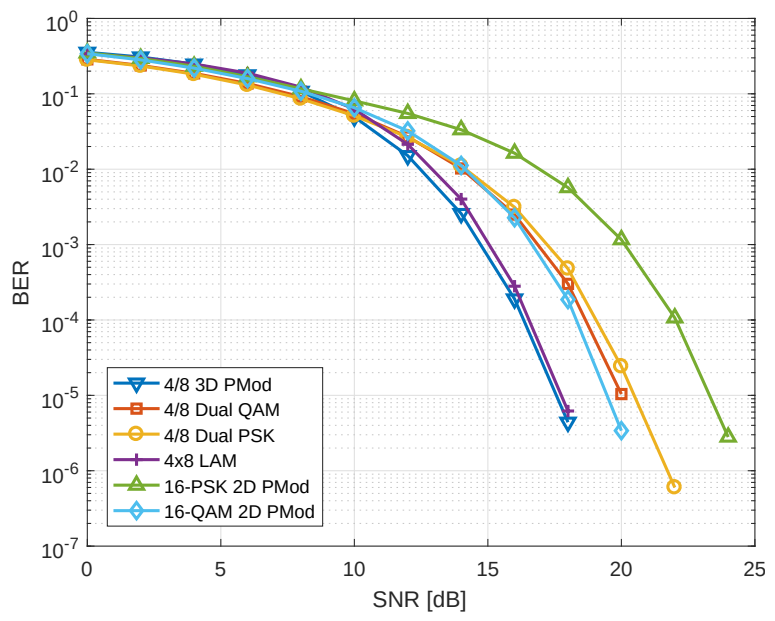

(B) SE 5 bps/Hz

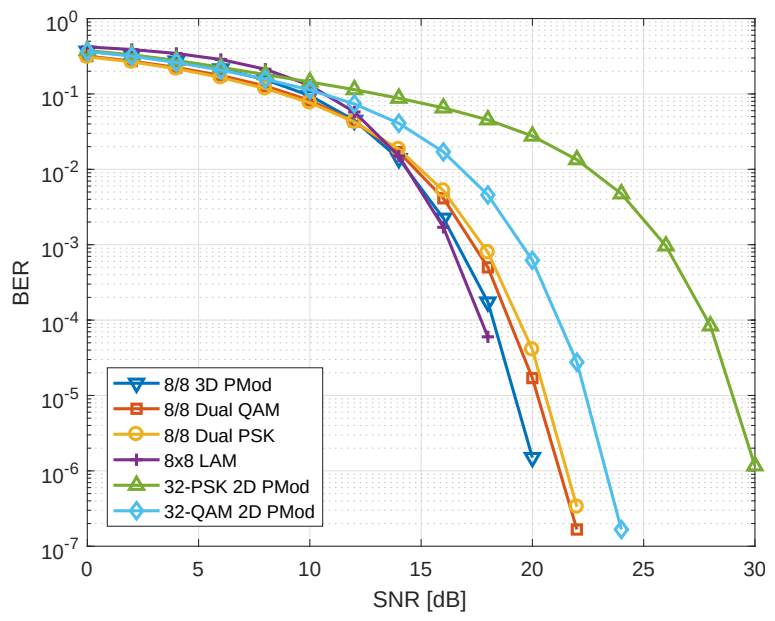

(C) SE 6 bps/Hz

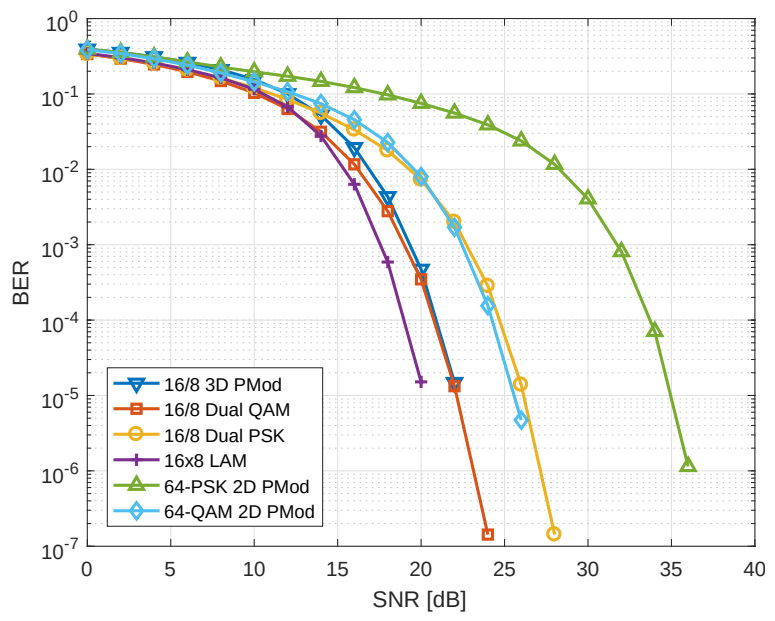

(D) SE 7 bps/Hz

FIGURE 5.17: BER of 2D PMod and 3D PMod with optimal mode for different spectral efficiencies.

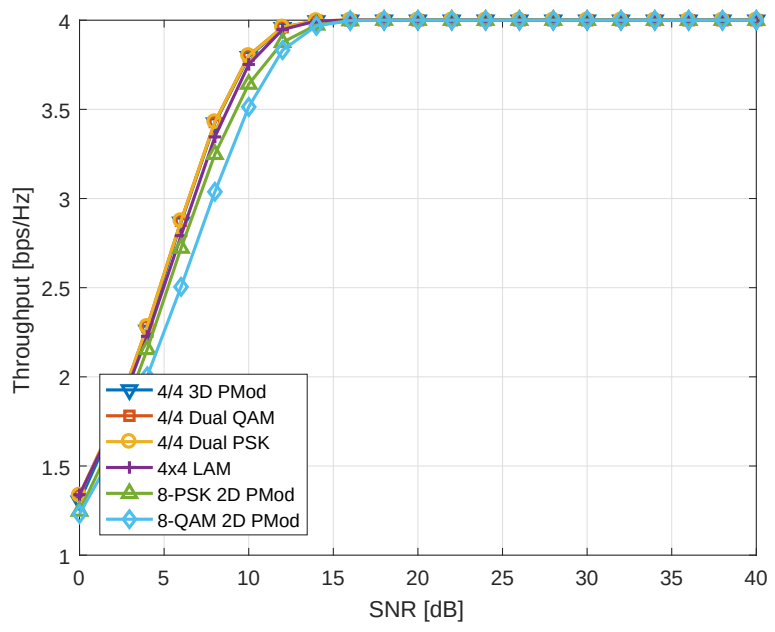

(A) SE 4 bps/Hz

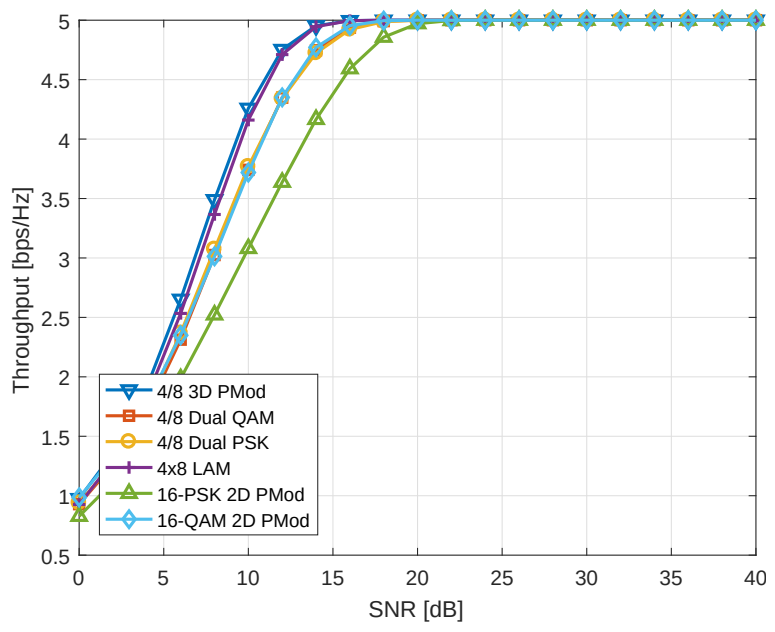

(B) SE 5 bps/Hz

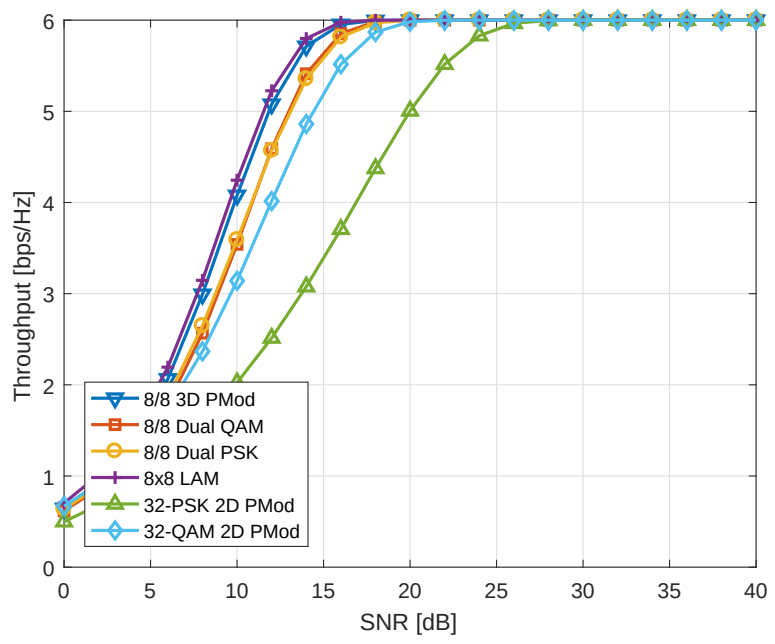

(C) SE 6 bps/Hz

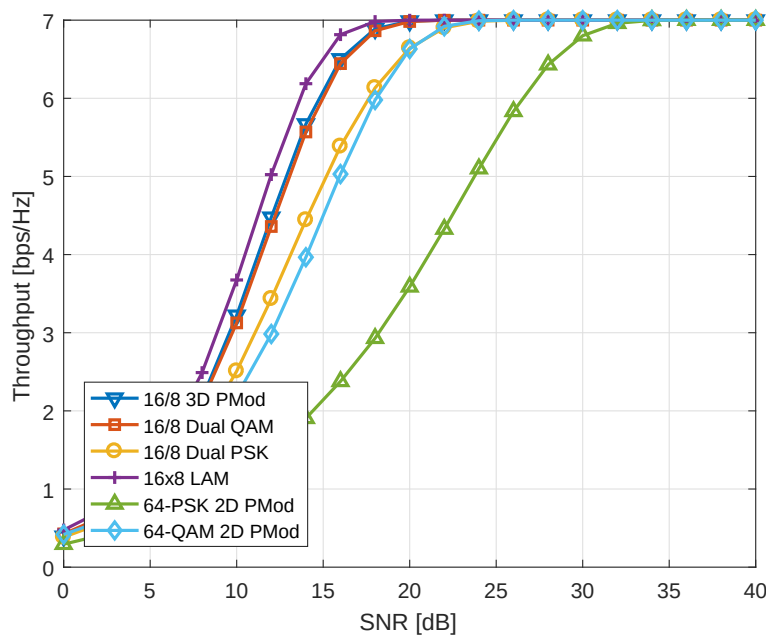

(D) SE 7 bps/Hz

FIGURE 5.18: Throughput of 2D PMod and 3D PMod with optimal mode for different spectral efficiencies.

# Chapter 6

# Adaptive Modulation, Coding and MIMO Schemes with Dual Polarization

> If quantum mechanics hasn't profoundly shocked you, you haven't understood it yet.
>
> N. Bohr

Dual polarization is introduced as a solution to double the throughput in low computational complexity communication systems. Although it has been used for many decades in fixed satellite communications, multiplexing was performed without any adaptation nor flexibility and waves dually polarized were not received simultaneously. However, it has been proven that dual polarization can also be applied to mobile satellite communications. In this way, it may be employed to increase the system capacity to increase the throughput of the individual links and increase the number of User Equipments (UE) connected to the network by taking the advantage of the partial decorrelation of the two polarizations. This approach is modeled as Multiple-Input Multiple-Output (MIMO) and it can be exploited by the MIMO signal processing techniques.

The first challenge of dual polarization systems is to provide a new communication mechanism where information can be modulated on the polarization state of the waveform whilst satisfying the specific constraints of the scenario. In order to achieve it, terminals must be able to adapt to the satellite channel. In addition, they must also be capable of feeding back information to the ground gateway regarding which modulation and coding scheme is the best for the session and which polarization scheme shall be used.

The second challenge is to implement the proposed algorithm in an existent standard-compliant system. Whilst in the previous chapter we implemented dual polarized in the BGAN standard, in this chapter we aim at implementing the proposed adaptive scheduling algorithms in the same standard. In order to achieve this, we aim at deploying the adaptive algorithms and use the BGAN standard, specified in [ETS], as a benchmark. This standard provides multimedia mobile satellite communications with low latency and high flexibility in terms of throughput. In order to test and validate the proposed schemes, we describe a Physical Layer Abstraction (PLA). Due to the long and slow shadowing, the simulation took too much time to compute the results. Thus, the PLA is a tool to model the PHY, obtain the parameters involved in the adaptation of the link and estimate the error rate without executing the entire coding and decoding chain.

## 6.1 PHY Layer Abstraction (PLA)

The goal of PLA is to obtain the instantaneous error rate in order to estimate the instantaneous throughput as a function of the radio channel coefficients. Hence, it is possible to speed up simulation time since it is not necessary to run the entire PHY signal processing chain. The model takes the modulation scheme, the coding rate, polarization scheme and other parameters to adjust bit loading depending on the magnitudes of the radio channel. PLA also offers the chance to study and analyze the impact of the feedback sent by UE.

Since most transmission schemes convey blocks of symbols that are convolved by the channel, each symbol experiences a different channel magnitude and, therefore, the Signal to Interference plus Noise Ratio (SINR) is different. Thus, a metric of effective SINR is needed. This metric maps the equivalent SINR of the transmitted block to the error rate and it is called effective SINR mapping (ESM). Hence, the ESM is defined as a function to obtain the error rate from a single value that represents the effective SINR.

From [MO06], the effective SINR is mathematically defined as

$$\bar{\gamma} = \Phi^{-1}\left(\frac{1}{N}\sum_{n}^{N}\Phi\left(\gamma_n\right)\right) \tag{6.1}$$

where $\gamma$ is the $N$-length vector of the SINR of each symbol. The function $\Phi(\cdot)$ defines the mapping of ESM. Two of them are described below.

In some cases, the representation of the error curves does not contain an analytical expression or becomes too complex. Thus, different approaches are proposed in the literature. The most relevant are:

- Exponential Effective SINR Mapping (EESM) [TW05]. This approach approximates the error curves with the Chernoff bound, which simplifies the expression as

$$\Phi(x) = 1 - e^{-x}. \tag{6.2}$$

In [ARM12], authors provide an approximation

$$\Phi(x) = 1 - \alpha_1 e^{-\beta_1 x} + (1 - \alpha_1)\, e^{-\beta_2 x} \tag{6.3}$$

where $\alpha_1$, $\alpha_2$, $\beta_1$ and $beta_2$ are tuned depending on the constellation.

- Mutual Information Effective SINR Mapping (MIESM). This approach takes the function of the capacity of the link and estimates the equivalent SINR. It is expressed as

$$\Phi(x) = \mathbb{E}_{XY} \left\{ \log_2 \frac{P\,(Y\,|X\,, x)}{\sum_{X'} P\,(X')\, P\,(Y\,|X'\,, x)} \right\} \tag{6.4}$$

where $X$ is the transmitted symbol, $Y$ is the received symbol and $\mathbb{E}\{\cdot\}$ is the expected value. Assuming that a symbol is transmitted with a $M$-ary constellation, (6.4) can be expressed as

$$\Phi(\gamma) = \log_2 M - \frac{1}{M} \sum_{x \in \mathcal{X}} \mathbb{E}_w \left\{ \log_2 \sum_{x' \in \mathcal{X}} e^{-\frac{|(x - x') + w|^2 - |w|^2}{\sigma^2}} \right\}, \tag{6.5}$$

where $\mathcal{X}$ is the set of the constellation and $w \sim \mathcal{CN}\left(0, \sigma^2\right)$ and $\sigma^2 = 1/\gamma$.

This expression can be computed off-line via Montecarlo simulations generating different realizations of the random variable. Nevertheless, in [Sri+08] different results are exposed by QPSK, 16QAM and 64QAM, for a range of $[-20, 27]$ dB of SINR. Although a closed expression is not provided, it is possible to compute this expression for different values and store the results in a lookup table (LUT) in order to find the values of $\Phi^{-1}(x)$ [RAM14].

- Received Bit Mutual Information Rate (RBIR) ESM. This approach is a normalization of the MIESM by the number of

received bits of each symbol. Thus,

$$\text{RBIR} = \frac{\sum_n^N \Phi(\gamma_n)}{\sum_n^N M_n}, \tag{6.6}$$

where $\Phi(\gamma_n)$ is defined in (6.5) and $M_n$ is the number of bits of $n$th symbol.

Whilst EESM is very attractive due to its simplicity and closed expression, MIESM and RBIR are better approximated in models where MIMO or hybrid automated repeat request (HARQ) are used. Finally, once the ESM method is defined, it is possible to use the error curves for the AWGN channel depending on the $\bar{\gamma}$ as a function of the channel to obtain the throughput of the system.

Fig. 6.1a and fig. 6.1b describe the curves of $\Phi(\gamma)$ and RBIR for QPSK, 16QAM, 32QAM and 64QAM defined in [ETS].

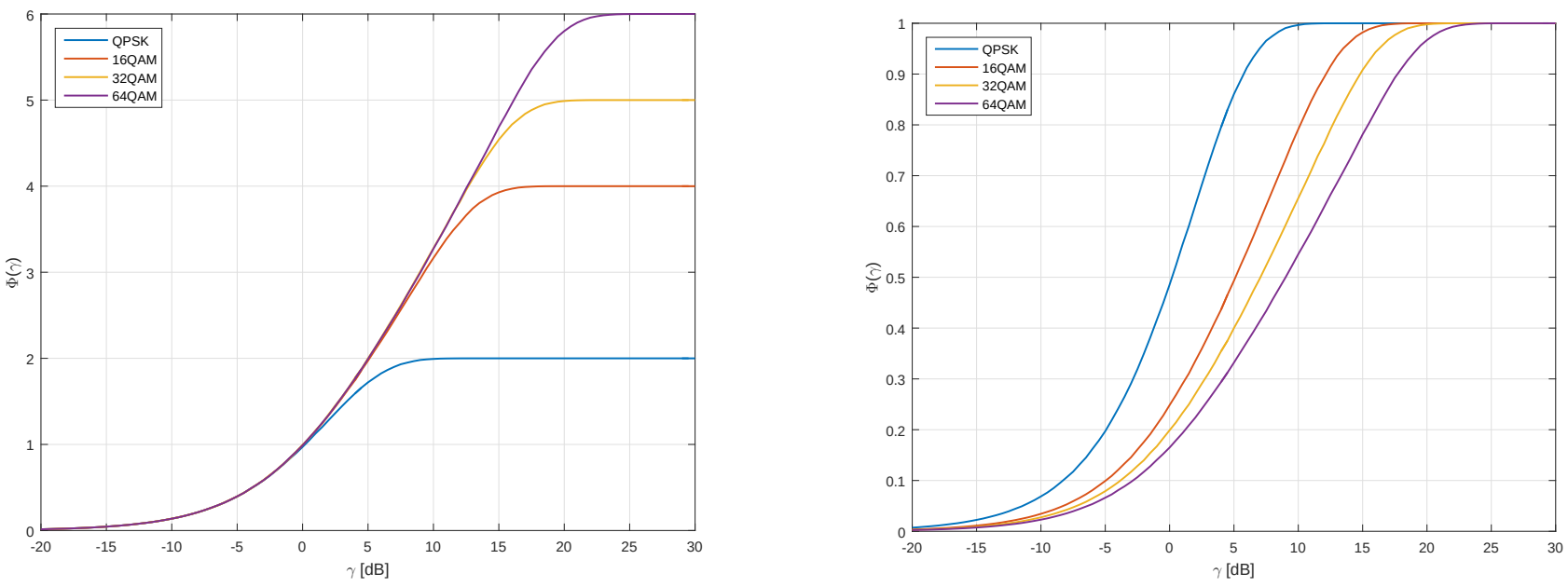

(A) MIESM

(B) RBIR

FIGURE 6.1: MIESM and RBIR curves for each constellation defined in [ETS].

## 6.2 PLA and MIMO

In the previous section we described the PLA for the Single-Input Single-Output (SISO) scenarios. In the case of dual polarized scenarios, the performance of the previous abstraction depends on the implementation of the receiver. In [Sri+08] two approaches are proposed depending on the receiver:

- Linear MIMO Receivers. The use of linear receivers allows low computational complexity implementations and offers the chance to suppress or mitigate the cross interference of the inputs. Thus, without loss of generality, the receiver can decouple both polarizations into two separate streams.
- Maximum Likelihood (ML) Receivers. In this approach, (6.6) is rewritten as a function of the probability of log-likelihood ratio (LLR). However, this approach requires much more computational complexity and additional LUTs, which enlarges the required memory.

### 6.2.1 RBIR for ML Receivers

In order to apply the SINR mapping in the case of Polarization Multiplexing (PM), hereinafter the metrics and parameters are referred to the $j$th stream. In contrast to (6.5), $\Phi(\gamma)$ is defined as

$$\Phi(\gamma) = \frac{1}{M} \sum_{m=1}^{M} \int_{-\infty}^{+\infty} f_{\Lambda_m}(x) \log_2 \left( \frac{M}{1 + e^{-\Lambda_m}} \right) \mathrm{d}x \tag{6.7}$$

where $\Lambda_i$ is the log-likelihood ratio (LLR) of the $i$th bit and $f_{\Lambda_m}(p)$ is its probability density function (pdf). The pdf can be approximated as a Gaussian and the integral can be reduced using the numerical integration of [Bec07]. Thus, (6.7) can be approximated

as

$$
\begin{aligned}
\Phi(\gamma) &\simeq \log_2 M - \frac{1}{\log_e 2} J \\
J &= \frac{J_a + J_b}{2} + \frac{J_a - J_b}{2} \operatorname{sign}(0.65 - J_b) \\
J_a &= \sqrt{\mathrm{VAR}} \left( \frac{-\eta}{2} \operatorname{erfc}\left( \frac{\eta}{\sqrt{2}} \right) + \frac{1}{\sqrt{2\pi}} e^{-\frac{\eta^2}{2}} \right) \\
\eta &= \frac{\mathrm{AVE}}{\sqrt{\mathrm{VAR}}} \\
J_b &= \frac{2}{3} f(\mathrm{AVE}) + \frac{1}{6} f\left(\mathrm{AVE} + \sqrt{3\mathrm{VAR}}\right) + \frac{1}{6} f\left(\mathrm{AVE} - \sqrt{3\mathrm{VAR}}\right) \\
f(x) &= \log_e\left(1 + e^{-x}\right) \\
\mathrm{AVE} &= d_{min}^2 \frac{\|\mathbf{h}_j\|^2}{\sigma_{\mathbf{h}}^2} - \mathbb{E}\{K_j\} \\
\mathbb{E}\{K_j\} &= \int_{-\infty}^{+\infty} \frac{1}{\sqrt{2\pi d_{min}^2 \frac{\|\mathbf{h}_j\|^2}{\sigma_{\mathbf{h}}^2}}} e^{-\frac{1}{2} \frac{x^2}{d_{min}^2 \frac{\|\mathbf{h}_j\|^2}{\sigma_{\mathbf{h}}^2}}} \log_e \left( 2e^{-x} + e^{-d_{min}^2 \frac{\|\mathbf{h}_j\|^2}{\sigma_{\mathbf{h}}^2}} e^{-2x} \right) \mathrm{d}x \\
\mathrm{VAR} &= \mathbb{E}\left\{K_j^2\right\} - \mathbb{E}\{K_j\}^2 \\
\mathbb{E}\left\{K_j^2\right\} &= \int_{-\infty}^{+\infty} \frac{1}{\sqrt{2\pi d_{min}^2 \frac{\|\mathbf{h}_j\|^2}{\sigma_{\mathbf{h}}^2}}} e^{-\frac{1}{2} \frac{x^2}{d_{min}^2 \frac{\|\mathbf{h}_j\|^2}{\sigma_{\mathbf{h}}^2}}} \left[ \log_e \left( 2e^{-x} + e^{-d_{min}^2 \frac{\|\mathbf{h}_j\|^2}{\sigma_{\mathbf{h}}^2}} e^{-2x} \right) \right]^2 \mathrm{d}x
\end{aligned}
\tag{6.8}
$$

where AVE and VAR are obtained via numeric integrations, $d_{min}$ is the minimum distance between the points of the constellation, $\sigma_{\mathbf{h}}^2$ is the variance of the channel entries.

However, this scheme presents high computational complexity compared with linear receivers. Although linear receivers may be inaccurate in low SNR regimes, the accuracy of the ML receiver might not be compensated by the low throughput.

### 6.2.2 RBIR for Linear Receivers

When linear receivers are used, the computation of RBIR (6.6) is performed by using $\gamma_n$, which varies depending on which MIMO scheme is adopted. Hence, the SNR can be computed depending on the MIMO scheme as follows:

#### SISO

This is the simplest case, where a single polarization is used. Thus, the system model is reduced to

$$y_n = h_n x_n + w_n \tag{6.9}$$

and therefore

$$\gamma_n = \frac{|h_n|^2}{\sigma_w^2}. \tag{6.10}$$

#### Orthogonal Polarization Time Block Codes

This case is an adaptation of the Orthogonal Space-Time Block Codes, introduced in [Ala98], replacing the spatial component by the polarization component. Since Orthogonal Polarization-Time Block Codes scheme exploits the full diversity of the channel [KLK13], the SINR can be expressed as

$$\gamma_n = \frac{\|\mathbf{H}_n\|^2}{2\sigma_w^2}, \tag{6.11}$$

where $\|\mathbf{H}_n\|^2 = \mathrm{tr}\left(\mathbf{H}^H\mathbf{H}\right)$ is the Frobenius norm.

#### Polarization Multiplexing

In this case, each polarization conveys a symbol. Thus, two symbols are transmitted in each channel access. Assuming that the receiver is able to cancel the interference between both streams, we obtain two equivalent SINR for each symbol of each polarization [KLK13; Lat+13].

Therefore, the equivalent SINR of the $m$th polarization using the Zero Forcer (ZF) receiver can be expressed as

$$\gamma_{n,m} = \frac{|\mathbf{h}_{n,m}|}{\sigma_w^2} \tag{6.12}$$

where $\mathbf{h}_{n,m}$ is the $m$th column of matrix $\mathbf{H}_n$. Note that in this case, we obtain $2N$ equivalent SNR instead of $N$, since we are conveying $2N$ symbols.

### Polarized Modulation

In this scheme, a single symbol is transmitted using a single polarization. However, the index of the used polarization is also a place for conveying bits. In the case where two polarizations are used, PMod conveys $M + 1$ bits ($M$ bits of the symbol and an additional bit of the polarization state index). Assuming that we employ a receiver with interference cancellation of cross-polarization and that the polarization selection bit is equiprobable, we use the average SNR in each polarization. Thus,

$$\gamma_n = \frac{|h_{00}|^2 + |h_{11}|^2}{2\sigma_w^2}. \tag{6.13}$$

## 6.3 Adaptive MODCOD and MIMO Scheme

After the introduction of the PLA for MIMO schemes, we aim at implementing it to the BGAN standard. This standard describes different modulation and coding schemes (MODCOD), called bearers. Each bearer defines a MODCOD, which has a different bitrate. Table H.1 describes the parameters of the small subset of all bearers. As shown in this table, the length of the block $N$ can be $640$, $1098$ or $941$, and the constellation size $M$ can be $2$, $4$, $5$ or $6$. It must be noted that since each MIMO scheme produces a different SINR, for the same channel realization each MIMO scheme produces a different error curve.

We can thus formulate the objective problem as

$$\begin{aligned} \max_{u_{m,d,c}} & \sum_{m\in\mathcal{M}} \sum_{d\in\mathcal{D}} \sum_{c\in\mathcal{C}} u_{m,d,c} r_{m,d,c}\left(\bar{\gamma}\right) \\ \text{s.t. } & \text{PER}\left(\bar{\gamma}\right) \leq 10^{-3} \\ & \sum_{m\in\mathcal{M}} \sum_{d\in\mathcal{D}} \sum_{c\in\mathcal{C}} u_{m,d,c} = 1 \end{aligned} \tag{6.14}$$

where $\mathcal{M}$ is the set of MIMO modes, $\mathcal{D}$ is the set of modulation orders and $\mathcal{C}$ is the set of available coding rates, $r_{m,d,c}\left(\bar{\gamma}\right)$ is the achievable rate given the effective SNR $\bar{\gamma}$ and the tuple $m, d, c$.

Compared with classic MODCOD adaptive algorithms, (6.14) introduces additional computational complexity since it has to optimize the MIMO mode. Fortunately, this additional increase is linearly proportional to the number of MIMO modes $\mathcal{M}$.

It is worth to mentioned that (6.14) is not constraint to any particular MIMO scheme or dimension. It can be applied to all MIMO schemes considered in the system. The requirement to apply (6.14) is to compute the SNR that this particular MIMO scheme achieves and include it to the set $\mathcal{M}$.

## 6.4 Channel Time Series Generator

Generating the channel realization is a key aspect. Although each realization must be independent for each trial, it must also be time-correlated during the whole transmission block. In order to implement a time series channel generator, we first generate independent realizations of channel coefficients for each polarization depending on the parameters of the scenarios. Then, we interpolate and filter them using a low pass band filter with a bandwidth equal to the Doppler spread. Thus, we can guarantee that

$$\mathbb{E}\left\{h^*(t)h(t-\tau)\right\} < 0.4,\ \tau > \tau_c, \tag{6.15}$$

where $\tau_c$ is the coherence time. The coherence time is defined in terms of Doppler Spectrum when the object is moving. The coherence time can be approximated by the inverse of Doppler spread

$$\tau_c \approx \frac{1}{D_s}, \tag{6.16}$$

where $D_s \leq v/\lambda = f_D$ and $f_D$ is the Doppler shift of an object moving at the speed of $v$ m/s. Assuming the Clarke's model [Rap96; Sha02], the coherence time can be approximated by

$$\tau_c = \frac{3\lambda}{4v\sqrt{\pi}}, \tag{6.17}$$

where $\lambda$ is the wavelength of carrier. If we assume a symbol rate of $F_s = 33600$ symbols/second, we can compute the number of correlated samples as $L = \lceil \tau_c F \rceil$ and the number of independent realizations as $Q = \lceil P/(\tau_c F) \rceil$, where $P$ is the number of samples to be generated.

### 6.4.1 Mobile Satellite Dual Polarized Channel Model

In [SGL06], the authors describe statistics of the mobile satellite dual polarized channel as the sum of a Line of Sight (LoS), specular and diffuse signals. LoS and specular components are modeled as Rice random variables and correspond to the direct link between satellite and UE and the specular ray produced by the sea effect, respectively. The diffuse component is produced by the scatters near the UE. Hence,

$$\mathbf{H} = \mathbf{L}\mathbf{K}_L + \mathbf{S}\mathbf{K}_S + \mathbf{D}\mathbf{K}_D. \tag{6.18}$$

Each matrix is $2 \times 2$ and are composed by the different components. $\mathbf{L}$ contains LoS components and is generated as follows

$$\mathbf{L} = \begin{pmatrix} \sqrt{1-\beta_1} & \sqrt{\beta_2} \\ \sqrt{\beta_1} & \sqrt{1-\beta_2} \end{pmatrix} (\cos\phi + j\sin\phi) = \boldsymbol{\beta} e^{j\phi}, \tag{6.19}$$

where $\phi$ is a uniform random variable between the range $[0, 2\pi)$. $\beta_i$ is related with the scenario and define the autocorrelation of each component of matrix $\mathbf{L}$. The K-factor matrix of the direct component is defined as

$$\mathbf{K}_L = \begin{pmatrix} \sqrt{\frac{K_1^L}{K_1^L+K_1^S+1}} & 0 \\ 0 & \sqrt{\frac{K_2^L}{K_2^L+K_2^S+1}} \end{pmatrix}, \tag{6.20}$$

where $K_i^L$ and $K_i^S$ are the K-factors of $i$th polarization of LoS and specular components, respectively.

In the same form, the specular component $\mathbf{S}$ is generated as

$$\mathbf{S} = \begin{pmatrix} \sqrt{1-\xi_1} & \sqrt{\xi_2} \\ \sqrt{\xi_1} & \sqrt{1-\xi_2} \end{pmatrix} (\cos\phi + j\sin\phi) = \boldsymbol{\xi} e^{j\phi}, \tag{6.21}$$

where $\xi_i$ is related with the scenario and define the autocorrelation of each component of matrix $\mathbf{S}$. The K-factor matrix of the specular component is defined as

$$\mathbf{K}_S = \begin{pmatrix} \sqrt{\frac{K_1^S}{K_1^L+K_1^S+1}} & 0 \\ 0 & \sqrt{\frac{K_2^S}{K_2^L+K_2^S+1}} \end{pmatrix}. \tag{6.22}$$

Finally, the diffuse matrix is composed by complex Gaussian random variables. Since it is a $2 \times 2$ matrix it can be envisaged as a zero mean multivariate random variable of order 4, $\mathbf{d} = \begin{pmatrix} d_{11} & d_{12} & d_{21} & d_{22} \end{pmatrix}^T$, where each $d_{ij}$ is the entry $ij$ of the matrix $\mathbf{D}$, with the following covariance matrix:

$$\mathbf{\Sigma} = \begin{pmatrix} 1-\alpha_1 & \rho_1^t\sqrt{(1-\alpha_1)\,\alpha_1} & \rho_1^r\sqrt{(1-\alpha_1)\,\alpha_2} & 0 \\ \rho_1^t\sqrt{(1-\alpha_1)\,\alpha_1} & \alpha_1 & 0 & \rho_2^r\sqrt{(1-\alpha_2)\,\alpha_1} \\ \rho_1^r\sqrt{(1-\alpha_1)\,\alpha_2} & 0 & \alpha_2 & \rho_2^t\sqrt{(1-\alpha_2)\,\alpha_2} \\ 0 & \rho_2^r\sqrt{(1-\alpha_2)\,\alpha_1} & \rho_2^t\sqrt{(1-\alpha_2)\,\alpha_2} & 1-\alpha_2 \end{pmatrix}, \tag{6.23}$$

where $\alpha_i$ and $\rho_i^{t,r}$ are specific parameters depending on the scenario that adjust the autocorrelation of each component as well as the cross-correlation between polarizations. The K-factor matrix of diffuse component is described as

$$\mathbf{K}_D = \begin{pmatrix} \sqrt{\frac{1}{K_1^L+K_1^S+1}} & 0 \\ 0 & \sqrt{\frac{1}{K_2^L+K_2^S+1}} \end{pmatrix}. \tag{6.24}$$

From [SGL06], for the specific case of maritime/aeronautical scenario where the specular component is relevant, the parameters are set as follows:

$$\begin{aligned} K_i^L &= 10 \\ K_i^S &= 5 \\ \beta_i &= 0.3 \\ \alpha_i &= 0.4 \\ \rho_i^{t,r} &= 0.5 \\ \xi_i &= 0.3. \end{aligned} \tag{6.25}$$

Fig. 6.2 summarizes the channel generation based on the described framework. Fig. 6.3 illustrates a random snapshot of the MIMO maritime satellite channel, which contains LoS components, specular and diffuse components.

### 6.4.2 Interference Model

Due to the non-isotropic reflectors at the satellite and the speed of the terminal, the difference of the SINR between the center and the edge of the beam spot may be significant. Additionally, the adjacent beams may introduce interferences. Whilst the gain of the intended beam decreases from the center to the edge, the interference increases from the center to the edge. Although the channel across the intended beam and the interferences is the same, the gains are different. Hence, assuming a linear interpolation in logarithmic scale, the gain of the intended beam is

$$C = C_c + \frac{(C_e - C_c)}{W} \frac{vn}{F_s}, \tag{6.26}$$

where $C_c = 40.83$ dB is the gain at the center of the beam, $C_e = 36.74$ dB is the gain at the edge of the beam, $W = 300$ km is the beam width, $v$ is the speed of the terminal, $n$ is the sample index, and $F_s$ is the sample frequency. The same is applied to the interference to obtain $I$ with the values $I_c = 17.49$ dB and $I_e = 25.1$ dB.

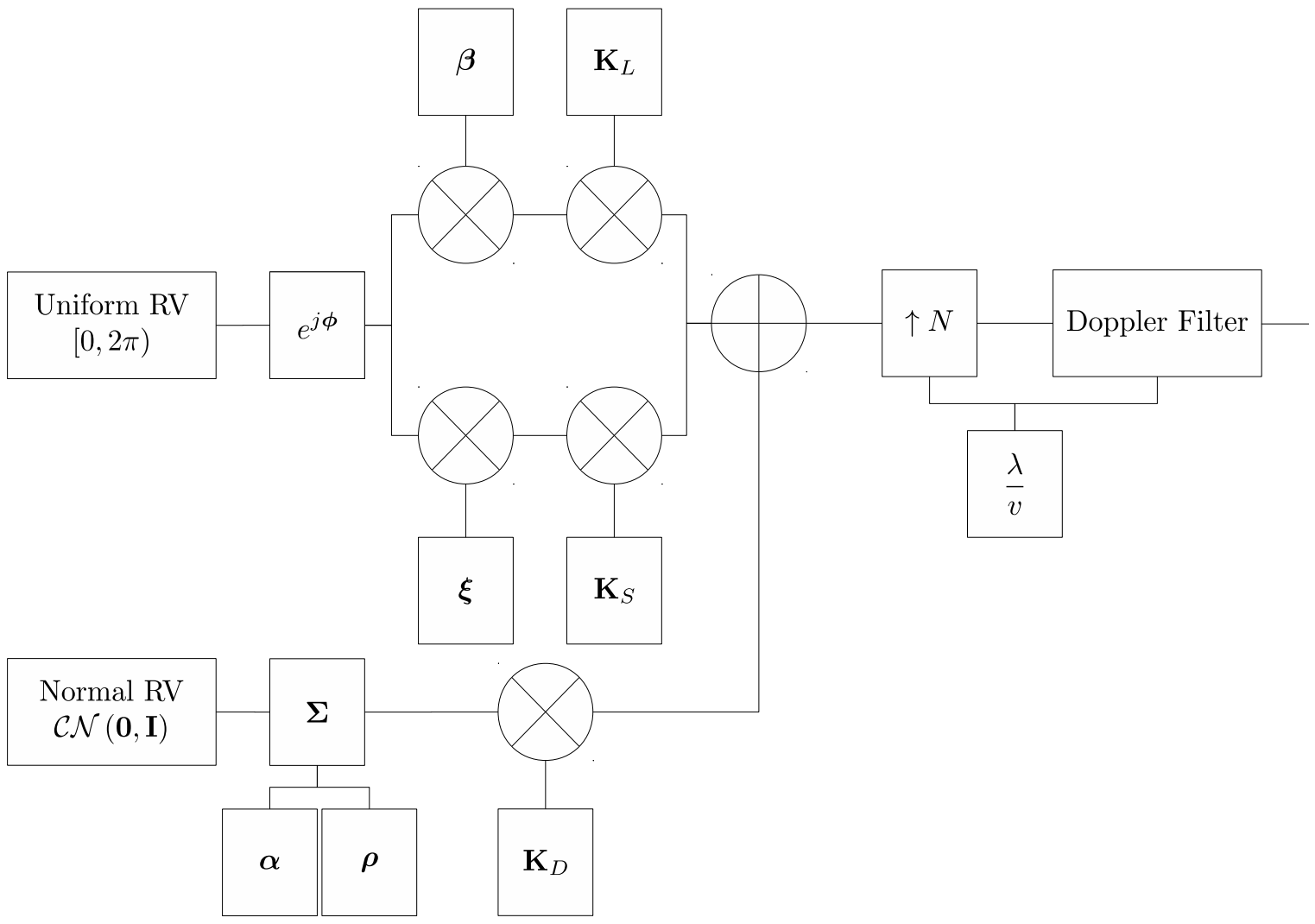

FIGURE 6.2: Block diagram of time series mobile polarized channel generator.

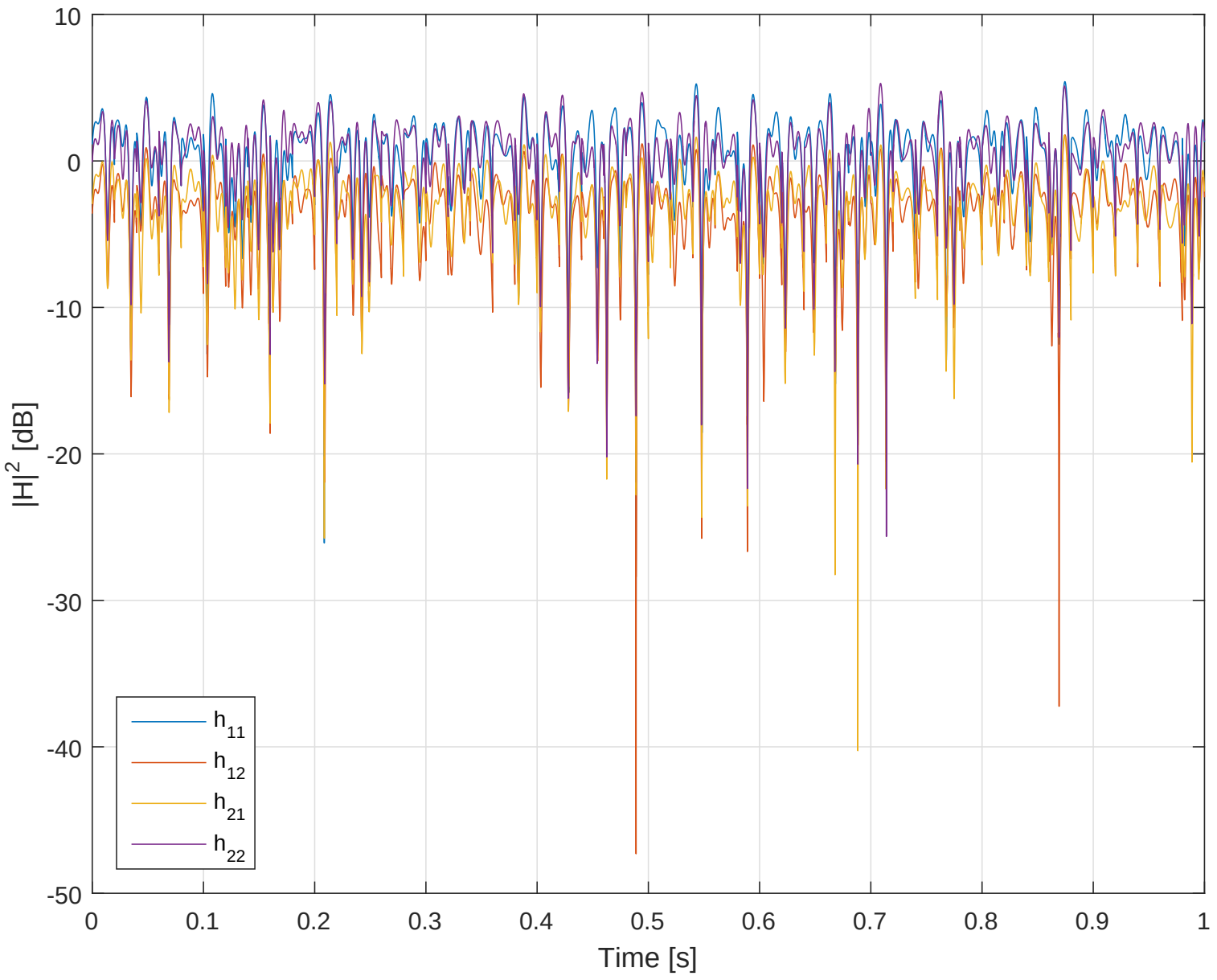

FIGURE 6.3: Snapshot of MIMO maritime mobile satellite channel magnitude.

Finally, (6.10)-(6.12) are replaced by $\gamma_n \leftarrow C\gamma_n \left(\sigma_w^2 \rightarrow I \, \mathrm{tr}\left(\mathbf{H}^H\mathbf{H}\right) + \sigma_w^2\right)$, where $\sigma_w^2$ is replaced by the sum of $\sigma_w^2$ and the interference $I \, \mathrm{tr}\left(\mathbf{H}^H\mathbf{H}\right)$.

## 6.5 Remarks

One of the major drawbacks of MIESM is the high resolution needed to compute $\Phi^{-1}$. Since the objective is PER $\leq 10^{-3}$, we work in high SNR regime, which corresponds to the flat upper part of MIESM curves in Fig. 6.1a. In this zone, a small increment in MIESM requires a significant increment in SNR. Thus, when we compute $\Phi^{-1}$, there is no reciprocal value in the SNR axis. In other words, for a certain precision and for SNR $> [10, 17, 20, 23]$ dB (QPSK, 16QAM, 32QAM and 64QAM, respectively) the $\Phi(x)$ function is not injective and thus, $\Phi^{-1}$ does not exist.

To deal with this issue, we replace (6.4) by

$$\bar{\gamma} = \breve{\Phi}^{-1}\left(\frac{1}{N}\sum_{n}^{N}\breve{\Phi}\left(\gamma_n\right)\right), \tag{6.27}$$

where

$$\breve{\Phi}(x) = \begin{cases} \Phi(x) & \text{if } \left|\frac{1}{N}\sum_n^N \Phi\left(\gamma_n\right) - M\right| \geq \epsilon \\ x & \text{otherwise} \end{cases} \tag{6.28}$$

and $M$ is the spectral efficiency of the used modulation order. With (6.27) and (6.28), we are able to split the MIESM curves between the linear and non-linear regions.

## 6.6 Results

In this section we describe the obtained results for different use cases. We perform the analysis in a maritime scenario, where the user terminal is located at the center of the beam and is moving to the edge with constant speed. During the journey, the terminal receives the blocks from the satellite and feedbacks the MODCOD and MIMO mode, which optimizes the throughput with a maximum PER $\leq 10^{-3}$. We assume a delay of $500$ ms. Additionally, we assume that we are in the high SNR margin and therefore we use the RBIR for MMSE as the PLA approach (as it obtains the same results as RBIR for ML).

To ensure a fair comparison, first we generate a time series channel snapshot corresponding to $300$ km trip. Later, we use this snapshot to run the different simulations for the different use cases. Hence, all simulations use the same channel realizations.

Table 6.1 describes the main parameters used in the system simulation. Next sections illustrate the results for different terminals, delays and MIMO modes. MIMO modes are denoted as $1$ (SISO), $2$ (OPTBC), $3$ (PMod) and $4$ (V-BLAST).

#### SISO Scenario: the benchmark

First, we include the simulation results with MODCOD adaptation but without MIMO mode adaptation. We use this scenario as the benchmark reference to evaluate the performance of the proposed techniques.

Introducing a delay during the feedback transmission may produce an outage on the constraint of PER. When the transmitter receives the feedback and applies it, the SINR of the terminal has decreased and the selected MODCOD and MIMO mode may require higher SNR. Thus, the result is an outage because the receiver cannot guarantee PER $\leq 10^{-3}$. To address this issue, we add a SINR margin of 1.8 dB. This value is obtained by running the simulations, where the SINR margin in the outage zones is always below than the proposed value.

TABLE 6.1: Scenario Main Parameters

| | |
|---|---|
| **Carrier** | 1.59GHz |
| **Beam Diameter** | 300 km |
| **Noise** | −204 dBW/Hz |
| **Bandwidth** | 32 KHz |
| **TX Power** | 34 dBm |
| **Symbols per Block ($N$)** | 640 dBW/Hz |
| **Block Length** | 20 ms |
| **Channel Profile** | Maritime |
| **Speed of Terminal** | 50 km/h |
| **G/T** | −10.5 dB/K |
| **Feedback Delay** | 500 ms |
| **PLA Scheme** | RBIR for MMSE |

Fig. 6.4 illustrates the adaptation of MODCOD without including the MIMO scheme. The SINR subplot includes the effective SINR ($\bar{\gamma}$) and the instantaneous SINR (a sample of $\gamma_n$). To avoid fast changes in $\bar{\gamma}$ we included a low-pass filter, which smooths the adaptation. This figure illustrates the SINR variations and the MODCOD adaptation. As seen in the figure, the throughput varies in small steps, thus guaranteeing a maximum PER.

Fig. 6.5 depicts a detailed curve of throughput and effective SINR. On the other hand, fig. 6.6 describes the throughput probability distribution. From this figure we can obtain the average and most probable throughput.

#### Fixed MIMO: the other benchmark

In the previous section, we compared the proposed framework with the benchmark reference, where a SISO transmission is performed without MIMO adaptation. In this section we compare the adaptive MIMO framework with the fixed V-BLAST scenario, i.e., where V-BLAST is always performed and the adaptation is done only through the MODCOD. Fig. 6.7, fig. 6.8 and fig. 6.9 depict the MODCOD adaptation, throughput result and pdf of this scenario.

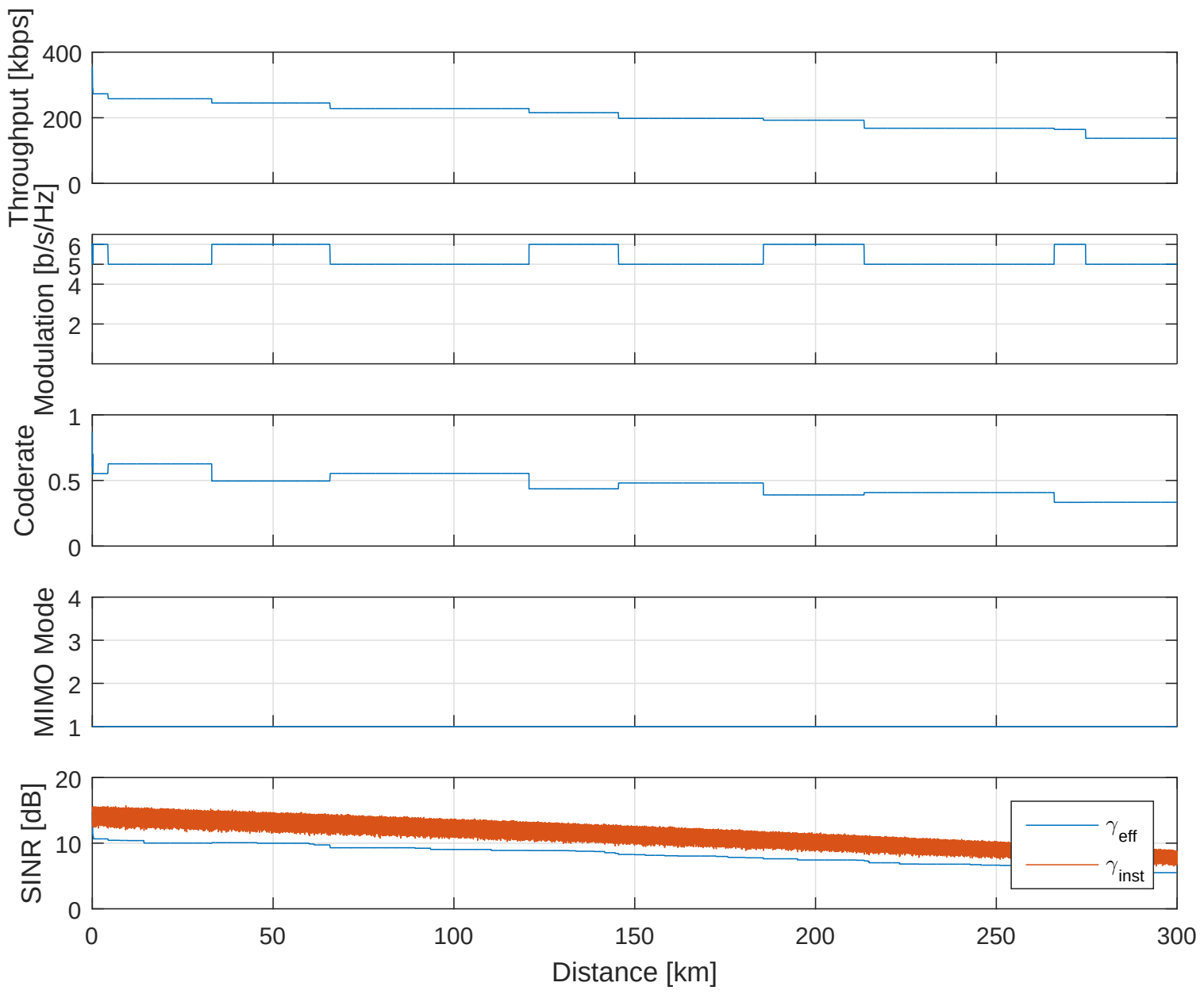

FIGURE 6.4: MODCOD adaptation without MIMO adaptation.

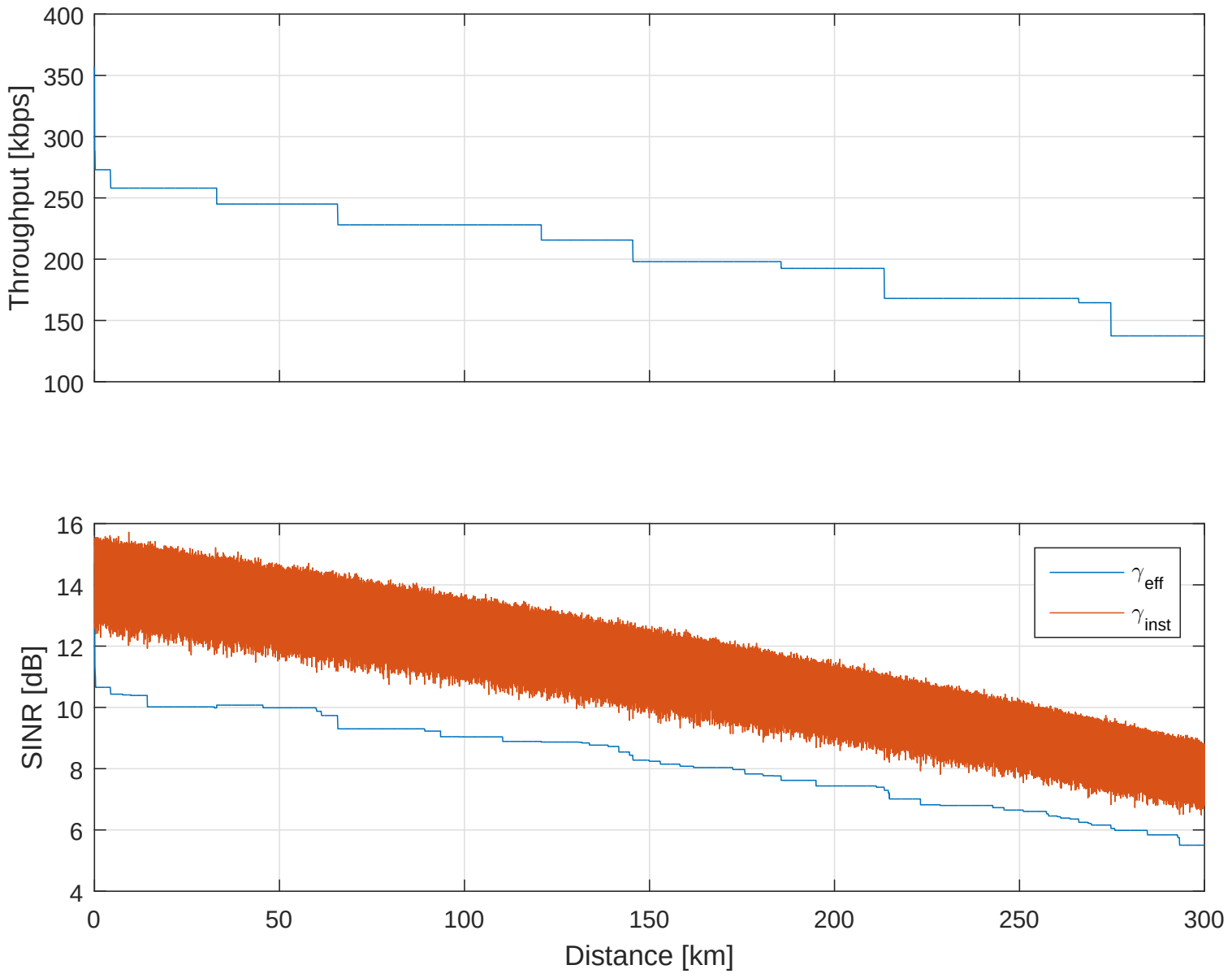

FIGURE 6.5: Detailed throughput performance of SISO scenario.

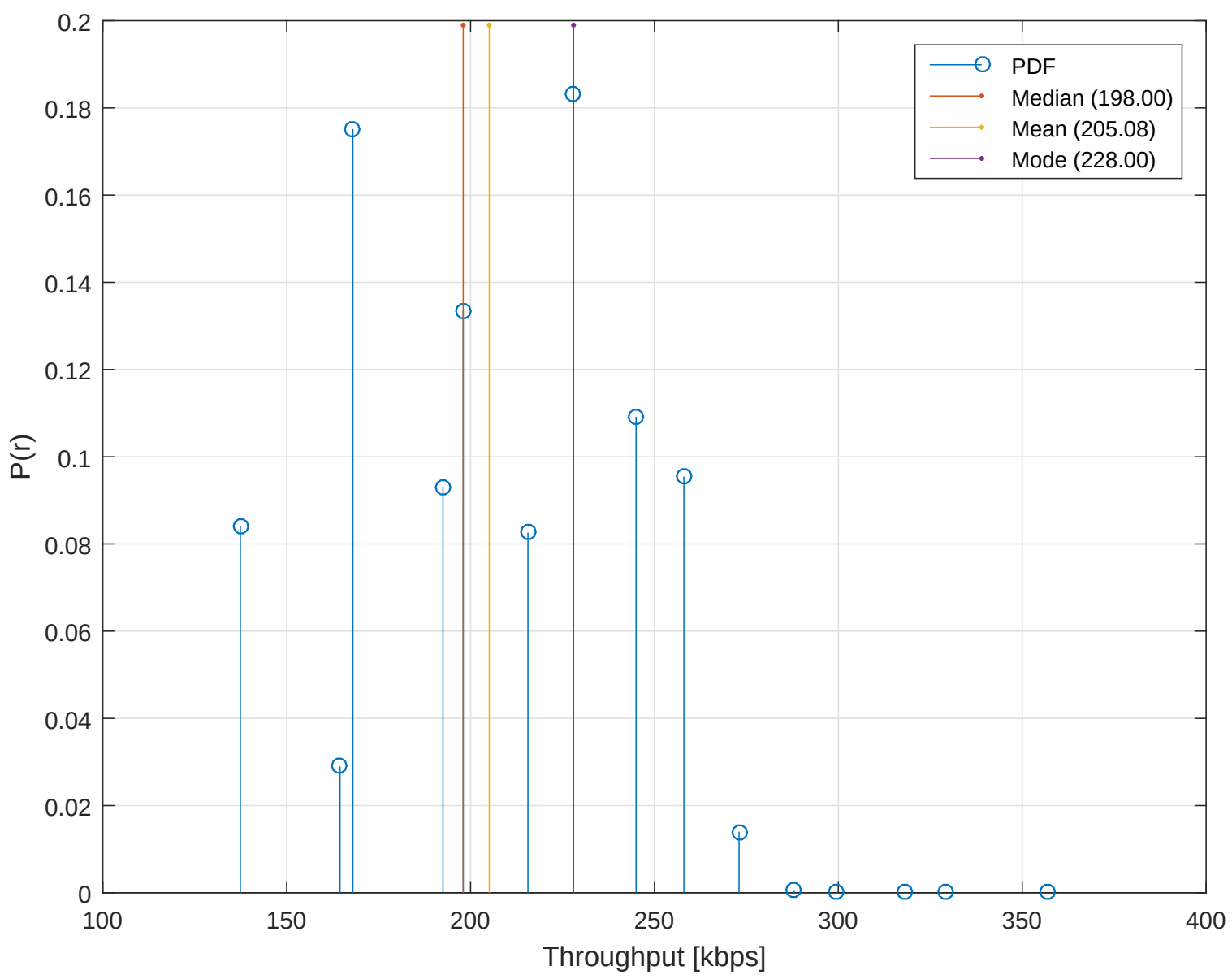

FIGURE 6.6: Throughput's distribution for the SISO scenario.

### MIMO Scenario: full adaptation

In this use case, the MIMO scheme is added to the adaptation. Hence, the terminal has three degrees of freedom for the optimization, namely modulation order, coding rate and MIMO scheme.

Fig. 6.10 summarizes the performance of the proposed adaptive techniques. The figure illustrates the adaptation between the modulation, code rate and MIMO mode. During the journey, the transmitter switches between V-BLAST and PMod in order to optimize the throughput. Fig. 6.11 describes the evolution of the throughput and SNR in detail.

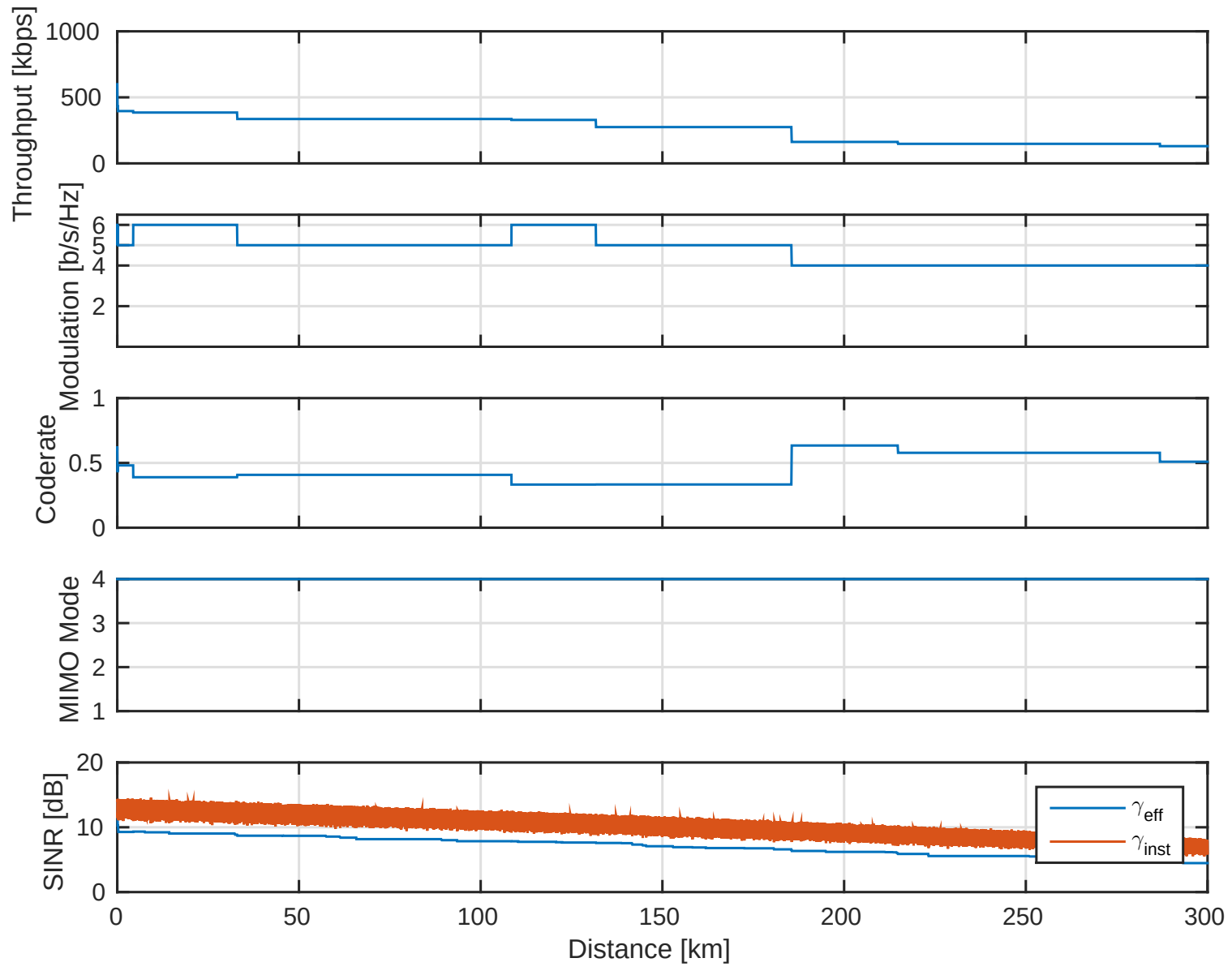

FIGURE 6.7: MODCOD adaptation of fixed MIMO scenario.

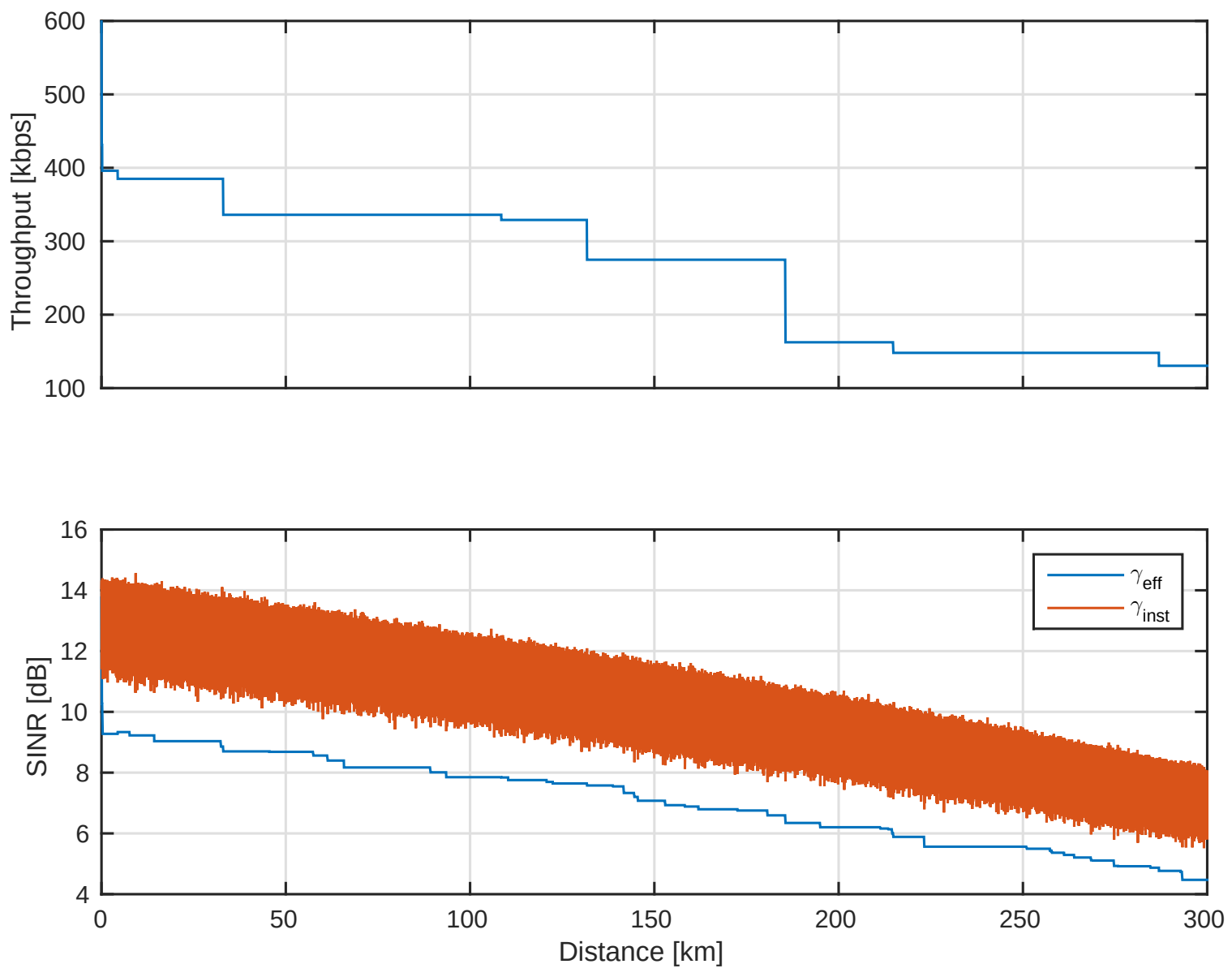

FIGURE 6.8: Detailed throughput performance of fixed MIMO scenario.

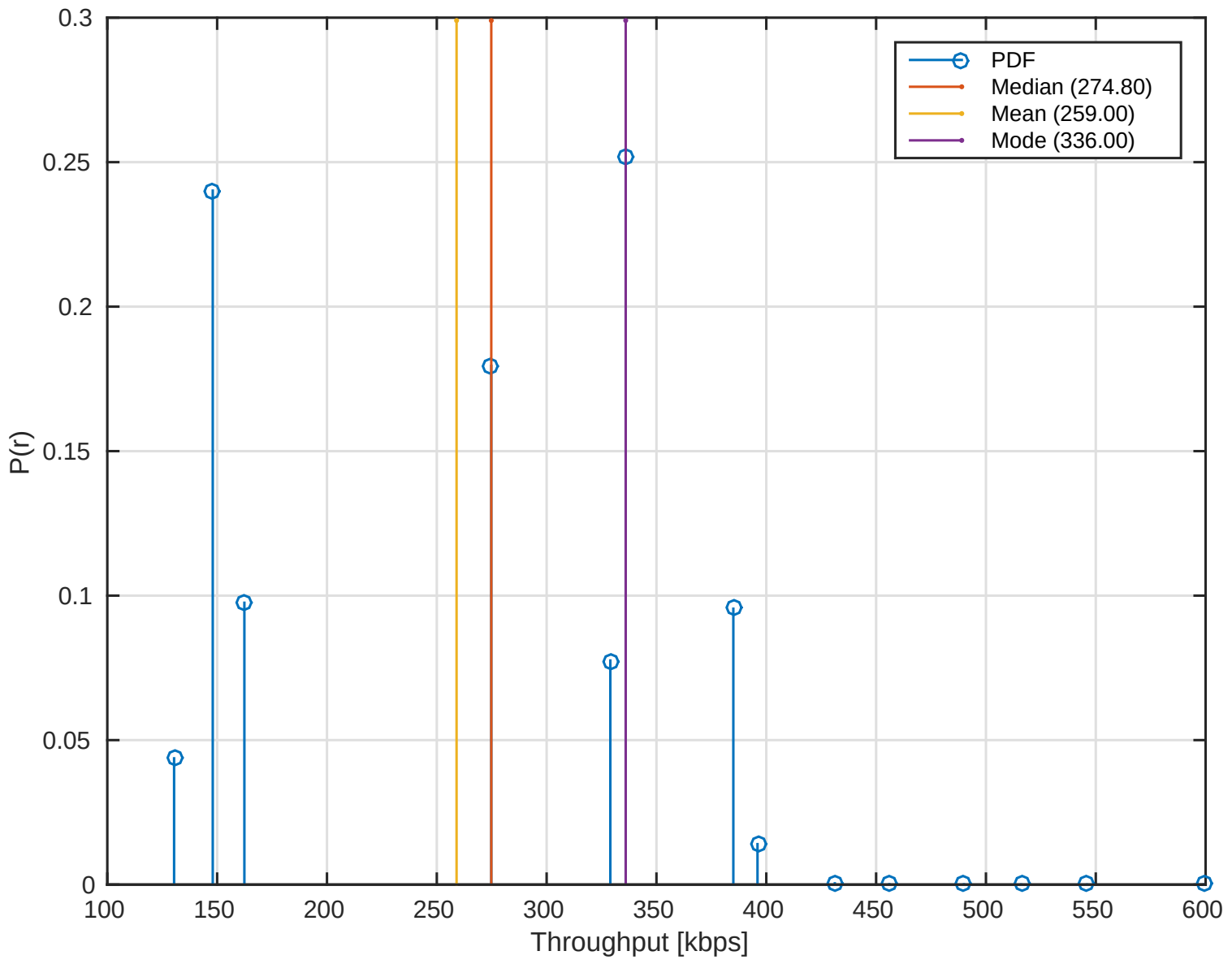

FIGURE 6.9: Throughput's distribution for the fixed MIMO scenario.

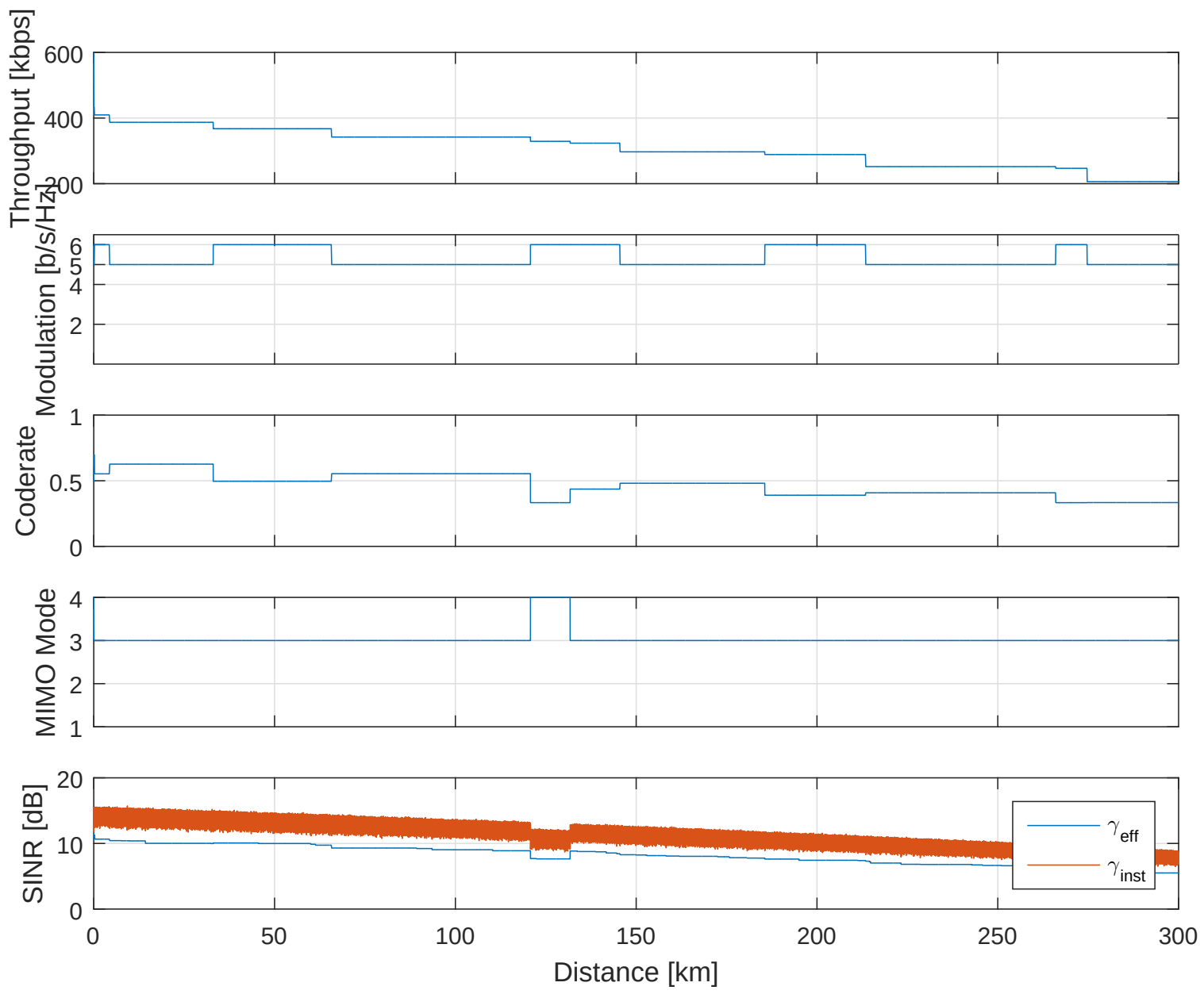

FIGURE 6.10: MODCOD adaptation with MIMO adaptation.

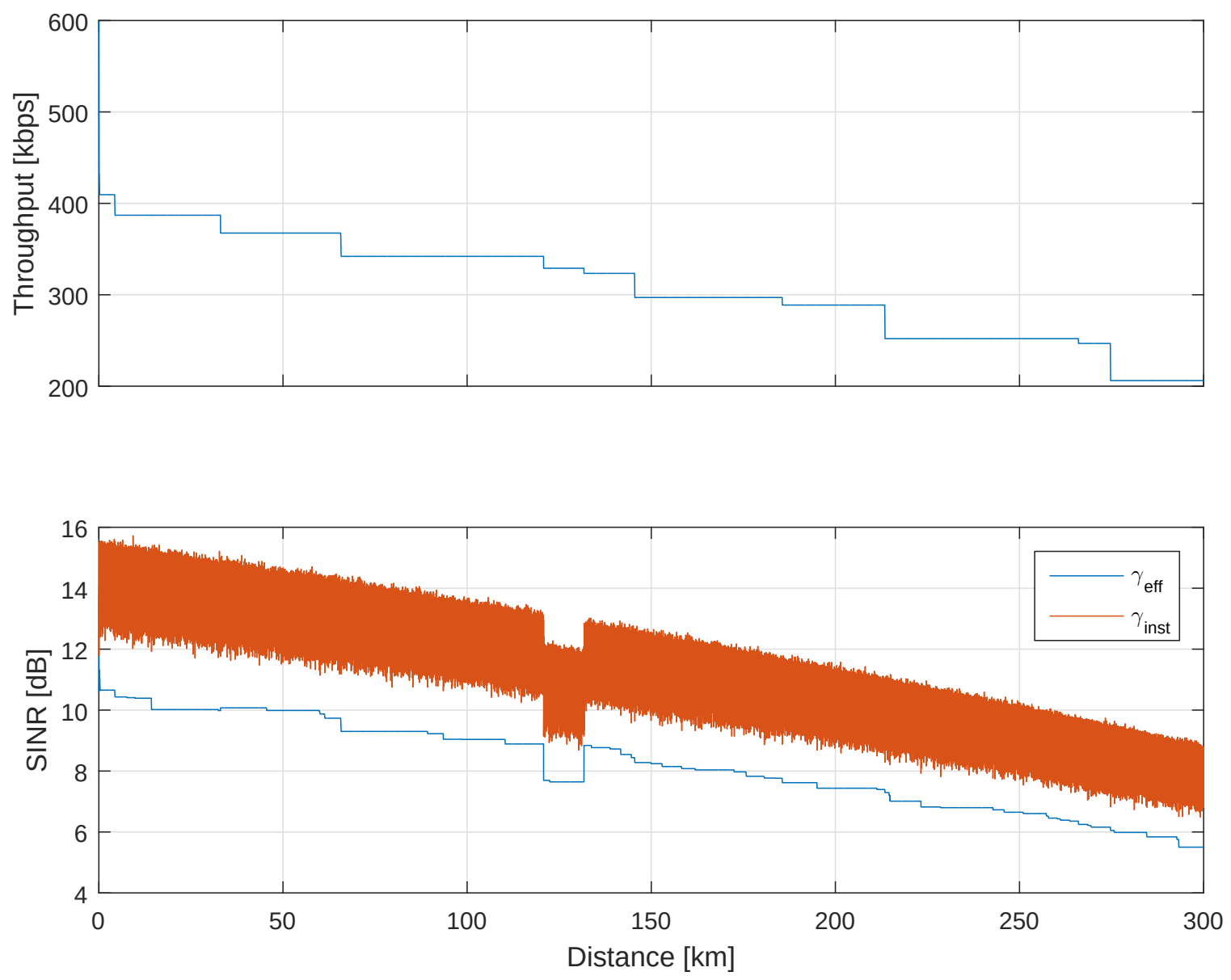

FIGURE 6.11: Detailed throughput performance of MIMO scenario.

Fig. 6.12 depicts the probability distribution of the throughput in this scenario. In this case, we observe an increase of throughput of $50\%$ compared with the SISO case. The terminal obtains an average throughput of $308$ kbps.

Fig. 6.13 describes the CDF for the SISO and MIMO cases with a delayed feedback. For example, the probability to obtain a throughput below $300$ kbps is $1$ in SISO and $0.5$ in MIMO. Hence, we can observe that introducing MIMO adaptation not only increases the throughput, but also increases the robustness of the transmission.

Finally, fig. 6.14 depicts the CDF for the fixed MIMO (V-BLAST scheme) and adaptive MIMO. In this case, it is also clear that using adaptive MIMO achieves higher performance compared with the fixed MIMO case. For instance, the probability to obtain a throughput below $330$ kbps is $0.9$ in fixed MIMO and $0.6$ in adaptive MIMO.

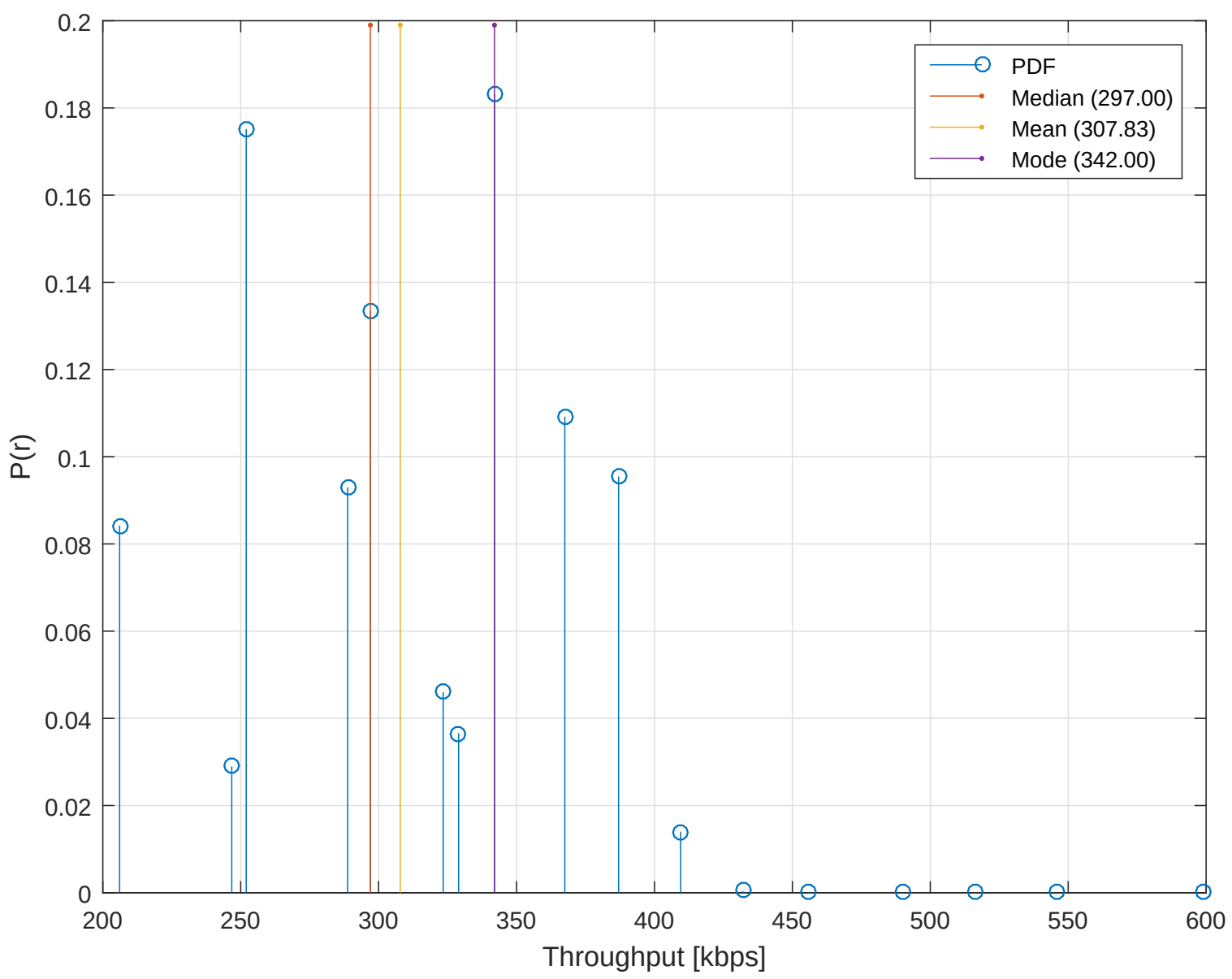

FIGURE 6.12: Throughput's distribution for the MIMO scenario.

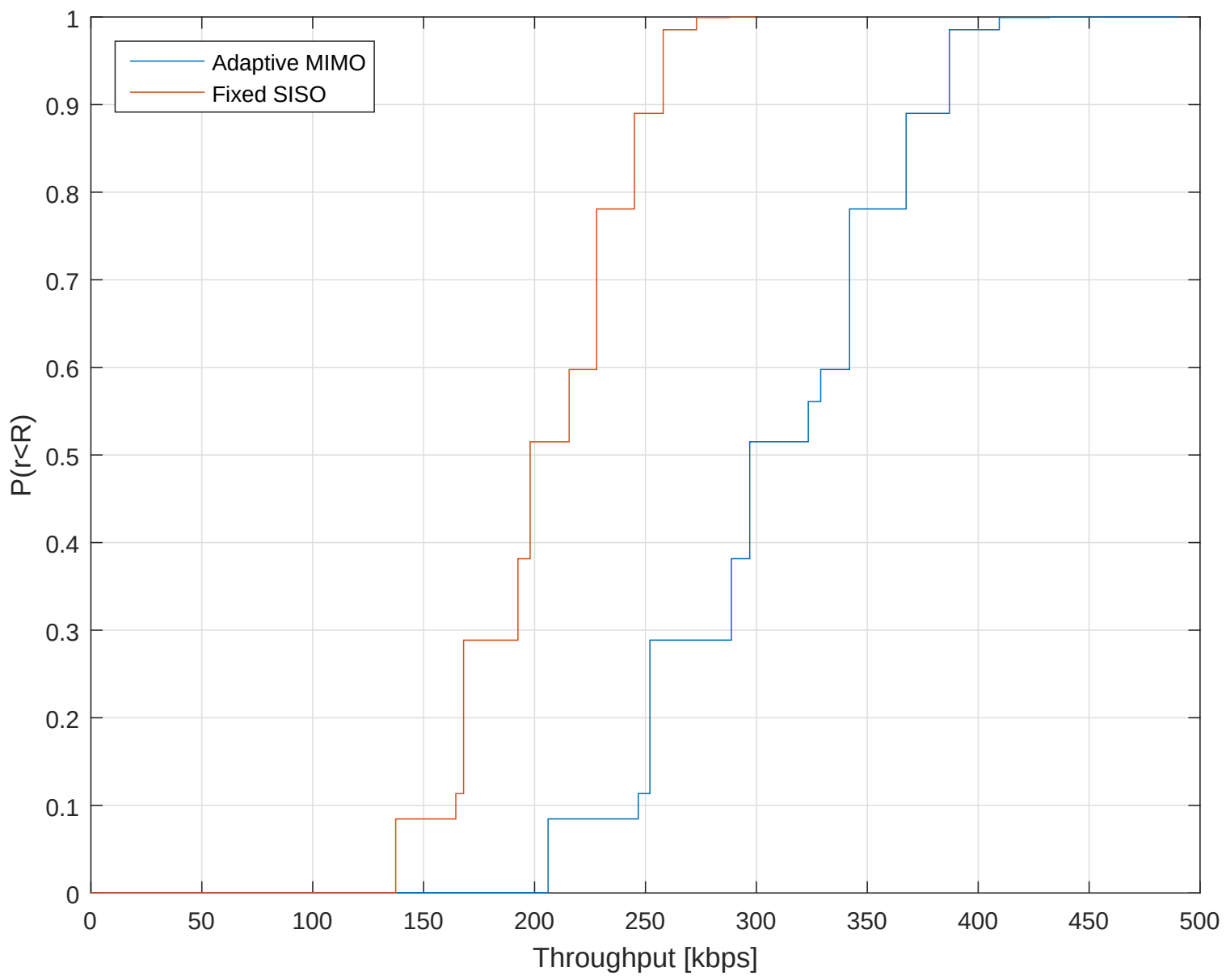

FIGURE 6.13: CDF of fixed SISO and adaptive MIMO.

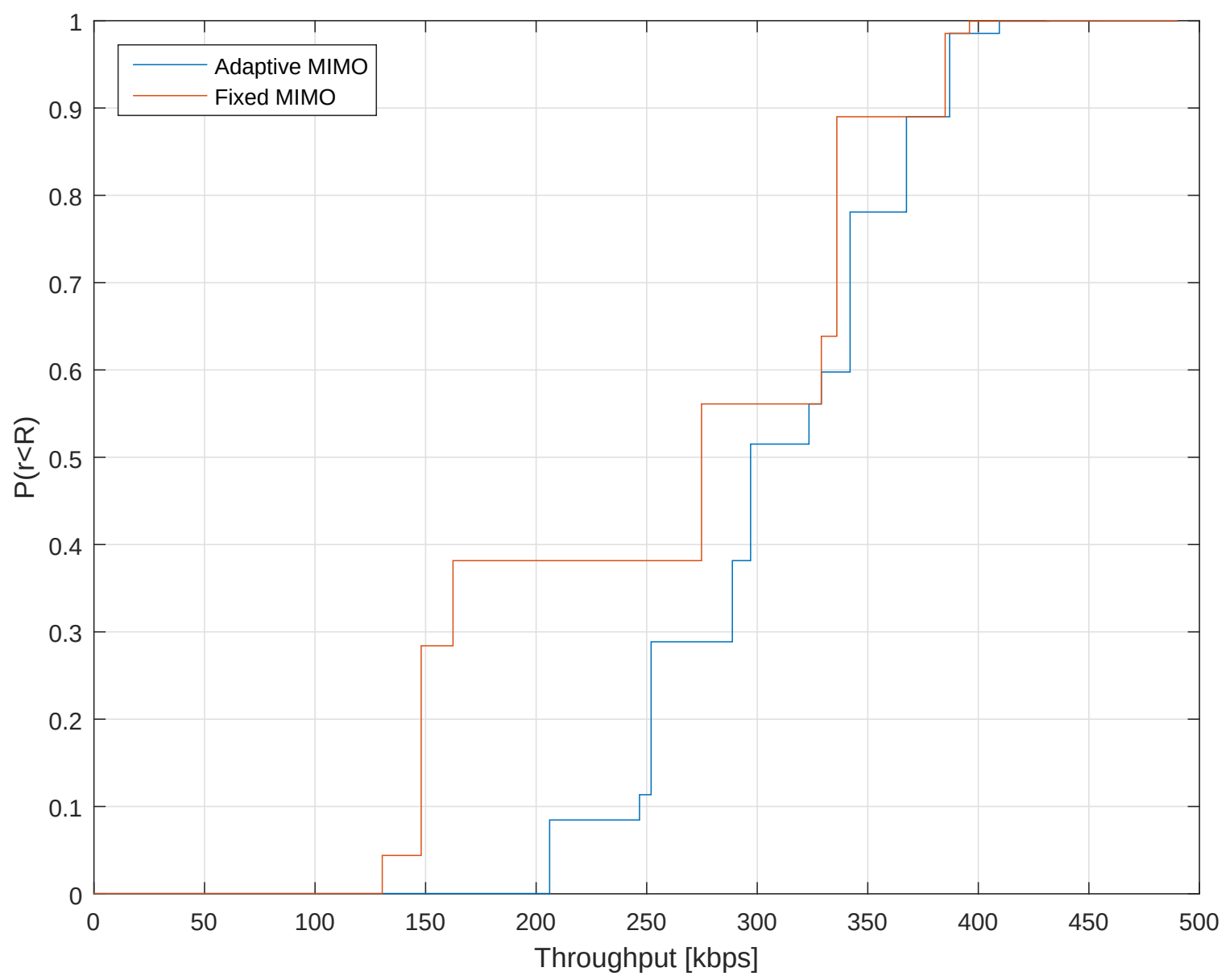

FIGURE 6.14: CDF of fixed MIMO and adaptive MIMO.

TABLE 6.2: Gain of average throughput comparison between adaptive MIMO and benchmarks.

| | **SISO Case** | **Fixed MIMO Case** |
|---|---|---|
| **Adaptive MIMO** | 102.75 | 48.83 |
| | 50.10% | 18.85% |

In summary, it becomes clear that using adaptive MIMO always achieves higher performance compared with the SISO case or fixed V-BLAST MIMO scheme. Table 6.2 compares the gain of using adaptive MIMO compared with the two benchmarks, its absolute value (in kbps) and relative percentage.

## 6.7 Conclusions

In this chapter we introduce the concept of PLA and we apply it to the MIMO case. Using the BGAN standard as a benchmark reference, we propose the use of PLA to model the adaptive downlink based on the effective SNR at the terminal. We implement the proposed adaptive scheme to the particular satellite mobile maritime scenario with MIMO schemes in polarization.

The precise dual polarized channel time series generator allows us the possibility of generating time-correlated channel snapshots which simulate the channel variations of a terminal moving from the center to the edge of the beam. This model also includes the effect of the interferences of the adjacent beams.

Finally, the addition of delayed feedback increases the accuracy of the simulations and the results are more realistic. Table 7 summarizes the obtained results for both terminals and the delayed feedback.

The use of MIMO technologies applied to dual and orthogonal polarizations are worth studying. They can provide gains near of $\sim 50\%$, without additional power. Furthermore, incorporating the MIMO scheme adaptation into the adaptive algorithms provides higher gains compared with the classic approach, where only the MODCOD is optimized. Hence, switching between different MIMO schemes depending on the channel conditions and optimizing the MODCOD jointly is particularly interesting in terms of throughput.

It is worth observing that the gain is more important in those terminals where the SINR is low. In these cases, the best option is to switch between the available MIMO schemes since it provides more rate diversity. On the other hand, in terminals where the SINR is high, the gain of using the proposed algorithm is relatively low. In these cases, the SINR is good enough to not switch between MIMO schemes and use the highest scheme instead.

# Chapter 7

# Final

In a time not distant, it will be possible to flash any image formed in thought on a screen and render it visible at any place desired. The perfection of this means of reading thought will create a revolution for the better in all our social relations.

N. Tesla

This chapter concludes this dissertation. Throughout all previous chapters we have introduced the concept of Index Modulations combined with the polarization dimension, spanning from the theoretical point view to the realistic implementation in actual standard scenarios [ETS]. The final goal is to study the usefulness of the polarization dimension in the communication schemes.

## 7.1 Conclusions

In this thesis, we depart from the theoretical fundamentals in terms of capacity of IM. After this, we apply the concept of IM to the polarization physical dimension and we introduce two communication systems based on 2D and 3D constellation mappings. Finally, we move one step forward to a more realistic scenario where the coexistence with other polarization schemes is presented and exploited.

**Chapter 2** in this dissertation introduces the polarization dimension, some schemes used in the polarization dimension, as well as the Index Modulations. The Stokes Vector and its parameters, as well as the Jones Vector and Poincaré Sphere, are also introduced and described in detail. These concepts are closely related to the polarization dimension and can be exploited in communication systems. Common polarizations are illustrated in the Stokes, Jones and Poincaré forms, and the reference system is defined. Various phenomena affecting the polarization during a reflection have also been described, as well as the implications of the Faraday Effect. Appendix C analyzes in more detail the Faraday Effect. In the second chapter, polarization is also presented as a channel diagonalizer. This approach allows to understand the reasons why polarization is a potential dimension in communication systems from the point of view of capacity. Finally, this chapter also discusses some common communication systems that use polarization dimension, as well as their context.

**Chapter 3** analyzes the mutual information of IM and proposes different approximations. The mutual information expression involves an integral that is particularly difficult to solve. This chapter develops this expression and introduces different approximations with various levels of accuracy. The approximations' error is analyzed in the limits and the ergodic capacity is also studied for different channel distributions. The expression is used to better understand IM and to compare different applications of IM, such as those in the spatial, polarization or frequency dimensions. Appendix G contains the proof of the theorems introduced in this chapter.

The application of PMod in a communication system is described in **Chapter 4**. This communication system implements realistic signal processing blocks. Furthermore, new problems arise when PMod is combined with other types of processing. Four receivers are discussed according to their accuracy and computational complexity. The chapter finishes with a complete analysis for a realistic satellite mobile scenario, i.e., considering typical impairments in satellite communications such as XPD or imperfect channel estimation.

Following the PMod philosophy, **Chapter 5** introduces a new PMod with an arbitrary number of polarizations, the so-called 3D PMod. Whilst the previous chapters analyze the PMod constraint with two polarizations, this chapter overcomes this assumption and describes a solution where an arbitrary number of polarizations can be used. The chapter describes why the minimum distance increases with traditional schemes. According to this proposition, communication systems can attain lower BER without an increase in power consumption.

Finally, going one step beyond, the system presented in the previous chapter is introduced in **Chapter 6** with a return link. The chapter introduces some concepts such as PLA and time-series channel generation, which is used to perform the AMC and polarization MIMO scheme optimization. The scenario emulates a ship journey from the center to the edge of a beam. During this journey, the system not only adapts the modulation and coding scheme, but also the polarization MIMO scheme. Results show that a joint optimization of the polarization MIMO scheme offers notably gains compared with AMC adaptation only.

The conclusions of all chapters are summarized as follows:

- The capacity expression of IM can be expressed without indefinite integrals and using approximations involving only arithmetic and harmonic means, which reduces computational complexity.
- IM can be applied to the spatial, polarization and frequency dimensions, considering different impairments depending

on the selected dimension. For instance, the polarization dimension is affected by the specular components. The spatial dimension is affected by the antenna correlation.

- PMod is a very strong candidate to cover the gap between orthogonal block codes and data multiplexing. PMod allows a smoother rate adaptation without requiring extra power consumption. In addition, it can be implemented in realistic systems.
- Hierarchical modulations are enabled with the use of PMod, since different bitrates can be multiplexed using PMod.
- Leveraging the polarization MIMO scheme on top of adaptation of the modulation and coding scheme offers substantial gains. For instance, adaptation of AMC and V-BLAST / PMod offers gains of more than $50\%$ compared with current deployments.
- PMod can be used with an arbitrary number of polarizations (3D PMod), exploiting the Stokes vector, the use of multiple polarizations and PSK.
- 3D PMod has a higher minimum distance than other modulation schemes and, therefore, the BER is reduced without additional power consumption.

Finally, this dissertation takes account of the following aspects:

- Although polarization was discovered many centuries ago (even before Maxwell's equations) it is currently being restricted to specific scenarios. This dissertation aims at presenting the polarization dimension as a more accessible dimension to be considered in upcoming communication systems paradigms.
- PMod and 3D PMod introduce interesting advantages that can be implemented in real-life systems. Additionally, they

increase the throughput without any penalties on extra power consumption and with low computational complexity deployments.

Although this dissertation covers many aspects related with polarization and IM, some open problems arose during the research carried out. These issues open interesting future research lines that can bring new advantages and improvements. The first problem deals with 3D PMod and the constraint of PSK modulation of the initial phase. The second problem deals with the angular momentum of electromagnetic waves.

## 7.2 Future Research

### 7.2.1 QAM & 3D PMod

One of the immediate questions that arises from 3D PMod is

> Can a generalized modulation (i.e., $M$-QAM) be used with 3D PMod?

The answer is still open and, at the moment of writing, there are no studies that focus on this problem. This question contains two major issues that are described in the next sections.

#### Constant Radius Issue

The radius of the Poincaré Sphere is characterized by the Stokes parameter $S_0$. Amongst the use of $S_1$, $S_2$ and $S_3$, it could be possible to map bits into the $S_0$ parameter. This produces different sphere radius, such as QAM with the 2D Cartesian plane. From the transmitter's point of view, this implementation does not present any issue. However, from the receiver perspective, there is a major issue with the Poincaré Sphere.

One of the constraints of using the Poincaré Sphere is that the radius of the sphere is normalized by 1. Indeed, the sphere spans from $S_1$, $S_2$ and $S_3$, but not with the $S_0$ parameter. This implies that the energy of the electromagnetic wave remains constant between in each symbol. Clearly, this assumption does not hold for QAM, since each symbol has different radius.

The system can detect $S_0$ first and after decoding $S_1$, $S_2$ and $S_3$ as described in the Chapter 5. However, the problem requires a very precise calibration of the receiving power, such as in the classical 2D QAM approach. The next issue is related with the use of different radius.

#### Maximizing the Distance

The problem of maximizing the minimum distance is studied in [HS95]. This analysis is valid for the constant unitary radius sphere. However, when two radius are involved the problem becomes different. In this case, there are two concentric spheres with different radius. Whilst the original problem is formulated in terms of maximizing the distance between points, in the second problem the points of the first sphere may pose an influence to the points of the second sphere. The problem is generalized for an arbitrary number of concentric spheres.

This problem has an analogue formulation with the 2D Cartesian constellation APSK. Intuitively, if one circumference is rotated with respect to the other, the BER is decreased. Fig. 7.1 illustrates the concept of rotation in concentric constellations to decrease the BER. This concept can be extended to three dimensions with concentric spheres instead of concentric circumferences. However, the solution is still open for an arbitrary number of spheres.

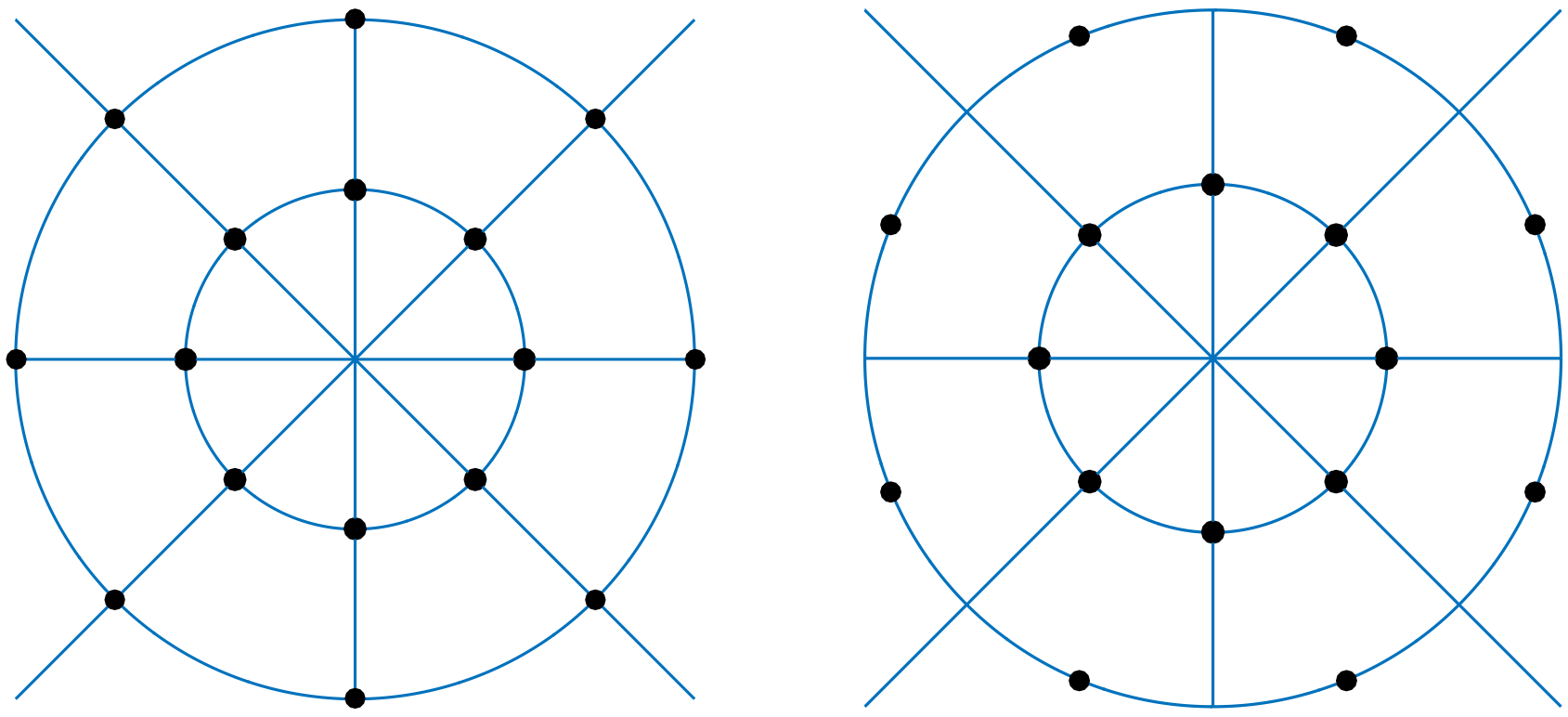

(A) Constellation without rotation.

(B) Constellation with rotation of $\frac{2\pi}{M}$.

FIGURE 7.1: APSK constellations.

Assuming that all symbols are equiprobable, the error probability is bounded by

$$\begin{aligned} P_e &\leq \frac{1}{M} \sum_{m=1}^{M} \sum_{n \neq m} P_{m \to n} \\ &= \frac{1}{M} \sum_{m=1}^{M} \sum_{n \neq m} Q\left(\sqrt{\frac{d_{mn}^2}{2\sigma_w^2}}\right) \\ &\leq (M-1) Q\left(\sqrt{\frac{d_{min}^2}{2\sigma_w^2}}\right), \end{aligned} \tag{7.1}$$

where $Q(x)$ is the Q-Function. Since it is a bound, it implies that the error probability can take lower values. Indeed, maximizing the pair-wise distance decreases the error probability. Hence, although fig. 7.1a and fig. 7.1b have the same minimum distance, which is determined by the distance of adjacent symbols in the inner circumference, the error probability of fig. 7.1b is less than the error probability of fig. 7.1a. This is due to the fact that the outer circumference has less influence into the inner circumference in the second picture compared to the first picture.

Apart from the aspects described above, there is still missing an exhaustive study of 3D PMod under non-Gaussian channels, such as Rayleigh or even with non-linear satellite channels. This could be achieved by performing the same analysis with the channel models described in the previous chapters. Hence, the study of 3D PMod for non-Gaussian channels deserves a future research line.

Finally, it is important to note that this concept is completely compatible with the proposed 3D PMod in Chapter 5. The phase of the QAM modulation scheme can be modulated using the initial phase as described, and the radius of the QAM symbol can be modulated using the $S_0$ Stokes parameter.

### 7.2.2 Orbital Angular Momentum Modulations

One of the inherent properties of electromagnetic waves is that they carry energy and momentum and both quantities are conserved. Angular momentum can be also carried in the wave. Indeed, the polarization is a type of angular momentum and it can be seen as the rotation of the electric field in time, which describes an ellipse. In other words, if we are able to give an angular momentum to the electromagnetic wave, it will be conserved along the propagation.

In all chapters we assumed that the centre of the polarization ellipse remains static and does not have any motion. But the centre of the ellipse can rotate and describe a translational motion. Like the Earth rotates and translates around the Sun, the electromagnetic wave can also rotate and translate. These motions are known as *spinning* and *twisting*. The first is defined by the polarization theory, described in Chapter 2, and named Spin Angular Momentum (SAM). The latter is defined by the wavefront shape and is named Orbital Angular Momentum (OAM). OAM is closely related with the spatial distribution of the electromagnetic field. It is worth mentioning that SAM and OAM are concepts completely independent. A wave carrying OAM can be polarized at the same time with an independent polarization, without affecting the OAM.

It is important to differentiate the non-planar wavefront from the planar electromagnetic waves. Planar electromagnetic waves are waves where the electric and magnetic fields are orthogonal to the direction of propagation. Also, this is a condition to the existence of the Maxwell's equations. The planar wavefront is a region of the space where the electric field is in phase in all points of this region and describes a plane.

OAM modifies the shape of the wavefront, thus changing the phase of certain points in the space. Hence, given a space region, the electric field presents a phase variation. Specifically, the wavefront of an electromagnetic wave that carries OAM describes an helical wavefront shape. Note that, in order to produce a stationary wave, not all angular frequency of OAM are valid. In particular, the orbital angular frequency is the same as the original wave and, therefore, the separation between in-phase points of the helical modes is also $\lambda$.

A particularly interesting aspect is that the OAM accepts different levels (or orbitals) and that all of them are orthogonal. This property implies that a wave that carries OAM can be constructed by infinite helices and, therefore, having infinite orthogonal dimensions. The sign of these levels defines the direction of rotation of the helices.

Nevertheless, generating an electromagnetic wave is a challenging task. In the next section we describe a method to generate waves carrying OAM.

**OAM Generation**

Albeit the OAM has been known in optics for many decades, the application to radio frequency is a relatively recent field with just a few years of work. In 2012, the authors of [EJ12] asked if OAM had a potential in radio communications. Works such as [Moh+10; MW12; MW13] analyze the feasibility using microwaves and describe the methodology to generate waves with OAM. Additionally, a complete communication system is described in [Zha+17]. Finally, [Yan+14] performs a proof-of-concept of SAM-OAM multiplexing, where up to $8$ data channels are multiplexed using the same wave.

In Chapter 2, we described the expressions of the electric field (2.17), where we decoupled the time-space dependency into a static component, $\mathbf{E}_0$. Since the OAM defines the spatial distribution, $\mathbf{E}_0$ must have spatial dependency, i.e., $\mathbf{E}_0(x, y, z)$. In particular, the aforementioned works use the Laguerre-Gaussian modes, which introduce OAM and has the following expression in cylindrical coordinates

$$E_0(r, \phi, z) = \sqrt{\frac{p!}{\pi\,(|l|+p)\,!}}\frac{1}{w(z)}\left(\frac{\sqrt{2}r}{w(z)}\right)^{|l|} e^{-\frac{r^2}{w^2(z)}} L_p^{|l|}\left(\frac{2r^2}{w^2(z)}\right) \times e^{-jk\frac{r^2 z}{2\left(z^2+z_0^2\right)}} e^{-jl\phi} e^{-j(|l|+2p+1)\arctan\left(\frac{z^2}{z^2+z_0^2}\right)}, \tag{7.2}$$

where $l$ is the orbital level, $p$ is the degree of the Laguerre polynomial, $w(z) = w_0\sqrt{1+(z/z_0)^2}$, $w_0$ is the beam waist size, $z_0 = \pi w_0^2/\lambda$, and $L_p^{|l|}(x)$ is the Generalized Laguerre polynomial. Note that for OAM generation, $p = 0$ and $L_p^{|l|}(x) = 1$. The term $e^{-jl\phi}$ defines the orbital shape corresponding to this mode.

Figs. 7.2, 7.3, 7.4, 7.5, 7.6, 7.7 and 7.8 describe different perspectives of the electric field at $z = 20\lambda$. The left plot illustrates the helical modes for each level. It is clear that each mode has $|l|$ helices. The figure only plots all points that have the same phase. The middle figure depicts the phase of the electric field from the $\mathbf{z}$ axis. This figure describes the direction of rotation of the electric field. The rightmost figure illustrates the intensity distribution.

Each orbital level is concentric to the others and, thus, orthogonal. Note that the distribution aperture increases as a function of $z$ and depicts a cone in 3D.

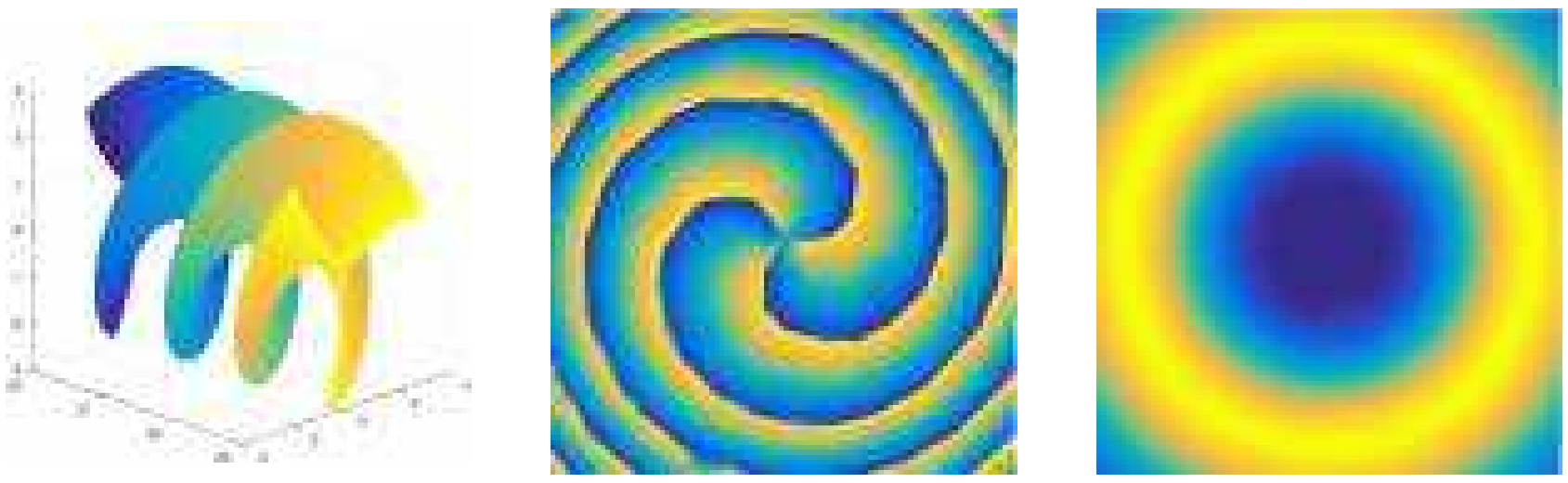

FIGURE 7.2: Orbital mode $l = -3$. Helical wavefront with the same phase, frontal phase and frontal intensity.

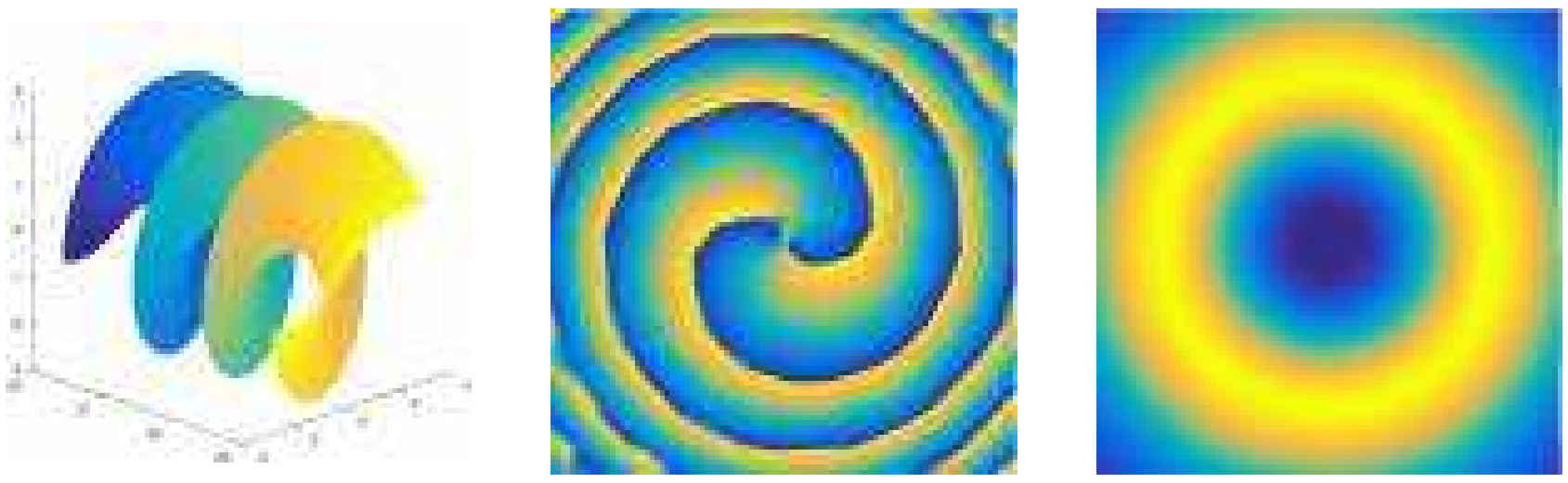

FIGURE 7.3: Orbital mode $l = -2$. Helical wavefront with the same phase, frontal phase and frontal intensity.

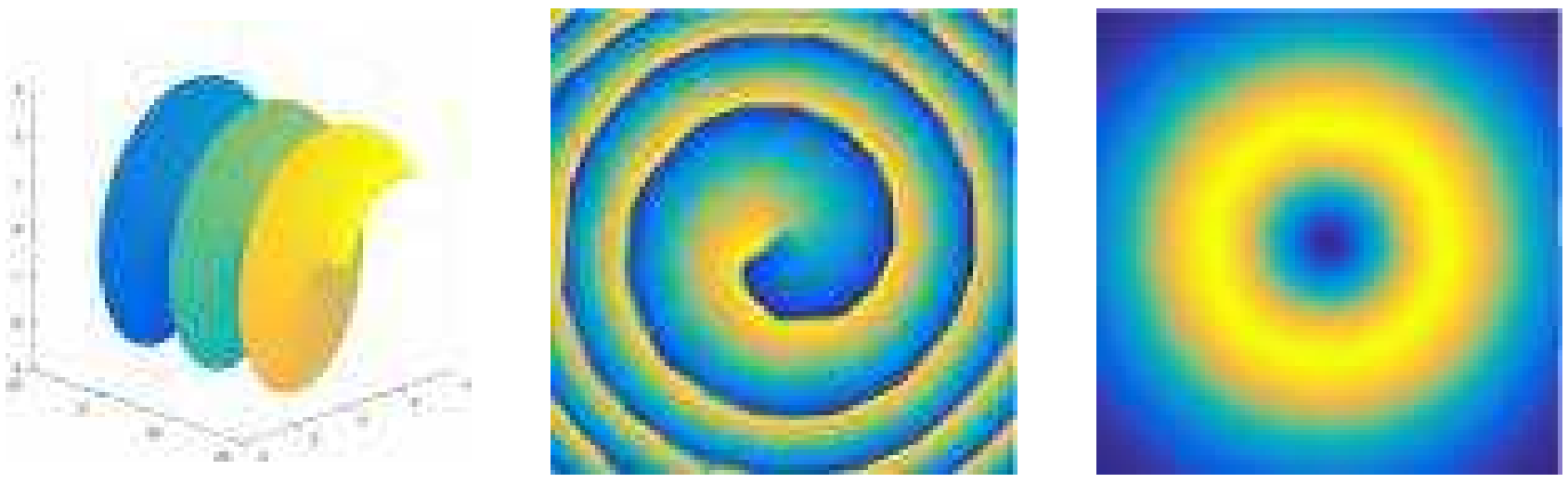

FIGURE 7.4: Orbital mode $l = -1$. Helical wavefront with the same phase, frontal phase and frontal intensity.

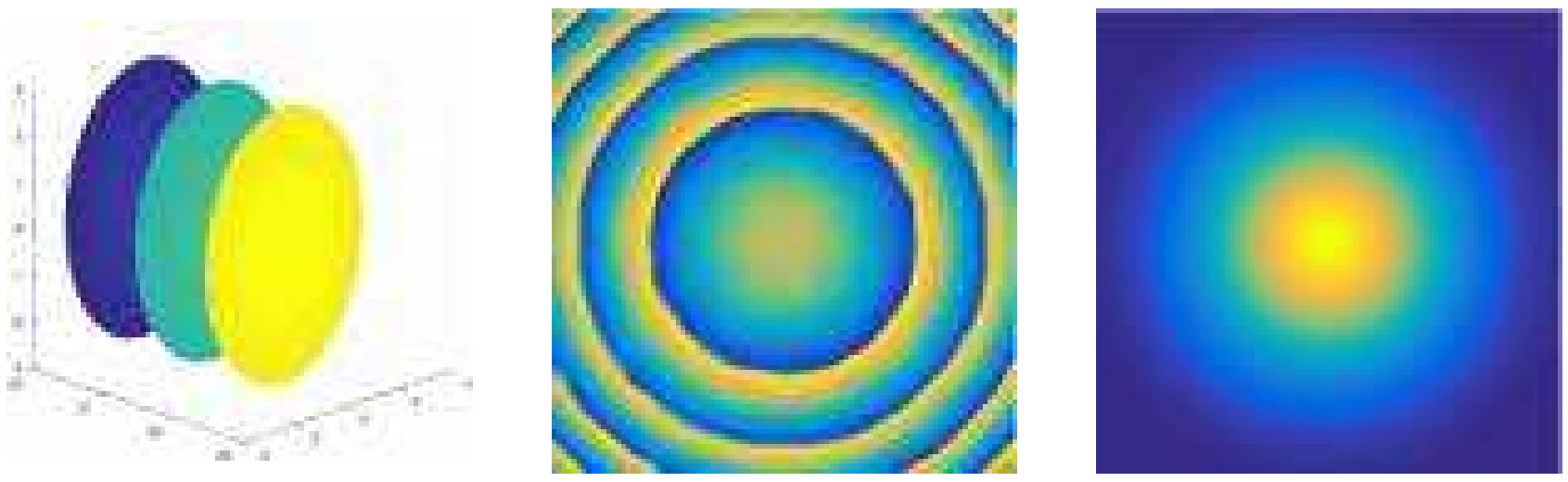

FIGURE 7.5: Orbital mode $l = 0$. Helical wavefront with the same phase, frontal phase and frontal intensity.

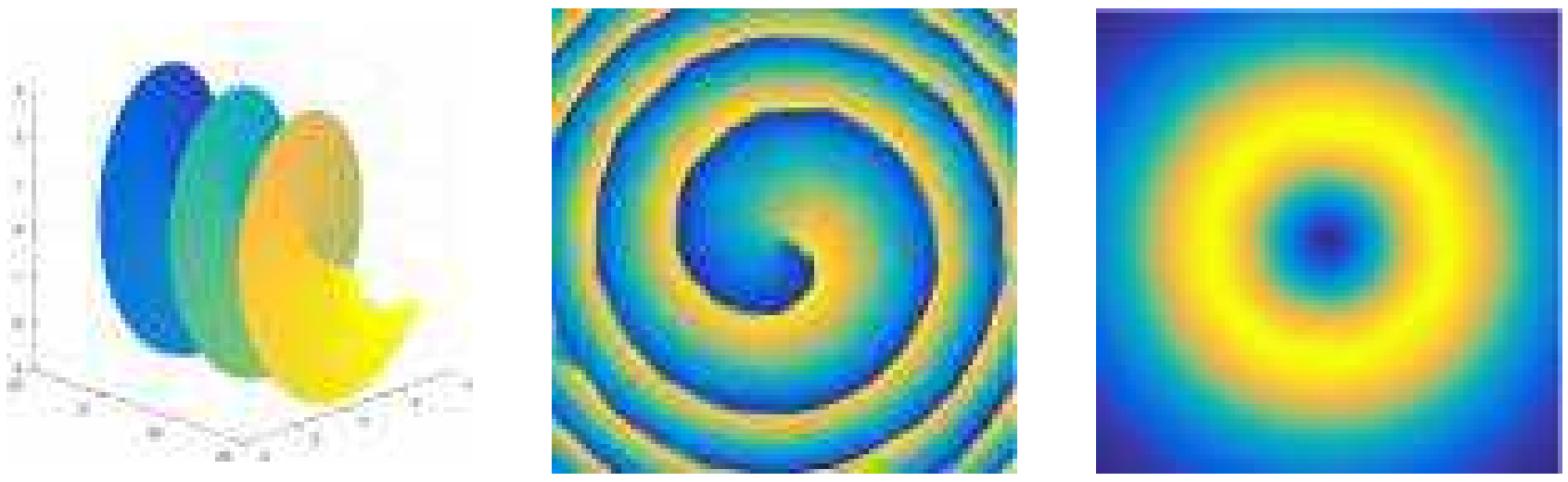

FIGURE 7.6: Orbital mode $l = 1$. Helical wavefront with the same phase, frontal phase and frontal intensity.

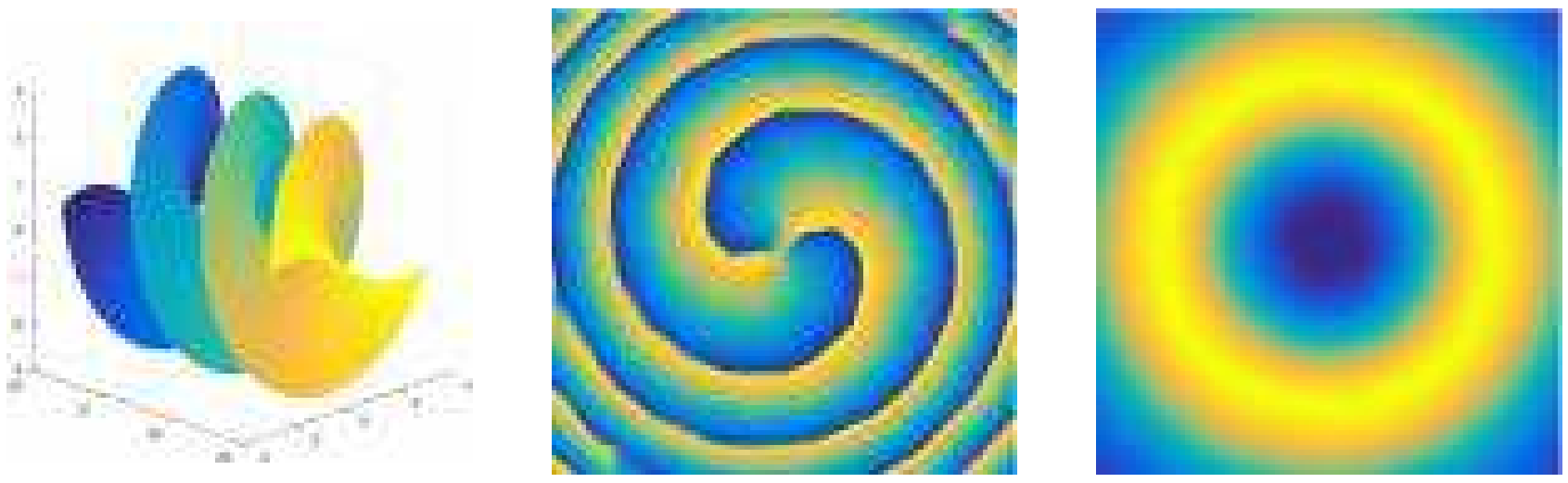

FIGURE 7.7: Orbital mode $l = 2$. Helical wavefront with the same phase, frontal phase and frontal intensity.

In optics, Laguerre-Gaussian modes can be generated either from the laser source or by means of Spiral Phase Plates (SPP). SPP are plates such that their height varies depending on the phase. The total height is a multiple of $\lambda$, and the direction of rotation is given by the plate. Hence, to generate the $m$th mode, the height of the plate must be $\pm m\lambda$. Fig. 7.9 depicts the SPP for $m = \pm 1$. Since the material of the plate has a lower wavenumber, the light leaves this medium at different speeds depending on the phase. The resulting waveform is twisted, carrying an OAM.

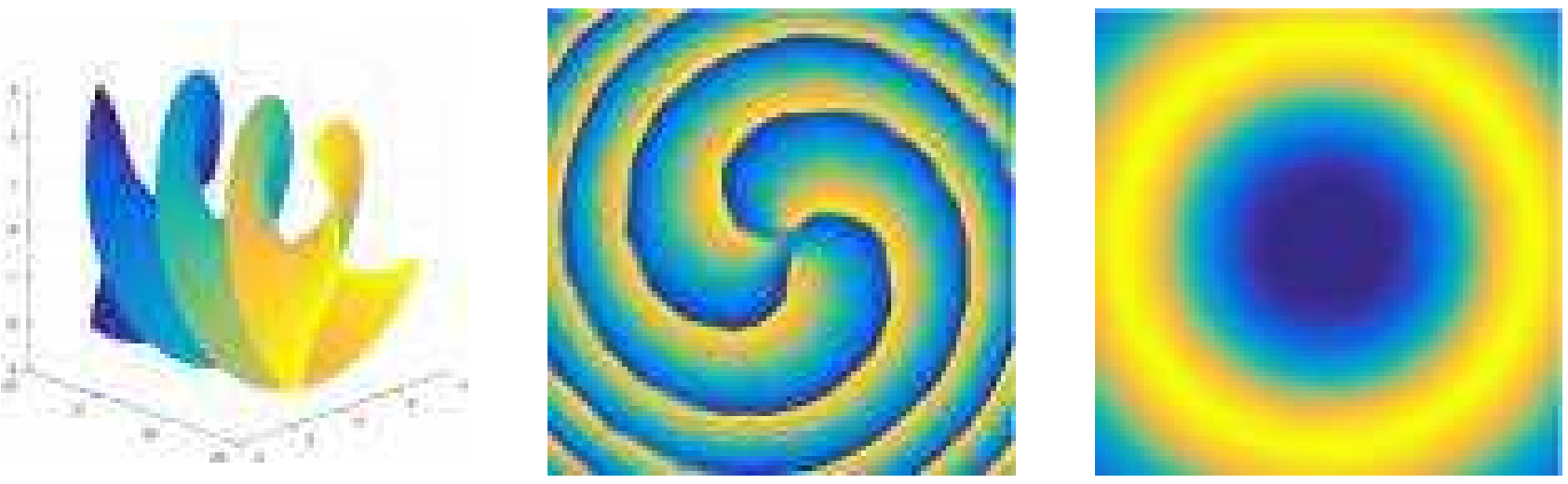

FIGURE 7.8: Orbital mode $l = 3$. Helical wavefront with the same phase, frontal phase and frontal intensity.

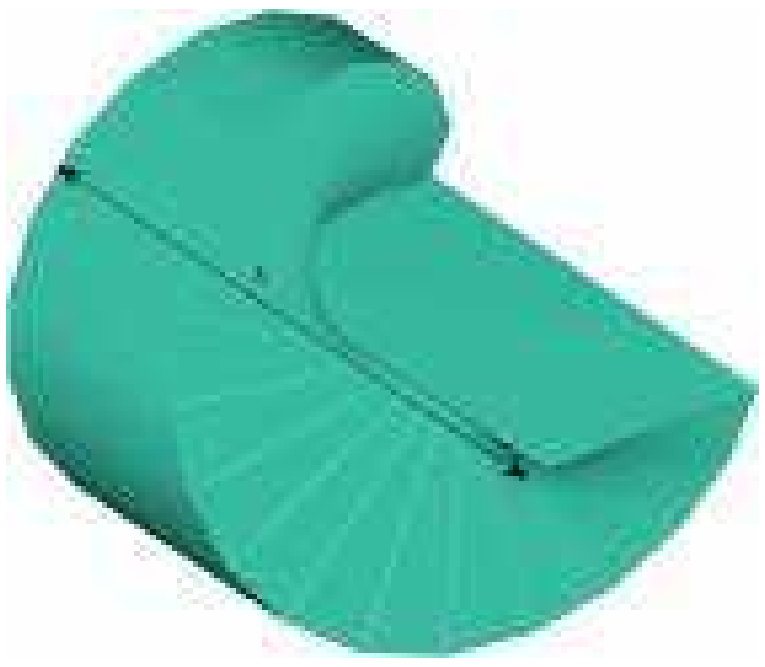

(A) $m = 1$

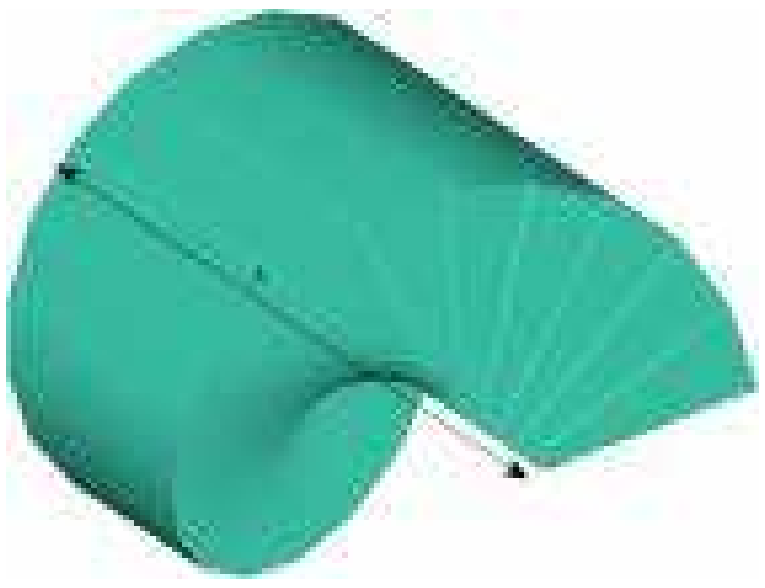

(B) $m = -1$.

FIGURE 7.9: SPP for $m = \pm 1$.

In [Yan+14], a SPP for 60 GHz is used for data multiplexing. However, using SPP lacks flexibility. Following the same approach, it is possible to generate OAM using circular arrays of antennas. In [BTA14; Bai+14], circular arrays are used to generate $m = \pm 1$ OAM. Varying the phase of each antenna generates OAM. Finally, [Bai+15] implements a novel scheme using horn arrays with Septum polarizers, described in 2.

### Beamforming and OAM

From the point of view of beamforming, this is particular interesting. OAM waveforms are, indeed, directive transmissions, such as in array theory. The difference between classic array theory and OAM is the use of the vortex to modulate information. The result can be envisaged to construct a classical beam with a vortex in the center. Depending on the shape of the vortex, it is possible to recover the OAM index. Moreover, it is possible to multiplex different beams jointly without interfering.

Hence, OAM enables to multiplex different sources of information orthogonally. In the same sense of PMod / 3D PMod, OAM can be applied in IM. By selecting which OAM index, it carries information. Following this approach, a primary system of Shift Keying using OAM is studied in [Kai+17].

However, the reception side is still a challenging problem. The sensitivity of the terminal plays a crucial role in the detection. Examining figs. 7.2b, 7.3b, 7.4b, 7.5b, 7.6b, 7.7b, 7.8b we can appreciate that the phase contains the helical mode. Thus, applying the gradient method we could estimate this parameter. In [CA15], this method is studied with realistic measurements.

In conclusion, OAM provides an additional degree of freedom to IM. It has been studied in different research areas and it can be applied to radiocommunications. It can be combined with polarization dimension, as an additional domain. Although it requires an exhaustive research, the results are courageous to persist in this line.

# Appendix A

# Decoupling Linear Polarizations

A wave linearly polarized implies that phases of $E_x$ and $E_y$ are the same (or shifted by $\pm\pi$). For simplicity, we demonstrate that the sum of two circular and opposite polarizations produces a linear polarization.

This statement can be demonstrated partially using Jones Vectors. Jones Vectors of RHCP and LHCP are described as

$$\begin{aligned} E_{\text{RHCP}} &= \frac{1}{\sqrt{2}} \begin{pmatrix} 1 \\ -j \end{pmatrix} \\ E_{\text{LHCP}} &= \frac{1}{\sqrt{2}} \begin{pmatrix} 1 \\ j \end{pmatrix} . \end{aligned} \tag{A.1}$$

Hence, the addition of both circular polarizations is

$$E = E_{\text{RHCP}} + E_{\text{LHCP}} = \frac{2}{\sqrt{2}} \begin{pmatrix} 1 \\ 0 \end{pmatrix} , \tag{A.2}$$

which corresponds to the linear polarization. However, Jones calculus does not contains the information of initial phase. To solve it for any initial phase, we use the complete formulation.

The electric field of RHCP is denoted by

$$\begin{aligned} E_x^{(R)} &= E_0 e^{j\phi_R} \\ E_y^{(R)} &= E_0 e^{j\left(\phi_R + \frac{\pi}{2}\right)} \end{aligned} \tag{A.3}$$

and for the LHCP,

$$\begin{aligned} E_x^{(L)} &= E_0 e^{j\left(\phi_L + \frac{\pi}{2}\right)} \\ E_y^{(L)} &= E_0 e^{j\phi_L}. \end{aligned} \tag{A.4}$$

The sum of both polarizations is

$$\begin{aligned} E_x &= E_x^{(R)} + E_x^{(L)} = E_0 e^{j\phi_R} + E_0 e^{j\left(\phi_L + \frac{\pi}{2}\right)} \\ E_y &= E_y^{(R)} + E_y^{(L)} = E_0 e^{j\phi_L} + E_0 e^{j\left(\phi_R + \frac{\pi}{2}\right)}. \end{aligned} \tag{A.5}$$

The objective is to demonstrate that $E_x$ and $E_y$ have the same phase. We note that

$$\begin{aligned} E_x &= E_0 \left(\cos\phi_R + j\sin\phi_R\right) + E_0 \left(\cos\left(\phi_L + \frac{\pi}{2}\right) + j\sin\left(\phi_L + \frac{\pi}{2}\right)\right) \\ E_y &= E_0 \left(\cos\phi_L + j\sin\phi_L\right) + E_0 \left(\cos\left(\phi_R + \frac{\pi}{2}\right) + j\sin\left(\phi_R + \frac{\pi}{2}\right)\right). \end{aligned} \tag{A.6}$$

Hence, the phase of $E_x$ and $E_y$ is

$$\begin{aligned} \arg\left(E_x\right) &= \arctan\left(\frac{\sin\phi_R + \sin\left(\phi_L + \frac{\pi}{2}\right)}{\cos\phi_R + \cos\left(\phi_L + \frac{\pi}{2}\right)}\right) \\ \arg\left(E_y\right) &= \arctan\left(\frac{\sin\phi_L + \sin\left(\phi_R + \frac{\pi}{2}\right)}{\cos\phi_L + \cos\left(\phi_R + \frac{\pi}{2}\right)}\right). \end{aligned} \tag{A.7}$$

Using the trigonometric identities $\sin\alpha + \sin\beta = 2\sin\left(\frac{\alpha+\beta}{2}\right)\cos\left(\frac{\alpha-\beta}{2}\right)$

and $\cos\alpha + \cos\beta = 2\cos\left(\frac{\alpha+\beta}{2}\right)\cos\left(\frac{\alpha-\beta}{2}\right)$, both arguments are reduced to

$$\begin{aligned}\arg\left(E_x\right) &= \arctan\left(\frac{\sin\left(\frac{\phi_R+\phi_L+\frac{\pi}{2}}{2}\right)}{\cos\left(\frac{\phi_R+\phi_L+\frac{\pi}{2}}{2}\right)}\right) = \frac{\phi_R+\phi_L}{2}+\frac{\pi}{4}\\ \arg\left(E_y\right) &= \arctan\left(\frac{\sin\left(\frac{\phi_L+\phi_R+\frac{\pi}{2}}{2}\right)}{\cos\left(\frac{\phi_L+\phi_R+\frac{\pi}{2}}{2}\right)}\right) = \frac{\phi_L+\phi_R}{2}+\frac{\pi}{4}.\end{aligned} \tag{A.8}$$

Since both arguments are the same, we prove that the result is a linear polarization.

# Appendix B

# Linear Block Codes

This section describes some of the most used Linear Block Codes. Note that these codes are not the unique solution, since can be rotated.

## B.1 Orthogonal Block Codes

For $t = 2$, $M = 2$, $N = 2$, with a rate of $R = 1$:

$$\mathbf{X} = \begin{pmatrix} s_1 & s_2^* \\ s_2 & -s_1^* \end{pmatrix} \tag{B.1}$$

For $t = 3$, $M = 4$, $N = 3$, with a rate of $R = 3/4$:

$$\mathbf{X} = \begin{pmatrix} s_1 & -s_2^* & s_3^* & 0 \\ s_2 & s_1^* & 0 & -s_3^* \\ s_3 & 0 & -s_1^* & s_2^* \end{pmatrix} \tag{B.2}$$

For $t = 4$, $M = 4$, $N = 3$, with a rate of $R = 3/4$:

$$\mathbf{X} = \begin{pmatrix} s_1 & 0 & s_2 & -s_3 \\ 0 & s_1 & s_3^* & s_2^* \\ -s_2^* & -s_3 & s_1^* & 0 \\ s_3^* & -s_2 & 0 & s_1^* \end{pmatrix} \tag{B.3}$$

For $t = 8$, $M = 8$, $N = 3$, with a rate of $R = 3/4$:

$$\mathbf{X} = \begin{pmatrix} s_1 & 0 & 0 & 0 & s_4 & 0 & s_2 & -s_3 \\ 0 & s_1 & 0 & 0 & 0 & s_4 & s_3^* & s_2^* \\ 0 & 0 & s_1 & 0 & -s_2^* & -s_3 & s_4^* & 0 \\ 0 & 0 & 0 & s_1 & s_3^* & -s_2 & 0 & s_4^* \\ -s_4^* & 0 & s_2 & -s_3 & s_1^* & 0 & 0 & 0 \\ 0 & -s_4^* & s_3^* & s_2^* & 0 & s_1^* & 0 & 0 \\ -s_2^* & -s_3 & -s_4 & 0 & 0 & 0 & s_1^* & 0 \\ s_3^* & -s_2 & 0 & -s_4 & 0 & 0 & 0 & s_1^* \end{pmatrix} \tag{B.4}$$

## B.2 Data Multiplexing

For $t = M = N$, with a rate of $R = 1$ (D-BLAST):

$$\mathbf{X} = \begin{pmatrix} s_1 & s_N & \cdots & s_2 \\ s_2 & s_1 & \cdots & s_3 \\ \vdots & \vdots & \ddots & \vdots \\ s_N & s_{N-1} & \cdots & s_1 \end{pmatrix} \tag{B.5}$$

For $t = M = N$, with a rate of $R = N$ (V-BLAST):

$$\mathbf{X} = \begin{pmatrix} s_1 & s_{N+1} & \cdots \\ s_2 & s_{N+2} & \cdots \\ \vdots & \vdots & \ddots \\ s_N & s_{2N} & \cdots \end{pmatrix} \tag{B.6}$$

For $t = M = N$, with a rate of $R = 1$ (Diagonal Code):

$$\mathbf{X} = \begin{pmatrix} s_1 & 0 & \cdots & 0 \\ 0 & s_2 & \cdots & 0 \\ \vdots & \vdots & \ddots & \vdots \\ 0 & 0 & \cdots & s_N \end{pmatrix} \tag{B.7}$$

## B.3 Other Block Codes

Golden Codes:

$$\mathbf{X} = \frac{1}{\sqrt{5}} \begin{pmatrix} \alpha\left(s_1 + \theta s_2\right) & \alpha\left(s_3 + \theta s_4\right) \\ j\bar{\alpha}\left(s_3 + \bar{\theta} s_4\right) & \bar{\alpha}\left(s_3 + \bar{\theta} s_4\right) \end{pmatrix} \tag{B.8}$$

where $\theta = \frac{1+\sqrt{5}}{2}$, $\bar{\theta} = \frac{1-\sqrt{5}}{2}$, $\alpha = 1 + j + j\theta$, $\bar{\alpha} = 1 + j + j\bar{\theta}$.

Silver Codes:

$$\mathbf{X} = \begin{pmatrix} s_1 + z_3 & -s_2^* - z_4^* \\ s_2 - z_4 & s_1^* - z_3^* \end{pmatrix} \tag{B.9}$$

where

$$\begin{pmatrix} z_3 & z_4 \end{pmatrix} = \frac{1}{\sqrt{7}} \begin{pmatrix} 1+j & -1+2j \\ 1+2j & 1-j \end{pmatrix} \begin{pmatrix} s_3 \\ s_4 \end{pmatrix}. \tag{B.10}$$

# Appendix C

# Demonstration of Faraday Effect

In 1892, Hendrik Lorentz, based on the works of Joseph J. Thomson and James C. Maxwell, derived the formula of the force applied to a moving charged particle in the presence of electric and magnetic fields. The expression takes the following form

$$\mathbf{F} = q\left(\mathbf{E} + \mathbf{v} \times \mathbf{B}\right) \tag{C.1}$$

where $q$ is the charge quantity of the particle and $\mathbf{v}$ is the velocity vector of the moving particle. Since the ionosphere can be modeled as a free plasma of electrons, $q$ is the charge of an electron. Thus, assuming that $\mathbf{v} = \frac{d\mathbf{r}}{dt}$,

$$m\frac{d^2\mathbf{r}}{dt^2} + K\mathbf{r} = q\left(\mathbf{E} + \frac{d\mathbf{r}}{dt} \times \mathbf{B}\right) \tag{C.2}$$

where $m$ is the mass of the particle, $K = m\omega_e^2$ is the restoring force constant, which restores the particle to the equilibrium point. For the sake of simplicity, we perform the following demonstration in the complex domain. The presence of an oscillating electric field induces a movement of the particle, at the same frequency of the oscillating electric field, whose movement is described by $\mathbf{r} = \mathbf{r}_0 e^{j(\omega t - kz)}$, where $\mathbf{r}_0 = r_{0x}\mathbf{x} + r_{0y}\mathbf{y}$. Additionally, we assume that the wave is propagated along the $\mathbf{z}$ axis and $\mathbf{B} = B_0\mathbf{z}$. Hence,

(C.2) is reduced to

$$m\left(\omega_e^2-\omega^2\right)\mathbf{r}_0 e^{j(\omega t-kz)} = q\left(\mathbf{E}_0 e^{j(\omega t-kz)} + j\omega\mathbf{r}_0 e^{j(\omega t-kz)} \times B_0\mathbf{z}\right). \tag{C.3}$$

Arranging terms, we have that

$$\Delta\omega\mathbf{r}_0 - j\omega\Omega\left(\mathbf{r}_0\times\mathbf{z}\right) = \frac{q}{m}\mathbf{E}_0, \tag{C.4}$$

where $\Delta\omega = \omega_e^2 - \omega^2$ and $\Omega = \frac{qB_0}{m}$ is the cyclotron frequency of the particle. Before to proceed, we first define a new basis $(\mathbf{e}_+, \mathbf{e}_-)$, such that $\mathbf{e}_+ = \mathbf{x} + j\mathbf{y}$ and $\mathbf{e}_- = \mathbf{x} - j\mathbf{y}$[1]. Note that $\mathbf{e}_\pm$ is an orthogonal basis that can describe circular polarizations using only one component of this basis. Hence, $\mathbf{e}_+$ is the basis component for the LHCP and $\mathbf{e}_-$ is for the RHCP. Therefore, we can rewrite (C.4) as

$$\begin{aligned}
\Delta\omega\left(r_{0x}\frac{\mathbf{e}_+ + \mathbf{e}_-}{2} + r_{0y}\frac{\mathbf{e}_+ - \mathbf{e}_-}{2j}\right) - j\omega\Omega\left(r_{0y}\frac{\mathbf{e}_+ + \mathbf{e}_-}{2} - r_{0x}\frac{\mathbf{e}_+ - \mathbf{e}_-}{2j}\right)&\\
= \frac{q}{m}\left(E_{0x}\frac{\mathbf{e}_+ + \mathbf{e}_-}{2} + E_{0y}\frac{\mathbf{e}_+ - \mathbf{e}_-}{2j}\right)&\\
\left(\Delta\omega + \omega\Omega\right)\left(r_{0x} - jr_{0y}\right)\mathbf{e}_+ + \left(\Delta\omega - \omega\Omega\right)\left(r_{0x} + jr_{0y}\right)\mathbf{e}_-&\\
= \frac{q}{m}\left[\left(E_{0x} - jE_{0y}\right)\mathbf{e}_+ + \left(E_{0x} + jE_{0y}\right)\mathbf{e}_-\right].&
\end{aligned} \tag{C.5}$$

Defining $r_{0\pm} = r_{0x} \pm jr_{0y}$ and $E_{0\pm} = E_{0x} \pm jE_{0y}$, we have the following equations per each basis component

$$\begin{aligned}
\left(\Delta\omega - \omega\Omega\right) r_{0+} &= \frac{q}{m}E_{0+}\\
\left(\Delta\omega + \omega\Omega\right) r_{0-} &= \frac{q}{m}E_{0-}
\end{aligned} \tag{C.6}$$

and the displacement $r_{0\pm}$ can be expressed with a single equation as

$$r_{0\pm} = \frac{q}{m}\frac{E_{0\pm}}{\Delta\omega \mp \omega\Omega}. \tag{C.7}$$

The polarization vector can be computed as $\mathbf{P} = \varepsilon_0\left(\varepsilon_r - 1\right)\mathbf{E}$, where $\varepsilon_0$ and $\varepsilon_r$ are the vacuum permittivity constant and relative

[1] It is shown that $\mathbf{x} = \frac{\mathbf{e}_+ + \mathbf{e}_-}{2}$ and $\mathbf{y} = \frac{\mathbf{e}_+ - \mathbf{e}_-}{2j}$

permittivity, respectively. But the polarization vector can also be computed as the sum of all contributions of each particle in the medium. The polarization vector of each particle is equivalent to the dipole moment, expressed as $\mathbf{p} = q\mathbf{r}$. Hence, assuming that the medium contains $N_e$ particles per volume unit describing the same movement, then the total polarization vector can be expressed $\mathbf{P} = N_e q\mathbf{r}$. Expressing both expressions in the circular basis $\mathbf{e}_{\pm}$, the relative permittivity can be expressed as

$$\begin{aligned} \varepsilon_{r\pm} &= 1 + \frac{N_e q^2}{\varepsilon_0 m \left(\Delta\omega \mp \omega\Omega\right)} \\ &= 1 + \frac{\omega_p^2}{\omega_e^2 - \omega\left(\omega \mp \Omega\right)}. \end{aligned} \tag{C.8}$$

where $\omega_p = \sqrt{\frac{N_e q^2}{m\varepsilon_0}} = 2\pi f_p$ is the plasma frequency of the medium.

The wave number $k$ can be computed from the relative permittivity as follows

$$k_{\pm} = k_0 n_{\pm} = k_0 \left(\sqrt{\varepsilon_{r\pm}}\right)^* \tag{C.9}$$

where $k_0 = \omega/c$ is the vacuum wave number. An important remark is the dependency with $\omega$. In the next section we analyze the function of $k_{\pm}$.

Having two different wave numbers depending on the basis component implies that the electric field radiated in one component is travelling with lower phase velocity. Hence, we can express the electric field through the medium as

$$\mathbf{E}(z,t) = \mathbf{E}_+ e^{j(\omega t - k_+ z)} + \mathbf{E}_- e^{j(\omega t - k_- z)} \tag{C.10}$$

where $\mathbf{E}_\pm = E_\pm \mathbf{e}_\mp$ is the electric field using the circular basis. Thus, expressing the electric field in Cartesian coordinates

$$\begin{aligned}\mathbf{E}(z,t) &= \left(E_{0+}e^{j(\omega t - k_+ z)} + E_{0-}e^{j(\omega t - k_- z)}\right)\mathbf{x}\\ &+ j\left(E_{0-}e^{j(\omega t - k_- z)} - E_{0+}e^{j(\omega t - k_+ z)}\right)\mathbf{y}\end{aligned} \tag{C.11}$$

For the sake of simplicity, we define the $\mathbf{x}$ axis in the direction of linear polarization, i.e., $E_{0x} = E_0$ and $E_{0y} = 0$. After some mathematical arrangements, we can describe the electric field as

$$\begin{aligned}\mathbf{E}(z,t) &= E_0\left[\left(e^{j(\omega t - k_- z)} + e^{j(\omega t - k_+ z)}\right)\mathbf{x} + j\left(e^{j(\omega t - k_- z)} - e^{j(\omega t - k_+ z)}\right)\mathbf{y}\right]\\ &= E_0 e^{j\left(\omega t - \frac{k_- + k_+}{2}z\right)}\left[\left(e^{-j\frac{k_- - k_+}{2}z} + e^{j\frac{k_- - k_+}{2}z}\right)\mathbf{x}\right.\\ &\left.+ j\left(e^{-j\frac{k_- - k_+}{2}z} - e^{j\frac{k_- - k_+}{2}z}\right)\mathbf{y}\right]\\ &= 2E_0 e^{j\left(\omega t - \frac{k_- + k_+}{2}z\right)}\left[\cos\left(\frac{k_- - k_+}{2}z\right)\mathbf{x} + \sin\left(\frac{k_- - k_+}{2}z\right)\mathbf{y}\right]\\ &= 2E_0 e^{j\left(\omega t - \bar{k}z\right)}\left[\cos\left(\Delta k z\right)\mathbf{x} + \sin\left(\Delta k z\right)\mathbf{y}\right]\end{aligned} \tag{C.12}$$

where $\bar{k} = \frac{k_- + k_+}{2}$ and $\Delta k = \frac{k_- - k_+}{2}$. The interesting aspect of this equation is twofold: 1) the wave number across the medium is the average individual wave numbers of each circular polarization component; and 2) the linear polarization rotates as a function of $z$. Hence, we can compute the rotation with respect the initial polarization axis $\mathbf{x}$ as

$$\theta = \Delta k z. \tag{C.13}$$

Although the demonstration is performed assuming that the external field is parallel to the direction of propagation, it is straightforward to show that, in any case, the results depends on the angle between $B_0$ and $\mathbf{z}$. The medium is still birefringence but the magnitude of such phenomena depends on this angle.

## C.1 Wave Number Analysis

In the previous section we formulated the form of the wave number of each circular polarization component as a function of the angular frequency. In this section we analyze the behaviour of the wave number for different interesting frequencies.

Since the electromagnetic waves make sense for positive and non-zero angular frequencies, we omit the study for $\omega = 0$. In the limit, $\omega \to +\infty$, it is straightforward to show that the wave number of each component tends to $1$ regardless the component:

$$\lim_{\omega \to +\infty} k_{\pm} = 1. \tag{C.14}$$

This is particularly interesting because it reveals that the medium does not affect to the electromagnetic wave and, therefore, it acts as in the vacuum. Therefore, high frequencies do not rotate the incident linear polarized wave.

However, there is a critical frequency such that the denominator is zero. Assuming that the angular frequency is always positive, it is easy to shown that each circular component has a critical angular frequency such that the wave number is not well defined. Hence,

$$\omega_{\pm}^{\dagger} = \frac{\mp\Omega + \sqrt{\Omega^2 + 4\omega_e^2}}{2} = 2\pi f_{\omega\pm}^{\dagger}. \tag{C.15}$$

Although both frequencies are always different ($\omega_{+}^{\dagger} \leq \omega_{-}^{\dagger}$), it implies that for such frequency, the circular component is suppressed. In this case, the incident linear polarized electromagnetic wave is transformed into a circular polarization.

The electron density in the ionosphere varies for many aspects. For instance, the interaction with the Sun winds affects the density of electrons. Thus, depending on the daylight, this quantity varies significantly. Moreover, the density also depends on the altitude and we can distinguish several sublayers in the ionosphere, each with different densities. To illustrate the concept of critical frequency, we use a typical value of $N_e = 10^{12}$ [$\mathrm{m}^{-3}$]. Since the terrestrial magnetic field is $B_0 = 50\ \mu$T, the cyclotron frequency of an electron is $\Omega = 2\pi f_\Omega$, where $f_\Omega = 1400$ MHz. The plasma frequency is $f_p = 9$ MHz, in the case of electron with the aforementioned density.

Examining (C.15), we can deduce that exists two frequencies such that LHCP and RHCP are vanished. Using the numbers described before, the critical frequency for LHCP is $f^\dagger_{\omega+} \approx 58$ kHz and for RHCP is $f^\dagger_{\omega-} \approx 1400$ MHz.

From the communications point of view, in one hand, the density of plasma is determinant for the LHCP critical frequency. However, the quantity $\sqrt{\Omega^2 + 4\omega_e^2} - \Omega$ is low, which produces a low frequency. In satellite communications, which are affected by the Faraday rotation, are defined from 1 GHz and above. See Fig. C.1. Hence, we assume that the LHCP component of a linear polarization is never affected by the Faraday rotation. However, on the other hand, the RHCP critical frequency can be approximated to $\Omega$, which is in the order of GHz. In this case, at this frequency, the RHCP component is vanished. The result is that the incident linear polarized electromagnetic wave leaves the ionosphere with a LHCP polarization.

Critical frequencies only depend on the cyclotron frequency $\Omega$ and the natural frequency of electron $\omega_e$. Moreover, note that there is the region in the critical frequency vicinity such that the wave number becomes complex. In this case, a complex wave number implies that the medium absorbs the incident wave. The other area near the critical frequency implies that the wave number tends to infinity. In this case, the incident wave is reflected. Hence, we distinguish three zones of frequencies:

1. Refractive zone: all frequencies such that the incident electromagnetic wave is propagated without reflections or absorptions.

2. Reflective zone: frequencies such that the ionosphere reflects the incident wave completely.

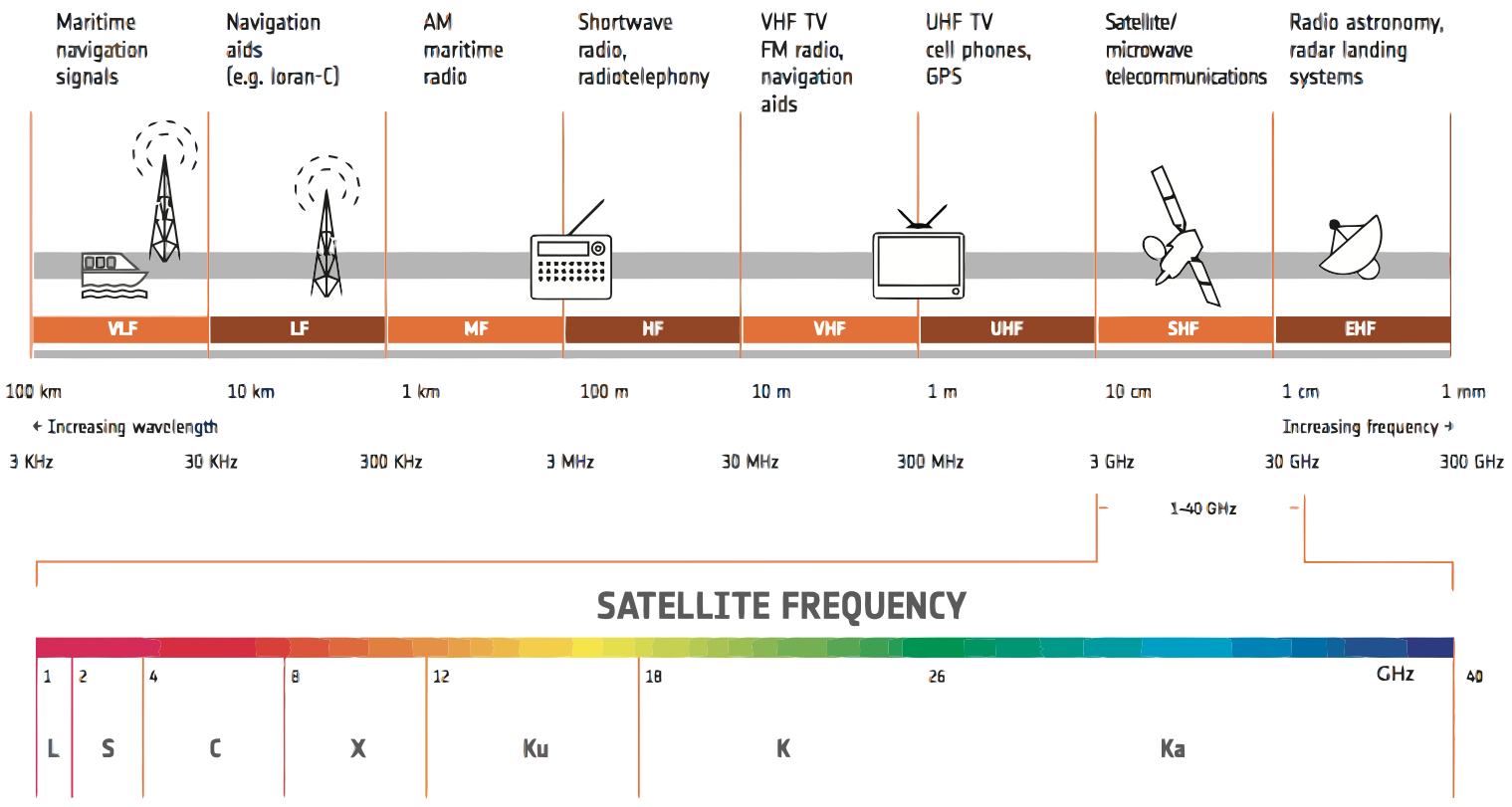

FIGURE C.1: Satellite frequencies. Credits to ESA.

3. Absorption zone: frequencies such that the ionosphere absorbs the incident wave.

All these areas are summarized in Fig. C.2. Note that frequencies in reflective and absorption zones causes that the incident wave is not transmitted through the ionosphere. In this case, the circular polarization component is not propagated and therefore, the resulting wave is circularly polarized.

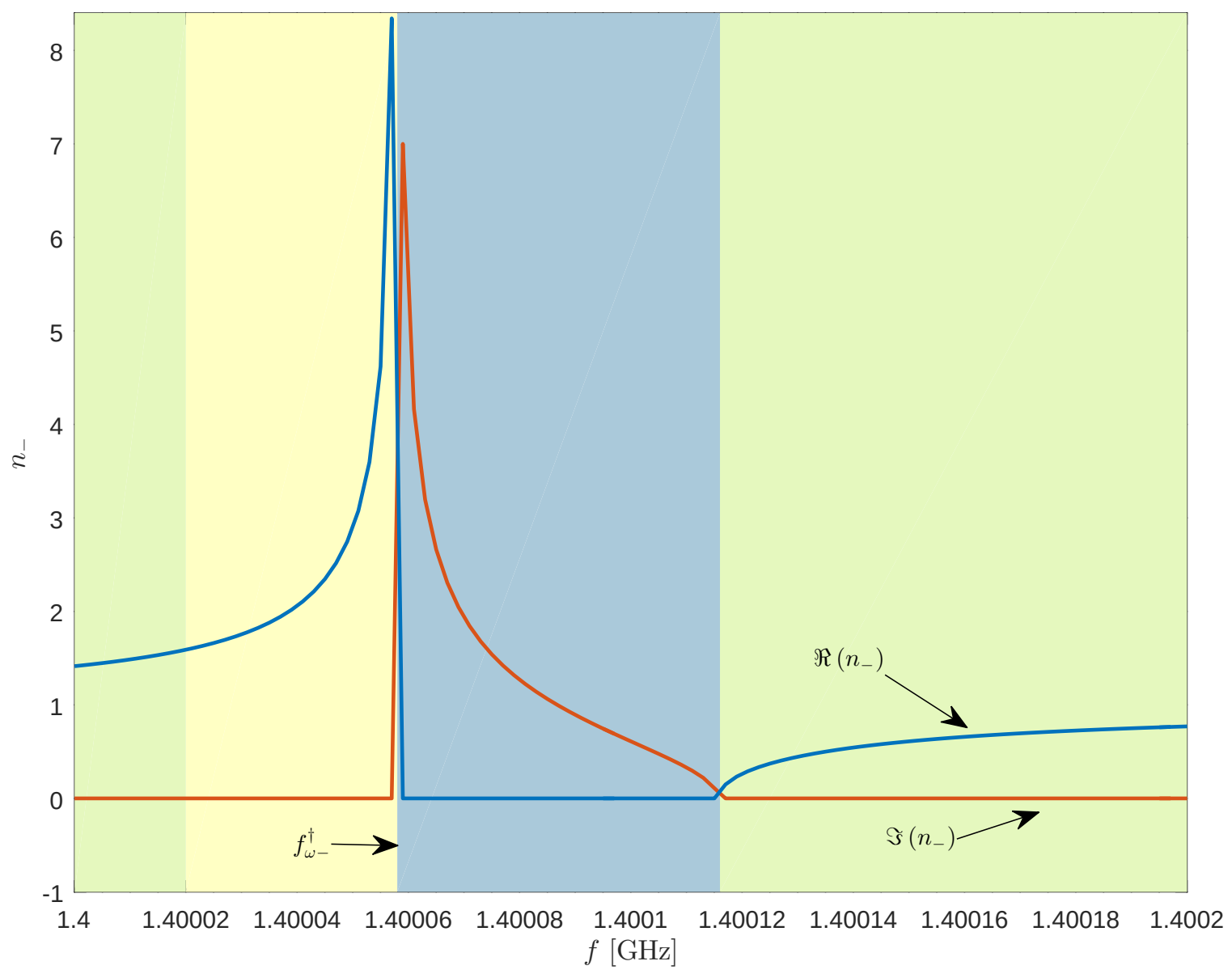

FIGURE C.2: Different regions for the frequencies near to the critical frequency. The green area corresponds to the full refraction. The yellow area is the reflected frequency range. The blue area is the absorbed frequency range.

# Appendix D

# Derivations of the Ergodic Capacity

Based on (3.40), we need to compute three expectations: $E_1 = \mathbb{E}\left[\log_2\left(\sigma_1^2\right)\right]$, $E_2 = \mathbb{E}\left[\log_2\left(\sigma_1^2+\sigma_2^2\right)\right]$ and $E_3 = \mathbb{E}\left[\frac{\sigma_1^2}{\sigma_2^2}\right]$.

The distribution of $\sigma_l^2$ is closely related to the distribution of $h_{il}$. In the following subsections we study the ergodic capacity by using the 2nd order approximation (3.27), particularized by $t = 2$, for different common channel distributions, Rice, Nakagami-$m$ and Rayleigh channel distributions, where the latter can be derived from Nakagami-$m$ with $m = 1$.

## D.1 Nakagami-$m$ Channel

In the Nakagami-$m$ channel the envelope of each component of the channel matrix $|h_{ij}| \sim$ Nakagami-$m(\Omega)$ follows the Nakagami-$m$ distribution with a variance of $\Omega$. In this case, $\|\mathbf{h}_l\|^2$ is a sum of $r$ Gamma distributed RV with parameters $(m, \Omega/m)$. Assuming that each component $h_{ij}$ follows the same distribution parameters, the sum of Gamma distributions with same parameters is also a Gamma distribution, where the shape parameter is the sum of the individual parameters. Thus, the pdf of $\|\mathbf{h}_l\|^2$ is denoted by

$$f_{\text{Gamma}(mr,\Omega/m)}(x) = \frac{m^{mr}}{\Gamma(mr)\Omega^{mr}} x^{mr-1} e^{-m\frac{x}{\Omega}}. \tag{D.1}$$

The transformation of $\sigma_l^2$ is also a RV with the following pdf

$$f_{\sigma_l^2}(x) = \frac{m^{mr}}{\Gamma(mr)\,(\rho\Omega)^{mr}}(x-1)^{mr-1}e^{-m\frac{x-1}{\rho\Omega}} \tag{D.2}$$

and moments $\mu_{\sigma_l^2} = r\rho\Omega + 1$, $\sigma^2_{\sigma_l^2} = r(\rho\Omega)^2$.

$E_1$ can be solved as follows

$$\begin{aligned}\mathbb{E}\left[\log_2\left(\sigma_l^2\right)\right] &= \int_0^\infty \log_2(1+\rho x)\, f_{\text{Gamma}(mr,\Omega/m)}(x)\,\mathrm{d}x \\ &= \frac{1}{\log(2)}\Upsilon(mr,\beta)\,. \end{aligned}\tag{D.3}$$

In order to compute $\mathbb{E}\left[\log_2\left(\sigma_1^2+\sigma_2^2\right)\right]$ we exploit the fact that $\log_2\left(\sigma_1^2+\sigma_2^2\right) = 1+\log_2\left(1+\frac{\rho}{2}\left(\|\mathbf{h}_1\|^2+\|\mathbf{h}_2\|^2\right)\right)$. Thus,

$$\begin{aligned}&\mathbb{E}\left[\log_2\left(\sigma_1^2+\sigma_2^2\right)\right] \\ &= 1+\int_0^\infty \log_2\left(1+\frac{\rho}{2}x\right) f_{\text{Gamma}(2mr,\Omega/m)}(x)\,\mathrm{d}x \\ &= 1+\frac{1}{\log(2)}\Upsilon(2mr,2\beta). \end{aligned}\tag{D.4}$$

Finally, to compute $E_3$, we recall that $\sigma_l^2$ are independent RV and thus

$$\mathbb{E}\left[\frac{\sigma_1^2}{\sigma_2^2}\right] = \mathbb{E}\left[\sigma_1^2\right]\mathbb{E}\left[\frac{1}{\sigma_2^2}\right] \tag{D.5}$$

where the second expectation is defined as follows

$$\begin{aligned}\mathbb{E}\left[\frac{1}{\sigma_2^2}\right] &= \int_0^\infty \frac{1}{1+\rho x} f_{\text{Gamma}(mr,\Omega/m)}(x)\,\mathrm{d}x \\ &= \beta^{mr}e^{\beta}\Gamma(1-mr,\beta)\,. \end{aligned}\tag{D.6}$$

Hence, $E_3$ is solved as

$$\mathbb{E}\left[\frac{\sigma_1^2}{\sigma_2^2}\right] = \left(1+mr\beta^{-1}\right)\beta^{mr}e^{\beta}\Gamma(1-mr,\beta)\,. \tag{D.7}$$

Finally, we can express the Ergodic Capacity for the Nakagami-$m$ Channel joining (D.3), (D.4) and (D.7) in (3.40).

Note that the Rayleigh channel is obtained when Nakagami-$m$ is particularized for $m = 1$ and $\Omega = 2\varrho^2$.

## D.2 Rice Channel

The Rice channel is such that each component $h_{ij} \sim \mathcal{CN}(\nu, \varrho^2)$. In this case, $\|\mathbf{h}_l\|^2$ is a Non-Central Chi-Squared distribution of $2r$ degrees of freedom $\chi^2_{2r}(\lambda)$ whose pdf is denoted by

$$f_{\chi^2_{2r}(\lambda)}(x) = \frac{e^{-\frac{\lambda}{2}}}{2\varrho^2} e^{-\frac{x}{2\varrho^2}} \left(\frac{x}{\lambda\varrho^2}\right)^{\frac{r-1}{2}} I_{r-1}\left(\sqrt{\frac{\lambda x}{\varrho^2}}\right) \tag{D.8}$$

where $\lambda = 2r\nu^2$ is the non-centrality parameter and $I_a(2x) = x^a \sum_{k=0}^{\infty} \frac{x^{2k}}{k!\Gamma(a+k+1)}$ is the modified Bessel function of first kind of $a$ degrees of freedom.

The transformation of $\sigma_l^2$ is also a RV with the following pdf

$$f_{\sigma_l^2}(x) = \frac{e^{-\frac{\lambda}{2}}}{2\rho\varrho^2} e^{-\frac{x-1}{2\rho\varrho^2}} \left(\frac{x-1}{\lambda\rho\varrho^2}\right)^{\frac{r-1}{2}} I_{r-1}\left(\sqrt{\frac{\lambda(x-1)}{\rho\varrho^2}}\right) \tag{D.9}$$

and moments $\mu_{\sigma_l^2} = 2r\rho\left(\nu^2 + \varrho^2\right) + 1$, $\sigma^2_{\sigma_l^2} = 4r\rho^2\varrho^2\left(2\nu^2 + \varrho^2\right)$.

In order to compute the expectation of $\log_2\left(\sigma_l^2\right)$, we decompose a Non-Central Chi-Squared RV as a mixture of Central Chi-Squared RV as follows

$$f_{\chi^2_{2r}(\lambda)}(x) = \sum_{k=0}^{\infty} \frac{e^{-\frac{\lambda}{2}}\left(\frac{\lambda}{2}\right)^k}{k!} f_{\chi^2_{2r+2k}}(x). \tag{D.10}$$

Thus, $E_1$ can be expressed as

$$\mathbb{E}\left[\log_2\left(\sigma_l^2\right)\right]$$
$$= \int_0^\infty \log_2(1+\rho x) \sum_{k=0}^{\infty} \frac{e^{-\frac{\lambda}{2}}\left(\frac{\lambda}{2}\right)^k}{k!} f_{\chi^2_{2r+2k}}(x)\,\mathrm{d}x$$
$$= \frac{1}{\log(2)} \sum_{k=0}^{\infty} \frac{e^{-\frac{\lambda}{2}}\left(\frac{\lambda}{2}\right)^k}{k!} \Upsilon(r+k,\beta) \tag{D.11}$$

To find $E_2$, we exploit the result obtained in (D.4), thus obtaining

$$\mathbb{E}\left[\log_2\left(\sigma_1^2+\sigma_2^2\right)\right] = 1 + \int_0^\infty \log_2\left(1+\frac{\rho}{2}x\right) f_{\chi^2_{4r}(\lambda)}(x)\,\mathrm{d}x$$
$$= 1 + \frac{1}{\log(2)} \sum_{k=0}^{\infty} \frac{e^{-\frac{\lambda}{2}}\left(\frac{\lambda}{2}\right)^k}{k!} \Upsilon(2r+k, 2\beta). \tag{D.12}$$

Similarly, we use the result in (D.7) to compute $E_3$ as

$$\mathbb{E}\left[\frac{\sigma_1^2}{\sigma_2^2}\right] \tag{D.13}$$
$$= \sum_{k=0}^{\infty} \frac{e^{-\frac{\lambda}{2}}\left(\frac{\lambda}{2}\right)^k}{k!} \left(1+\frac{r+k}{\beta}\right) \beta^{r+k} e^{\beta} \Gamma\left(1-r-k,\beta\right). \tag{D.14}$$

Finally, we can express the Ergodic Capacity for a Rice Channel joining (D.11), (D.12) and (D.14) in (3.40).

Note that Rayleigh distribution can also be obtained from the Rice distribution imposing $\lambda = 0$. In this case, the mixture of Central Chi-Squared RV degenerates into an indetermination for $k = 0$. This problem can be solved examining $\lim_{\lambda\to 0} \bar{C}$, whose solution agrees with the analytical expression in Table 3.1.

# Appendix E

# CASTLE: the Platform where Researchers can Develop Standards

During the process of this dissertation, a tool for developing standards has been created. This tool is known as CASTLE Platform® and stands for Cloud Architecture for Standards Development. CASTLE is a platform where researchers, students and industry can develop new enhancements, test features and deploy new algorithms using different telecommunication standards.

CASTLE is conceived as Platform as a Service (PaaS). It is maintained by Centre Tecnològic de Telecomunicacions de Catalunya and is being upgraded all time with new enhancements and features. The contribution of this thesis to CASTLE is in the form of implementing and prototyping of the proposed techniques. See [CAS] for further details.

Although the BGAN and PMod implementation are part of CASTLE, it is not limited to this standard. Currently implements the physical layer of the following standards:

### LTE

LTE physical procedures (3GPP TS 36.211, 36.212, 36.213, 36.310, 36.321, 36.322, 36.331) are implemented up to the recent Release

13. In more detail:

- Uplink and Downlink
- Transmission Modes 1 10
- Transmit Diversity, Beamforming, Precoding and Spatial Multiplexing
- Resource Allocation Formats 0 3
- DCI Formats 0, 1, 1A-1D, 2, 2A-2D, 3, 3A-3B, 4
- Constellations QPSK, 16QAM, 64QAM and 256QAM with soft demapping
- PHY, MAC, RLC and RRC layers
- Multiple eNB and multiple UE
- Synchronization, time and frequency offset estimation
- Channel estimation and equalization
- MAC heading
- RLC: TM and UM modes
- RRC: ASN.1 encoding
- ITU, ETSI channel models (EVA, EPA, ETU) with different Doppler shifts

**BGAN**

This standard implements the specifications of ETSI TS 102.744-2-1. In detail:

- Forward link
- All supported bearers

- Polarization Diversity, Polarized Modulation, Polarization Multiplexing
- Land Mobile Satellite and Maritime scenarios
- Synchronization and time offset estimation
- Channel estimation and equalization
- Constellations QPSK, 16QAM, 32QAM and 64QAM with soft demapping

**Li-Fi**

This standard implements the specifications of IEEE 802.15.7. In particular:

- OOK and VPPM dimming
- PHY I, PHY II and PHY III
- Synchronization
- Channel estimation and equalization

CASTLE is composed by these components:

- Core Standards: implement the majority of procedures defined in the standards. This module contains LTE / LTE-Advance / LTE-A Pro and newest 5G technology candidates; Broadband Global Area Network (BGAN) standard, which defines bidirectional multimedia communications with very low latency; and Li-Fi standard defined by IEEE 802.15.7, which describes schemes to transmit information using the visible light.
- Simulation & Emulation: this module constructs scenarios where several eNB and UE are configured with different parameters and run long time simulations to obtain useful metrics for performance evaluation, via software or over the air.

- Interface: the end user may interact with CASTLE via web or an API for Matlab/Octave and C++.

CASTLE is the first cloud-based architecture to run and test communication standards.

Researchers, students and industry professionals can develop new features, test improvements and deploy novel algorithms using multiple communications standards without the need to download and install additional software. With CASTLE, end users can test their designs directly from the cloud.

With CASTLE, the users can benefit from:

- Parallel programming and GPU computing. Users can experience the power of 40 CPU cores and more than 6500+ GPUs.

- Over the Air. Synthesize and analyze waveforms using state-of-the-art USRP devices. Enable over-the-air MIMO capabilities using CASTLE's SDR component.

- High sampling frequency. Thanks to a dedicated 10 Gigabit per stream, CASTLE can achieve a sampling rate of 100 Msps (mega samples per second) with a 12-bit resolution using the most advanced USRP hardware. Experiment with cloud-based Software Defined Radio.

- Analyze intermediate signals. Debug intermediate signal processing blocks and collect inputs, outputs and variable states. This feature enables you to probe outputs from the Turbodecoder, the constellation points or coded symbols, amongst other.

- Export data to MATLAB / Octave. Export collected data in MATLAB file format (.mat). Process output files with MATLAB to analyze and plot results.

- Public API. Create MATLAB scripts, write your own C++ code and embed different standard procedures of using CASTLE's public API.

## E.1 Architecture

CASTLE is constructed with an heterogeneous architecture. Fig. E.1 illustrates the components of CASTLE.

- Standards: this services manages the implementations of the different standards.
  - 4G/5G: current and future communications are controlled by this component.
    - LTE Pro: it implements the LTE Release 13.
  - Satellite: this component controls satellite-oriented standards.
    - ETSI TS 102 744: BGAN standard, which allows multimedia bidirectional low latency communications.
  - Li-Fi: Visible Light Communications (VLC) are controlled by this component.
    - IEEE 802.15.7: standard that defines the specifications of VLC.
- Simulation & Emulation: this service manages the simulation framework that allows the end users to configure and run their customized scenarios.
  - The simulation is performed using a cloud of dedicated servers for intensive simulation of large scale scenarios.
  - The emulation can be performed at small scale by radiating and receiving real waveforms Over the Air (OTA).

* Software Defined Radio (SDR): the OTA emulation is performed by means of SDR using 4 Universal Software Radio Peripherals. These devices allows to transmit and receive digital baseband signals to / from analog RF signals.

- Interfaces: this service manages the communication between the end-user and CASTLE.
  - API: the communication can be stablished using an API, which allows end users to communicate with CASTLE using their respective programs. Users call different orders that execute different tasks remotely in CASTLE Platform. This interface is intended for advanced users.
    * C++: this API is suited for programs written in C++. It is a class that implements several methods and instances to communicate with CASTLE transparently.
    * MATLAB: this API is suited for scripts executed in MATLAB. It is a class that implements several methods to interact with CASTLE. These procedures return MATLAB objects such as matrices, cells or arrays.
  - Web: this access is more user-friendly. It allows to configure customized scenarios using a graphical interface by means of Web App.
  - VPN: this is the full access to CASTLE Platform. It is intended for admin users.

- Certification Authority (CA): it is the Certification Authority service that signs the certificates.
  - Licensing: Intermediate CA that signs the certificates of licenses.
  - Web: Intermediate CA that signs the certificates of web and web Apps.

- VPN: Intermediate CA that signs VPN connections.

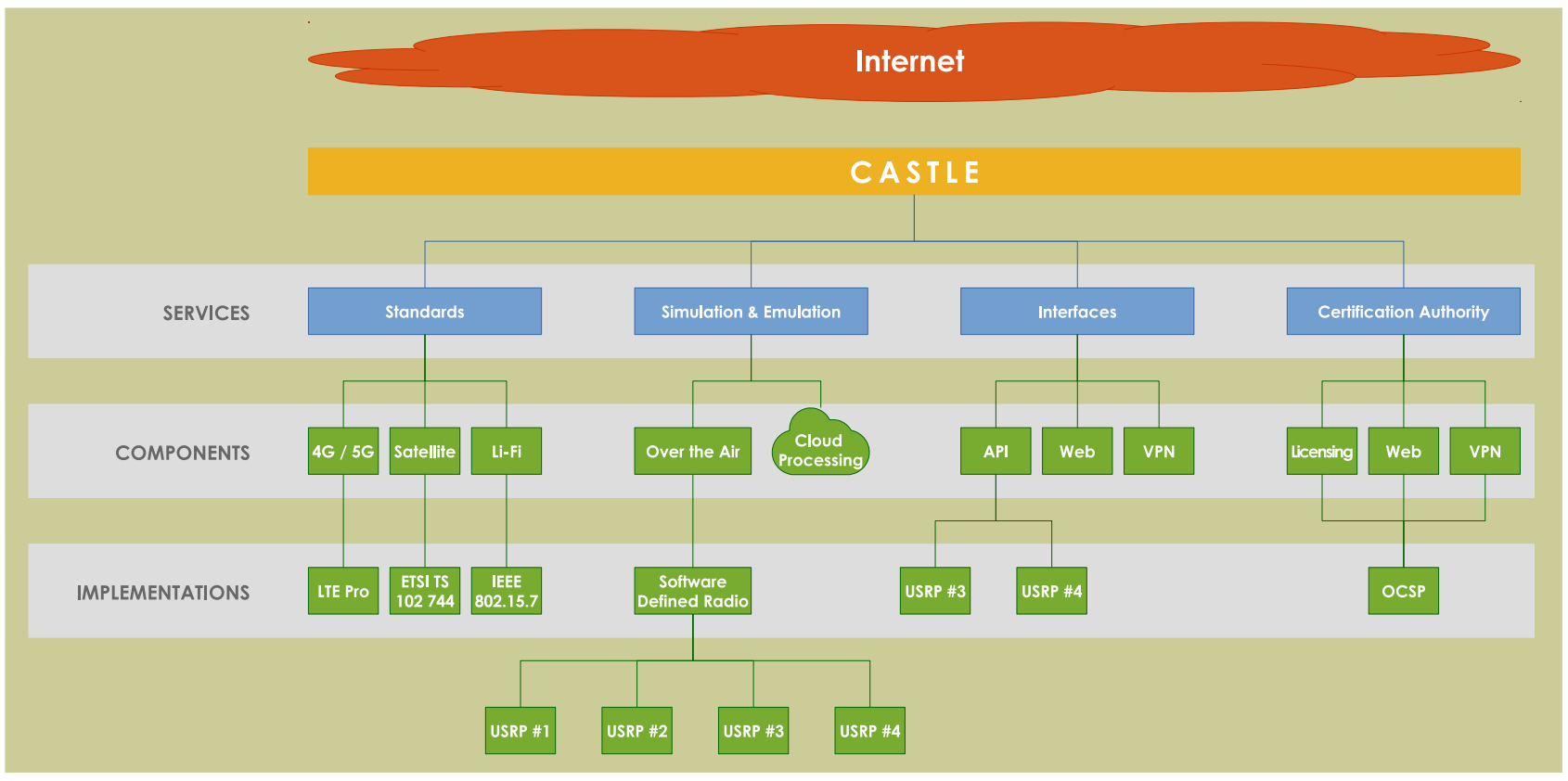

FIGURE E.1: CASTLE Architecture

## E.2 Handshaking & Session

The remote connection with CASTLE is ciphered by robust algorithms. Fig E.2 illustrates an example of session used to convey and process LTE signals. Before communicating with CASTLE, the client and the server performs a handshaking, where different pair of keys are exchanged.

The connection is encrypted using two 256-bit AES keys for uplink and downlink. The AES keys are stored in the cache of the session and destroyed if the session is closed. The exchange of AES keys is using RSA encryption and decryption. Each link (downlink and uplink) has a pair of public/private RSA keys, only valid for the current session. These RSA keys are destroyed when the session is closed. The process can be summarized as follows:

1. Client requests a connection.
2. Server accepts or denies the connection. The denial may occur for several reasons (IP blacklisting, high concurrency, etc.). At this point, the client is still not identified.

3. Client and server exchange public RSA keys.
4. Client encrypts its private AES key using the server public RSA key and sends it to the server.
5. Server decrypts the encrypted client AES key using its private RSA key.
6. Server encrypts its private AES key using the client public RSA key and sends it to the client.
7. Client decrypts the encrypted server AES key using its private RSA key. At this point, client and server have the respective AES key. Hereinafter all the content is ciphered using AES keys.
8. Client sends the license.
9. Server accepts or denies the license and closes the connection if it is necessary.
10. Client sends the XML scenario configuration.
11. ...

The AES implementation is based on AES-NI, the Intel implementation. This is embedded into the processor unit. Therefore, the implementation is not a software part but a hardware procedure, which produces a $5x$ faster results.

The configuration of scenario is performed using a XML file. This file contains all the parameters involved in the standard. It can be modified easily by the client to be adjusted to its needs.

## E.3 Licensing

CASTLE implements a novel system of licensing based on X.509 certificate system. CASTLE CA creates a certificate request with special fields that identify the client. These fields contain information about:

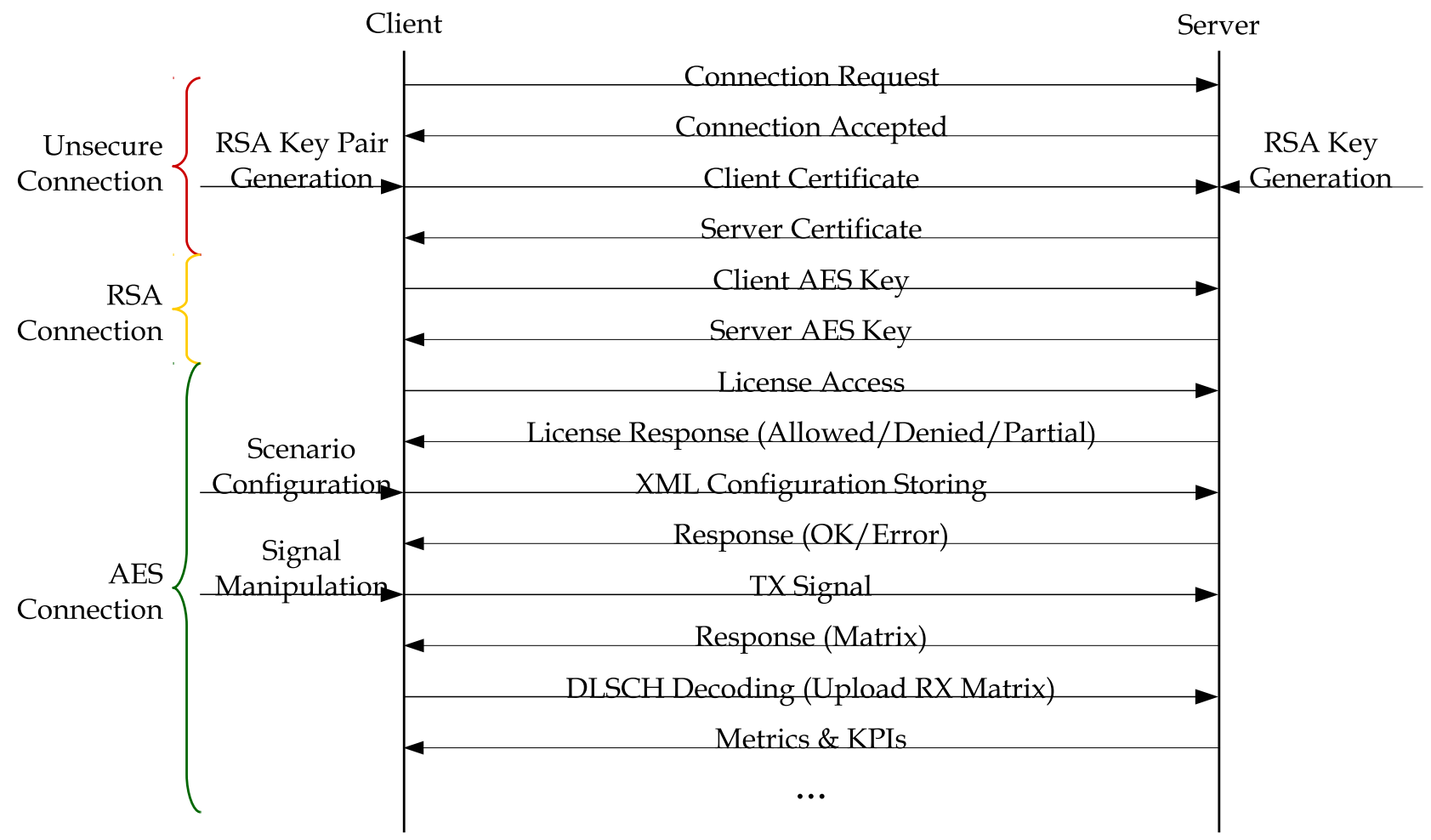

FIGURE E.2: Example of session with CASTLE to send and process a LTE signal.

- Period of granted access: range of dates where the license is valid.
- IP source: IP, range of IP or subnetwork from where the client is connected.
- Floating-point precision: defines the accuracy of floating decimal numbers (quad, half, single or double precision) in bytes.
- Granted commands: which commands the user may use.
- ID of license
- Contact point of license

All this information is encapsulated using X.509 certificate and signed by the CASTLE License CA. The resulting certificate is delivered to the end user. This certificate is the license file and it is sent to CASTLE during the handshake.

# Appendix F

# Broadband Global Area Network Standard

BGAN standard is used in this dissertation to implement some of the results to contextualize the research in a realistic scenario. This standard aims to be full duplex and capable to transmit multimedia communications bidirectionally. In contrast to standards, such as [DVBb] or [DVBa], where long interleaving blocks are used, BGAN implements fast interleaving to reduce dramatically the latency. This dissertation implemented the downlink of this standard. The features of BGAN can be summarized as follows (for the downlink):

- Maximum data rate: 858 kbps.
- TTI duration: 80 ms.
- Minimum latency: 10 ms.
- Modulation types: QPSK, 16QAM, 32QAM and 64QAM.
- Channel coding: Turboencoding.
- Interleaver and puncturing.

Fig. F.1 describes the PHY block diagram of BGAN (downlink).

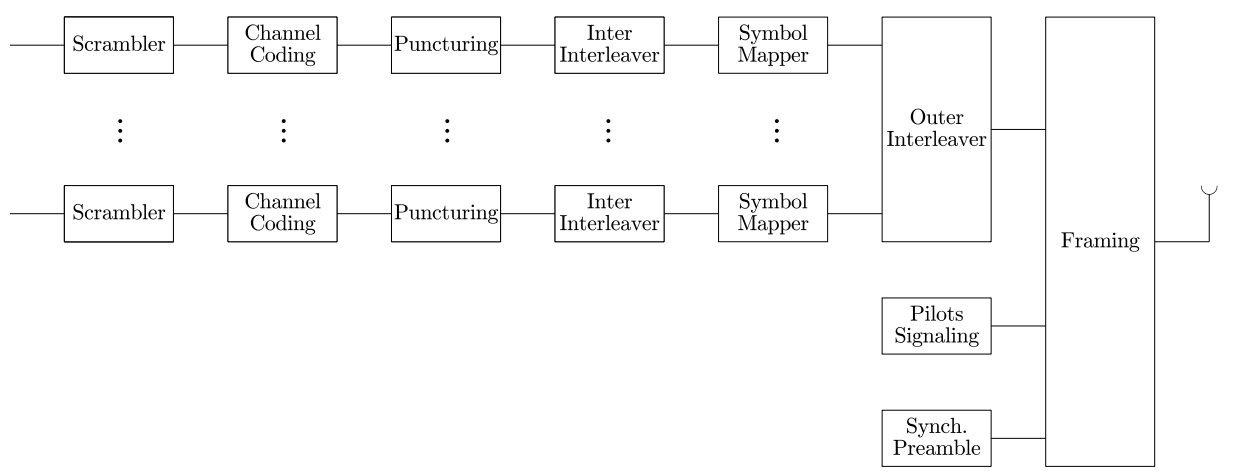

FIGURE F.1: BGAN block diagram.

# Appendix G

# Proofs of Theorems (1) and (2)

In this appendix we describe the second and fourth derivatives of $g(y)$, involved in chapter 3. We recall that

$$g\left(y\right) = \log_2\left(\sum_{l=1}^{t}\frac{1}{\pi\sigma_l^2}\,e^{-\frac{y_1^2+y_2^2}{\sigma_l^2}}\right)$$
$$g(y) \doteq \log_2(f_2(y)), \tag{G.1}$$

where we define $f_n(y) \doteq \sum_{l=1}^{t} \frac{1}{\pi \sigma_l^n} e^{-\frac{y_1^2+y_2^2}{\sigma_l^2}}$. We first compute the first four derivatives of $g(y)$ as a function of $f_n(y)$:

$$\begin{aligned}
\frac{\partial g}{\partial y_1}(y) &= \frac{1}{\log 2} \frac{\frac{\partial f_2}{\partial y_1}(y)}{f_2(y)} \\
\frac{\partial^2 g}{\partial y_1^2}(y) &= \frac{1}{\log 2} \left( \frac{\frac{\partial^2 f_2}{\partial y_1^2}(y)}{f_2(y)} - \frac{\left(\frac{\partial f_2}{\partial y_1}(y)\right)^2}{f_2^2(y)} \right) \\
\frac{\partial^3 g}{\partial y_1^3}(y) &= \frac{1}{\log 2} \left( \frac{\frac{\partial^3 f_2}{\partial y_1^3}(y)}{f_2(y)} - 3\frac{\frac{\partial^2 f_2}{\partial y_1^2}(y)\frac{\partial f_2}{\partial y_1}(y)}{f_2^2(y)} + 2\frac{\left(\frac{\partial f_2}{\partial y_1}(y)\right)^3}{f_2^3(y)} \right) \\
\frac{\partial^4 g}{\partial y_1^4}(y) &= \frac{1}{\log 2} \left( \frac{\frac{\partial^4 f_2}{\partial y_1^4}(y)}{f_2(y)} - \frac{4\frac{\partial^3 f_2}{\partial y_1^3}(y)\frac{\partial f_2}{\partial y_1}(y) + 3\left(\frac{\partial^2 f_2}{\partial y_1^2}(y)\right)^2}{f_2^2(y)} \right. \\
&\qquad \left. +12\frac{\frac{\partial^2 f_2}{\partial y_1^2}(y)\left(\frac{\partial f_2}{\partial y_1}(y)\right)^2}{f_2^3(y)} - 6\frac{\left(\frac{\partial f_2}{\partial y_1}(y)\right)^4}{f_2^4(y)} \right)
\end{aligned} \tag{G.2}$$

The derivatives of $f_2(y)$ are denoted as follows:

$$\begin{aligned}
\frac{\partial f_2}{\partial y_1}(y) &= -2y_1 f_4(y) \\
\frac{\partial^2 f_2}{\partial y_1^2}(y) &= 2\left(2y_1^2 f_6(y) - f_4(y)\right) \\
\frac{\partial^3 f_2}{\partial y_1^3}(y) &= 4y_1\left(3f_6(y) - 2y_1^2 f_8(y)\right) \\
\frac{\partial^4 f_2}{\partial y_1^4}(y) &= 4\left(3f_6(y) - 12y_1^2 f_8(y) + 4y_1^4 f_{10}(y)\right)
\end{aligned} \tag{G.3}$$

Thus, combining (G.2) and (G.3) and after some mathematical arrangements, we encounter that

$$\begin{aligned}
\frac{\partial g}{\partial y_1}(y) &= -\frac{2y_1}{\log 2}\frac{f_4(y)}{f_2(y)} \\
\frac{\partial^2 g}{\partial y_1^2}(y) &= \frac{2}{\log 2}\left(\frac{2y_1^2 f_6(y) - f_4(y)}{f_2(y)} - \frac{2y_1^2 f_4^2(y)}{f_2^2(y)}\right) \\
\frac{\partial^3 g}{\partial y_1^3}(y) &= \frac{4y_1}{\log 2}\left(\frac{3f_6(y) - 2y_1^2 f_8(y)}{f_2(y)} + 3f_4(y)\frac{2y_1^2 f_6(y) - f_4(y)}{f_2^2(y)} - 4y_1^2\frac{f_4^3(y)}{f_2^3(y)}\right) \\
\frac{\partial^4 g}{\partial y_1^4}(y) &= \frac{4}{\log 2}\left(\frac{3f_6(y) - 12y_1^2 f_8(y) + 4y_1^4 f_{10}(y)}{f_2(y)}\right. \\
&+ \frac{8y_1^2 f_4(y)\left(3f_6(y) - 2y_1^2 f_8(y)\right) - 3\left(2y_1^2 f_6(y) - f_4(y)\right)^2}{f_2^2(y)} \\
&\left. +24y_1^2 f_4^2(y)\frac{2y_1^2 f_6(y) - f_4(y)}{f_2^3(y)} - 24y_1^4\frac{f_4^4(y)}{f_2^4(y)}\right)
\end{aligned} \tag{G.4}$$

To prove the theorem 1 we compute the limit of the fourth derivative of $g(y)$.

*Proof.* The equality $H(\boldsymbol{\sigma}^n) = A(\boldsymbol{\sigma}^n)$ holds when $\sigma_1^n = \ldots = \sigma_t^n = S$. Thus, we compute the limit of the fourth derivative assuming this equality. Computing the limit is equivalent to compute the limit of each $f_n$ term of the fourth derivative[1]. In any case, it is straightforward that

$$\lim_{\boldsymbol{\sigma}^2 \to S\mathbf{1}} f_n(y) = \frac{t}{\pi S^{\frac{n}{2}}} e^{-\frac{y_1^2 + y_2^2}{S}} = S^{-\frac{n}{2}} F(y). \tag{G.5}$$

Hence, we have that

$$\lim_{\boldsymbol{\sigma}^2 \to S\mathbf{1}} f_n^p(y) f_m^q(y) = S^{-\frac{pn+qm}{2}} F^{p+q}(y). \tag{G.6}$$

[1] We denote $\boldsymbol{\sigma}^2 \to S\mathbf{1}$ is equivalent to $\sigma_l^2 \to S, \forall l \in [1, t]$.

After few mathematical manipulations using (G.6), we can ensure that

$$\lim_{\boldsymbol{\sigma}^2 \to S\mathbf{1}} \frac{\partial^4 g}{\partial y_1^4}(y) = 0. \tag{G.7}$$

□

To prove the theorem 2 we proceed as previously but in the limit with $\gamma \to \infty$.

*Proof.* It can be seen that $\lim_{\gamma\to\infty} \frac{\partial^4 g}{\partial y_1^4}(\xi) = 0$ and $\lim_{\gamma\to\infty} A\left(\boldsymbol{\sigma}^n\right) = \gamma^{\frac{n}{2}} A\left(\boldsymbol{\kappa}^n\right)$, where $\boldsymbol{\kappa}^n = (\|\mathbf{h}_1\|^n \ldots \|\mathbf{h}_t\|^n)^T$. Hence, the remainder takes the form of the indetermination $\infty \cdot 0$. To solve it we first compute the following limit:

$$\lim_{\gamma\to\infty} f_n(\xi) = \gamma^{-\frac{n}{2}} \sum_{l=1}^{t} \frac{1}{\pi \|\mathbf{h}_l\|^n} e^{-\frac{\xi_1^2+\xi_2^2}{\sigma_l^2}} = \gamma^{-\frac{n}{2}} \frac{t}{\pi} H\left(\boldsymbol{\kappa}^n\right)^{-1} \tag{G.8}$$

Hence, we have that

$$\begin{aligned} \lim_{\gamma\to\infty} A\left(\boldsymbol{\sigma}^k\right) f_n^p(\xi) &= \gamma^{\frac{k-np}{2}} A\left(\boldsymbol{\kappa}^n\right) \left(\frac{t}{\pi}\right)^p H\left(\boldsymbol{\kappa}^n\right)^{-p} \\ &= \begin{cases} 0 & \text{if } k < pn \\ A\left(\boldsymbol{\kappa}^k\right) \left(\frac{t}{\pi}\right)^p H\left(\boldsymbol{\kappa}^n\right)^{-p} & \text{if } k = pn \\ \infty & \text{if } k > pn \end{cases}. \end{aligned} \tag{G.9}$$

Applying (G.9) to the fourth remainder, it is reduced to

$$\begin{aligned} \lim_{\gamma\to\infty} \frac{A\left(\boldsymbol{\sigma}^4\right)}{32} \left(\frac{\partial^4 g}{\partial y_1^4}(\xi) + \frac{\partial^4 g}{\partial y_2^4}(\xi)\right) &= \lim_{\gamma\to\infty} \frac{3}{4\log 2} A\left(\boldsymbol{\sigma}^4\right) \left(\frac{f_6(\xi)}{f_2(\xi)} - \frac{f_4^2(\xi)}{f_2^2(\xi)}\right) \\ &= \frac{3}{4\log 2} A\left(\boldsymbol{\kappa}^4\right) \left(\frac{H\left(\boldsymbol{\kappa}^2\right)}{H\left(\boldsymbol{\kappa}^6\right)} - \frac{H\left(\boldsymbol{\kappa}^2\right)^2}{H\left(\boldsymbol{\kappa}^4\right)^2}\right). \end{aligned} \tag{G.10}$$

Hence, the expectation of the remainder is a constant that does not depend on $\gamma$, regardless the $\xi$ value. We can conclude that

$\mathbb{E}\left\{R_3\left(g, y, \mu_y\right)\right\} = o(\gamma)$ since

$$\lim_{\gamma \to \infty} \frac{\mathbb{E}\left\{R_3\left(g, y, \mu_y\right)\right\}}{\gamma} = 0, \tag{G.11}$$

which is the definition of the $o(\gamma)$. □

Note that (G.10) also holds to the theorem (1) when $\boldsymbol{\sigma}^2 \to S\mathbf{1}$, which in this case the limit is $0$.

# Appendix H

# Reference Tables

Table H.1 describes the parameters of each bearer of BGAN standard.

Tables H.2, H.3, H.4 and H.5 describe the exact symbol mapping on the surface of Poincareé Sphere for $L = 2, 4, 8, 16$.

TABLE H.1: BGAN Bearers Specification

| Bearer Name | Symbol Rate | Modulation | FEC Duration | FEC Size | FEC Size | FEC Data | FEC Data | Coding Rate | Data Rate | Required C/No | Es/No | Eb/No |
|---|---|---|---|---|---|---|---|---|---|---|---|---|
| **Type-subtype** | **Ksps** | **bits/sy** | **s** | **symbols** | **bits** | **bits** | **bytes** | **R** | **kbps** | **dBHz** | **dB** | **dB** |
| **F80T1Q4B-L8** | 33600 | 2 | 0,02 | 640 | 1280 | 432 | 54 | 0,34 | 21,6 | 46,1 | 0,8 | 2,8 |
| **F80T1Q4B-L7** | 33600 | 2 | 0,02 | 640 | 1280 | 512 | 64 | 0,40 | 25,6 | 46,9 | 1,6 | 2,8 |
| **F80T1Q4B-L6** | 33600 | 2 | 0,02 | 640 | 1280 | 608 | 76 | 0,48 | 30,4 | 47,9 | 2,6 | 3,1 |
| **F80T1Q4B-L5** | 33600 | 2 | 0,02 | 640 | 1280 | 704 | 88 | 0,55 | 35,2 | 48,8 | 3,5 | 3,3 |
| **F80T1Q4B-L4** | 33600 | 2 | 0,02 | 640 | 1280 | 800 | 100 | 0,63 | 40,0 | 49,7 | 4,4 | 3,7 |
| **F80T1Q4B-L3** | 33600 | 2 | 0,02 | 640 | 1280 | 896 | 112 | 0,70 | 44,8 | 50,6 | 5,3 | 4,1 |
| **F80T1Q4B-L2** | 33600 | 2 | 0,02 | 640 | 1280 | 984 | 123 | 0,77 | 49,2 | 51,6 | 6,3 | 4,7 |
| **F80T1Q4B-L1** | 33600 | 2 | 0,02 | 640 | 1280 | 1056 | 132 | 0,83 | 52,8 | 52,6 | 7,3 | 5,4 |
| **F80T1Q4B-R** | 33600 | 2 | 0,02 | 640 | 1280 | 1112 | 139 | 0,87 | 55,6 | 53,6 | 8,3 | 6,1 |
| **F80T1X4B-L3** | 33600 | 4 | 0,02 | 640 | 2560 | 856 | 107 | 0,33 | 42,8 | 50,6 | 5,3 | 4,3 |
| **F80T1X4B-L2** | 33600 | 4 | 0,02 | 640 | 2560 | 1000 | 125 | 0,39 | 50,0 | 51,6 | 6,3 | 4,6 |
| **F80T1X4B-L1** | 33600 | 4 | 0,02 | 640 | 2560 | 1152 | 144 | 0,45 | 57,6 | 52,5 | 7,2 | 4,9 |
| **F80T1X4B-R** | 33600 | 4 | 0,02 | 640 | 2560 | 1304 | 163 | 0,51 | 65,2 | 53,5 | 8,2 | 5,4 |
| **F80T1X4B-H1** | 33600 | 4 | 0,02 | 640 | 2560 | 1480 | 185 | 0,58 | 74,0 | 54,5 | 9,2 | 5,8 |
| **F80T1X4B-H2** | 33600 | 4 | 0,02 | 640 | 2560 | 1624 | 203 | 0,63 | 81,2 | 55,6 | 10,3 | 6,5 |
| **F80T1X4B-H3** | 33600 | 4 | 0,02 | 640 | 2560 | 1800 | 225 | 0,70 | 90,0 | 56,5 | 11,2 | 7,0 |
| **F80T1X4B-H4** | 33600 | 4 | 0,02 | 640 | 2560 | 1960 | 245 | 0,77 | 98,0 | 57,6 | 12,3 | 7,7 |
| **F80T1X4B-H5** | 33600 | 4 | 0,02 | 640 | 2560 | 2104 | 263 | 0,82 | 105,2 | 58,7 | 13,4 | 8,5 |
| **F80T1X4B-H6** | 33600 | 4 | 0,02 | 640 | 2560 | 2184 | 273 | 0,85 | 109,2 | 59,5 | 14,2 | 9,1 |
| **F80T2.5X32-6B-L3** | 84000 | 5 | 0,013 | 1098 | 5490 | 1832 | 229 | 0,33 | 137,4 | 56,2 | 7,0 | 4,9 |
| **F80T2.5X32-6B-L2** | 84000 | 5 | 0,013 | 1098 | 5490 | 2240 | 280 | 0,41 | 168,0 | 57,5 | 8,3 | 5,3 |
| **F80T2.5X32-6B-L1** | 84000 | 5 | 0,013 | 1098 | 5490 | 2640 | 330 | 0,48 | 198,0 | 58,9 | 9,7 | 6,0 |
| **F80T2.5X32-6B-R** | 84000 | 5 | 0,013 | 1098 | 5490 | 3040 | 380 | 0,55 | 228,0 | 60,0 | 10,8 | 6,5 |
| **F80T2.5X32-6B-H1** | 84000 | 5 | 0,013 | 1098 | 5490 | 3440 | 430 | 0,63 | 258,0 | 61,1 | 11,9 | 7,0 |
| **F80T2.5X32-6B-H2** | 84000 | 5 | 0,013 | 1098 | 5490 | 3840 | 480 | 0,70 | 288,0 | 62,4 | 13,2 | 7,8 |
| **F80T2.5X32-6B-H3** | 84000 | 5 | 0,013 | 1098 | 5490 | 4240 | 530 | 0,77 | 318,0 | 63,6 | 14,4 | 8,6 |
| **F80T2.5X32-6B-H4** | 84000 | 5 | 0,013 | 1098 | 5490 | 4520 | 565 | 0,82 | 339,0 | 64,8 | 15,6 | 9,5 |
| **F80T2.5X32-6B-H5** | 84000 | 5 | 0,013 | 1098 | 5490 | 4760 | 595 | 0,87 | 357,0 | 65,8 | 16,6 | 10,3 |
| **F80T2.5X32-6B-H6** | 84000 | 5 | 0,013 | 1098 | 5490 | 4960 | 620 | 0,90 | 372,0 | 67,1 | 17,9 | 11,4 |
| **F80T2.5X64-7B-L3** | 84000 | 6 | 0,011 | 941 | 5646 | 1880 | 235 | 0,33 | 164,5 | 57,3 | 8,1 | 5,2 |
| **F80T2.5X64-7B-L2** | 84000 | 6 | 0,011 | 941 | 5646 | 2200 | 275 | 0,39 | 192,5 | 58,5 | 9,3 | 5,7 |
| **F80T2.5X64-7B-L1** | 84000 | 6 | 0,011 | 941 | 5646 | 2464 | 308 | 0,44 | 215,6 | 59,6 | 10,4 | 6,3 |
| **F80T2.5X64-7B-R** | 84000 | 6 | 0,011 | 941 | 5646 | 2800 | 350 | 0,50 | 245,0 | 60,7 | 11,5 | 6,9 |
| **F80T2.5X64-7B-H1** | 84000 | 6 | 0,011 | 941 | 5646 | 3120 | 390 | 0,55 | 273,0 | 61,8 | 12,6 | 7,5 |
| **F80T2.5X64-7B-H2** | 84000 | 6 | 0,011 | 941 | 5646 | 3424 | 428 | 0,61 | 299,6 | 63,0 | 13,8 | 8,3 |
| **F80T2.5X64-7B-H3** | 84000 | 6 | 0,011 | 941 | 5646 | 3760 | 470 | 0,67 | 329,0 | 64,3 | 15,1 | 9,2 |
| **F80T2.5X64-7B-H4** | 84000 | 6 | 0,011 | 941 | 5646 | 4080 | 510 | 0,72 | 357,0 | 65,3 | 16,1 | 9,8 |
| **F80T2.5X64-7B-H5** | 84000 | 6 | 0,011 | 941 | 5646 | 4400 | 550 | 0,78 | 385,0 | 66,4 | 17,2 | 10,6 |
| **F80T2.5X64-7B-H6** | 84000 | 6 | 0,011 | 941 | 5646 | 4720 | 590 | 0,84 | 413,0 | 67,7 | 18,5 | 11,6 |

TABLE H.2: Packing for $L = 2$

| Bit Mapping | $\phi$ | $\vartheta$ |
|---|---|---|
| 0 | $0$ | $0$ |
| 1 | $0$ | $\pi$ |

TABLE H.3: Packing for $L = 4$. $\alpha = \arccos\left(-\frac{1}{3}\right)$

| Bit Mapping | $\phi$ | $\vartheta$ |
|---|---|---|
| 00 | $\frac{\pi}{2}$ | $\pi$ |
| 01 | $0$ | $\alpha$ |
| 10 | $\frac{2\pi}{3}$ | $\alpha$ |
| 11 | $\frac{4\pi}{3}$ | $\alpha$ |

TABLE H.4: Packing for $L = 8$

| Bit Mapping | $\phi$ | $\vartheta$ |
|---|---|---|
| 000 | $0$ | $\frac{\pi}{3}$ |
| 001 | $\frac{\pi}{2}$ | $\frac{\pi}{3}$ |
| 010 | $\frac{3\pi}{2}$ | $\frac{\pi}{3}$ |
| 011 | $\pi$ | $\frac{\pi}{3}$ |
| 100 | $\frac{\pi}{4}$ | $\frac{2\pi}{3}$ |
| 101 | $\frac{3\pi}{4}$ | $\frac{2\pi}{3}$ |
| 110 | $\frac{7\pi}{4}$ | $\frac{2\pi}{3}$ |
| 111 | $\frac{5\pi}{4}$ | $\frac{2\pi}{3}$ |

TABLE H.5: Packing for $L = 16$. $\alpha = \frac{2}{3}$

| Point no. | $\phi$ | $\vartheta$ |
|---|---|---|
| 0000 | $\frac{\pi}{4}$ | $\alpha$ |
| 0001 | $\frac{3\pi}{4}$ | $\alpha$ |
| 0010 | $\frac{7\pi}{4}$ | $\alpha$ |
| 0011 | $\frac{5\pi}{4}$ | $\alpha$ |
| 0100 | $0$ | $2\alpha$ |
| 0101 | $\frac{\pi}{2}$ | $2\alpha$ |
| 0110 | $\frac{3\pi}{2}$ | $2\alpha$ |
| 0111 | $\pi$ | $2\alpha$ |
| 1000 | $0$ | $\pi - \alpha$ |
| 1001 | $\frac{\pi}{2}$ | $\pi - \alpha$ |
| 1010 | $\frac{3\pi}{2}$ | $\pi - \alpha$ |
| 1011 | $\pi$ | $\pi - \alpha$ |
| 1100 | $\frac{\pi}{4}$ | $\pi - 2\alpha$ |
| 1101 | $\frac{3\pi}{4}$ | $\pi - 2\alpha$ |
| 1110 | $\frac{7\pi}{4}$ | $\pi - 2\alpha$ |
| 1111 | $\frac{5\pi}{4}$ | $\pi - 2\alpha$ |

# Index